\documentclass[12pt]{report}
\usepackage[titletoc]{appendix}
\usepackage{verbatim}
\usepackage{cite}
\usepackage{graphicx}
\usepackage{color}
\usepackage[fleqn]{amsmath}
\usepackage{amssymb}
\usepackage{slashed}
\usepackage{epsfig}
\usepackage{textcomp}
\usepackage{hyperref}
 \usepackage{geometry}
\begin{document}
\pagenumbering{roman}
\thispagestyle{empty}

\begin{center}
      {\bf {\Large Structural Properties of Finite and Infinite Nuclear Systems
and Related Phenomena} }
\end{center}

\vspace{0.3 cm}
\begin{center}
    {\it \large By}
\end{center}

\begin{center}
    {\bf {\Large Subrata Kumar Biswal} \\ PHYS07201104002}
\end{center}

\begin{center}
\bf {{\large Institute of Physics, Bhubaneswar }}\\
\bf {{\large INDIA}}
\end{center}

\vskip 2.0 cm
\begin{center}
\large{
{ A thesis submitted to the }  \\
 {Board of studies in Physical Sciences }\\
In partial fulfillment of requirements \\
For the Degree of } \\
{\bf  DOCTOR OF PHILOSOPHY} \\
\emph{of} \\
{\bf HOMI BHABHA NATIONAL INSTITUTE }
\vskip 1.0 cm
\epsfig{file=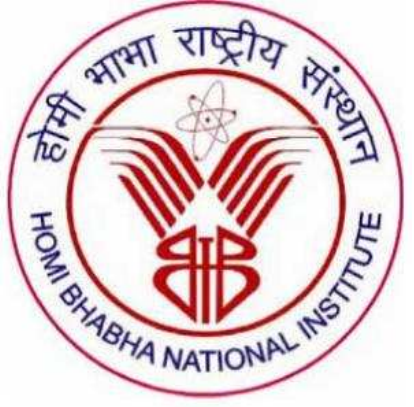, width=4.0cm, height= 4.0cm}\\
\vskip 0.2 cm
{\bf February, 2017}
\end{center}
\newpage
\begin{center}
\large \bf 
STATEMENT BY AUTHOR
\end{center}
This dissertation has been submitted in partial fulfillment of requirements
for an advance degree at Homi Bhabha National Institute (HBNI) and deposited 
in the Library to be made available to borrowers under rules of the HBNI. 

    Brief quotations from this dissertation are allowed without special 
permission, provided that accurate acknowledgement of source is made. 
Requests for permission for extended  quotation from or reproduction    
of this manuscript in whole or in part may be granted by the competent 
Authority of HBNI when in his or her judgment the proposed use of the 
meterial is in the interests of scholarship. In all other instances, 
however, permission must be obtained from the author.\\
\bigskip
\\
\\
\\
\\
\\
Date:  \hspace*{8.3cm} {(Subrata Kumar Biswal)}\\
\newpage
\thispagestyle{empty}
\begin{center}
\large \bf
DECLARATION
\end{center}

I, {\bf Subrata Kumar Biswal}, hereby declare that the investigations
presented in the thesis have been carried out by me. The matter
embodied in the thesis is original and has not been submitted earlier
as a whole or in part for a degree/diploma at this or any other
Institution/University.

\vskip 3.0cm
\noindent
\hspace{10.5cm}\underline{Signature}\\
\hspace*{10.5cm} (Subrata Kumar Biswal)\\
\noindent
Date :

\newpage
\begin{center}
{\bf List of Publications arising from the thesis}
\end{center}
{\large\bf Journal}
\begin{enumerate}
\item{``The Effects of Self Interacting Isoscalar-Vector Meson
on Finite Nuclei and Infinite Nuclear Matter'',
{\bf S. K. Biswal},  S. K. Singh, M. Bhuyan and  S. K. Patra, 
{\it Braz. J. Phys}, {\bf 2014}, {45}, 347-352}.
\item  { ``Isoscalar giant monopole resonance for drip-line and
super heavy nuclei in the framework of relativistic
mean field formalism with scaling calculation'', 
{\bf S. K. Biswal} and S. K. Patra, 
{\it Cent. Eur. J. Phys.}, {\bf 2014}, {12}, 582-596 }.
\item { ``Search of double shell closure in the superheavy nuclei
using a simple effective interaction'', {\bf S. K. Biswal}, M. Bhuyan, 
S. K. Sing and S. K. Patra, {\it International Journal of Modern Physics E}, 
{\bf 2014}, {23}, 1450017(1-14).}
\item { ``Effects of N-N potentials on p Nuclides in the A$\sim$100-120 region'', C. lahiri, {\bf S. K. Biswal} and S. K. Patra, {\it International Journal of Modern Physics E}, {\bf 2016}, { 25}, 1650015(1-15). }
\item{  ``Softness of Sn isotopes in relativistic semiclassical approximation''' {\bf S. K. Biswal}, S. K. Singh, and S. K. Patra, {\it Modern Physics Letters A}, {\bf 2015}, {30}, 1550097(1-13).}
\item 
 {``Effects of isovector scalar meson on neutron star both with and without 
 hyperon'', 
 {\bf S. K. Biswal}, Bharat Kumar, and S. K. Patra, }
{\it International Journal of Modern Physics E}, {\bf 2016}, {25}, 1650090(1-22).\\

\item {``New constrained calculation of excitation energy of Isoscalar
giant monopole resonance and Isovector giant dipole resonance in
relativistic Thomas-Fermi formalism'', {\bf S. K. Biswal},
X. Vi\~nas, M. Centelles, and S. K. Patra.}\\
(Communicated to a Journal)\\

\end{enumerate}
\newpage
\bigskip
{\large \bf Conferences }
\begin{enumerate}
\item
``Isovector giant dipole resonance in Relativistic Thomas-Fermi formalism'',  
{\bf S. K Biswal}, S. K Patra and X. Vi\~nas, {\it Proceedings of the DAE-BRNS Symp. on Nucl. Phys.}, {\bf 2015}, {60}, 106-107. 

\item ``Effects of different NN potentials on $^{120}$Te nucleus'', 
C. Lahiri, {\bf S.K. Biswal} and  S.K. Patra, {\it Proceedings of the 
DAE-BRNS Symp. on Nucl. Phys.}, {\bf 2015}, {60}, 214-215.

\item
``Effects of the isovector-scalar meson on the softness of the Sn isotopes'', 
{\bf S. K. Biswal}, S. K. Singh, S. K. Patra and B. K. Agrawal, 
{\it Proceedings of the DAE Symp. on Nucl. Phys.}, {\bf 2014}, {59}, 84-85.

\item 
``Effect of Isospin on compressibility of drip line and superheavy nuclei'', 
{\bf S. K. Biswal} and  S. K. Patra,
{\it Proceedings of the DAE Symp. on Nucl. Phys.}, (2013), {\bf 58}, 146-147. 

\end{enumerate}
\vskip 8.0cm
\hspace*{11cm} {(Subrata Kumar Biswal)}\\

\newpage
\thispagestyle{empty}
\vspace*{3.0in}
\hspace*{2.0in}
\begin{center}
{\Large \emph{ To My Parents}\\}
\hrule
\end{center}

\newpage
\thispagestyle{empty}
\begin{center}
\large \bf
ACKNOWLEDGMENTS
\end{center}
\vspace*{-0.1in}
{ \noindent
With my deep respect, I like to thank my supervisor Prof. S. K. Patra for his 
constant support and guidance during my research work. His experienced advice 
has helped me in many ways, without which it would not have been possible to 
complete this thesis. He has been always available for me, which 
helped me to accelerate my research work. Not only nuclear physics,
he has also taught me the fact of life and philosophies, which 
will constantly guide me throughout my life. I take this opportunity 
to thanks, Prof. L. Satpathy, whose inspirational words made me more 
versatile during my Ph.D carrier. Prof. Satpathy is a man of 
physics and philosophies. His words make me to think beyond my capacity.

          Special thanks to Prof. Xavier Vi\~nas for his help. I will 
remain always greatfull to him for his kindness. I would like to express 
my sincere gratitude to Prof. Sudhakara Panda, Director, Institute of Physics. 
His helpful attitude has made my rigorous Ph.D. life more enjoyable and 
worth living. Thanks for his constant support.

                 Institute of Physics  always  provided 
us a good academic and friendly atmosphere. I would like to thank  all the 
office and library staffs of the Institute. Special thanks 
to Rajesh Mohapatra, Ramchandra Hansda, Ravana and others. for their 
co-operation during my stay  at IOP. I like to thank my junior Bharat Kumar 
and my senior S. K. Singh, M. Ikram and M. Bhuyan for their 
support.

        After all, I thank, Department of Atomic Energy for 
giving me financial support to conduct my Ph.D. work.

                     I will always remain grateful to my parents and my friends for 
their patience and support.

\vskip 2.0cm
{\bf Date:}
\hspace{3.2in} {\bf Subrat Kumar Biswal}
 \hspace*{0.32cm}



\numberwithin{equation}{chapter}
\numberwithin{figure}{chapter}
\numberwithin{table}{chapter}
\tableofcontents
\addcontentsline{toc}{chapter}{Synopsis}
\newpage
\begin{flushleft}
{\large{\bf Synopsis }}\\
\end{flushleft}
\vskip 2.0cm

The atomic nucleus is a strongly interacting many-body quantum mechanical 
system that exhibits a fascinating variety of shape and excitation modes, 
starting from spherical to super deformed and from single particle 
excitation to collective excitation like vibration and rotation of 
nucleus as a whole. The study of nuclear structure attempts to 
elucidate the unified mechanism by which these rich patterns of behavior 
emerge from a common underlying strong nuclear force between the nucleons.
 Nucleons are the building blocks of the nucleus. It is the nature of 
nucleon-nucleon interaction, that decides the each and every characteristic of a
nucleus. Nuclear structure is  a consequence of the nuclear interaction. A thorough  study of the nuclear structure gives prerequisite information about the interaction of  a nucleon with another nucleon. Not only the study of a finite nucleus but also the study of an infinite nuclear  matter (INM) system like neutron star has attracted nuclear physicists and many information can be
 accumulated  from the study of INM. A neutron star is a very complex  
and dense system. It also provides a well-riched laboratory to test the nuclear theory under extreme condition of density and asymmetry, which can not be created in a terrestrial laboratory.  Our primary aim is to study nuclear structure for both finite and infinite nuclear matter. 
In the present thesis, we explore  both finite nuclei and neutron star structure using relativistic mean field (RMF) formalism.   

                          In finite nuclear structure physics, the magic number has a special place. The magic combinations of protons and neutrons give the extra stability to the nucleus in comparison with the neighboring one. We have searched a magic combination of  protons and neutrons in the super heavy region. We found the proton number Z= 120 and corresponding neutron number 
N=182/184 \cite{1} can be a possible combination of the next magic numbers beyond $^{208}Pb$ in the super heavy region. We have shown  that our newly developed  non-relativistic interaction, simple effective interaction (SEI ) along  relativistic RMF model  with  various parameter sets  predict similar results. Our prediction of magic combination based on the various signature of the 
magic nuclei like the sharp decrease in the two-neutron separation energy $S_{2n}$ and 
two-proton separation energy $S_{2p}$ - energy,  zero pairing gaps $\Delta_n$ and $\Delta_p$ and the large gap in protons and neutrons single particle energy levels. All the evidences show a clear signature of the presence of magic number at Z= 120 and N= 182 or 184 \cite{1}. 

               In  recent years with the advent of the radio active ion 
beam (RBI) facility drip-line and super-heavy nuclei are on the spot 
light. We have studied the collective 
excitation of drip-line and super-heavy nuclei. As we know that nucleus is a many-body quantum system, collective excitation happens quite often instead of single particle excitation. In collective excitation, nucleons are  excited collectively ( all protons and neutrons excited together ) instead of single particle excitation. There are various types of collective excitation present like Isoscalar giant monopole resonance (ISGMR) , isovector giant dipole resonance (IVGDR), isovector giant monopole resonance (IVGMR) etc . 
ISGMR and IVGDR play very important role in nuclear structure physics. 
Both are known as the squeezing mode of vibration.  In ISGMR the protons and neutrons vibrate in the same phase to each other. Either 
both the proton and neutron Fermi sphere expand or compress at the same time.  It is related to the incompressibility of a nucleus \cite{2}. From the finite nuclear incompressibility, we can calculate the infinite nuclear matter incompressibility ($K_\infty$) by leptodermous expansion \cite{2}.  In  the leptodermous expansion, the finite nuclear incompressibility ($K_A$) can be 
expanded into  various terms like volume, surface and asymmetry as : 
\begin{equation}
K_A = {K_\infty}+ {K_s} A^{-1/3} + K_{curv} A^{-2/3} + K_{\tau} \frac{(N-Z)^2 }
{A^2} + K_c Z^2 A^{-4/3}.
\end{equation}
Volume term of the finite nuclear incompressibility gives the value of  $K_\infty$, which has a very imperative role in deciding the nature of the equation of state (EOS). Softness and stiffness of an EOS are decided by its $K_\infty$ value. Usually higher $K_\infty$  corresponds to the stiff EOS, while low $K_\infty$ gives soft EOS. In other words, $K_\infty$  controls the curvature of EOS at saturation density.  But incompressibility is not a directly measurable quantity. In experiment, we measured excitation energy of ISGMR, which can be related to the incompressibility of finite nucleus by the formula : 
\begin{equation}
E_x=\hbar \sqrt\frac {A {K_A}}{m <r^2>}.
\end {equation}
There is also another method to calculate  $K_\infty$  from $K_A$. Theoretical strength function distribution is calculated with the microscopic model like random phase approximation (RPA) with various parameter sets. The $K_\infty$  of the parameter set which reproduces the experimental strength distribution exactly  is considered as the right $K_\infty$ value.

                                           We have calculated the incompressibility ($K_A$) and the excitation energy of the dripline and super heavy nuclei \cite{3}. The collective excitation like ISGMR is a smooth function of the mass number (A).  So we can apply semiclassical  approximation  like relativistic Thomas-Fermi (RTF) and extended Thomas-Fermi (RETF) approximation \cite{4} to study the excitation energy as well as the incompressibility of the finite nucleus. In the present thesis, we have applied RETF approximation along with the constrained and scaling approximation to calculate excitation energy ($E_x$) of ISGMR. 
In order to compare the excitation energy  of macroscopic model with the microscopic model, three average energy are defined :  
constrained energy ($\sqrt{\frac{m_1}{m_{-1}}}$) , centroid energy  (${\frac{m_1}{m_{0}}}$) and   scaling energy  ($\sqrt{\frac{m_3}{m_{1}}}$).  We focused to calculate the constrained and scaling energy. The constrained energy is calculated by minimizing the constrained energy functional (${\cal H} - \lambda Q$), where Q is the excitation operator and $\lambda $ is the parameter.  In scaling approach, we scaled  the density and solved the scaled equation of motion to calculate the restoring force : 
\begin{equation}
C_M = \bigg[ \frac{\partial^2 }{\partial \lambda^2}\int d(\lambda r) 
\frac{{\cal H}_\lambda (r) } {\lambda^3} \bigg]_{\lambda=1}.
\end{equation}
The restoring force and incompressibility are connected by the formula 
\begin{equation}
E_x=\sqrt{\frac{C_M}{B_M}}, 
\end{equation}
where $C_m$ is the restoring force and $B_M$ is the mass parameter. In relativistic case, mass  parameter is defined like 
\begin{equation}
{B_M} =\int {\cal H} {r^2} {dr}
\end{equation}
and in the non-relativistic case ${B_M}^{nr} = m {A}<{r^2}>$.

                    Along with the constrained method developed in the recent past,  we have developed also a different constrained technique, which is based on the Taylor  series expansion around the equilibrium. As we know that the giant resonance can be viewed as a small amplitude vibration of density or shape around the equilibrium, we can expand the constrained energy functional around the equilibrium in Taylor series.  We have applied the new method to calculate the excitation energies of both ISGMR and IVGDR. We have compared our results with other theoretical models like random phase approximation (RPA). Our results well matched with other theoretical calculations  as well as experimental data.

                  Apart from the  drip-line and super-heavy region the  medium-heavy nuclei (A$\sim$100) have also very crucial in the study of  giant resonance. In recent years, softness of Sn isotopes have gained a lot of attention for nuclear theorists.  The excitation energy of isoscalar giant monopole resonance  of Sn series ($Sn^{112-124})$ shows a low value with respect to the experimental excitation energy, whereas excitation energy of $^{208}{Pb}$ and $^{90}{Zr}$ well matched with the experimental data. Now the question arises why Sn is so floppy ? Lots of literature have been devoted to explain this softness of Sn isotopes. But still, it is an open problem. We have discussed this problem within our RETF formalism. We elaborately discussed the contribution of various terms of RMF Lagrangian \cite{5}. A large number of nonlinear relativistic force parameters are used in these calculations. We find that a parameter set is
capable of reproducing the experimental monopole energy of Sn isotopes when its nuclear matter compressibility lies within 210-230 MeV, however, it fails to reproduce the GMR energy of other nuclei. That means, simultaneously a parameter set cannot reproduce
the GMR values of Sn and other nuclei. 
            
                The nuclear force is the central theme of the nuclear 
physics ever since from the starting point (1932: discovery of neutron). It is the aim of every nuclear physicist to understand the atomic nuclei on the basis of bare nucleon-nucleon interaction. The interaction between the nucleon is the strong interaction in nature, so 
in principle, one should use quantum chromo dynamics (QCD) to study the atomic nuclei. But it is not easy to use QCD starting from quark interaction to calculate various properties of nucleus. In fact, we can not use perturbative formalism like QCD approach in a case of nuclear physics due to the energy range involved. In this energy range, the QCD coupling constants become large so the perturbative approach is not valid. In nuclear physics, it is a tradition to use nucleons as the degrees of freedom and massive bosons are exchanged between the nucleon to generate the nuclear force. Patra et al. have calculated nucleon-nucleon potential starting from the effective RMF Lagrangian. This is known as the R3Y interaction \cite{7}. We have used this R3Y interaction to calculate various properties of finite nuclei. We have also introduced a new cross-coupling of omega meson into the interaction and analyzed its effects on the finite and infinite nuclear system.  We found that self-interacting $\omega$-meson term has a significant influence on the finite as well as an infinite nuclear system. A strong but an attractive components of nuclear force is  generated due to $\omega^4$ term at very short distance (r$<$0.2 fm) \cite{6}. We have applied R3Y as well as the M3Y interaction to obtain the folding potential to study the (p, $\gamma$) and (p, n) types of reaction for proton-rich nuclei. These proton-rich nuclei are very less in number but has great astrophysical implication to study r- and p- processes. Effects of linear and non-linear interaction terms of RMF Langrangian on s-factor calculation is discussed \cite{9}. 
We have shown that the non-linear $\sigma$ meson coupling has strong effect 
on the S-factor calculation on the p-nuclei.

After discussion of various aspects of finite nuclear structure, we shifted our attention to the infinite nuclear matter, which is a hypothetical system of an infinite number of nucleons interacting through nuclear force only (uncharged), having infinite volume (no surface). Due to the heavy mass, the contribution of $\delta$-meson has been neglected for many years in RMF theory. But in a neutron star, a highly asymmetric and dense system, we can not overlook the 
contribution of $\delta$-meson. It affects the transport properties of  
asymmetric nuclear matter in a significant way.  Using the effective field theory approach, like RMF, we discussed the dominance of isovector-scalar $\delta$-meson on the neutron stars as well as hyperon star \cite{10}. From simplification point of view, we started with an assumption that neutron star is a static and uncharged body, constituted with a maximum number of neutrons and little amount of proton and electrons, which are necessary for the $\beta$-equilibrium.  Then extended our investigation to a more realistic situation, where a neutron can rotate with hyperon at the core of the star. Neutron star is a system of degenerate fermions, so simple energetic consideration implies the presence of hyperon in the high density (9-10 times of $\rho_0$)  environment of a neutron star. Further the time scale associated with the neutron star is much larger than the time scale associated with the weak interaction, which favors the production of hyperons inside the neutron star. However, in the standard picture, the inclusion of hyperon degrees of freedom leads a considerable softening of EOS, which consequently leads a maximum allowed the mass of neutron star lower than the current observation. So there is no question of hesitance to re-investigate the EOS of neutron star with new degrees of freedom in different physical condition.  We focused on the effects of $\delta$-meson on :  (a) static proton-neutron and hyperon star. (b) rotating proton-neutron and hyperon star.  All  the calculation  have done with RMF formalism with G2+$\delta$ parameter set. G2 is considered as one of the efficient parameter set, where maximum number of interactions have been taken care. We have added an extra $\delta$-degrees of freedom with existing $\sigma$, $\omega$ and $\rho$-mesons. It is not conceptually right to vary the $\delta$-meson strength ($g_\delta$) as a free parameter to see the effects of the $\delta$-meson on the nuclear system. Due to the isovector nature of both $\rho$ and $\delta$-meson, they contribute to the isovector channel simultaneously. It is customary to take into account the contribution of $\rho$-meson, while discussing the effects of the $\delta$-meson \cite{11}. There are various procedure to fit $g_\rho$ and $g_\delta$ to see the effects. We have adopted two of them. In one case, the $g_\rho$ and $g_\delta$  are fitted in such way a that the symmetry energy of the original G2 parameter (36.4 MeV) remain fixed and in another case the coupling 
constants are adjusted to fix the binding energy and charge radius of finite nucleus fixed.  We calculated the maximum allowed mass of the neutron start with these pairs of $g_\rho$ and $g_\delta$. With the inclusion of $\delta$-meson, the maximum allowed mass value of mass for proton-neutron star increases due to the stiff equation of state at higher density. Results are same for both rotating and static case. Quantitatively we get maximum mass of static and rotating proton-neutron star $\sim $ 2 $M_\odot$ and $\sim$2.4 $M_\odot$ respectively in G2 +$\delta$ parameter, which are close to the current observations. But in hyperon star, maximum mass decreases with the inclusion of $\delta$-meson, this is due to the increase of strangeness in the system. The $\delta$-meson interaction affects the production yield of hyperon and hyperon produced at a lower density than in non-$\delta$-system. To see a consistent effects of $\delta$-meson on the nuclear system, the more appropriate way to form a parameter set by reshuffling all the parameters and readjusting with various finite and infinite nuclear matter properties. Work in this direction is in progress.

\addcontentsline{toc}{chapter}{List of Figures}
\listoffigures 
\listoftables
\addcontentsline{toc}{chapter}{List of Tables}

\newpage
\setcounter{page}{1}
\pagenumbering{arabic}



\newpage
\chapter{Introduction }
\label{chapter1}
Nuclear Physics has been providing a platform to test all four types of basic interactions unlike to any other branch of physics. It is not an easy task to deal with nuclear physics because it needs a sound knowledge in all four fundamental interactions starting from the electromagnetic interaction between protons of finite nuclei to gravitational interaction in the case of the neutron star, which can be viewed as a big nucleus. The most mysterious and puzzling thing in  nuclear physics is the nucleon-nucleon (N-N) interaction, which has consumed maximum manpower to fix its nature, but still a debatable subject. In 1911, Rutherford \cite{mars09} discovered nucleus from the famous $\alpha$-particle scattering experiment. Factual nature of a nucleus came to known after the discovery of neutron by James Chadwick in 1932. A nucleus consists of protons and neutrons, both the individuals are known as nucleon. Nucleons are the quanta of  the nucleus. In 1935 Yukawa purposed the meson theory \cite {yakawa35,brink65} of nuclear force. According  to this theory, nucleons interact with each other through the exchange of mesons. The nuclear force is similar to the Van der Waals force between atoms and molecules. Both Van der Waals and Nuclear forces are residual forces of fundamental  interaction. The former one is the residual interaction of electromagnetic interaction, while later one is same as the strong interaction. In some sense,  nuclear force is not a fundamental force rather it is the residual interaction of  strong interaction between the quarks. In principle,  we can study the  interaction between two nucleons starting from the interaction between the quarks,  which are the constituent particles of nucleons. But for  practical  purpose, it is very difficult to deal with a theory starting from  quark -quark  interaction i.e. QCD (Quantum Chromo Dynamics). In the QCD level, things become very complicated to study the properties of finite and infinite nuclear systems. So in nuclear,  physics it is customary to use effective mean filed theories, like SHF 
( Skyrme Hartree-Fock ), RMF (Relativistic Mean Field), DBHF (Dirac- Bruckner  Hartree Fock ). In effective theories, the total interaction is not the  sum of  all possible two-body interactions but  
each nucleon feels as if it moves in a common (mean) potential/ field generated  by rest of the  nucleons. Suppose in a nucleus there are A number of nucleons  then every nucleon will feel a mean field generated by A-1 number  of nucleons. Basically, there are two types of effective theories mostly  adopted in  nuclear physics. In one case nucleons are treated like  non-relativistic particles (SHF, Gogny) \cite{skyrme56,skyrme58,gogny80} and in other case nucleons are  treated like relativistic particles (RMF)\cite{miller72,walecka74,walecka85}. In the non-relativistic 
case nucleon's motion are governed by Schrodinger equation (SC), while in the case  of relativistic one nucleons motion are governed by Dirac equation . There are some merits of  relativistic  formalism over the non-relativistic formalism. Relativistic formalism  is one step ahead of  non-relativistic formalism. In relativistic  formalism, we deal with Dirac equations, so the spin-orbit interaction  comes in a more natural way unlike to the non-relativistic model.  No extra parameter is used to fit the spin-orbit interaction term.  In relativistic mechanics, there is the presence of both  positive and  negative energy solution. We avoid the negative solution by assuming  Dirac no sea approximation.     

\section{Effective Theory }

Self-consistent models are efficacious tools for the 
investigation of the nuclear structure and low energy dynamics. 
Mean field models are based on the effective energy-density functional 
theory. Interaction between nucleons is formulated in terms of density 
dependent Lagrangian. Energy- density functional contains many free parameters
 whose values are fitted to reproduce empirical and experimental data. Theses 
parameters are fitted based on the constraints related with 
\begin{itemize}
\item Experimental data on the static properties of the finite nucleus 
( Binding energy and charge radius ).

\item Characteristic properties of the nuclear matter 
( saturation density, binding energy per nucleon at saturation density,
symmetry energy).
\item Excitation of giant monopole and dipole resonances.

\item Observational information of the neutron star and supernovae
( mass and radius of the neutron star).
\end{itemize}

\section{Non relativistic self consistent theory }

Basically, there are two types of non-relativistic theories widely used 
, one is zero range Skryme Hartree-Fock (SHF) and other is finite range 
Gogny interaction.  In  both the cases, basic assumption is that nucleons are 
moving at a speed much less than the speed of light. This assumption is validated  taking into account the average binding energy per nucleon in an nucleus is 
$\sim$ 8 MeV comparing with the mass of the nucleon $\sim$938 MeV. So no relativistic 
formalism is needed. In Skryme Hartree Fock approach the effective 
potential between the  nucleons can be written as the sum of two bodies 
and three bodies interactions. That is  
\begin{equation}
V_{eff} = \sum_{ij} {V_{ij}}^{ (2)} + \sum _{i j k } {V_{ijk}}^{(3)},
\end{equation}
where $V_{ij}$ and  $V_{ijk}$ are  two body and three body  
interactions respectively.  It is a tedious work to deal with 3-body 
and many-body interactions. Thus for the practical purpose, the 
leading order interaction i.e the 2-body and for very specific case 
the 3-body is sufficient. In short  range expansion of two body 
interaction. The Skyrme energy functional can be written as 

\begin{eqnarray}
{\cal{E}} &=&\frac{1}{2} {t_0}\bigg[  (1+ \frac{x_0}{2}) {\rho}^2 -
 (x_0 + \frac{1}{2}) \sum_q {\rho_q}^2] + {t_1}\bigg[ 
\bigg [ (1+ \frac{x_1}{2} ) \bigg] \bigg[ \rho \tau + 
\frac{3}{4} (\Delta \rho) ^2 \bigg] \nonumber\\
&-& (x_1 +\frac{1}{2}) \sum_q \bigg[ {\rho_q} {\tau_q} + 
\frac{3}{4} (\Delta {\rho_q}) ^2 \bigg] \bigg]+\frac{t_2}{4}  
( 1+\frac{x_2}{2}) \bigg[ \rho \tau -\frac {1}{4} (\Delta \rho)^2 \bigg] 
\nonumber\\
&+& ( \frac{x_2}{ 2}+\frac{1}{2} ) 
\sum_q  \bigg[ \rho_q \tau_q - \frac{1}{4} 
 (\Delta \rho)^2 \bigg]- \frac{1}{16} (t_1x_1 + t_2 x_2) J^2 + \frac{1}{16} 
(t_1 - t_2 ) \sum_q {J_q}^2 \nonumber\\
&+& \frac{1}{12} t_3 {\rho}^{\gamma} 
\bigg [ (1+ \frac{x_3}{2}) {\rho^2} - (x_3 +
 \frac{1}{2} ) \sum_q {\rho_q}^2 \bigg]+ \frac{1}{2} {W_0} (J \Delta \rho + \sum_q J_q \Delta \rho_q ).
\end{eqnarray} 

The terms proportional to  $\rho^2$ represent two body attractive potential. It is counter balanced by the terms proportional to the ${\rho}^{\gamma+2}$, 
which are the repulsive terms. These two terms counter balance each other and 
give saturation properties of nuclear force. The terms proportional to the 
$\rho \tau$ give kinetic energy contribution. The surface term 
$(\Delta \rho)^2$  is essential for finite nuclei. Spin-orbit 
interaction has also a vital role in finite nuclei, which is represented by 
$J \Delta \rho$.

\begin{itemize}
\item $\rho$ = Local baryon density
\item $\tau $ = Kinetic energy density of the baryon 
\item J= Spin-orbit density 
\end{itemize}
 
In the Skyrme energy functional there are both T=0 ( isoscalar ) and 
T=1 (isovector) part. The densities without subscript represent 
isoscalar part and  densities with subscript q represent 
isovector part.

\section{Relativistic self consistent theory }

Fundamentally RMF is different from SHF and Gogny, by including the 
relativistic effects in the formalism. In simple words, one can say RMF is nothing but the relativistic generalization of  non-relativistic 
mean field formalism. Unlike SHF, the RMF is a finite range interaction. Nucleons are interact with each other through the exchange of  various effective mesons-like $\sigma$-, $\omega$-, $\rho$- and $\delta$- mesons. In nature, there are so many effective mesons are present, but why few of them are taken into account ? This is due to the symmetry of nuclear potential. Except the symmetry, there are some meson's contribution are excluded due to 
their heavy masses and negligible contributions. They do not give any 
significant contribution neither qualitatively nor quantitatively for the 
improvement of the model, except the mathematical complexity. So it is wise to take the minimal set of mesons, which can describe the nucleon-nucleon interaction up to a desired level. In our work we have taken $\sigma$-, $\omega$-  and $\rho$- mesons in most of cases. $\delta $-meson is also added in some cases. The starting point of the RMF theory is the nucleon-meson interacting Hamiltonian, which is given by

\begin{eqnarray}
{\cal H}&= &\underbrace{\sum_i \varphi_i^{\dagger}
\bigg[ - i \vec{\alpha} \cdot \vec{\nabla} 
\bigg]\varphi_i}_\text{a}+ \underbrace{\sum_i \varphi_i^{\dagger} 
\beta m^* \varphi_i}_\text{b}
+ \underbrace{\sum_i \varphi_i^{\dagger} g_{v} V \varphi_i}_\text{c}
+ \underbrace{\sum_i \varphi_i^{\dagger}\frac{1}{2} g_{\rho} R \tau_3 \varphi_i}_\text{d}
+\underbrace{\frac{1}{2} e {\cal A} (1+\tau_3)  \varphi_i}_\text {e}\nonumber\\ 
&+&\underbrace{ \frac{1}{2} \left[ (\vec{\nabla}\phi)^2 
+ m_{s}^2 \phi^2 \right]}_\text{f}
-\underbrace{\frac{1}{2} \left[ (\vec{\nabla} V)^2 + m_{v}^2 V^2 \right]}_\text{g}
- \underbrace{\frac{1}{2} \left[ (\vec{\nabla} R)^2 + m_\rho^2 R^2 \right]}_\text{h}
- \underbrace{\frac{1}{2} \left(\vec{\nabla}  {\cal A}\right)^2}_\text{i} -
\underbrace{\frac {\zeta_0}{24} g_{v}^4{V}^4 }_{j}\nonumber\\
&+&\underbrace{\frac{1}{3} b \phi^3
+ \frac{1}{4} c \phi^4}_\text{k}
- \underbrace{{\Lambda_V} {g_v^2}{g_\rho}^2{R^2}{V^2}}_\text{l} 
\end{eqnarray} 
The meaning of the terms in the Hamiltonian are as follow:
\begin{itemize}
\item a:  The kinetic energy contribution of the  nucleons. 
\item b: Rest energy of the nucleon plus interaction between the nucleons 
         and  sigma meson. 
\item c,d: Interaction between the omega  and rho-mesons with nucleons 
           respectively.
\item e: Interaction of photon with the proton. 
 
\item f,g, h: Free meson contribution of $\sigma$, $\omega$, and $\rho$-meson 
respectively. First term represents the kinetic part and second term is
the rest energy part for each case. 

\item i: Kinetic energy contribution of the photon. Rest energy part is zero 
because of zero rest mass. 

\item j,k: Self coupling of the $\omega$ and $\sigma$-mesons respectively. 

\item l: Cross coupling between the $\omega$ and $\rho$-mesons. 
\end{itemize}

\section{Drip-line and Super heavy nuclei}
Both drip-lines and super-heavy regions are most venerable areas of 
nuclear physics. Many exotic phenomena can be found in these areas of 
landscape. Physics inside  light nuclei is not exactly same as  
in super-heavy nuclei. As the nuclear force is density depended, 
it changes with  number of nucleons. There are lots of 
discrepancy  in the drip-line and $\beta$- stable nuclei. The experimental and theoretical investigations of  nuclei far from the valley of  $\beta$- stability  is the main aim of modern nuclear structural research. Radio active ion beam (RIB) facilities have disclosed a wealth of structural phenomena in exotic nuclei having the extreme value of $N/Z$ ratio. The super-heavy and drip-line nuclei regions involve many exotic and anomalous phenomena like
\begin{itemize}
\item Halo Nucleus 

\item Neutron skin

\item Quenching of the shell effect in neutron-rich nuclei and 
the resulting deformation

\item New magic number (N, Z)

\item Shape co-existence 

\item Ground-state proton radioactive 

\item Synthesis of the heavy-element 

\item Onset of the exotic collective modes.

\end{itemize}
New upgraded experimental facilities will improve and extend  our 
understanding of physics of nuclei up to  drip-line and super-heavy 
region. With the modern technology, we can find out the exact location of  proton and neutron drip-line and possible to  synthesis of  super heavy nuclei having a longer life time so that spectroscopy of the nuclei can also be studied. More new important information will be accumulated on masses, life times and reaction cross-sections, which are urgently needed in astrophysical calculations for better understanding of nuclear process occurring in the universe. Proton-rich nuclei play a vital role in understanding the astrophysical process like r- and p- process ( rapid neutron capture process and rapid proton capture process).

 Unraveling of the physics of drip-line and super-heavy 
 regions not only due to the lack of experimental up-gradation, but also the lack of the proper theoretical understanding. All most all theoretical models converge in the light nuclei and $\beta$-stable region with acceptable error. But if we  start the journey towards the drip-line and super-heavy region they diverge with an unacceptable difference.  In drip line region the exotic phenomena mainly due to  (a) pronounced effects of the coupling between bound state and particle continuum (while in the $\beta $ stable nuclei system is mostly bound state ) (b) weak binding energy of  outer most neutron (c) region of neutron halo with very diffuse  neutron density (d ) major modification in shell structure. The vital problem of  theoretical investigation in the drip-line region is the closeness of  Fermi level to the particle continuum. Self-consistent theories (relativistic and non-relativistic ) can describe the coupling between the bound state and particle continuum. But  both relativistic and non-relativistic lead to a huge discrepancy in  the drip-line region, while they move in hand  to hand  in the  $\beta-$ stability line. Particularly in the way they describe  spin-orbit are very different from each other  and its iso-spin dependence also. This is due to the fitting of the parameter, which mostly belongs to the $\beta$- stability region.

       In general, $A> 210 $ is considered as  super-heavy region in the
   nuclear chart. The most challenged task in the super heavy nuclear physics is  its production in the laboratory or available in nature. Its reaction cross-section goes inversely with 
 atomic number in the super heavy region. Most of the super heavy elements have a half-life 
in the order of few  $\mu$-seconds, which is too small to characterize its chemical and structural 
properties. This is the main reason of worried for a nuclear physicists to find a combination of Z and N, 
which gives element with an appreciable half-life in nuclear scale. Due to the lack of  technical up-gradation, 
experimentalists always depend upon the theoretical confirmation. Theorists also face compulsion due to the 
diverge behavior of the various models. They always face problem to conclude with an unique 
combination of N and Z. Condition become more worst, when a model with different parameter sets lead 
different conclusion. Basic reason of this is the non-transparent nature of nuclear force, which is a 
fundamental problem in the nuclear physics for all  time. Then the validity and applicability  of the 
different nuclear force and the model becomes a debatable subject, which consumed maximum man power in the nuclear theory.

 In the present thesis, we are particularly concerned about  
nuclear structure. Nuclear structure is the result of  an interplay between the surface tension and elector-magnetic interaction between  protons. 
Surface tension originates from  nuclear force between the nucleons 
and tries to maintain a spherical shape, while Coulomb force provokes 
the nucleus to be deformed. Most of the heaviest elements were found in three " heavy elements factories ": Lawrence Berkely Laboratory in Berkely,  Joint Institute for Nuclear Research (JINR), Dubuna (Russia) and GSI near Darmstadt, Germany. Probably the first theoretical prediction of  the super heavy element was by Nilson et.al \cite{nilson69} with the help of  liquid drop model plus Strutinsky shell correction,and the prediction was  Z=114 and N=184. The magic number Z=118  was 
predicted in earliest macro-microscopic calculation \cite{patyk91} and later confirmed in \cite{nix92, miller72}. The fully microscopic approaches predict the proton shell closure at Z=120 \cite{74}, Z=126, or Z= 114,120 126 \cite{rutz97} depending upon the chosen  
nucleon-nucleon interaction in  mean field theories. The neutron magic 
number N=184 is almost firmly predicted by different theoretical models.   With the advent of radioactive ion beam(RIB) facilities, last twenty years have remained as a golden era for heavy elements production. New super-heavy elements are now produced with both cold and hot combination of colliding nuclei. The heaviest yet discovered element is Z=118 , synthesized in hot fusion reaction of $^{48}{Ca}$ beam and $^{248}Cf$ is the heaviest available target, which  has been used in this Z=118\cite{oga11}. Thus to get super heavy (SH)  elements with Z $>118$  fusion reaction, one should be processed to heavier than $^{48}Ca$ projectile  ( $^{50}Ti$, $^{54} Cr$). The corresponding cross-section for the production  of  elements Z=119 and Z=120 are predicted to be smaller by about two order  of magnitude as compared $^ {48} Ca $- induced fusion reaction.  Another  limitation of the fusion reaction (both hot and cold ) for producing SH  elements consists in the fact that they lead to neutron deficient isotopes having rather a short lifetime.  In Chapter 2, we  discuss the prediction  of magic number at Z=120 and N=182/184 with our new non-relativistic SEI model and  compared with other theoretical calculations.

\section {Giant resonances}

Giant resonance is a collective vibration in which the protons and neutrons vibrate in a collective manner instead of single particle vibration. In microscopic formalism, it can be viewed as a coherent superposition of particle-hole excitation. Macroscopic formalism describes it as a vibration of shape and density of nucleus around the equilibrium shape and density. On the basis of their quantum numbers like multi-polarity (L), spin (S) and isospin (T), giant resonances can be divided like isoscalar giant monopole resonance ( ISGMR), isovector giant monopole resonance(IVGMR), isovector giant dipole resonance (IVGDR) etc\ldots . Some of the resonances 
like ISGMR and IVGDR are  crucial in the nuclear structure 
physics. ISGMR is also refereed as the  breathing mode of oscillation. 
In ISGMR the proton and neutron vibrate in a phase to each other, 
either they are expanding or compressing.   It is a density oscillation just like human breathing, expansion, and compression of the nucleus.  In another word, it corresponds to the radial oscillation, in which the radius oscillates around the equilibrium radius. It has a small amplitude of vibration, only 1-2 \%  of the original radius.  As it is related to  breathing mode, the excitation energy of the ISGMR gives a way to calculate the incomprehensibility (K ) of the finite nucleus and consequently the incompressibility of infinite nuclear matter ($K_\infty$).  Infinite nuclear matter incompressibility is a key quantity in the calculation of equation of state (EOS) of a neutron star. In IVGDR, the protons and neutrons vibrate in  opposite  phase to each other. It related to the symmetry energy of a nuclear system, which is  another most important quantity to calculate the EOS of a neutron star.  In the present thesis, we have developed a slightly different  formalism to calculate the excitation energy $E_x$ of ISGMR and IVGDR. Our formalism is based on  constrained method calculation. This constrained  formalism is based on Taylor series expansion, which is different from the constrained Hartree-Fock formalism, where we have to minimize the constrained Hartree-Fock constrained Hamiltonian.  Generally, we defined three mean energy like constrained energy  ($\sqrt\frac{m_1}{m_{-1}}$), centroid energy ($\frac{m_1}{m_{0}}$) and scaling energy ($\sqrt\frac{m_3}{m_{1}}$). We will discuss more the excitation energy and incomprehensibility in Chapter IV.

\begin{figure} 
\includegraphics[width=0.8\columnwidth,clip=true]{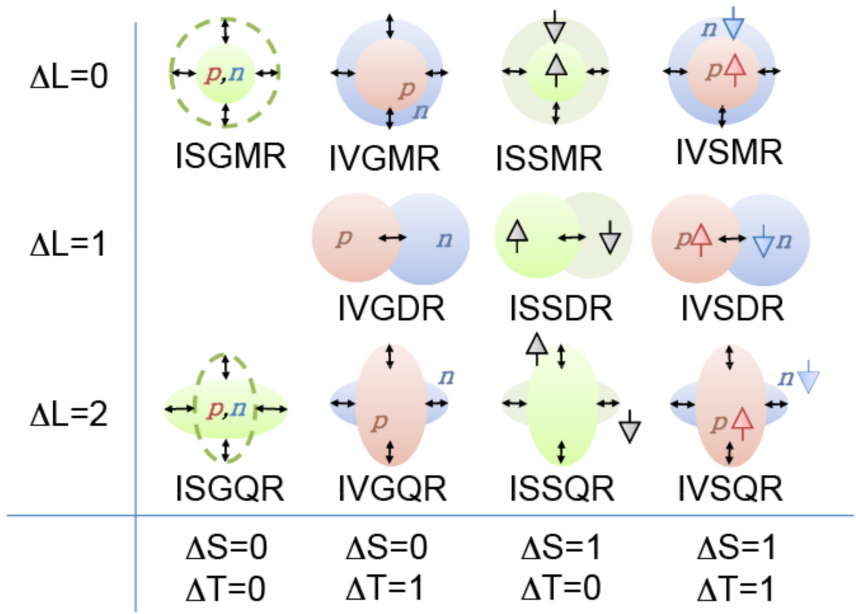}
\caption{ Various electric and magnetic giant resonance, which classified 
on the basis of their multi polarity(L), spin (s)and isospin(T) 
quantum number.}
\label{Fig.2}
\end{figure}

\section{Infinite Nuclear matter (INM) }

The universe contains a remarkable  wide variety of atomic nuclei with the
mass number up to A$\sim$250. There are many interesting properties, which 
differentiate these nuclei from  each other, while there is also a 
powerful set of systematic trend and general properties that provide an 
important and useful frame work for understanding the basic structure 
of  nuclei. We can classified the properties of  nucleus  
mainly into two categories: local and  global properties. These global 
properties are well studied with the help of a hypothetical system i.e., 
infinite nuclear matter (INM). Properties of this system do not depends
 on the local parameters like a number of the nucleon and structure 
of the nucleus. We can define the INM as " it is a hypothetical 
interacting system of infinite number of nucleons, with no surface effect 
and in an absence of Coulomb interaction". The nucleons interact through 
strong interaction only. As we are considering A$\rightarrow \infty$ , 
so there must not be any boundary (zero surface effect). Then a common 
question arises "How can we address the infinite nuclear matter"? 
For that, we have to attribute some properties to the infinite nuclear matter. 
These are (a) saturation density ($\rho_0$ ) (b) binding energy per particle 
at saturation density ($E/A$)  (c) symmetry energy (J) (d) L-coefficient 
(e) incompressibility ($K_\infty$). In semi-empirical mass formula, the binding 
energy  per particle of a finite nucleus can be written as 
\begin{equation}
E/A= \underbrace{{a_V}}_\text{Global (bulk)} - 
\underbrace{\frac{{a_s}}{A^{1/3}}-{a_c} \frac {Z (Z-1)}{A^{2/3}}-
a_{sym} \frac{(N-Z)^2}{A^{2}}\pm\delta }_\text{local} 
\end{equation}  
Here $a_v$, $a_s$, $a_c$ and $a_{sym}$ carry their usual meaning of volume, 
surface, Coulomb and symmetry coefficient respectively. In binding energy per
 particle expression, first term is independent of  A, which represent the 
bulk properties  and same for all nucleus. Second part differentiate 
nucleus from each other. The first term (volume ) represents the INM properties 
with the condition of symmetric nuclear matter (N=Z). Thus we 
will get rid of the asymmetry and Coulomb terms. The surface contribution 
goes on decreasing in comparison with the volume term with an increase of the 
mass number and Coulomb term is switched off by definition. The pairing 
term can be neglected, which is very small in comparison to the volume term. 
So finally binding energy expression of INM left with $a_v$ term, which 
has value $a_v \sim -16 MeV $.

      Another characteristic property of the INM is the  saturation density, 
which has a typical value of $\rho_0 = 0.16 fm^{-3}$. It is the maximum 
density, which can be achieved inside the finite nuclei. In other words 
, one can say, it is not possible to go beyond this density under the 
nuclear force only. This property is the bi-product of the short range 
repulsive nature of the nucleon-nucleon interaction. We can not compress 
the nuclear system after certain limit due to its repulsive nature at 
short range. We need some other force to increase the density of a 
nuclear system. This is the gravity which helps the system of the neutron 
star to increase 9-10 time of the saturation density. INM is a homogeneous  and 
isotropic infinite system, the position of the nucleons do not depend only 
the relative distance between the nucleon matter. Due to the homogeneity 
nature, we can write the nucleon wave function as plane wave function $e^{ikx}$. The particle 
density is obtained by the sum over all the occupied state inside the
 phase space volume $(2\pi)^3 $. 

\begin{equation}
\rho= \frac{N}{V}= \sum_{K,\lambda} {\psi_{k,\lambda}}^* (\vec{r}) 
{\psi_{k,\lambda}} (\vec{r}) \rightarrow \frac{\gamma}{(2\pi)^3} 
\int d^3K n(\vec{k})
\end{equation}    
 Here $\gamma$ is the degeneracy factor, which has value $\gamma =4 $ for the 
symmetric nuclear matter (N=Z ) and $\gamma =2$ for the pure neutron matter. 
The nuclear matter symmetry energy $ E_{sym} (\rho)$ is an essential tool 
to characterize the isospin dependent part of the equation of state (EOS)
 of the asymmetric nuclear matter. Knowledge about the symmetry energy is 
unavoidable in understanding many aspects of the nuclear physics and 
astrophysics\cite{dani02, latt07, stei05}. The symmetry energy of the 
nuclear matter can be defined as the difference between the energies 
of pure neutron matter and symmetric nuclear matter as function 
of the density. Energy density can be function of density and asymmetry 
of the system $E (\rho, \alpha=\frac{N-Z}{A})$.      

\begin{equation}
E(\rho, \alpha)= E(\rho,0) + S(\rho) {\alpha}^2 + O({\alpha}^4) + .......
\end{equation}
\begin{equation}
S(\rho) = \frac{1}{2} \frac{\partial^2 E(\rho,\alpha) }
{\partial \alpha^2}\bigg|_{\alpha=0}= {S_v} + \frac{P_0}{{\rho_s}^2}
(\rho-\rho_s) + \frac{\Delta K}{18 {\rho_s}^2}(\rho-\rho_s)^2 +....... 
\end{equation}

$S_v$ is the infinite nuclear matter symmetry energy, whose empirical value lies 
$30\pm 5 $ MeV.

Another most important quantity of the infinite nuclear matter is its 
incomressibility ($K_\infty$). The nuclear matter incompressibility ($K_\infty$) is 
the measure of the stiffness of the equation of state (EOS). 
It can be calculated from the curvature of the EOS at saturation 
density. $K_\infty$ puts a stringent constraint on the
theoretical description of effective nuclear interaction and density
dependence of the nuclear interaction. The correct value of the infinite
nuclear matter incompressibility ($K_\infty$) is a debatable subject in
the current scenario. It has been remained a long standing technique to
constraint the $K_\infty$ with the help of isoscalar giant monopole
resonance\cite{blaizot}.  There are two ways to calculate the infinite 
nuclear matter incompressibility ($K_\infty$). One is the direct way, in which 
the monopole excitation energy is calculated for heavy nucleus with 
different nucleon-nucleon interaction and the correct nucleon-nucleon 
interaction is pointed out which gives the exact excitation energy. 
With this nucleon-nucleon interaction, the incompressibility of infinite 
nuclear matter is calculated. The second method is little indirect, first, the 
monopole excitation energy of different nuclei is measured and the 
incompressibility of finite nuclei is calculated by the formula 
${E_A}=\sqrt{\left [ \frac{\hbar^2 {K_A}}{m \langle r^2 \rangle}\right]}$.
Like the semi-empirical mass formula, the finite nuclear incompressibility can 
be expanded in terms of volume, surface, asymmetry and Coulomb expression. 

\begin{equation}
\label{ser}
{K_A}= {K_V}+ {K_s}A^{-1/3}+ {K_c} A^{-2/3} +[ K_{\delta v}
+K_{\delta s}A^{-1/3}]\delta^2+ K_{coul}\frac{Z^2}{A^{4/3}}+......
\end{equation}  

In the limit of infinite nuclear matter ($A\rightarrow \infty$ ) all other 
terms go to zero except $K_V$. So $K_V $ gives the infinite nuclear matter 
incompressibility ($K_\infty$). The various coefficient in the expression 
(\ref{ser}) can estimated from the chisqaure fitting with the experimental data.
But there is a huge uncertainty in the value of $K_\infty$ due to very 
limited number of excitation energy data. It is also  model sensitive 
quantity. In current scenario both the relativistic and non-relativistic 
formalism come to a conclusion with a value of $240\pm 20 $ of $K_\infty$.   
There is various application of the concept of infinite nuclear matter 
in understanding nuclear physics and astronuclear physics, mainly the 
physics of the neutron star.

\section {Neutron Star}
  A neutron star is probably the densest object found in the visible 
universe, which has the density around 9-10 time the density of the 
infinite nuclear matter (density of infinite nuclear matter $\rho_0 \approx 0.16  fm^{-3}$). Neutron star 
is the compact remnants of a massive star after it undergoes core collapse. 
During lifetime of a star, a hydro-dynamical equilibrium is 
maintained between the radiation pressure created from the fusion process, 
which is  outward and the gravitational attraction towards the core of the 
star. Fusion process starts with the fusion of hydrogen to form Helium 
with a huge amount of energy production. This process continues till the 
production of $^{56}Fe$, after which it become an endothermic process 
and not favorable for further processed. With the absence of the 
outward pressure gravity, contract and core collapse starts.     
Fate of a star after core collapse usually decided by the mass of the star 
before the core collapse. The star whose mass below $4 M_\odot$  become the 
white dwarf after the core collapse. Neutron stars and black hole are 
believed to originate from a more massive star. However, the dividing 
line between those stars that from neutron star and those that form a black hole 
is very uncertain because the final stage of the evolution of massive star 
are poorly understood. The formation of neutron star from a massive star is in the mass range  
$4-10 M_\odot$ and assume that all star with masses greater than 
$10 M_\odot$ end up as a black hole. One should not misunderstood from the 
name "Neutron star", that is composed of the only neutron. Neutron star derive 
their name from the predominance of neutrons in their interior, following 
the mutual elimination of electron and proton by inverse $\beta$-process. 
Neutron star composed of neutrons, protons and electrons. 
One can simply shows that $n_e : n_p : n_n = 1:1: 8$  in the limit of 
very large density is a trivial consequence of charge neutrality, $\beta$-
equilibrium and extremely relativistic degeneracy. Due to the high density, 
there is the probability of production of strange baryon (hyperon ). 
It is also assumed that quarks become unconfined due to extremely high 
density in the inner core of the neutron star.

\section{Hyperon Star}

 As we know that neutron star is one of the most dense object of the universe.
Neutron star has a core of density 9-10 times denser than the saturation 
density ($0.16 fm^{-3}$). This high density is the main cause of the 
many anomalous physics inside the neutron star. As the density increases, 
new hadronic degree of freedom may appear in addition to the neutron and 
proton. One such degree of freedom is hyperon, baryon with strangeness 
quantum number. We can not deny the presence of hyperon inside the 
core of neutron star. So we need to study the EOS and mass-radius profile 
of the hyperon star for complete knowledge of the neutron structure. 
In Chapter V we will discuss about the effect on EOS with inclusion of 
the baryon octet.

\section{Plan of the thesis}
In this  thesis we have outlined the structural properties  of 
finite and infinite nuclear matter. The thesis is organized as follows:

\subsection{In chapter 2} 

We have discussed  the magic number in the super heavy region. The  familiar 
 magic numbers Z= 2, 8, 20, 28, 50, 82 and N= 2, 8, 20, 28, 50, 82, 126 
in the light and medium 
heavy region. However the magic number in super heavy region beyond $^{208}Pb$ 
is not clear. Magic number has a great importance in the nuclear 
structure physics. It is a basic assumption that both proton and 
neutron double magic nuclei are spherical. In this chapter we will 
discuss the prediction of magic number with help two proton and 
neutron separation energy, $\Delta_n$ and $\Delta_p$ gaps, single particle energy levels and chemical potential. All the calculation have done with both relativistic (RMF) and non-relativistic (SEI) formalism.      
\subsection{In chapter 3} 
In this chapter, we will discuss about the excitation energy of the 
isoscalar giant monopole resonance (ISGMR) energy. Excitation 
energy of ISGMR provides a crucial tool to calculate infinite 
nuclear matter (INM) incompressibility ($K_\infty$). This $K_\infty$ 
has a major role in the EOS of infinite nuclear matter. 
So ($K_\infty$)  plays an imperative role in both structure of finite 
and neutron star. The study of nuclear structure physics will remain 
incomplete without the study of giant resonance excitation. In this 
chapter, we have discussed only one type of resonance i.e., ISGMR. 
Mostly we discussed the excitation energy of ISGMR in super-heavy region.
In recent years the excitation energy measurement reveals the softness behavior of the Sn isotopes. Last part of this chapter is dedicated  to explain the softness of Sn nuclei in our formalism, relativistic extended Thomas-Fermi the (RETF) formalism. 

\subsection{In chapter 4} This chapter is fully dedicated to develop a 
new formalism to calculate the giant resonance excitation energy 
of nucleus. Here we have developed a new technique based on Taylor series expansion to calculate the constrained energy ($ E_1 = \sqrt\frac{m_1}{m_1}$) for isoscalar giant monopole resonance (ISGMR) and isovector giant dipole resonance (IVDGR). We have given a comprehensive analysis of our new model and compared with other theoretical model as well as with the experimental data.
\subsection { In chapter 5}
In this chapter, the effects of nucleon-nucleon interaction on the finite and infinite nuclear system are discussed.  All the results are obtained with R3Y interaction, which was purpose recently by our group. The effects of self-interaction of $\omega $ meson on various properties of nuclear system is also extensively discussed. 

\subsection{ In chapter 6}
We have investigated the cross-section and astrophysical S-factor for the proton rich nuclei in the mass range A$\sim$100-120, using the R3Y and density dependent M3Y interaction. The effect of self-interaction of 
$\sigma$ and $\omega $ meson are also discussed extensively.  

\subsection{In chapter 7} 

After the discussion of finite nuclear structure, we move forward to the 
infinite nuclear matter. The best example of INM is neutron star. We have  calculated mass-radius profile of the neutron star using Tollmann-Volkoff- Oppenhimer (TOV) equation. The EOS for the TOV equation has taken from the RMF formalism with an extra degree of freedom $\delta $ -meson, which generally never included  in the most of the RMF interaction. We have analyzed the effect of $\delta$-meson on the EOS of both neutron and hyperon stars. Both static and rotating systems have been taken into consideration.

\setcounter{equation}{0}
\setcounter{figure}{0}


\newpage
\chapter{Double shell closure in the super-heavy region}
\label{chapter2}

Magic number is one of the most important tool to study the nuclear structure. Magic numbers in the $\beta$-stability regions are well known but in the drip-line and super heavy area it is still a debatable subject. Various theoretical models proposed different combination of protons and neutrons as magic combinations. In this chapter, an attempt to search for spherical double shell closure nuclei beyond {\it Z}=82, {\it N}=126 is discussed. We have used the non-relativistic (SEI) and relativistic (RMF) calculations for our analysis. This will help us to reduce the theoretical uncertainty in the prediction of  magic number. Here the calculations and results are based on a newly developed approach entitled simple effective interaction (SEI) and well known RMF formalism .  Our results predict the combination of magic nucleus occurs at {\it N}=182 and {\it Z}=114,120,126. All possible evidence for the occurrence of magic nuclei is discussed systematically.

\section{Introduction}
Starting from the discovery of nucleus, the formation of new element is
an interesting topic in Nuclear Physics. So far the synthesis of heaviest
element in laboratory is {\it Z}=118 in the hot fusion reaction process at
JINR Dubna \cite{ogan06,ogan12}. The possibility of the existence (synthesis)
of these super-heavy elements is mainly due to the attractive shell corrections
against the destructive Coulomb repulsion. Although atomic number Z=114 was
predicted to be the next magic number after {\it Z}=82 and neutron number
{\it N}=184, recently attention has shifted to the nucleus {\it Z}=120
with {\it N}=182/184 \cite{patra97,rutz97,sil04}. The experimental
discovery  of the super-heavy elements also support this prediction to some
extent. Thus, the synthesis of {\it Z}=120 is in full swing at the worlds'
most laboratories like, Dubna (Russia), RIKEN (Japan), GSI (Germany).

Using cold fusion reaction, elements from $Z=107-112$ are synthesized
at GSI~\cite{hofman00,hofman81,hofman1995,hofman95,hofman96,hofman98,hofman09}.
At the production time of {\it Z} = 112 nucleus at GSI, the fusion cross
section was extremely small ($1$ pb) \cite{hofman96}, which led to the
conclusion that reaching still heavier elements will be very difficult
by this process. The element {\it Z}=113 was also synthesized in cold-fusion
reaction at RIKEN with a very low cross section $\sim 0.03$ pb \cite{morita07}
confirming the limitation of cold-fusion synthesis. To overcome this problem
in hot fusion evaporation reaction with  deformed actinide targets and
 neutron-rich doubly magic spherical projectile like $^{48}Ca$ are used
in the production of super-heavy nuclei $Z = 112 - 118$ at Dubna
\cite{ogni98,ogni01,ogni04,ogni07,ogni10,ogni11}.

It is thus a matter of challenge for every theoretical prediction in
nuclear  physics to find suitable combination of proton and neutron,
which gives double closure shell nuclei beyond $^{208}$Pb and will be
the next element of epicenter for experimental synthesis. 
Our aim is to look for a suitable combination of proton and neutron
in such a way that the resultant combination will be the next magic nucleus
after $^{208}$Pb. This work is not new, but a revisit
of our earlier prediction with in a new simple effective interaction (SEI).
The SEI interaction is recently developed in Ref. \cite{bhuyan13} and
given a parameter set which is consistent with both nuclear matter and
finite nuclei. Here, we have used this
SEI interaction. A systematic investigation of
the nuclear structure is done and reconfirmed the double closed nucleus
as Z=120 with N=182/184.

This chapter is organized as follows: In Sec.~\ref{seisec2}, the theoretical formalism
of the SEI is presented. The procedures for numerical calculations to estimate
the binding energy and root mean square radii are outlined. The results
and discussions are given in Sec.\ref{seisec3}. The characteristics of magic structure
of nucleus using two neutron separation energy, pairing gap of proton and
neutron are analyzed for super-heavy region. In this section stability of
such nuclei are also studied in terms of the chemical potentials. Finally
 summary and  concluding remarks are given in Sec.~\ref{seisec4}.

\section{The Theoretical Framework}
\label{seisec2}
\subsection{Simple Effective Interaction}

The present formalism is based on a simple way to make a consistent
parametrization for both finite nucleus and infinite nuclear matter
with a momentum dependence finite range term of conventional form,
such as {\it Yukawa}, {\it Gaussian} or {\it exponential} to the
standard Skyrme interaction \cite{trr98,trr02,bhuyan13}. We have used
the technique of Refs. \cite{trr98,trr02,bhuyan13} considering a Gaussian
term as the momentum dependence finite range interaction which simulate
the effect of Gogny type interaction \cite{gogny80,gogny84}. Then it is
applied to nuclear equation of state as well as to finite nuclei through
out the periodic table \cite{bhuyan13}. The Hartree-Fock (HF) formalism is
adopted to calculate the wave-function of the nuclear system which then used
to evaluate the nuclear observable, such as binding energy, root mean square
radii etc. The detail formalism and numerical procedure can be found in
\cite{bhuyan13}. The form of the simple effective interaction (SEI) is given
by \cite{bhuyan13}:
\begin{eqnarray}
\label{3.1}
v_{eff}({\bf r})&=&t_0 (1+x_0P_{\sigma})\delta({\bf r}) 
+t_3(1+x_3 P_{\sigma})\left(\frac{\rho({\bf R})}{1+b\rho({\bf R})}\right)^{\gamma} 
\delta({\bf r})\nonumber\\
&&+ \left(W+BP_{\sigma}-HP_{\tau}
-MP_{\sigma}P_{\tau}\right)f({\bf r}) \nonumber \\
&&+i W_{0}({\bf \sigma}_i+{\bf \sigma}_j)({\bf k'}\times{\delta}({\bf r}_i+{\bf r}_j){\bf k}).
\end{eqnarray}
Where, $f({\bf r})$ is the functional form of the finite
range interaction containing a single range parameter $\alpha$. The
finite range Gaussian form is given as $f({\bf r})=e^{-r^2/\alpha^2}$. The
other terms having their usual meaning \cite{bhuyan13}.
To prevent the supra luminous behavior of the nuclear matter, the
usual value of $b$ \cite{trr05,trr07,trr09,trr11} is taken.
There are $11$-parameters in the interaction, namely
$t_0$, $x_0$, $t_3$, $x_3$, $b$, $W$, $B$, $H$, $M$, $\gamma$ and $\alpha$.
The expression for energy density, single particle energy and other
relevant quantities are obtained from Eq. (~\ref{3.1}) for Gaussian $f(r)$ 
defined in Ref. \cite{bhuyan13}. The numerical values of the parameter
set, SEI and RMF(NL3) are given in Table \ref{tab:table1a}.

\begin{table}
\caption{\label {tab:table1a}{The value of interaction parameters 
for simple effective interaction (SEI) and RMF (NL3) [33] sets and 
their nuclear matter properties at saturation.}}
\renewcommand{\tabcolsep}{0.6cm}
\renewcommand{\arraystretch}{1.0}
\begin{tabular}{|cc|cc|cccccccc}
\hline
\multicolumn{2}{|c|}{SEI} & \multicolumn{2}{c|}{RMF (NL3)} \\
\hline
$\gamma$                    & $\frac{1}{2}$ & ${M} $ (MeV)         & 939 \\
$b$ ($fm^3$)                & 0.5914        & ${m}_{\sigma} $ (MeV)& 508.1941 \\
$t_0$ ($MeVfm^3$)           & 437.0         & ${m}_{\omega} $ (MeV)& 782.6010 \\
$x_0$                       & 0.6           & ${m}_{\rho}   $(MeV) & 7630.0 \\
$t_3$ ($MeVfm^{3(\gamma+1)})$ & 9955.2      & ${g}_{\sigma} $      & 10.2169 \\
$x_3$                       & -0.1180       & ${g}_{\omega} $      & 12.8675 \\
W (MeV)                     & -589.09       & ${g}_{\rho}  $       & 8.9488 \\
B (MeV)                     & 130.36        & ${g}_2$  ($fm^{-1}$) & -10.4307 \\
H (MeV)                     & -272.42       & ${g}_3$              & 28.8851 \\
M (MeV)                     & -192.16       &                      &          \\
$\alpha$ (fm)               &  0.7596       &                      &          \\ 
$W_0$ (MeV)                 & 115.0         &                      &           \\
\hline
\multicolumn{4}{|c|}{Nuclear matter}                                           \\
\hline
$\rho_o$ ($fm^{-3}$)       & 0.157           & $\rho_o$ ($fm^{-3}$)& 0.148 \\
$e(\rho_0)$ (MeV)          & -16.0           & $e(\rho_0)$ (MeV)   & -16.24  \\
$E_{s}$ (MeV)              & 35.0            & $E_{s}$ (MeV)       & 37.4  \\
$K_0$ (MeV)                & 245             & $K_0$ (MeV)         & 271.5  \\
\hline
\hline
\end{tabular}
\end{table}

\subsection{Relativistic mean field (RMF) formalism}

The starting point of the RMF theory is the basic Lagrangian containing
nucleons interacting with $\sigma-$, $\omega-$ and $\rho-$meson fields.
The photon field $A_{\mu}$ is included to take care of the Coulomb
interaction of protons. The relativistic mean field Lagrangian density
is expressed as~\cite{boguta97,walecka},
\begin{eqnarray}
L & =&  \overline{\psi_{i}}\{i\gamma^{\mu}\partial_{\mu}-M\}\psi_{i}+{\frac12}\partial^{\mu}
\sigma\partial_{\mu}\sigma-{\frac12}m_{\sigma}^{2}\sigma^{2}
 -g_{\sigma}\overline{\psi_{i}}\psi_{i}
\sigma-{\frac14}\Omega^{\mu\nu}\Omega_{\mu\nu}
+{\frac12}m_{w}^{2}V^{\mu} V_{\mu}\nonumber \\
&-&g_{w}\overline\psi_{i}\gamma^{\mu} \psi_{i}V_{\mu}-{\frac14}
\vec{B}^{\mu\nu}.\vec{B}_{\mu\nu}
+{\frac12}m_{\rho}^{2}{\vec R^{\mu}} .{\vec{R}_{\mu}}
-g_{\rho}\overline\psi_{i}\gamma^{\mu}\vec{\tau}\psi_{i}.\vec{R^{\mu}}-{\frac12}m_{\delta}^{2}
\delta^{2}+g_{\delta}\overline\psi_{i}\delta\vec{\tau}\psi_{i}.\nonumber\\
\end{eqnarray}
Here M, $m_{\sigma}$, $m_{\omega}$ and $m_{\rho}$ are the masses for nucleon,
${\sigma}$-, ${\omega}$- and ${\rho}$-mesons and ${\psi}$ is the Dirac spinor.
The field for the ${\sigma}$-meson is denoted by ${\sigma}$, ${\omega}$-meson
by $V_{\mu}$ and ${\rho}$-meson by $R_{\mu}$.
The parameters $g_s$, $g_{\omega}$, $g_{\rho}$ and $e^2/4{\pi}$=1/137 are the coupling
constants for the ${\sigma}$, ${\omega}$, ${\rho}$-mesons and photon
respectively. $g_2$ and $g_3$ are the self-interaction coupling constants
for ${\sigma}$ mesons. By using the classical variational principle we
obtain the field equations for the nucleons and mesons.
A static solution is obtained from the equations of motion to describe
the ground state properties of nuclei.
The set of nonlinear coupled equations are solved self-consistently \cite{estal01}.
The total energy of the system is given by
\begin{equation}
 E_{total} = E_{part}+E_{\sigma}+E_{\omega}+E_{\rho}+E_{c}+E_{pair}+E_{c.m.},
\end{equation}
where $E_{part}$ is the sum of the single particle energies of the nucleons and
$E_{\sigma}$, $E_{\omega}$, $E_{\rho}$, $E_{c}$, $E_{pair}$, $E_{cm}$ are
the contributions of the meson fields, the Coulomb field, pairing energy
and the center-of-mass motion correction energy, respectively.
We have used the well known NL3 parameter set~\cite{lala97} in our
calculations for RMF formalism.

\subsection{Pairing Correlation}

To take care of the pairing correlation for open shell nuclei the
constant gap, BCS-approach is used in our calculations.
The pairing energy expression is written as
\begin{equation}
E_{pair}=-G\left[\sum_{i>0}u_{i}v_{i}\right]^2,
\end{equation}
with $G$ is the  pairing force constant. The quantities $v_i^2$ and $u_i^2=1-v_i^2$ are the occupation
probabilities \cite{patra93,sero861,pres82}. The variational approach with
respect to $v_i^2$ gives the BCS equation
\begin{equation}
2\epsilon_iu_iv_i-\triangle(u_i^2-v_i^2)=0,
\end{equation}
using $\triangle=G\sum_{i>0}u_{i}v_{i}$.
The occupation number is defined as
\begin{equation}
n_i=v_i^2=\frac{1}{2}\left[1-\frac{\epsilon_i-\lambda}{\sqrt{(\epsilon_i-\lambda)^2+\triangle^2}}\right].
\end{equation}

The chemical potentials $\lambda_n$ and $\lambda_p$ are determined
by the particle number for protons and neutrons. The pairing energy
is computed as $E_{pair}=-\triangle\sum_{i>0}u_{i}v_{i}$. For a particular
value of $\triangle$ and $G$, the pairing energy $E_{pair}$ diverges if it
is extended to an infinite configuration space. In fact, in all realistic
calculations with finite range forces, $\triangle$ decreases with state for
large momenta near the Fermi surface. In the present case, we assume equal
pairing gap for all states $\mid\alpha>=\mid nljm>$ near the Fermi surface.
We use a pairing window, where the equations are extended up to the level
$\epsilon_i-\lambda\leq 2(41A^{1/3})$. The factor 2 has been determined
so as to reproduce the pairing correlation energy for neutrons in $^{118}$Sn
using Gogny force \cite{gogny80,patra93,sero861}.

\section{Results and Discussions}
\label{seisec3}
The quasi local Density Functional Theory (DFT) is used in this work, which is similar to
the one used by Hoffman and Lenske in Ref. \cite{hofman98}.
The total energy is nothing but the sum of the
energy density contribution from different components of the
interaction along with spin-orbit and Coulomb term.
The energy density ${\cal H}_0$ for SEI set can be expressed as
\begin{eqnarray}
{\mathcal H}_0&=&\frac{\hbar^2}{2m}\left(\tau_n+\tau_p\right)+
{\mathcal H}_{d}^{Nucl}+{\mathcal H}_{exch}^{Nucl}+ {\mathcal H}^{SO} 
+{\mathcal H}^{Coul}.
\end{eqnarray}
From this effective Hamiltonian $\tilde{H}$ we obtain the quasi local energy functional as:
\begin{eqnarray}
\varepsilon_0\left[\rho^{QL}\right]=\int{\mathcal H}_0d^3R.
\label{eq12}
\end{eqnarray}
The equations solved self-consistently to get the solution for
nucleonic system. Here we have taken only spherical solution for both RMF and SEI.

\begin{table}
\caption{\label{tab:table1}{The binding energy (BE) obtained from SEI
calculation is compared with the RMF(NL3) [33], finite range droplet 
model (FRDM) [39] and with experimental data of some of the known 
super-heavy nuclei. The BE is in MeV.}}
\renewcommand{\tabcolsep}{0.6cm}
\renewcommand{\arraystretch}{1.0}
\begin{tabular}{|c|c|c|c|c|c|c|c|c|}
\hline
Element &\multicolumn{4}{c|}{BE}  \\
\hline
&SEI& RMF(NL3) &FRDM & Expt. \\
\hline
$^{258}{Md}$&1896.19&1897.70&1911.53&1911.69 \\
$^{258}{Rf}$&1884.95&1890.86&1905.25 &1904.69 \\
$^{261}{Rf}$&1906.38&1911.04&1924.28& 1923.93 \\
$^{259}{Db}$& 1886.94&1894.58 & 1907.00&1906.33  \\
$^{260}{Db}$& 1894.31& 1901.4 & 1913.34 & 1912.82\# \\
$^{260}{Sg}$& 1888.62& 1897.9 & 1909.90&1909.07  \\
$^{261}{Sg}$&1896.17&1905.02&1916.27& 1915.68 \\
$^{264}{Hs}$& 1906.86&1915.5 & 1927.62&1926.77 \\
$^{265}{Hs}$&1914.59&1922.9 &1934.40 &1933.50  \\
$^{269}{Ds}$&1932.81&1941.21&1952.06&1950.290\\
$^{285}{Fl}$&2029.41&2039.19&2044.12& 2040.03\# \\
$^{286}{Fl}$&2036.74&2046.17&2051.59& 2047.474\# \\
$^{287}{Fl}$&2043.36&2052.50&2057.65& 2053.19\# \\
$^{288}{Fl}$&2050.14&2058.73&2065.01& 2060.64\# \\
$^{289}{Fl}$&2056.80&2064.87&2071.04& 2066.06\#  \\
\hline
\hline
\end{tabular}
\end{table}

\subsection{Ground state binding energy}

The main objective of the present study is to find the double
shell closure in the superheavy valley. In this context, we
have concentrated on few observable such as separation energy $S_{2n}$,
chemical potential $\mu_n$, single-particle levels ${\cal E}_{n,p}$
and pairing energy $E_{pair}$. Before going to this unknown region
(super-heavy valley), it is important to test our model for known
 nuclei, which are experimentally and theoretically well
established. We calculate the binding energy of few known super-heavy
nuclei using SEI. The obtained results are compared with RMF,
finite-Range-Droplet-Model (FRDM) \cite{moll97} and experimental
data \cite{audi12} in Table~\ref{tab:table1}. The \# marks in the experimental
column are for the extrapolated data from Ref\cite{audi12}. From the table, we find that the SEI and NL3 results are slightly
overestimated to the experimental values. A close observation of
the table shows the superiority of FRDM over SEI or NL3 for
lighter masses of the super-heavy nuclei. In contrast to the lighter
region, the SEI predicts better results
for heavier isotopes. For example,  binding energy of $^{289}{Fl}$
is 2056.80 $MeV$ in SEI, whereas the values are 2064.87, 2071.04
and 2066.06 $MeV$ in RMF(NL3), FRDM and experiment (or systematic),
respectively. Based on this trend, one can expect that the prediction
of SEI gives us better insight about the magic structures
of super-heavy nuclei in heavier mass region, which is the main objective
of this present investigation.


\begin{figure}
\includegraphics[scale=0.40]{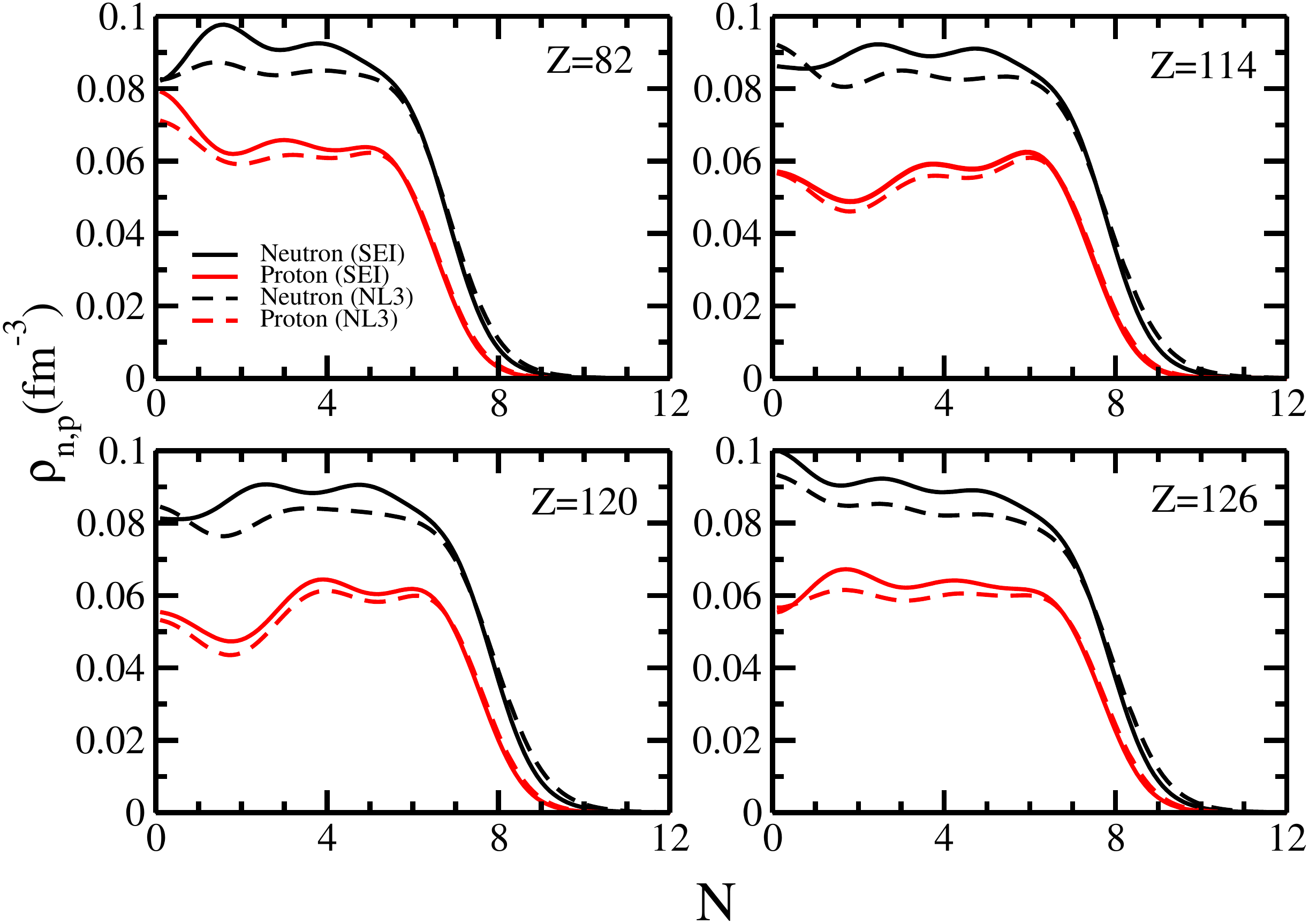}
\caption{\label{sefi1}The ground state densities with SEI
for $^{208}$Pb, $^{298}$114, $^{304}$120 and $^{310}$126 are compared
with the RMF(NL3) results.}
\label{seifi1}
\end{figure}

\subsection{Density distribution of Neutrons and Protons}

After convinced with the binding energy of the super-heavy nuclei,
we present the density distribution of protons and neutrons in
Fig.~\ref{seifi1}. The densities are compared with the RMF(NL3) calculations.
In general, the RMF and SEI densities are  almost similar with
each other. However, a proper inspection reveals that the SEI
densities slightly over estimate the RMF(NL3) densities. This
overestimation is mostly at the middle region of the nucleus.
The humps at the central region for both the densities show shell
effect for all nuclei shown in the figure.

\subsection{Two neutron separation energy and location of closed shell}
\begin{figure}
\includegraphics[scale=0.40]{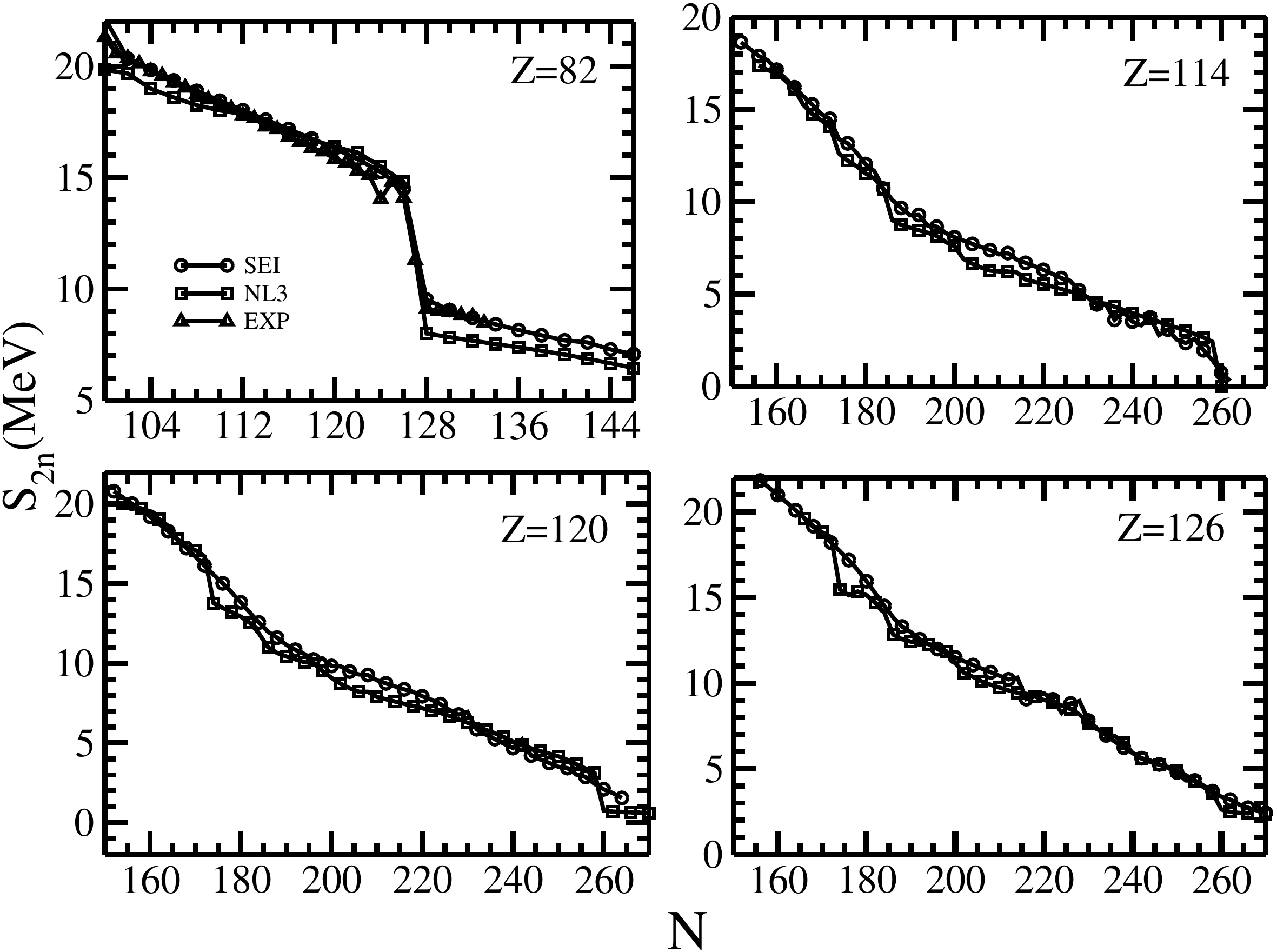}
\caption{\label{seifi2}The two neutron separation energy obtained from NL3 and
SEI for $^{208}$Pb, $^{298}$114, $^{304}$120 and $^{310}$126.}
\end{figure}
From the binding energy, we have calculated the two neutron
separation energy using the relation $S_{2n}(N,Z)=BE(N,Z)-BE(N-2,Z)$.
The $S_{2n}$ for all the four isotopic chains are shown in Fig.~\ref{seifi2}
as a function of neutron numbers. In case of Pb isotopes, the sudden
decrease of $S_{2n}$  at neutron number {\it N}=126,
is the well known neutron magic number for the largest known {\it Z}=82
magic nucleus. The analysis is extended to the recently predicted
proton magic  numbers like Z=114, 120 and 126, which are currently
under scrutiny for their confirmation.

It is important to mention that, the next proton magic
number beyond {\it Z}=82 would be {\it Z}=126 considering the traditional proton and neutron magic numbers for known closed shell nuclei
\cite{goldhaber57,myers66}.  However, several microscopic calculations
\cite{nilsson78,nilsson59,andersson76,meldner67,meldner78a,kohler71}
suggest a shift of this number to 114. One of the cause of the shift is the
Coulomb effect on the spherical single particle levels. The use of shell
correction by V. M. Strutinsky \cite{strutinsky67} to the liquid-drop
calculation of binding energy (BE) opens a more satisfactory
exploration towards the search of double closed nucleus beyond $^{208}$Pb.
Using this approach, {\it Z}=114 is supported to be the proton magic after 82
\cite{nix94,sob94,sob07,smo95}, which was regarded as the magic
number in the super-heavy valley \cite{kumar89} with {\it N}=184
as the corresponding neutron magic number. However, the recent relativistic
mean field calculations using various force parameters \cite{bhuyan12},
predict {\it Z}=120 as the next magic number with {\it N}=172/182 as the
neutron closed shell. Contrary to all these predictions, some non-relativistic
calculations reported {\it Z}=126 as the next magic proton in the super-heavy
valley. The microscopic calculations using Skyrme Hartree-Fock
formalism predict {\it N}=182 as the next neutron closed shell after
N=126, which differs by 2 unit from other predictions\cite{bhuyan12}.


\begin{figure}
\includegraphics[scale=0.40]{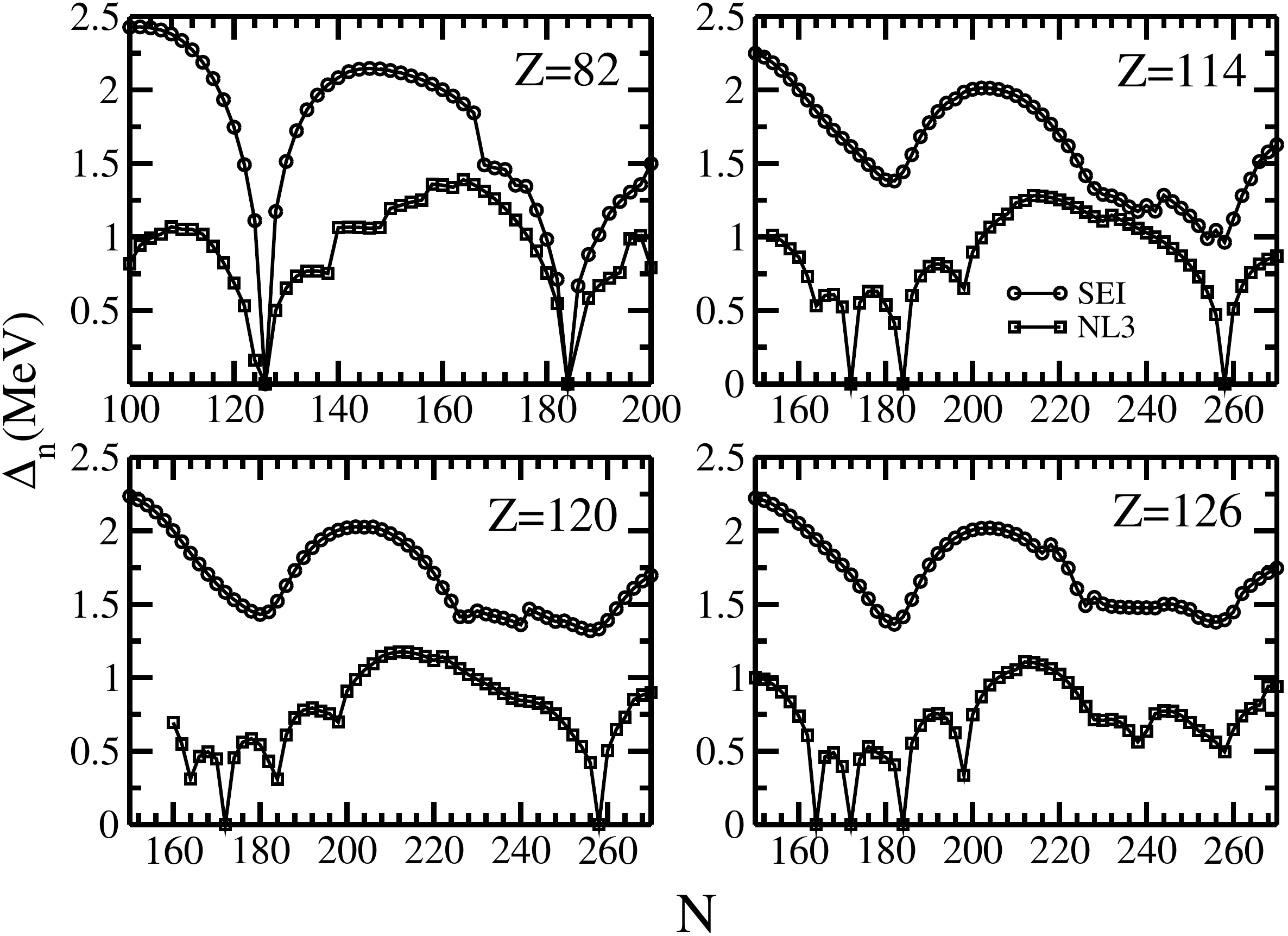}
\caption{\label{seifi3}The calculated pairing gap of neutron $\triangle_n$
with SEI for the isotopic series Z=82, 114, 120 and 126 are
compared with the NL3 results.}
\end{figure}

Analyzing  the $S_{2n}$ energy for the isotopic chain
of {\it Z}=82, 114, 120 and 126, the sharp fall of $S_{2n}$ at
{\it N}=126 is a clear evidence of magic combination of {\it Z}=82 and
 {\it N}=126. Our newly developed SEI model and previously
existing NL3 follow the  same trend as experiment.
But whenever we analyzed the plots of {\it Z}=114, 120 and 126,
we find a slight difference in these two models (SEI and RMF).
In RMF(NL3), when we go
from one magic neutron number to the next one, the $S_{2n}$ energy suddenly
decreases to a lower value, which reflect in Fig.~\ref{seifi2}. In SEI, the
$S_{2n}$ energy follows same pattern but the magnitude of decreseness
some how less.


\begin{figure}
\includegraphics[scale=0.40]{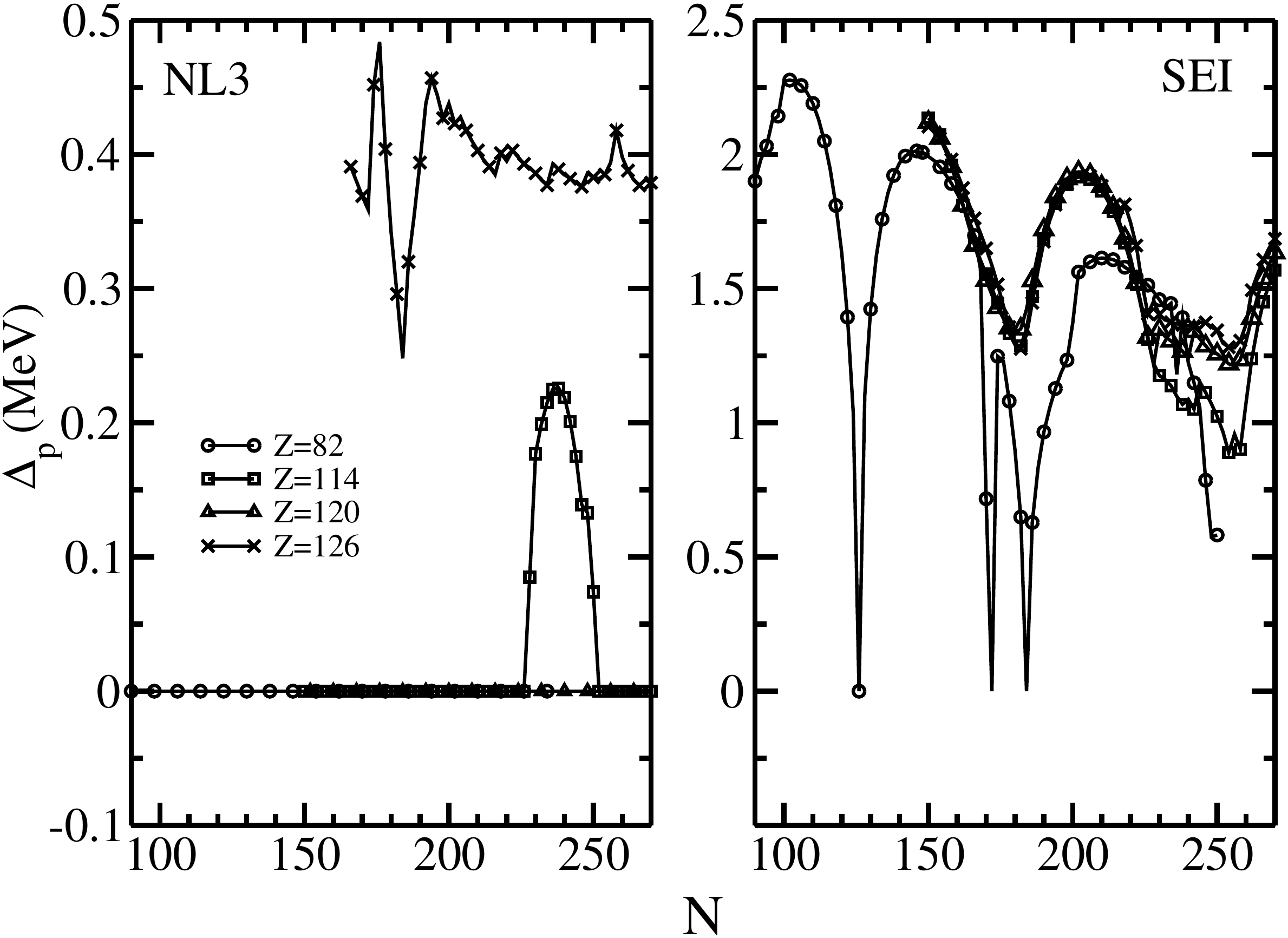}
\caption{\label {seifi4}(a)The pairing gap of proton  $\triangle_p$ with SEI
for the isotopic series {\it Z}=82, 114, 120 and 126 compared with
the NL3 results.}
\label{fig:4}
\end{figure}

\subsection{Pairing gaps and pairing energy}

Another important quantity, which helps us to locate the closed shell is
the pairing gaps of proton and neutron in a constant force BCS
calculation. Here, we calculate the pairing gap for the isotopic chain
of {\it Z}=82, 114, 120 and 126 and locate the minimum values of $\triangle_n$
and $\triangle_p$. The results are depicted in Figs.~\ref{seifi3} and ~\ref{seifi4} and also
compared with the RMF(NL3) force. It is well known that NL3 force
satisfies this criteria for the location of magicity \cite{sil04,bhuyan12}.
Although, SEI overestimates the paring gaps of  $\triangle_n$,
$\triangle_p$, the trend for both  NL3 and SEI are found to be similar.
Consistence with NL3 results as well as with earlier calculations with
a variety of force parameters, our present SEI reproduces minima at
{\it N}=182/184 and {\it Z}=120 and to some extent at {\it Z}=114.


\begin{figure}
\vspace{1.0cm}
\includegraphics[scale=0.40]{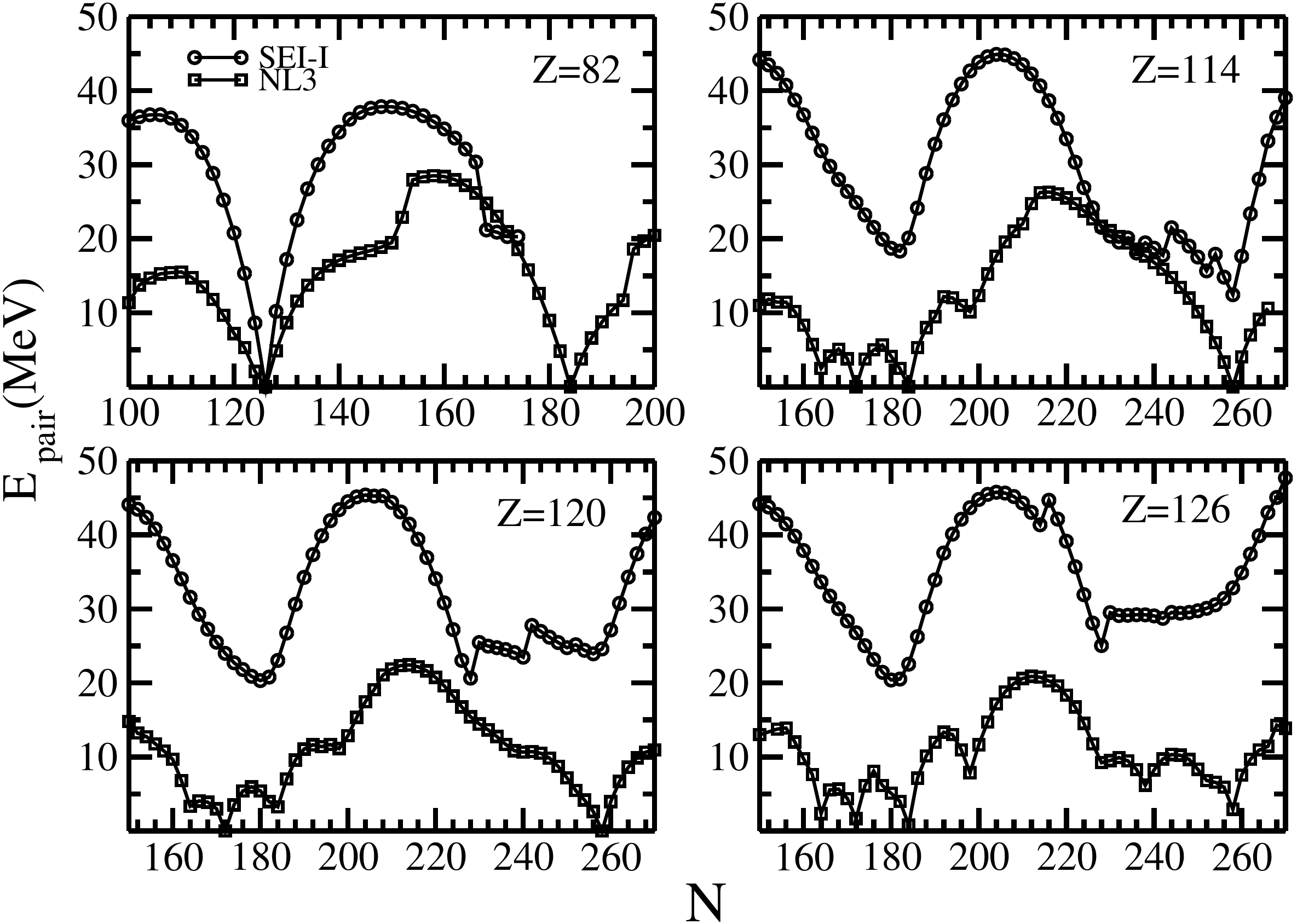}
\caption{\label{seifi5}The pairing energy as a function of neutron number
for {\it Z}=82, 114, 120 and 126 with SEI and NL3 forces.}
\end{figure}

To see the trend of pairing energy $E_{pair}$ at the discussed neutron
number {\it N}=184, we plot $E_{pair}$ as a function of neutron number
N in Fig.~\ref{seifi5}. Surprisingly, we get almost zero pairing energy at N=126
for {\it Z}=82 isotopic case. The formalism is extended to {\it Z}=114,
120 and 126 cases. We find minimum or zero $E_{pair}$ at {\it N}=182/184
confirming the earlier predictions of this neutron magic number at {\it N}
=182/184 \cite{bhuyan12}.Qualitatively, the SEI interaction follows the
trend of RMF(NL3) as shown in Fig.~\ref{seifi5}, but fails when we have a quantitative
estimation. For example, the $\triangle_n$ or $E_{pair}$ at {\it N}=182/184
is minimum but has a finite value unlike to the NL3 prediction, which has 
zero value. As a matter
of fact, the validity of pairing scheme to this region of nuclei may not be
100 {\% } applicable. The improvement of pairing is needed to keep the value
of $\triangle _n$ and $\triangle_p$ zero at the appropriate magic number.


\begin{figure}
\includegraphics[scale=0.40]{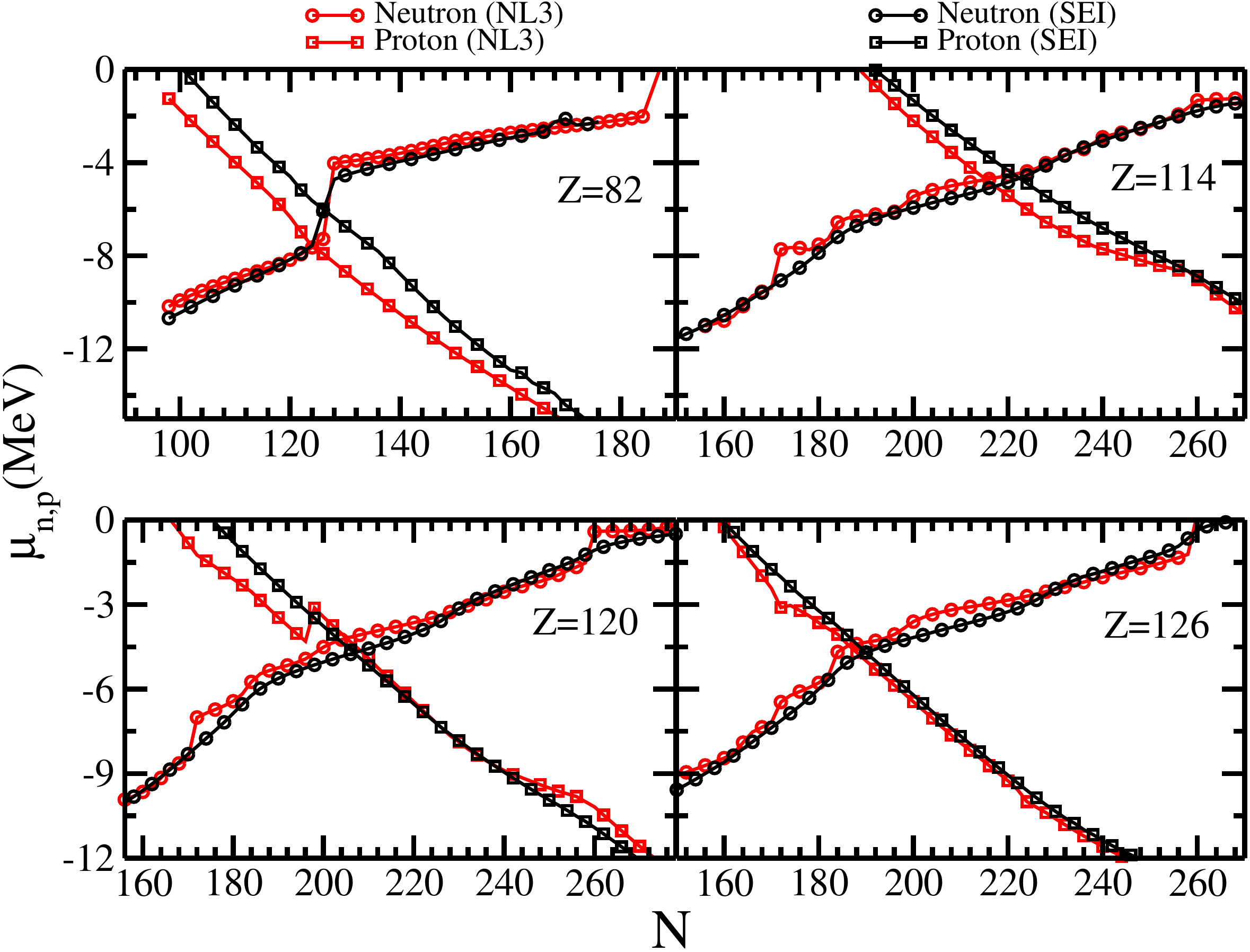}
\caption{\label{seifi6}Systematic of chemical energy $\mu_n$ and $\mu_p$
as a function of neutron number for {\it Z}=82, 114, 120 and
126 with SEI and NL3 sets.}
\end{figure}

\subsection{Chemical energy and stability}
The stability of an element not only depends on its binding energy and
shell structure, but also very much affected by the chemical potential $\mu$.
For a bound nucleus, both the chemical potentials of protons $\mu_p$
and neutrons $\mu_n$  must be negative. To realize the relative stability
from chemical point of view, we have plotted $\mu_p$ and $\mu_n$
with neutron number in Fig.~\ref{seifi6}. The results are also compared with the
$\mu-$value of NL3 set. In both the cases, we find similar chemical
potential. In some previous papers it is suggested that we can take
{\it N}=172 as magic number for neutron. But our SEI model shows that the
combination {\it Z}=120 and {\it N}=172 is strictly not allowed. Because
in this case $\mu_p=0.69$ MeV, which gives proton instability. However
NL3 result shows this combination is a loosely bound system having
$\mu_n=-1.240$ MeV and $\mu_p=-7.007$ MeV. Although the BE/A curve shows
a local maximum at {\it Z}=114 and N=172 in SEI model, we can not take
this as a stable system because of $\triangle_n$ and $\triangle_p$ value, which
does not show any signature of stability. The SEI model gives a clear picture
that the isotope $^{302}120$ can be a suitable combination for the next double
closed nucleus. One can justify it by analysis of BE/A data of $^{302}120$.
For example $BE/A=7.007$ MeV which create a local maxima in its neighbor-hood for
$^{302}120$. In the same time, the optimum negative value of chemical potential
energies of $\mu_n$and $\mu_p$ gives a sign of maximum stability. A similar
analysis of numerical data for $\mu_p$ of isotopes of {\it Z}=126 shows that
there is no reason of taking {\it Z}=126 and {\it N}=182/184 as a stable
combination. This is because of the positive value of $\mu_p$ (1.36) MeV.


\begin{figure}
\includegraphics[scale=0.40]{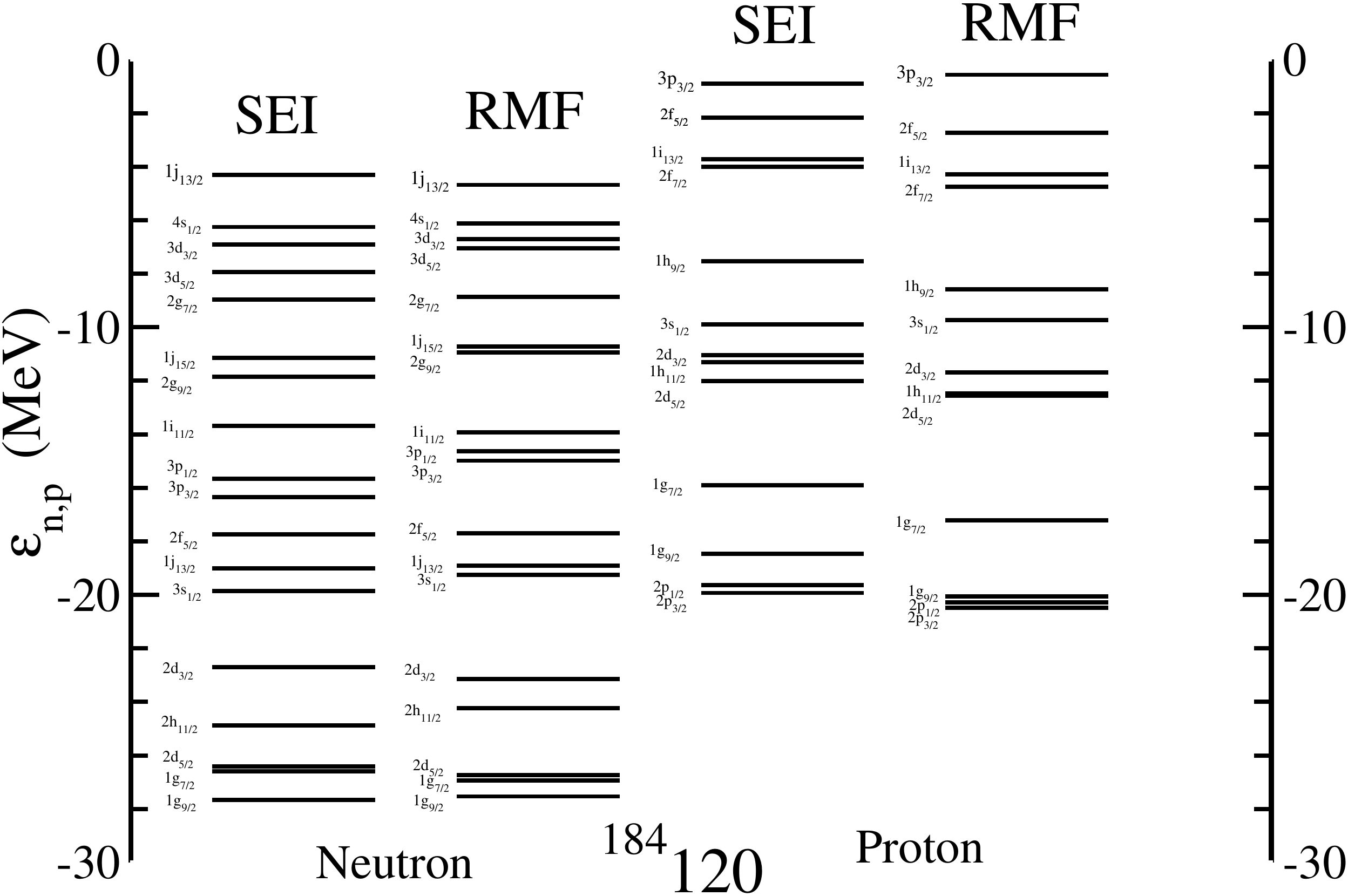}
\caption{ \label{seifi7}The single particle energy levels for $^{304}$120 with
NL3 and SEI parametrization.}
\end{figure}

\subsection{Single particle energy}
The single particle energies for $^{304}$120 with NL3 and SEI
for proton and neutron are shown in Fig.~\ref{seifi7}. The single particle
solutions are obtained without including the pairing correlation
into account to intact the degeneracy of the levels. The calculation
of single particle energies of SEI with pairing shows that the degeneracy
of the energy levels are not invariant. The basic cause of this discrepancy
is the over estimation of our pairing strength in SEI model which may
be an interesting analysis for pairing in future. The filling up single
particle energy levels for neutrons in SEI with pairing is different from
that of without pairing. The energy levels without pairing are given by
[178] $({{3d}_{3/2}})^4$, $({{4s}_{1/2}})^2$
while the same with pairing are [178] $({{3d}_{3/2}})^3$, $({{4s}_{1/2}})^1$,
$({{1j}_{11/2}})^2$. That means an empty orbital is created at 4$s_{1/2}$ and
occupied in 1$j_{11/2}$. We have also analyzed the single particle levels for
$^{302}$120, which is not given in the figure. From the anatomy of single particle energies for neutron and proton $\epsilon_n$
and $\epsilon_p$, we find large gaps at neutron number {\it N}=184 and proton
number {\it Z}=120.  The value of neutron gap at {\it N}=184 is 1.949 $MeV$
and that of proton is 1.275 $MeV$ for the last occupied and first unoccupied
nucleon. On the other hand the neutron and proton gap for $^{302}$120 are
respectively  $\sim 0.6$ and $\sim 1.663$$ MeV$. The above data say the
energy gaps for the neutron and proton in $^{304}$120 are greater than
the  gap in $^{302}$120. This give us an indication to take the combination
{\it N}=184 and {\it  Z}=120 as the next magic nucleus. From the  analysis
of single particle energy level of $^{304}$120 with NL3 parameter set one
can see the neutron and proton gaps are $1.4503$ MeV and $2.1781$ MeV
respectively for the last occupied and first unoccupied nucleon.
 The RMF(NL3) and SEI data are comparable with each other.

\section{Summary and Conclusions}
\label{seisec4}
In summery, we have calculated the binding energy, $S_{2n}$ energy, single
particle levels, pairing gaps and chemical potential, in the isotopic chain of {\it Z}=82,
114, 120 and 126. All our calculations are done in the frame work of SEI
interaction. We have compared our results with standard RMF (NL3) parameter.
Over all discussion and analysis of all possible  evidences of shell closure
 property show that, one can take {\it Z}=120 and {\it N}=182 as the next magic combination beyond $Z=82$ and $N=126$, which is different from Skyrme, Gogny and RMF(NL3) by two unit. However on the basis of single particle energy levels, the preferred gap is at
{\it N}=184 which is consistent with these (Skyrme, Gogny and RMF) force
parameters. This happens due to the overestimation of pairing strength.


\setcounter{equation}{0}
\setcounter{figure}{0}
\newpage
\chapter{Monopole resonance in  drip-line
and super heavy  region.}
\label{chapter3}

In last chapter we discussed the magic properties of the super heavy and drip-line nuclei, which is prominently  based on single particle nature of the nuclear system.  Like the single particle state,   the collective  states are also equally responsible for the structure of the finite nucleus.
Many information about the nuclear structure can be obtained from the
collective excitation like the giant monopole, dipole, and quadrupole oscillations. Out of all collective excitation, isoscalar monopole and isovector dipole are most important for the nuclear structure physics. These are also known as the squeezing mode of oscillation.  In isoscalar monopole resonance (ISGMR) the protons and neutrons vibrate in a phase to each other.  It is also known as the breathing mode of oscillation. This collective mode is related to the incompressibility of finite  nuclear system and using leptodermous expansion one can get the information about the infinite nuclear matter incompressibility ($K_\infty$) from the finite one. $K_\infty$  is one of the fundamental properties of the equation of state, which decides the mass and radius of the neutron star. The general assumption is that a heavy finite nuclear system has a great resembles with the infinite nuclear matter. So the study of the incompressibility and excitation energy of heavy and drip line nuclei become important.   We study the isoscalar giant monopole resonance for drip-lines and super heavy nuclei in the framework of relativistic mean field theory with a scaling approach. The well-known extended Thomas-Fermi approximation in the nonlinear $\sigma$- the $\omega$ model is used to estimate the giant monopole excitation energy for some selected light spherical nuclei starting from the region of a proton to neutron drip-lines. The application is extended to the super  heavy region for Z=114 and 120, which are predicted by several models as  the next proton magic numbers beyond Z=82.  We compared the excitation energy obtained by four successful force parameters NL1, NL3, NL3$^*$, and FSUGold. 


\section{Introduction}

The study of nuclei far away from the drip-lines has a current research
interest due to their very different properties from nuclei at the
$\beta-$stability valley. New properties of these nuclei like the soft giant
resonance, change of magic number, halo and skin structures, and new decay modes
 stimulate strongly research using radioactive ion beams (RIB)
\cite{tan85,han89}. On the other hand, super heavy nuclei which are on the 
stability line, but extremely unstable due to excessive Coulomb repulsion, 
attract much theoretical attention for their resemblance to the highly 
asymmetric nuclear matter limit \cite{og,kumar89}. These nuclei possess 
a large amount of collective excitation and their study
along an isotopic chain is more informative for the structural evaluation 
of astrophysical objects like neutron stars \cite{aru04a}.  Also, the nuclear 
symmetry energy, and consequently the proton to neutron ratio, are
crucial factors in constructing the equation of state (EOS) for asymmetric 
nuclear matter \cite{manisa10}.

The incompressibility $K_A$ of a nuclear system depends on its neutron-proton 
asymmetry. It is also well known that the EOS of a highly asymmetric and dense 
object like a neutron star is substantially influenced by its incompressibility. 
Although the incompressibility at various asymmetries is an important quantity, 
it is not a direct experimental observable. Thus, one has to determine
the $K_A$ from a linked experimental quantity (which is directly or 
indirectly related to $K_A$) like isoscalar giant monopole resonance (ISGMR)
\cite{young99,bohigas79}. The ISGMR is a well-defined experimental 
observable, which can be measured precisely through various experimental 
techniques.  The drip-lines and super heavy nuclei are vulnerable
and unstable in nature, because of the presence of excess neutrons and 
large number of protons, respectively. Thus, it is instructive to know the 
giant monopole resonance, incompressibility modulus, and other related 
quantities for both drip-lines and super heavy nuclei. In this context, 
our aim is to study the giant monopole excitation energy and the 
incompressibility of finite nuclei near the drip-line \cite{han89} as 
well as for recently discussed super heavy nuclei with proton numbers Z=114 
and 120, which are predicted to be the next magic numbers beyond Z=82 with 
various models \cite{sil04,rutz95}. In addition, the calculations of 
Refs. \cite{sobi,patra} suggest that these nuclei possess
spherical ground state or low-lying spherical excited solutions.
More specifically, we aimed to study the following within the frame-work 
of an extended relativistic Thomas-Fermi approximation:

\begin{itemize}
\item
How the isoscalar excitation energy and the finite nuclear incompressibility vary in an isotopic
chain in drip-lines and super heavy nuclei within a well-tested model like the relativistic
extended Thomas-Fermi approximation using scaling and constrained approaches which were developed
by some of us recently \cite{patra01,patra02}.
\end{itemize}
\begin{itemize}
\item
A comparative study of ISGMR obtained with various parameter sets, originated
from several interactions, such as
NL1, NL3, NL3$^*$, and FSUGold for the same drip-lines and super heavy nuclei.
The large variation in nuclear matter incompressibility starting from 
$K_{\infty} \sim 211.7- 271.76$ MeV will give us an idea about the prediction
of ISGMR with different values of $K_{\infty}$. 
\end{itemize}
\begin{itemize}
\item
The resonance widths $\Sigma$, which are mostly the difference between the 
scaling and constraint excitation energies are analyzed in the isotopic 
chains of light and super heavy nuclei.

\end{itemize}
\begin{itemize}
\item
The relation between the finite nuclear incompressibility with the infinite nuclear matter values in various force parameter sets are looked for.
\end{itemize}
\begin{itemize}
\item 
Finally, We applied the scaling and constrained method for Cd and Sn isotopes 
 and compared with the excitation energy with moments ratio $\sqrt{\frac{m_3}{m_1}}$ and $\sqrt{\frac{m_1}{m_{-1}}}$ obtained  from multipole-decomposition  analysis (MCA). 

\end{itemize}

In relativistic mean field (RMF) formalism, the NL1 parameter set
\cite{rufa86} has been considered for a long time to be one of the best interactions
to predict the experimental observable. The excessively large 
value of the asymmetry coefficient $J\sim 43.6$ MeV brings into question the
accuracy of the prediction of neutron radius near the drip-line. As a 
result, the discovery of the NL3 parameter set \cite{lala97} complements the 
limitations of the NL1 force and evaluates the ground state properties of finite 
nuclei in excellent agreement with experiment \cite{boguta87,lala97,
patra91,gamb90,sumi93,toki94}. It reproduces the proton  or charge radius 
$r_{ch}$ precisely along with the ground state binding energy. Unfortunately, 
the experimental data for neutron radius has a large error bar \cite{batty}, 
which covers most of the prediction of all relativistic and non-relativistic 
models \cite{brow00}. The FSUGold parameter set \cite{fsu,fattoyev10} 
reproduces the ISGMR pretty well with the experimental data for $^{90}$Zr 
and $^{208}$Pb. There is also a possibility to solve the problem of the uncertainty in 
neutron radius  \cite{roca11a} using this interaction. The NL3$^*$ 
force parametrization \cite{lala09} is claimed to be an improved version of NL3 
to reproduce the experimental observable. We used all these forces and 
made a comparison of their predictive power for various experimental data. 
Then we selected NL3 as a suitable parameter set for our further investigations for ISGMR and related quantities. This chapter is organized as follows: In 
section \ref{cejsec2}, we outline in brief the formalism used in the present work. In section \ref{cejsec3}, we discuss our results for the ground 
state properties and isoscalar giant monopole resonance (ISGMR) for drip-lines and 
super heavy nuclei. The isoscalar monopole excitation energy $E_x$ 
and the incompressibility modulus of finite nuclei $K_A$ are also 
analyzed. We give the summary and concluding remarks in section \ref{cejsec4}

\section{The Formalism}
\label{cejsec2}

We use the principle of scale invariant
to obtain the virial theorem for the relativistic mean field  
theory in the relativistic Thomas--Fermi (RTF) and relativistic 
extended Thomas-Fermi (RETF) approximations \cite{patra01,mario93,mario92,
spei98,mario98,mario93a}. Although the scaling and constrained calculations 
are not new, the present technique was developed first by Patra et al 
\cite{patra01} and not much has been explored for various regions of 
the periodic chart. Thus, it is interesting to apply the model specially for 
drip-lines and super heavy nuclei. The calculations will explore the 
region ranging from Z=8 to Z=114, 120, where we can simulate the
properties of neutron matter from the neutron-rich finite nuclei.
For this purpose, we compute moments and average excitation energies of the 
isoscalar giant monopole resonance (ISGMR) through scaling and constrained 
self-consistent calculations for the ground state.

The detailed formalism of the scaling method is given in Refs. \cite{patra01,patra02}.
For completeness, we have outlined briefly some of the essential expressions
which are needed for the present purpose. We have worked  with the non-linear
Lagrangian of Boguta and Bodmer \cite{boguta77} to include the many-body correlation which arises from the non-linear terms of the $\sigma-$meson self-interaction 
\cite{schiff51} for a nuclear many-body system. The nuclear matter incompressibility $K_{\infty}$ is also reduced dramatically by the introduction of 
these terms, which motivate us to work with this non-linear Lagrangian. We have
also included the self-coupling of the vector $\omega-$meson ($aV^4$) and
the cross-coupling of the $\omega-$ and $\rho-$mesons $\Lambda{R^2}V^2$ in
the Lagrangian. The terms $aV^4$  and $\Lambda{R^2}V^2$ are very important
in the equations of state \cite{toki94} and symmetry energy \cite{fsu} 
for nuclear systems. 
The relativistic mean field Hamiltonian for a nucleon-meson interacting system
is written as \cite{patra01}:
\begin{eqnarray}
{\cal H}&= &\sum_i \varphi_i^{\dagger}
\bigg[ - i \vec{\alpha} \cdot \vec{\nabla} +
\beta m^* + g_{v} V + \frac{1}{2} g_{\rho} R \tau_3 
+\frac{1}{2} e {\cal A} (1+\tau_3) \bigg] \varphi_i 
+ \frac{1}{2} \left[ (\vec{\nabla}\phi)^2 + m_{s}^2 \phi^2 \right]
\nonumber\\
&+&\frac{1}{3} b \phi^3
+ \frac{1}{4} c \phi^4
-\frac{1}{2} \left[ (\vec{\nabla} V)^2 + m_{v}^2 V^2 \right]
- \frac{1}{2} \left[ (\vec{\nabla} R)^2 + m_\rho^2 R^2 \right]
- \frac{1}{2} \left(\vec{\nabla}  {\cal A}\right)^2+a{{V}}^4
\nonumber\\
&+&\Lambda{R^2}{V^2}. 
\end{eqnarray}
%
All the terms in the above Hamiltonian represent their usual meaning, 
which have already discussed in the chapter 2. Here we have added two new 
coupling constants, one is  self-coupling of the vector meson $\omega$  and 
other is the cross-coupling of the $\omega$ and $\rho-$meson, 
which  are represented by $a=\frac{\zeta_0}{24}{g_v}^4$, 
and $\Lambda=\Lambda_V{g_\rho}^2{g_v}^2$, respectively. By using the classical 
variational principle we obtain the 
field equations for the nucleons and mesons. In semi-classical approximation all the terms containg single particle wave-function converted to their 
corresponding density form and the  above Hamiltonian can written in term of density as:
\begin{eqnarray}
{\cal H}&=&{\cal E}+g_v V {\rho}+g_{\rho}R{\rho}_3+e{\cal A}{\rho}_p+{\cal H}_f,
\end{eqnarray}
where
\begin{eqnarray}
{\cal E}&= &\sum_i \varphi_i^{\dagger}
\bigg[ - i \vec{\alpha} \cdot \vec{\nabla} +
\beta m^*\bigg]\varphi_i={{\cal{E}}_0}+{{\cal{E}}_2},
\end{eqnarray}
\begin{eqnarray}
{\rho}_s&=&\sum_i \varphi_i^{\dagger}{\varphi}={\rho_{s0}}+{\rho_{s2}},
\end{eqnarray}
\begin{eqnarray}
{\rho}&=&\sum_i {\bar \varphi_i}{\varphi},
\end{eqnarray}
\begin{eqnarray}
{\rho_3}&=& \frac {1}{2}\sum_i{\varphi}_i^{\dagger}{\tau_3}{\varphi}_i,
\end{eqnarray}
and ${\cal H}_f$ is the free part of the Hamiltonian, contains the free 
meson contribution. ${\cal{E}}_2$ and ${\cal{\rho}}_{s2}$ correspond to the ${\hbar}^2$ correction to the energy and scalar density, respectively. 
These terms are considered as the extension of Thomas-Fermi approximation 
and known as the extended Thomas-Fermi (ETF) approach. In Thomas-Fermi 
approach, the density is considered as locally constant ( only depends on 
the  position co-ordinate) but in extended Thoms-Fermi approach, the 
density contains the terms, which are function of position as well as the 
derivative at that point. This ETF is considered as one step forward 
to TF approach to explain the real nuclear system, where the variation 
of density takes place on the surface of a finite nucleus.  
The complete expression for these quantities is found in \cite{patra01,patra02}. The total density $\rho$ is the sum of the proton $\rho_p$ and neutron
$\rho_n$ densities.
The semi-classical ground-state meson fields are obtained
by solving the Euler-Lagrange equations $\delta {\cal H}/\delta \rho_q = \mu_q$
($q=n, p$).
\begin{equation}
\label{cejeq7}
(\Delta- m_s^2)\phi = -g_{s} \rho_{s} +b\phi^2 +c\phi^3 ,
\end{equation}
\begin{equation}
(\Delta - m_{v}^2) V =  -g_{v} \rho-4a{{V}}^4-2\Lambda{R^2}{V},
\label{eq18} 
\end{equation}
\begin{equation}
\label{cejeq9}
(\Delta - m_\rho^2)  R  =   - g_\rho \rho_3-2\Lambda{R}{V^2},
\end{equation}
\begin{equation}
\label{cejeq10}
\Delta {\cal A}    =   -e \rho_{p}.
\end{equation}
The above field equations are solved self-consistently in an iterative method.
The pairing correlation is not included in the evaluation of the equilibrium 
properties including the monopole excitation energy. The Thomas-Fermi approach is 
a semi-classical approximation and pairing correlation has a minor role in 
giant resonance. It is shown in \cite{xavier11,baldo13} that the pairing has a marginal effect on the ISGMR energy, and only for open-
shell nuclei.  As far as pairing 
correlation is concerned, it is a quantal effect and  can be included in a 
semi-classical calculation as an average, as is adopted in semi-empirical 
mass formula.  In Ref.\cite{baldo13},  perturbative calculation on top of a 
semi-classical approach is done, and it suggests that pairing
correlation is not important in such approaches like relativistic Thomas-Fermi
(RTF) or relativistic extended Thomas-Fermi (REFT) approximations.  In our 
present calculations, the scalar density ($\rho_s$) and energy density 
($\cal{E}$) are calculated using RTF and RETF formalism. The RETF is the 
${\hbar}^2$ correction to the RTF, where the gradient of density is taken 
into account. This term of the density takes care of the variation of the density 
and involves more of the surface properties. Now transforming the term
$\left(\bigtriangledown V \right)^2+ {m^2}_v {V^2}$ into
$V\left(-\Delta+{m_v}^2\right)V=-g_vV\rho$ (similarly for other fields),
we can write the Hamiltonian as 
\begin{multline}
{\cal H} = {\cal E} + \frac{1}{2}g_{s}\phi \rho^{eff}_{s}
+ \frac{1}{3}b\phi^3+\frac{1}{4} c \phi^4
+\frac{1}{2} g_{v}  V \rho +\frac{1}{2} g_\rho R \rho_3 
+\frac{1}{2} e {\cal A} \rho_{p}-a{V}^4-\Lambda{{R}^2}{{V}^2},
\label{eqFN8c}
\end{multline}
with
\begin {eqnarray}
{\rho}_s^{eff}&= & g_s{\rho}_s-b{\phi}^2-c{\phi}^3.
\end{eqnarray}
In order to study the monopole vibration of the nucleus, we have scaled the baryon density
\cite{patra01}.
The normalized form of the scaled baryon density is given by
\begin {eqnarray}
{\rho}_{\lambda}\left(\bf r \right)&=&{\lambda}^3{\rho}\left(\lambda r \right),
\end{eqnarray}
where ${\lambda}$ is the collective co-ordinate associated
 with the monopole vibration. The other quantities are scaled like  
\begin {eqnarray}
K_{Fq{\lambda}}=\left[3{\pi}^2{\rho}_{q
\lambda}\left(\bf r \right) \right]^{\frac{1}{3}}= \lambda K_{F
q}\left(\lambda r \right),
\end {eqnarray}
\begin {eqnarray}
\tilde m^*_{{\lambda}}(r)=m-g_s{\phi_\lambda}(r)
{m^*}_{{\lambda}}(r)\equiv\lambda \tilde m^*(\lambda r) ,
\end {eqnarray}
where $\tilde m^**$ carries implicit dependence of $\lambda$ apart from the 
parametric dependence of $\lambda r$.
\begin{eqnarray}
{\cal E_{\lambda}}(\bf r)&\equiv&{\lambda}^4 {\tilde{\cal E}}(\lambda \bf r)
= \lambda^4[\tilde {\cal {E}}_0 (\lambda \bf r)+
\tilde{\cal {E}}_2(\lambda \bf r)],
\end {eqnarray}
\begin{eqnarray}
\rho_{{s\lambda}}(\bf r) \equiv {\lambda^3}{\tilde{\rho}}_{s}{(\lambda{\bf r})}.
\end{eqnarray}

Similarly, the $\phi$, $V$, $R$ and Coulomb fields are scaled due to the self-consistency
Eq.~\ref{cejeq7}-\ref{cejeq10} . But the $\phi$ field can not be scaled simply like the density 
and momentum, because its source term contains the $\phi$ 
field itself. Putting in all of the scaled variables one can get the scaled Hamiltonian:
\begin{eqnarray}
\frac{{\cal{H}}_{\lambda}}{\lambda^3} & = & {\lambda}
{\tilde{\cal{E}}}+\frac{1}{2}{g_s}
\phi_{\lambda}{{\tilde{\rho}}_s^{eff}}\nonumber
+\frac{1}{3}\frac{b}{\lambda^3}\phi_\lambda^3
+\frac{1}{4}\frac{c}{\lambda^3}{\phi_\lambda}^4
+\frac{1}{2}{g_v}V_\lambda{\rho}
+\frac{1}{2}g_{\rho}R_\lambda {\rho_3}+\frac{1}{2}e{A}_\lambda{\rho}_p-
a\frac{{V_\lambda}^4}{\lambda^3}
\nonumber\\
&-&\frac{\Lambda}{\lambda^3}{{R_\lambda}^2}
{{V_\lambda}^2}.
\end{eqnarray}
We know $\tilde{\rho_s}=\frac{\partial \tilde{\cal {E}}}{\partial 
\tilde m^*}$, $\frac{\partial \tilde {\cal {E}}}{\partial \lambda}=
\frac{\partial \tilde {\cal {E}}}{\partial\tilde m^*} 
\frac{\partial \tilde m^*}{\partial \lambda}=\rho_s 
\frac{\partial \tilde m^*}{\partial \lambda}$, and
$\frac{\partial m^*_\lambda}{\partial \lambda}=\lambda 
\frac{\partial \tilde  m^*}{\partial \lambda}+{\tilde m^*}=
-g_s\frac{\partial \phi_\lambda}{\partial \lambda}$.
%
Here we are interested in calculating the monopole excitation energy, which is 
defined as ${E}^{s}={\sqrt{\frac{C_m}{B_m}}}$, where ${C_m}$ is the restoring 
force and $B_m$ is the mass parameter. In our calculations, $C_m$ is
obtained from the double derivative of the scaled energy with respect to 
the scaled co-ordinate $\lambda$ at ${\lambda}=1$ \cite{patra01}. The first 
derivative is given by  
\begin {eqnarray}
\label{cejeq19}
&&\bigg[\frac {\partial}{\partial \lambda}
\int{d \lambda r}\frac {{\cal H}
_\lambda\left(r \right)}{\lambda^3}\bigg]_{\lambda =1}
=\int {d\lambda r}\bigg[\tilde{\cal E}-\tilde m^*\tilde \rho_s
-\frac{1}{2}g_s{\tilde {\rho_s}^{eff}}\frac 
{\partial \phi_{\lambda}}{\partial \lambda}+\frac {1}{2}g_s
{\phi_\lambda}\frac{\partial{\tilde {\rho_s}^{eff}}}
{\partial \lambda}- \frac{b}{\lambda^4}{\phi_\lambda}^3 
\nonumber\\
&&-\frac{3}{4}\frac{c}{\lambda^4}{\phi_\lambda}^4  
+\frac{1}{2}g_\rho\rho_3\frac{\partial R_\lambda}{\partial \lambda}
+\frac{1}{2}g_v\rho\frac{\partial V_\lambda}{\partial \lambda}
+\frac{1}{2}e\rho_p\frac{\partial A_\lambda}{\partial \lambda}
-4a\frac{{v_\lambda}^3}{\lambda^3}\frac{\partial V_\lambda}
{\partial \lambda}+3a\frac{{V_\lambda}^4}{\lambda^4}+\frac
{3\Lambda}{\lambda^4}{R_\lambda}^2{V_\lambda}^2\nonumber\\
&&-\frac{2\Lambda}{\lambda^3}{R_\lambda}^2{V_\lambda
}\frac{\partial V_\lambda}{\partial \lambda}-
\frac{2\Lambda}{\lambda^3}{R_\lambda}{V_\lambda
}^2\frac{\partial R_\lambda}{\partial \lambda}\bigg]_{\lambda=1}.
\end{eqnarray}
Now consider the field equation for the omega field 
\begin{eqnarray}
\left( \triangle -{m_v}^2 \right)V=-g_v\rho-4aV^3-2\Lambda{R^2}{V}.
\end{eqnarray}
The scaled equation is  
\begin{eqnarray}
\left( \triangle_u -\frac{{m_v}^2}{\lambda^2} \right){V_\lambda}=
-\lambda g_v\rho-\frac{4a{V_\lambda}^3}{\lambda^2}-\frac{2{\Lambda}}{\lambda^2}
{{R_\lambda}^2}
V_\lambda,
\end{eqnarray}
where $u=\lambda r$. Taking the first derivative with respect to $\lambda$, we
have 
\begin{eqnarray}
\left( \triangle_u -\frac{{m_v}^2}{\lambda^2} \right)
\frac{\partial V_\lambda}{\partial\lambda}&=&
-g_v\rho-\frac{12a{V_\lambda}^2}{\lambda^2}\frac{\partial V_\lambda}{\partial \lambda}
+\frac{8a{V_\lambda}^3}{\lambda^3}-\frac{2{m_v}^2V_\lambda}{\lambda^3}
+4\frac{\Lambda}{\lambda^3}{R_\lambda}^2{V_\lambda}
\nonumber\\
&&-4\frac{\Lambda}{\lambda^2}{R_\lambda}{V_\lambda}\frac{\partial 
{R_\lambda}}{\partial \lambda}
-2\frac{\Lambda}{\lambda^2}{R_\lambda}^2
\frac{\partial {V_\lambda}}{\partial \lambda}.
\end{eqnarray}

Multiplying $V_\lambda$ on both the sides and integrating, one can get

\begin{eqnarray} 
\int{d (\lambda r)}\frac{\partial V_\lambda}{\partial\lambda}
\left( \triangle_u -\frac{{m_v}^2}{\lambda^2} \right)V_\lambda&=&
\int{d\lambda r}\bigg[-g_v\rho V_\lambda-\frac{12a{V_\lambda}^3}{\lambda^2}
\frac{\partial V_\lambda}{\partial \lambda}
+\frac{8a{V_\lambda}^4}{\lambda^3}-\frac{2{m_v}^2{V_\lambda}^2}{\lambda^3}
\nonumber\\
&&+4\frac{\Lambda}{\lambda^3}{R_\lambda}^2{V_\lambda}^2
-4\frac{\Lambda}{\lambda^2}{R_\lambda}{V_\lambda}^2\frac{\partial 
{R_\lambda}}{\partial \lambda}-2\frac{\Lambda}{\lambda^2}{R_\lambda}^2
\frac{\partial {V_\lambda}}{\partial \lambda}V_{\lambda}\bigg].
\nonumber\\
\end{eqnarray}

Putting the field equation for omega field in the above equation, we get 
\bigskip
\begin{eqnarray}
\label{cejeq24}
\int{dr}\bigg[\frac{1}{2}\rho g_v\frac{\partial V_\lambda}{\partial \lambda}-
\frac{4a{V_\lambda}^3}{\lambda^3}\frac{\partial V_\lambda}{\partial \lambda}
\bigg]_{\lambda=1}&=&
\int{dr}\bigg[\frac{g_vV_\lambda\rho}{2\lambda}-\frac{4a{V_\lambda}^3}{\lambda^4}+\frac{{m_v}^2{V_\lambda}^2}{\lambda^4}
-2\frac{\Lambda}{\lambda^4}{R_\lambda}^2{V_\lambda}^2
\nonumber\\
&+&2\frac{\Lambda}{\lambda^3}{R_\lambda}{V_\lambda}^2
\frac{\partial R_\lambda}{\partial \lambda}\bigg]_{\lambda=1}.
\end{eqnarray}
Now, consider the field equation for $R$ field, where we have
\begin{equation}
(\Delta - {m_\rho^2})R  =- {g_\rho} {\rho_3}-2{\Lambda}R{V}^2.
\label{eq19} 
\\[3mm]
\end{equation}
By scaling the whole equation with the scaling parameter $\lambda$, 
we get the scaled equation as:
\begin{equation}
(\Delta_u - \frac{m_\rho^2}{\lambda^2})R_\lambda  =- g_\rho{\lambda} \rho_3-
2\frac{{\Lambda}R_{\lambda}{V_\lambda}^2}{\lambda^2}.
\label{eq19} 
\\[3mm]
\end{equation}
From the first and second derivative with respect to $\lambda$ and using
a similar procedure for $\omega-$field, one
can get the following equation
\begin {eqnarray}
\label{cejeq27}
(\Delta_u-\frac{{m_\rho}^2}{\lambda^2})\frac
{\partial R_\lambda}{\partial \lambda}=-{g_\rho}{\rho_3}-
2R_\lambda\frac{{m_\rho}^2}{\lambda^3}+
4\frac{\Lambda}{\lambda^3}{V_\lambda}^2{R_\lambda}
-4\frac{\Lambda}{\lambda^3}{R_\lambda}{V_\lambda}\frac
{\partial V_\lambda}{\partial \lambda}
-2\frac{\Lambda}{\lambda^2}{V_\lambda}^2\frac
{\partial R_\lambda}{\partial \lambda}.\nonumber\\
\end{eqnarray}

Substituting the relation of Eq.(~\ref{cejeq9}) in (~\ref{cejeq27}) at $\lambda=1$, we get,
\begin{eqnarray}
\label{cejeq28}
\int{dr}g_\rho\frac{1}{2}{\rho_3}\frac{\partial R_\lambda}
{\partial \lambda}\bigg|_{\lambda=1}
=\int{dr}\bigg[\frac{1}{2\lambda}g_\rho{\rho_3} R_\lambda
+\frac{{R_\lambda }^2{m_\rho}^2}{\lambda^4}
+2\frac{\Lambda}{\lambda^4}{R_\lambda}^2{V_\lambda}
\frac{\partial V_\lambda}{\partial \lambda}
-2\frac{\Lambda}{\lambda^4}{V_\lambda}^2{R_\lambda}^2\bigg]_{\lambda=1}.
\nonumber\\
\end{eqnarray}

With the help of this expression,  Eq.(\ref{cejeq19}) becomes 
\begin {eqnarray}
\bigg[\frac {\partial}{\partial \lambda}
\int\left({d \lambda r}\right)\frac {{\cal H}
_\lambda\left(r \right)}{\lambda^3}\bigg]_{\lambda =1}
=\int {d\lambda r}\bigg[\tilde{\cal E}-\tilde m^*\rho_s
-\frac{1}{2}g_s{\tilde {\rho_s}^{eff}}\frac 
{\partial \phi_{\lambda}}{\partial \lambda}+\frac {1}{2}g_s
{\phi_\lambda}\frac{\partial{\tilde {\rho_s}^{eff}}}
{\partial \lambda}\nonumber\\
-\frac{b}{\lambda^4}{\phi_\lambda}^3 
-\frac{3}{4}\frac{c}{\lambda^4}{\phi_\lambda}^4  
+g_\rho\rho_3\frac{1}{2\lambda}{R_\lambda}
+g_v\rho\frac{1}{2\lambda} V_\lambda
+e\rho_p\frac{1}{2}\frac{\partial A_\lambda}{\partial \lambda}
-a\frac{{v_\lambda}^3}{\lambda^4}+\frac{{m_v}^2{V_\lambda}^2}{\lambda^4}\nonumber\\
+\frac{{m_\rho}^2}{\lambda^4}{R_\lambda}^2-\frac{\Lambda}{\lambda^4}
{R_\lambda}^2{V_\lambda}^2\bigg]_{\lambda=1}.
\end{eqnarray}

Again, differentiating w.r.t $\lambda$ and substituting $\lambda=1$, 
the restoring force $C_m$ becomes:

\begin {eqnarray}
{C}_m&=&\bigg[\frac {\partial^2}{\partial \lambda^2}
\int{d \lambda r} \frac {{\cal {H}}
\lambda\left(r \right)}{\lambda^3}\bigg]_{\lambda =1}
 =\int {d\lambda r}\bigg[-\tilde {m}^*\frac{\partial \tilde {\rho_s}}{\partial \lambda}
-\frac{1}{2}g_s{\tilde {\rho_s}^{eff}}
\frac {\partial^2 \phi_{\lambda}}{\partial \lambda^2}
+\frac {1}{2}g_s {\phi_\lambda}
\frac{\partial^2{\tilde {\rho_s}^{eff}}}{\partial \lambda^2}
\nonumber\\
&&+4\frac{b}{\lambda^5}{\phi_\lambda}^3 
+3\frac{c}{\lambda^5}{\phi_\lambda}^4   
 -\frac{3}{\lambda^4} \left(b{{\phi_\lambda}^2}+c{\phi_\lambda}^3\right)
\frac{\partial \phi_\lambda}{\partial \lambda} 
+\frac{1}{2\lambda}g_\rho {\rho_3}\frac{\partial R_\lambda}{\partial \lambda}
+\frac{2}{\lambda^4}{m_\rho}^2{R_\lambda}\frac{\partial R_\lambda}{\partial \lambda}\nonumber\\
&-&\frac{2\Lambda}{\lambda^4}\left(R_\lambda {V_\lambda}^2\frac{\partial 
R_\lambda}{\partial \lambda}\right)+{R_\lambda}^2 {V_\lambda}\frac{\partial 
V_\lambda}{\partial \lambda}+\frac{4\Lambda}{\lambda^5}{R_\lambda}^2{V_\lambda}^2
+\frac{1}{2} e \rho_p \frac{\partial^2 {\cal{A_\lambda}}}{\partial \lambda^2}
-\frac{g_v V_\lambda\rho}{2\lambda^2}\nonumber\\
&+&\left[\frac{g_v\rho}{2\lambda}
-\frac{4a{V_\lambda}^3}{\lambda^4}
 +2\frac{{m_v}^2{V_\lambda}}{\lambda^4}
\right]
\frac{\partial V_\lambda}{\partial \lambda}
+\frac{4a{V_\lambda}^4}{\lambda^5} -4\frac{{m_v}^2{V_{\lambda}^2}}{\lambda^5}
-4\frac{{m_\rho}^2{R_{\lambda}^2}}{\lambda^5}\bigg]_{\lambda=1}.
\end{eqnarray}

The $\omega-$meson field equation (\ref{cejeq24}) at $\lambda=1$ can be  
written as
\begin{eqnarray}
\label{cejeq31}
\int{dr}\bigg[\frac{1}{2}g_v \rho-4aV^3\bigg]
{\frac{\partial V_\lambda}{\partial \lambda}}\bigg|_{\lambda=1}=
\int{dr}\bigg[\frac{g_vV\rho}{2}-4aV^3+{m_v}^2V^2-2{\Lambda}
{R}^2{V}^2
+2{\Lambda}R{V}^2
\frac{\partial R_\lambda}{\partial \lambda}\bigg]_{\lambda=1}.\nonumber\\
\end{eqnarray}
Now consider the scaled equation for the sigma field, 
\begin{eqnarray}
\label{cejeq32}
\left( \triangle_u -\frac{{m_s}^2}{\lambda^2} \right){\phi_\lambda}=
-\lambda g_s\tilde{\rho_s}^{eff}.
\end{eqnarray}
The double derivative of Eq.(\ref{cejeq32}) with respect to $\lambda$ is given by:
\begin{eqnarray}
\left( \triangle_u -\frac{{m_s}^2}{\lambda^2} \right)\frac{{\partial^2\phi_\lambda}}
{\partial \lambda^2}=
-2 g_s\frac{\partial\tilde{\rho_s}^{eff}}{\partial \lambda}-
2\frac{{m_s}^2}{\lambda^3}\frac{\partial \phi_\lambda}{\partial \lambda}
+\frac{6{m_s}^2}{\lambda^4}\phi_\lambda-g_s\lambda\frac{\partial^2{\tilde
\rho^{eff}}}{\partial \lambda^2}.
\end{eqnarray}

Multiplying by $\phi_\lambda$ and then integrating both sides, we get

\begin{eqnarray}
\label{cejeq34}
\int{dr} \bigg[ -\frac{1}{2}g_s \tilde{{\rho_s}}^{eff}
\frac{\partial^2 \phi_\lambda}{\partial \lambda^2} \bigg]
+\frac{1}{2} g_s\phi_\lambda\tilde{\rho_s}^{eff}
\frac{\partial^2 \tilde{\rho_s}^{eff}}{\partial \lambda^2}
=\int{dr}\bigg[-g_s\phi\frac{\partial \tilde{\rho_s}^{eff}}{\partial \lambda}
-2{m_s}^2\phi\frac{\partial \phi_\lambda}{\partial \lambda}+3{m_s}^2{\phi^2} 
\bigg],\nonumber\\
\end{eqnarray}
where
\begin{eqnarray}
\label{cejeq35}
-g_s\frac{\partial \tilde{\rho}^{eff}}{\partial \lambda}\bigg|_{\lambda=1}=
-g_s\phi\frac{\partial \tilde{\rho_s}}{\partial \lambda}\bigg|_{\lambda=1}
-3(b\phi^2+c\phi^3)-2\left(2b\phi+3c\phi^2\right)\frac{\partial 
\phi_\lambda}{\partial \lambda}\bigg|_{\lambda=1}.
\end{eqnarray}

{
Substituting the value of the scaled parameter $\lambda=1$, the restoring 
force $C_m$ can be written as:
\bigskip
\begin{align}
\label{cejeq36}
{C}_m&=\bigg[\frac {\partial^2}{\partial \lambda^2}
\int{d \lambda r} \frac {{\cal {H}}
_\lambda\left(r \right)}{\lambda^3}\bigg]_{\lambda =1}\nonumber\\
& =\int {d r}\bigg[-\tilde {m}^*
\frac{\partial \tilde {\rho_s}}{\partial \lambda}
\bigg|_{\lambda=1}-\frac{1}{2}g_s{\tilde {\rho_s}^{eff}}
\frac {\partial^2 \phi_{\lambda}}{\partial \lambda^2}\bigg|_{\lambda=1}
+\frac {1}{2}g_s {\phi} 
\frac{\partial^2{\tilde {\rho_s}^{eff}}}{\partial \lambda^2}
\bigg|_{\lambda=1}+4 {b}{\phi}^3 +3{c}{\phi}^4 \nonumber\\  
&-{3} \left(b{{\phi}^2}+c{\phi}^3\right)
\frac{\partial \phi_\lambda}{\partial \lambda}\bigg|_{\lambda=1} 
+\frac{1}{2}g_\rho {\rho_3}\frac{\partial R_\lambda}
{\partial \lambda}\bigg|_{\lambda=1}
+{2}{m_\rho}^2{R}\frac{\partial R_\lambda}{\partial \lambda}\bigg|_{\lambda=1}
-\frac{2\Lambda}{\lambda^4}(R {V}^2\frac{\partial V_\lambda}{\partial \lambda}\bigg|_{\lambda=1
}\nonumber\\
&+{R}^2 {V}\frac{\partial R_\lambda}{\partial \lambda}\bigg|_{\lambda=1})
+{4\Lambda}{R}^2{V}^2
+\frac{1}{2} e \rho_p \frac{\partial^2 {\cal{A_\lambda}}}{\partial \lambda^2}\bigg|_{\lambda=1}
-\frac{g_v V\rho}{2}
+\left(\frac{g_v\rho}{2}
-{4a{V}^3}
 +2{{m_v}^2{V}}\right)\nonumber\\
&\frac{\partial V_\lambda}{\partial \lambda}\bigg|_{\lambda=1}
+4a{V}-4{m_v}^2{V^2}-4{m_\rho}^2{R}^2 \bigg].
\end{align}
\bigskip
We put  $\frac{\partial^2 {\cal{A}}}{\partial \lambda^2}=0$, as
the photon has zero mass. Finally, substituting Eqs. (\ref{cejeq28}),(\ref{cejeq31} ), (\ref{cejeq34}), (\ref{cejeq35}) into Eq. (\ref{cejeq36}) and rearranging the terms, one can get 
the following expression for the restoring force  
\bigskip
\begin {eqnarray}
{C}_m&=&\int{dr}\bigg[-m\frac{\partial{\tilde{\rho_s}}}{\partial{\lambda}}
+3\bigg({m_s}^2{\phi}^2+\frac{1}{3}b{\phi}^3
- {m_v}^2{V^2}
-{m_{\rho}}^2R^2\bigg)-(2{m_s}^2{\phi}
+b{\phi}^2)\frac{\partial{\phi_\lambda}}{\partial{\lambda}}\nonumber\\
&+&2{m_v}^2V\frac{\partial{V_\lambda}}{\partial{\lambda}}
+2{m_{\rho}}^2R
\frac{\partial R_\lambda}{\partial \lambda}\bigg]_{\lambda=1}.
\end{eqnarray}
The mass parameter $B_{m}$ of the monopole vibration can be expressed as the
double derivative of the scaled energy with the  collective velocity $\dot{\lambda}$
as
\begin{eqnarray}
B_{m}=\int{dr}{U(\bf r)}^2{\cal {H}},
\end {eqnarray}
where $U(\bf r)$ is the displacement field, which can be determined from the relation
between collective velocity $\dot{\lambda}$ and velocity of the moving frame,
\begin {eqnarray}
U(\bf r)=\frac{1}{\rho(\bf r){\bf r}^2}\int{dr'}{\rho}_T(r'){r'}^2,
\end {eqnarray}
where ${\rho}_T $ is the transition density defined as
\begin {eqnarray}
{{\rho}_T(\bf r)}=\frac{\partial{\rho_\lambda(\bf r)}}{\partial{\lambda}}\bigg|_{\lambda = 1}
=3 {\rho}(\bf r)+r \frac{\partial{\rho(\bf r)}}{\partial r}.
\end {eqnarray}
Taking $U(\bf r)=r $ the relativistic mass parameter can be written as:
\begin{equation}
\label{mas1}
B_m=\int{dr}{r}^2{\cal H}.
\end{equation}
Similarly in the non-relativistic limit,
the mass parameter is defined as 
\begin{equation}
\label{mas2}
{B_m}^{nr}=\int{dr}{r^2}m{\rho}. 
\end{equation}
The scaled energy in terms of the moments of the strength function can be written as ${E_m}^{s}= \sqrt{\frac{m_3}{m_1}}$. The expressions for ${m_3}$ and 
${m_1}$ can be found in \cite{bohigas79}. In simple classical approximation 
the scaling excitation energy can be written $E_x^s = \sqrt{\frac{C_m}{B_M}}$.

                 Along with the scaling calculation, the monopole vibration can
also be studied with a constrained approach \cite{bohigas79,maru89,boer91,stoi94,stoi94a}. In the constrained method, one has to minimize the constrained 
energy functional: 
\begin{eqnarray}
\label{coneq}
\int{dr}\left[{\cal H}-{\eta}{r}^2 {\rho}\right]=E(\eta)-\eta\int{dr}{r}^2\rho,
\end{eqnarray}
 with respect to the variation of  densities and meson fields. Here 
$\eta$ is the constrained parameter. The densities, field and energy 
obtained from the solution of the above constrained equation 
( Eq. ~\ref{coneq} ) are the function of parameter $\eta$. The rms 
radius of the nucleus is given in terms of the parameter $\eta$: 
\begin{equation}
{R_\eta}=  \bigg[ \frac{1}{A} \int dr r^2 \rho \bigg]^{1/2}, 
\end{equation}
where A is the mass number of the nucleus. We are assuming that the parameter 
$\eta$ is very small, so E($\eta$) can be expanded around $\eta$=0, which 
corresponds to the ground state energy of the nucleus and $R_0$ gives 
the rms radius of the nucleus in ground state. The expansion of $E(\eta)$  
around $\eta=0$ is given by: 
\begin{equation}
E(\eta)= E(0) + {R_0} \frac{\partial E(\eta)}{\partial \eta}\bigg|_{\eta=0}
+ {R_0} \frac{\partial E(\eta)}{\partial \eta}\bigg|_{\eta=0}.
\end{equation}
The second order derivative of the expansion gives the effective 
incompressibility of finite nucleus and it is defined as : 
\begin{equation}
{K_A}^c = \frac{1}{A} {R_0} \frac{\partial^2 E(\eta)}{\partial \eta^2}\bigg|_{\eta=0}.
\end{equation}
We can calculate the incompressibility using this simple formula. In classical 
approximation the average constraint excitation energy is given by : 
${E_x}^c = \sqrt{\frac{A{K_A}^c}{B_m^c}}$, where $B_m^c$ is the mass parameter. The expression for the mass parameter both in relativistic and non-relativistic  are given in Eq. (~\ref{mas1}) and Eq. (~\ref{mas2}), respectively. Again, in the  
sum rule approach constrained excitation energy can be expressed as the 
ratio of moments of strength distribution. Constrained energy is 
defined as : $E_x^c = \sqrt{\frac{m_1}{m_{-1}}}$, where $m_1$  and 
$m_{-1}$ are the energy weighted and inverse energy weighted moments, 
respectively. $m_{-1}$ is interpreted as the polarizability of the system 
ie. nothing but the measure of energy changed due to excitation of 
the nucleus. Expression for the $m_{-1}$ is given by : 
\begin{equation}
m_{-1}= -\frac{A}{2} \frac{\partial {R_\eta}^2}{\partial \eta}\bigg|_{\eta=0} =
\frac{1}{2}\frac{\partial^2 E(\eta)}{\partial \eta^2}\bigg|_{\eta=0}.
\end{equation}
One can estimate $m_{-1}$ by calculating first derivative of ${R_\eta}^2 $ 
with respect to $\eta$ or second derivative of $E(\eta)$ with respect to the 
$\eta$. We have calculated $m_{-1}$ using both 5-point and 3-point formula. $m_1$ is proportional to the mass parameter, which can expressed as : $m_1= \frac{2}{m} A <r^2>= \frac{2}{m^2} {B_M}^{nr}$. From the inequality relations 
satisfied by the moments of the strength distribution, we get
 $\frac{m_1}{m_{-1}}\leq \frac{m_3}{m_{1}} $. This implies that the scaling 
energy gives the upper limit of the resonance, while the constrained 
energy gives the lower limit of the resonance spectra. The  resonance  width  \cite{bohigas79,mari05} is defined as:
\begin{eqnarray}
{\Sigma} &=& \sqrt{\left({E_m}^s\right)^2-\left({E_m}^c\right)^2} 
= \sqrt{({\frac{m_3}{m_1}})^2-{(\frac{m_1}{m_{-1}}})^2}. 
\end{eqnarray}

\section{Results and Discussions}
\label{cejsec3}
\begin{table}
\caption{The calculated binding BE and charge radius $r_{ch}$ obtained from
relativistic extended Thomas-Fermi (RETF) approximation is compared with 
relativistic Hartree (with various parameter sets) and experimental results 
\cite{wang12,angeli13}.The  RETF results are given in the parentheses. 
The empirical values \cite{bethe71,blaizot} of nuclear matter saturation
density $\rho_0$, binding energy per nucleon BE/A, incompressibility modulus 
$K$, asymmetry parameter $J$, and ratio of effective mass to the nucleon mass
$M^*/M$ are given in the lower part of the table. Energies are in MeV and 
radii are in fm.
}
\renewcommand{\tabcolsep}{0.3cm}
\renewcommand{\arraystretch}{0.8}
\begin{tabular}{|ccccccc|}
\hline
\hline
Nucleus & Set     & BE (calc.)  & BE (Expt.) & $r_{ch}$ (calc.) & $r_{ch}$ (Expt.)& \\
\hline
$^{16}$O & NL1 & 127.2(118.7) &127.6  & 2.772(2.636) &2.699  & \\
         & NL3 & 128.7(120.8) &  & 2.718(2.591) &  & \\
         & NL3$^*$& 128.1(119.5) &  & 2.724(2.603) &  & \\
         & FSUGold& 127.4(117.8) &  & 2.674(2.572) &  & \\
$^{40}$Ca& NL1 & 342.3(344.783) &342.0  & 3.501(3.371) & 3.478 & \\
         & NL3 & 341.6(346.2) &  & 3.470(3.343) &  & \\
         & NL3$^*$& 341.5(344.2) &  & 3.470(3.349) &  & \\
         & FSUGold& 340.8(342.2) &  & 3.429(3.327) &  & \\
$^{48}$Ca& NL1 & 412.7(419.5) & 416.0 & 3.501(3.445) & 3.477 & \\
         & NL3 & 414.6(422.6) &  & 3.472(3.426) &  & \\
         & NL3$^*$& 413.5(420.3) &  & 3.469(3.429) &  & \\
         & FSUGold& 411.2(418.0) &  & 3.456(3.418) &  & \\
$^{90}$Zr& NL1 & 784.3(801.1) &783.9 & 4.284(4.232) & 4.269 & \\
         & NL3 & 781.4(801.7) &  & 4.273(4.219) &  & \\
         & NL3$^*$& 781.6(798.7) &  & 4.267(4.219) &  & \\
         & FSUGold& 778.8(797.3) &  & 4.257(4.214) &  & \\
$^{116}$Sn&NL1 & 989.5(1013.7) &988.7  & 4.625(4.583) & 4.625 & \\
          &NL3 & 985.4(1014.6) &  & 4.617(4.571) &  & \\
         & NL3$^*$& 986.4(1011.0) &  & 4.609(4.569) &  & \\
         & FSUGold& 984.4(1010.7) &  & 4.611(4.569) &  & \\
$^{208}$Pb&NL1 & 1638.1(1653.7)&1636.4  & 5.536(5.564) & 5.501 & \\
          &NL3 & 1636.9(1661.2)&  & 5.522(5.541) &  & \\
         & NL3$^*$& 1636.5(1655.2)&  & 5.512(5.538) &  & \\
         & FSUGold& 1636.2(1661.4)&  &5.532(5.541)  &  & \\
\hline
\hline
Set &NL1& NL3&NL3$^*$&FSUGold &empirical& \\
\hline
\hline
$\rho_0$ &0.154 & 0.150 &0.148 &0.148&0.17 &  \\
$E/A$ &16.43&16.31&16.30&16.30&15.68&\\
$K$ &211.7&271.76&258.27&230.0&$210\pm 30$&\\
$J$ &43.6&38.68&37.4&32.597&$32\pm 2$&\\
$M^*/M$ &0.57& 0.594&0.60&0.61&0.6&\\
\hline
\end{tabular}
\label{table1}
\end{table}

\subsection{Force parameters of relativistic mean field formalism}

First of all, we examined the predictive power of various parameter sets. 
In this context we selected NL1 as a successful past set, and a few 
recently used forces like NL3, NL3$^*$, and FSUGold with varying
incompressibilities as shown in Table \ref{table1} (lower part of the table). 
The ground state 
observable obtained by these forces are depicted in Table \ref{table1}. 
Along with the relativistic extended Thomas-Fermi (RETF) results, the
relativistic Hartree values are also compared with the experimental data
\cite{wang12,angeli13}. The calculated RMF results obtained by all the 
force parameters considered in the present thesis  are very close to the 
experimental data \cite{wang12,angeli13}. A detailed analysis of the binding 
energy and charge radius clearly show that NL1 and FSUGold have superior 
predictive power for $^{16}$O in RMF level. The advantage of FSUGold 
decreases with increased mass number of the nucleus. Although the
predictive power of the relatively old NL1 set is very good for binding energy, 
it has a large asymmetry coefficient $J$, which may mislead the prediction 
in unknown territories, like the neutron drip-line or super heavy regions. The 
RETF prediction of binding energy and charge radius (numbers in the 
parenthesis) is very poor with the experimental data as compared to the RMF 
calculations. However, for relatively heavy masses, the RETF results can 
be used within acceptable error.  In general, taking into account the
binding energy BE and root mean square charge radius $r_{ch}$, one may prefer to use either of the NL3 or NL3$^*$ parametrization.
\begin{table}
\caption{The results of isoscalar giant monopole resonance with
various parameter sets for some known nuclei are compared with recent experimental
data \cite{young13}. The calculations are done with
relativistic extended Thomas-Fermi (RETF) approximation using both
scaling and constrained schemes. The values of $\Delta_1$,
$\Delta_2$, and $\Delta_3$ are obtained by subtracting the
results of (NL3, NL3$^*$), (NL3$^*$, FSUGold), and (NL3, FSUGold), respectively.
The monopole excitation energies with scaling $E^s$ and constrained $E^c$
are in MeV.
}
\renewcommand{\tabcolsep}{0.10cm}
\renewcommand{\arraystretch}{1.0}
\begin{tabular}{|c|c|c|c|c|c|c|c|c|c|c|c|c|c|}
\hline
\hline
Nucleus
& \multicolumn{2}{c|}{NL1}
& \multicolumn{2}{c|}{NL3}
&\multicolumn{2}{c|}{NL3$^*$}
&\multicolumn{2}{c|}{FSUGold}
&\multicolumn{1}{c|}{Expt.}
&{$\Delta_1$}&{$\Delta_2$}&{$\Delta_3$} \\
\hline
& ${E}^s$& ${E}^c$&${E}^s$& ${E}^c$& ${E}^s$& ${E}^c$& ${E}^s$& ${E}^c$& &
 && \\
\hline
$^{16}{O}$&23.31&21.75&27.83&25.97&26.86&25.20&26.97&25.17&21.13$\pm$0.49&0.97&0.11&0.86 \\
$^{40}{Ca}$&20.61&19.77&24.01&23.16&23.32&22.48&22.98&22.30&19.20$\pm$0.40&0.69&0.43&1.03\\
$^{48}{Ca}$&19.51&18.67&22.69&21.73&22.01&21.11&21.72&20.88&19.90$\pm$0.20&0.68&0.29&0.97\\
$^{90}{Zr}$&16.91&16.41&19.53&19.03&18.97&18.50&18.60&18.21&$17.89\pm 0.20$&
0.56&0.37&0.97\\
$^{110}{Sn}$&15.97&15.50&18.42&17.94&17.90&17.44&17.52&17.13&&0.52&0.38&0.90\\
$^{112}{Sn}$&15.87&15.39&18.29&17.81&17.78&17.32&17.42&17.02&16.1$\pm$0.10&0.51&0.36&0.86\\
$^{114}{Sn}$&15.76&15.28&18.16&17.67&17.65&17.18&17.31&16.90&15.9$\pm$0.10&0.51&0.34&0.85\\
$^{116}{Sn}$&15.63&15.19&18.02&17.52&17.51&17.04&17.19&16.77&15.80$\pm$0.10&0.51&0.32&0.83\\
$^{118}{Sn}$&15.51&15.03&17.87&17.36&17.37&16.89&17.07&16.63&15.6$\pm$0.10&0.50&0.30&0.80\\
$^{120}{Sn}$&15.38&14.90&17.72&17.20&17.22&16.73&16.94&16.49&15.4$\pm$0.20&0.50&0.28&0.78\\
$^{122}{Sn}$&15.24&14.76&17.56&17.03&17.07&16.57&16.81&16.34&15.0$\pm$0.20&0.49&0.24&0.77\\
$^{124}{Sn}$&15.11&14.61&17.40&16.85&16.91&16.40&16.67&16.19&14.80$\pm$0.20&0.48&0.24&0.72\\
$^{208}{Pb}$&12.69&12.11&14.58&13.91&14.18&13.55&14.04&13.44&14.17$\pm$0.28&0.40&0.14&0.54\\
$^{286}{114}$&11.32&10.60&13.00&12.14&12.64&11.83&12.55&11.79&&0.36&0.09&0.45\\
$^{298}{114}$&11.05&10.31&12.68&11.80&12.33&11.50&12.29&11.53&&0.35&0.04&0.37\\
$^{292}{120}$&11.28&10.53&12.96&12.07&12.60&11.76&12.48&11.69&&0.36&0.12&0.27\\
$^{304}{120}$&11.04&10.28&12.67&11.77&12.33&11.47&12.25&11.47&&0.34&0.08&0.42\\
\hline
\hline
\end{tabular}
\label{table2}
\end{table}
Before accepting NL3 or NL3$^*$ as the working parameter set for our 
further calculations, in Table \ref{table2}, we have given the excitation 
energy of some selective nuclei both in light and super heavy regions 
with various parameter sets for some further verification. The isoscalar 
giant monopole energies $E^s$ and $E^c$ are evaluated using both scaling and
constraint calculations, respectively. The forces like NL1, NL3, Nl3$^*$ and 
FSUGold  have a wide range of incompressibility $K_{\infty}$ starting from 
211.7 to 271.7 MeV (see Table \ref{table1}). Because of the large variation 
in $K_{\infty}$ of these sets, we expect various values of $E^s$ 
and $E^c$ with different parametrization. From Table \ref{table2}, it is 
noticed that the calculated results for $^{16}$O and $^{40}$Ca differ
substantially from the data. Again this deviation of calculated results 
continues decreasing with increasing mass number, irrespective of the 
parameter set. This may be due to the  use of semi-classical approximations 
like Thomas-Fermi  and extended Thomas-Fermi. In these approaches, quantal 
corrections are averaged out. When we are going from  light to
the heavy and then super heavy nuclei, the surface correction decreases 
appreciably. Consequently, the contribution to monopole excitation energy 
decreases with mass number A. In column 11, 12 and 13 of Table \ref{table2}, 
the differences in $E_x$ obtained from various parameter sets are given, 
namely, $\triangle_{1}$ is the difference in monopole excitation energy 
obtained by NL3 and NL3$^*$. Similarly, $\triangle_{2}$ and 
$\triangle_{3}$ are the ISGMR difference with  (NL3$^*$, FSUGold) and 
(NL3, FSUGold), respectively. The values of $\triangle_{1}$, $\triangle_{2}$ 
or  $\triangle_{3}$ go on decreasing with increasing mass number of 
the nucleus without depending on the parameter used. In other words, we may 
reach the same conclusion in the super heavy region irrespective of the 
parameter set. However, it is always better to use a successful
parameter set to explore an unknown territory. In this context, it is 
safer to choose the NL3 force for our further exploration. The second 
observation is also apparent from the Table. It is commonly believed that mostly
the incompressibility of the force parameter affects the excitation energy of 
ISGMR of the nucleus. That means, forces having different
$K_{\infty}$ have different excitation energy for the same nucleus. For example,
$^{208}{Pb}$ has excitation energy 14.58 and 14.04 MeV with NL3 and FSUGold, 
respectively. Although, the ground state binding energy of $^{208}$Pb, 
either with Hartree (RMF) or REFT approximation matches well with  NL3 
and FSUGold parameter sets (see Table \ref{table1}), their ISGMR differ by 
0.54 MeV, which is quite substantial.
The reason behind this difference in $E^s$ with various parameter sets is 
not yet clear. As we have stated, the incompressibility $K_{\infty}$ is not
the only controlling key to tune the monopole excitation energy. 
A lot of effort has been devoted to show that the ISGMR excitation 
energy can be altered by modifying other variables of the force parameter like 
effective mass $M^*$ and the symmetry energy coefficient $a_4$ \cite{toro05}. 
Thus, the relation $E^M=\sqrt{\frac{K_A}{M<r^2>}}$ needs modification
with the inclusion of some other variables including the nuclear matter
incompressibility, where $<r^2>$ is the rms matter radius and $M$ the 
mass of the nucleon. 
Actually, it is a long running debate and not yet clear, and the factors
responsible for the ISGMR are invite more work in this direction. 
\begin{table*}
\caption{The predicted proton and neutron drip-lines PDL and NDL for 
$O$, $Ca$, $Ni$, $Sn$, $Pb$,
Z=114, and Z=120 in relativistic mean field formalism (RMF) with various parameter
sets are compared
with experimental (wherever available) and Finite Range Droplet
Model (FRDM) predictions \cite{nix95}.
}
\renewcommand{\tabcolsep}{0.15cm}
\renewcommand{\arraystretch}{1.0}
\begin{tabular}{|c|c|c|c|c|c|c|c|c|c|c|c|c|c|}
\hline
\hline
Nucleus
& \multicolumn{8}{c|}{RMF}
&\multicolumn{2}{c|}{FRDM}
&\multicolumn{2}{c|}{Expt. \cite{wang12}}
\\
\hline
& \multicolumn{2}{c|}{NL1}
& \multicolumn{2}{c|}{NL3}
&\multicolumn{2}{c|}{NL3$^*$}
&\multicolumn{2}{c|}{FSUGold}
&\multicolumn{2}{c|}{}
&\multicolumn{2}{c|}{}
\\
\hline
& PDL& NDL&PDL& NDL & PDL & NDL & PDL &NDL &PDL &NDL &PDL&NDL \\
\hline
O& 12& 29& 13&  30&12 &  30& 12&  27& 12&  26& 12& 28\# \\
Ca&  34& 69& 33&  71& 34& 71& 34& 66& 30& 73& 35\#& 58 \#\\
Ni& 49& 94& 50&  98& 50& 98& 51& 94& 46& 99& 48& 79 \\
Sn& 99& 165& 100&  172& 100& 172& 99&1 64&94&169& 99\# & 138\# \\
Pb& 178&275& 180&  281& 180& 280& 179&269&175&273& 178 & 220\# \\
114& 267& 375& 271& 392& 274& 390&  271& 376& 269& 339& 285\#&289\# \\
120& 285&376& 288&  414& 288& 410& 289& 396& 287& 339&-&- \\
\hline
\hline
\end{tabular}
\label{table3}
\end{table*}
\subsection{Proton and neutron drip-lines}

In Table \ref{table3} we have shown the proton and neutron drip-lines (PDL 
and NDL) for various parameter sets. The neutron (or proton) drip-line of an 
isotope is defined when the neutron (or proton)
separation energy $S_n$ (or $S_p) \leq 0$, where $S_n=BE(N,Z)-BE(N-1,Z)$ or 
$S_p=BE(N,Z)-BE(N,Z-1)$ with BE(N,Z) is the binding energy of a nucleus
with N neutron and Z proton. From the table, it is seen that all the interactions
predict almost similar proton and neutron drip-lines. If one
compares the drip-lines of NL3 and NL3$^*$, then their predictions are almost 
identical, explicitly for lighter mass nuclei.  Thus, the location of the drip-line with various forces does not depend on its nuclear matter incompressibility 
or asymmetry coefficient. For example, the asymmetry coefficient $J=43.6$ 
MeV and $K_{\infty}=211.7$ MeV for the NL1 set and these are 38.68 and 271.76
 MeV in the NL3 parametrization. The corresponding proton drip-lines for
O isotopes are 12 and 13, and the neutron drip-lines are 29 and 30, 
respectively.  Similar effects are noticed for other isotopes of the 
considered nuclei (see Table \ref{table3}).
\begin{figure}
\includegraphics[scale=0.55]{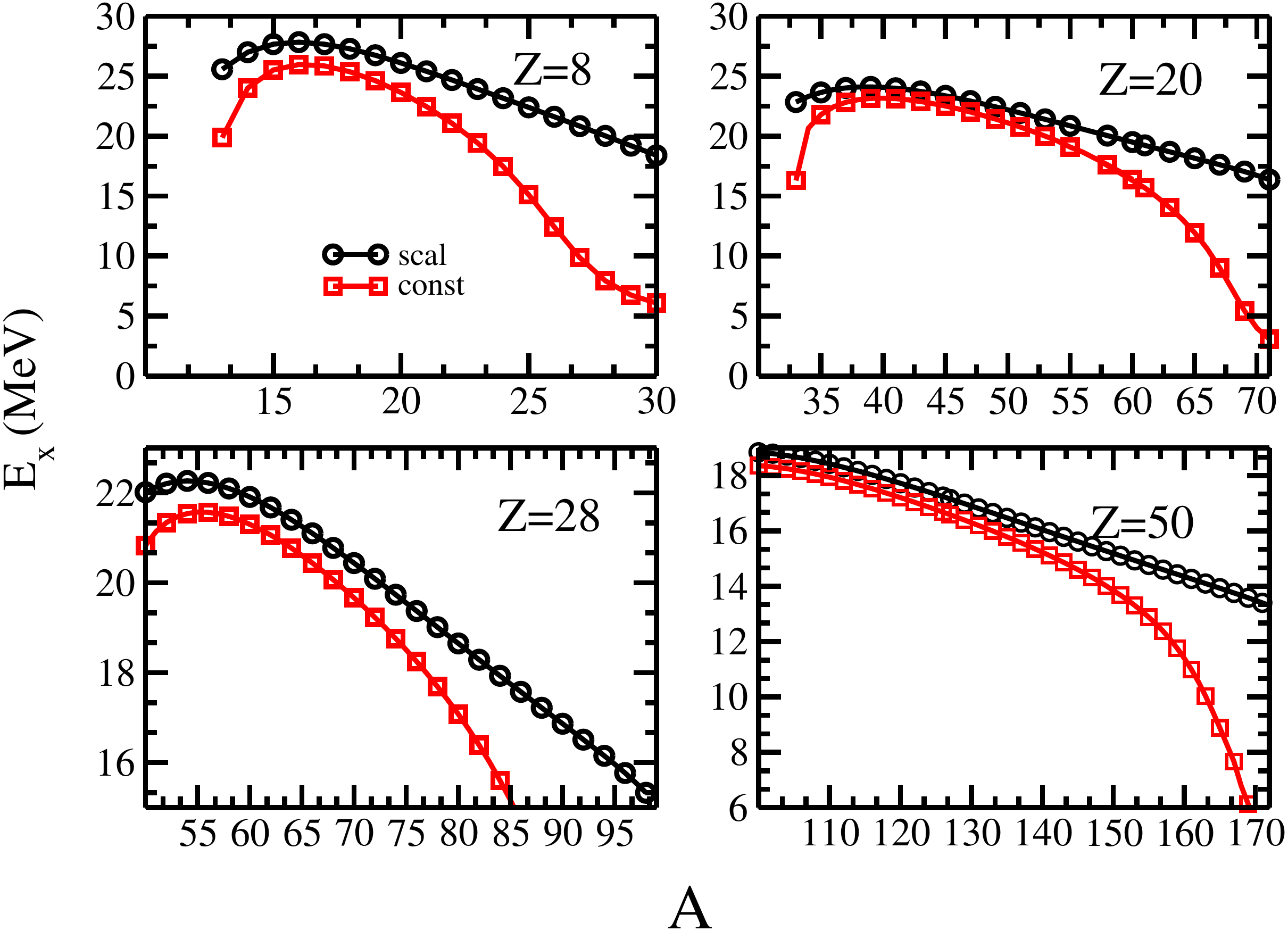}
\caption{\label{fig1:gmr}The excitation energy isoscalar giant 
monopole resonance (ISGMR) for O, Ca, Ni, and Sn isotopes from proton 
to neutron drip-lines as a function of mass number. }
\end{figure}

\begin{figure}
\includegraphics[scale=0.55]{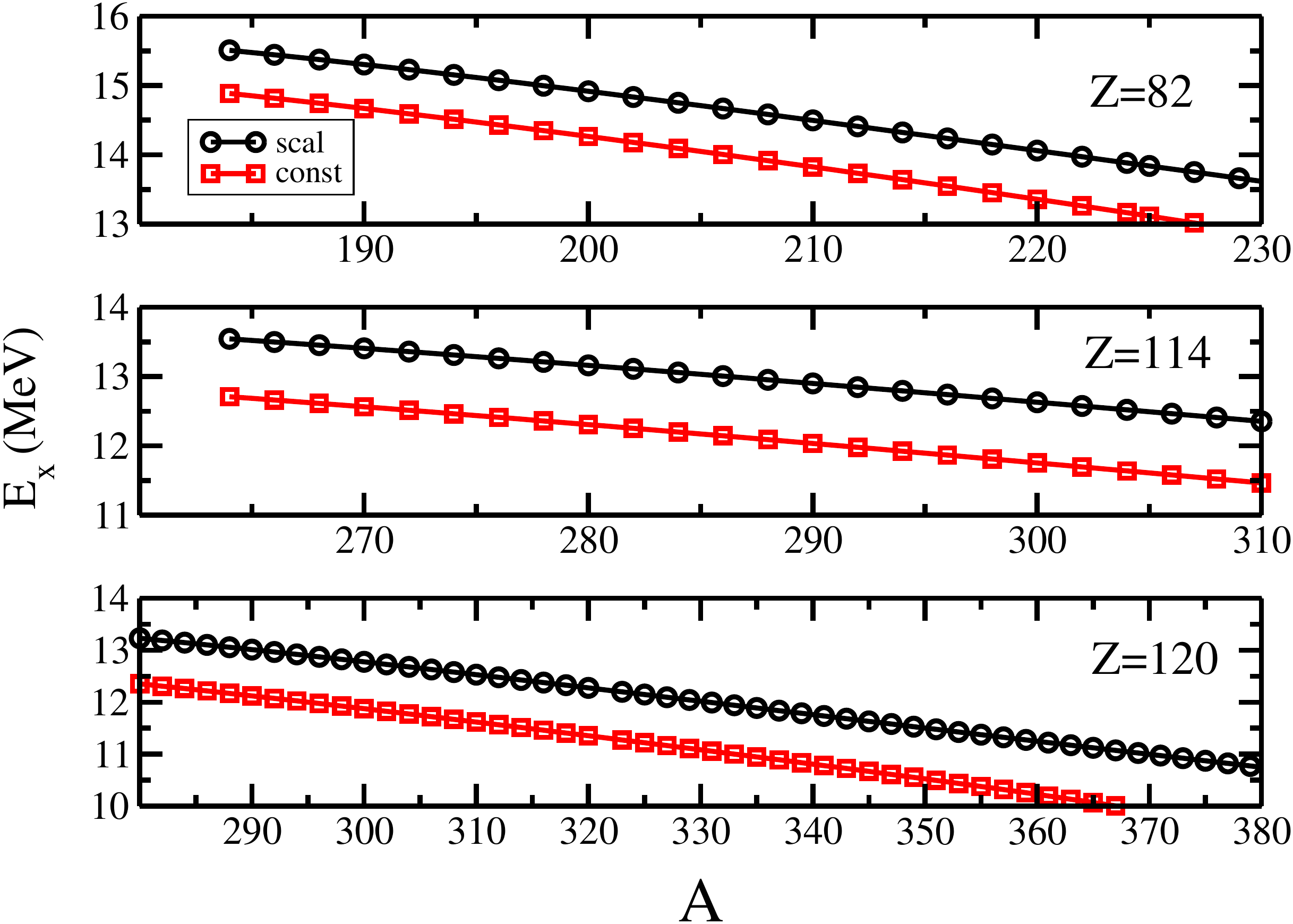}
\caption{\label{fig2:gmr}The excitation energy isoscalar giant monopole 
resonance (ISGMR) for Pb, Z=114, and Z=120 isotopes starting from proton 
to neutron drip-lines as a function of mass number. }
\end{figure}
\subsection{Isoscalar giant monopole resonance}

It is well understood that the ISGMR
 has a direct relation to the incompressibility of nuclear
matter, which decides the softness or stiffness of an equation of
state \cite{blaizot}. This EOS also estimates the structure of neutron stars, such as
mass and radius.
Thus, the ISGMR is an intrinsic property of finite nuclei as well
as nuclear equations of state and needed to be determined to shine
some light into nuclear properties. The excitation energies of ISGMR for 
O, Ca, Ni, Sn, Pb, Z=114, and Z=120 isotopic series are given in Figs. 
\ref{fig1:gmr} and \ref{fig2:gmr}. The results are calculated by using 
both constrained and scaling approaches in the isotopic chain, starting 
from proton to neutron drip-lines. We use the relation 
$E_m^s=\sqrt{\frac{AK_A^s}{B_m}}$ with the mass
parameter $B_m=\int d{\bf r}r^2{\cal H}$. The figure shows that 
excitation energy  obtained from scaling calculation is always
greater than the constrained value. The difference
between the monopole excitation of scaling and constrained calculations,
generally gives the resonance width $\Sigma = \frac{1}{2}\sqrt{E_3^2-E_1^2}$,
with $E_3=\sqrt{\frac{m_3}{m_1}}$ and $E_1=\sqrt{\frac{m_1}{m_1}}$ in terms
of the ratios of the integral moments
$m_k=\int_0^{\infty}d\omega\omega^KS(\omega)$ of the random phase 
approximation (RPA) strength function
$S(\omega)$ \cite{mario92}. It is also equivalent to
$m_1=\frac{2}{m}A<r^2>$ and from dielectric theorem, we have
$m_{-1}=-\frac{1}{2}A(\frac{\partial R_{\eta}^2}{\partial\eta}){\bigg |_{\eta=0}}$.

Now consider Fig. \ref{fig1:gmr}, where the excitation energies of giant 
isoscalar monopole resonance $E_x$ for lighter mass nuclei are plotted. 
For Z=8 the excitation energy decreases towards both the proton (A=12, 
${E_x}^s$= 22.51 MeV) and neutron drip-lines (A=26, ${E_x}^s$ =21.22 MeV). 
Excitation energy has a maximum value near N=Z (here it is a double 
closed isotope with Z=8, N=8, ${E_x}^s$= 27.83 MeV). Similar trend
is followed in the isotopic chain of Ca with Z=20. We find the maximum excitation
energy at $^{40}$Ca (${E_x}^s={24.07}$ MeV), whereas ${E_x}^s$ is found to be smaller both in the proton (A=34, ${E_x}=23.31$ MeV) and neutron drip-lines (A=71, 
${E_x}^s=16.80$ MeV). However, the trends are different for isotopic chains 
of higher Z like Z=50, 82, 114, and 120. In these series of nuclei, the 
excitation energy monotonically decreases starting from proton drip-line 
to neutron drip-line. For example, $^{180}$Pb and $^{280}$Pb are the proton and neutron drip nuclei having excitation energy ${E_x}=15.63$ and  
${E_x}=11.45$ MeV, respectively. Fig. \ref{fig2:gmr} shows clearly the 
monotonic decrease of excitation energy for super heavy nuclei. This 
discrepancy between super heavy and light nuclei may be due to the Coulomb 
interaction and the large value of isospin difference. For lighter values of Z, 
the proton drip-line occurs at a combination of proton and neutron where 
the neutron number is less than or near to the proton number. But for
higher Z nuclei, the proton drip-line is exhibited at a larger isospin.
As the excitation energy of a nucleus is a collective property, it varies 
smoothly with its mass number, which is also reflected in the figures. Consider 
the isotopic chain of Z=50: the drip-line nucleus (A=100) has  excitation 
energy 18.84 MeV and the neutron drip nucleus A=171 has $E_x=13.39$ MeV. 
The difference in excitation energy between these two isotopes is 5.32 MeV. 
This difference in proton and neutron drip nuclei is 4.31 MeV for Z=82 and 
is 2.37 MeV in Z=114. In summary, for higher Z nuclei, the variation 
of excitation energy in an isotopic chain is less than for a lighter Z nucleus. 
Again, by comparing with the empirical formula of $E_x=CA^{-1/3}$, our 
predictions show similar variation throughout the isotopic chains. 
Empirically, the value of $C$ is found to be 80 \cite{bert76}.   However, if 
we select $C=70-80$ for lighter mass isotopes and $C=80-86$ for the super heavy 
region, then that fits well with our results, which are slightly different 
than C=80 obtained by fitting the data for stable nuclei \cite{bert76}.

\begin{table}
\caption{The calculated excitation energies in the RETF formalism with
FSUGold parameter set  are  compared with other theoretical formalisms
and experimental data \cite{young99,young01,younga01}. }
\renewcommand{\tabcolsep}{0.02cm}
\renewcommand{\arraystretch}{1.0}
\begin{tabular}{|cccc|}
\hline
\hline
Nucleus &    Formalism (parameter) &   Excitation energy&
\\
\hline
$^{16}$O & RPA(FSUGold)& 23.09 & \\
         & CRMF(FSUGold) & 22.89& \\
         & RETF(FSUGold)& 25.17  & \\
        & Expt. & 21.13$\pm$0.49          &  \\
$^{40}$Ca& RPA(FSUGold) & 20.67  & \\
         & CRMF(FSUGold) & 20.67 &  \\
         & RETF(FSUGold)& 22.30  &   \\
         & Expt. & 19.18$\pm$0.37          &    \\
$^{90}$Zr& RPA(FSUGold) &17.44   &     \\
         & CRMF(FSUGold) & 17.70 &      \\
         & RETF(FSUGold)& 18.21  &       \\
         & Expt. & 17.89$\pm$0.20          &        \\
$^{208}$Pb& RPA(FSUGold) &13.76  &         \\
         & CRMF(FSUGold) & 13.50 &          \\
         & RETF(FSUGold)& 13.44   &           \\
         &Pairing+MEM &14.0  &            \\
         &Expt.& 13.5$\pm$0.1            &             \\
$^{204}$Pb& RETF(FSUGold) & 13.6&              \\
         & Pairing+MEM & 13.4&              \\
         & Expt.       & 13.7$\pm$0.1    &                \\
$^{206}$Pb& RETF(FSUGold) &13.51 &                 \\
       & Pairing+ MEM& 13.4&                 \\
         & Expt.     &   13.6$\pm$0.1       &                  \\
\hline
\end{tabular}
\label{table4}
\end{table}

\begin{figure}
\includegraphics[scale=0.55]{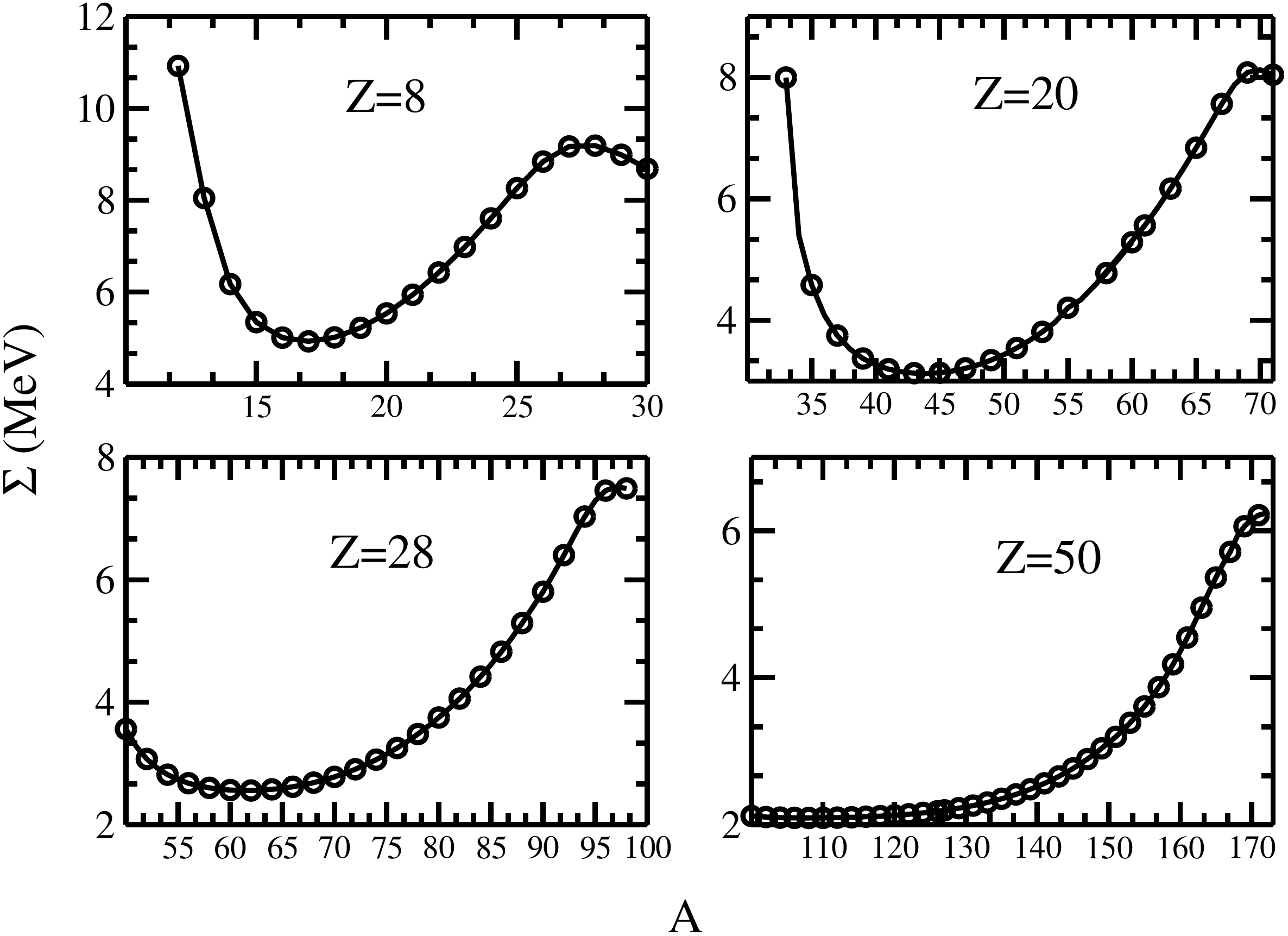}
\caption{\label{fig3:sigma} The difference between the monopole excitation
energies of scaling and constrained calculations $\Sigma = \frac{1}{2}\sqrt{E_3^2-E_1^2}$
as a function of mass number A for O, Ca, Ni, and Sn. }
\end{figure}

\begin{figure}
\includegraphics[scale=0.62]{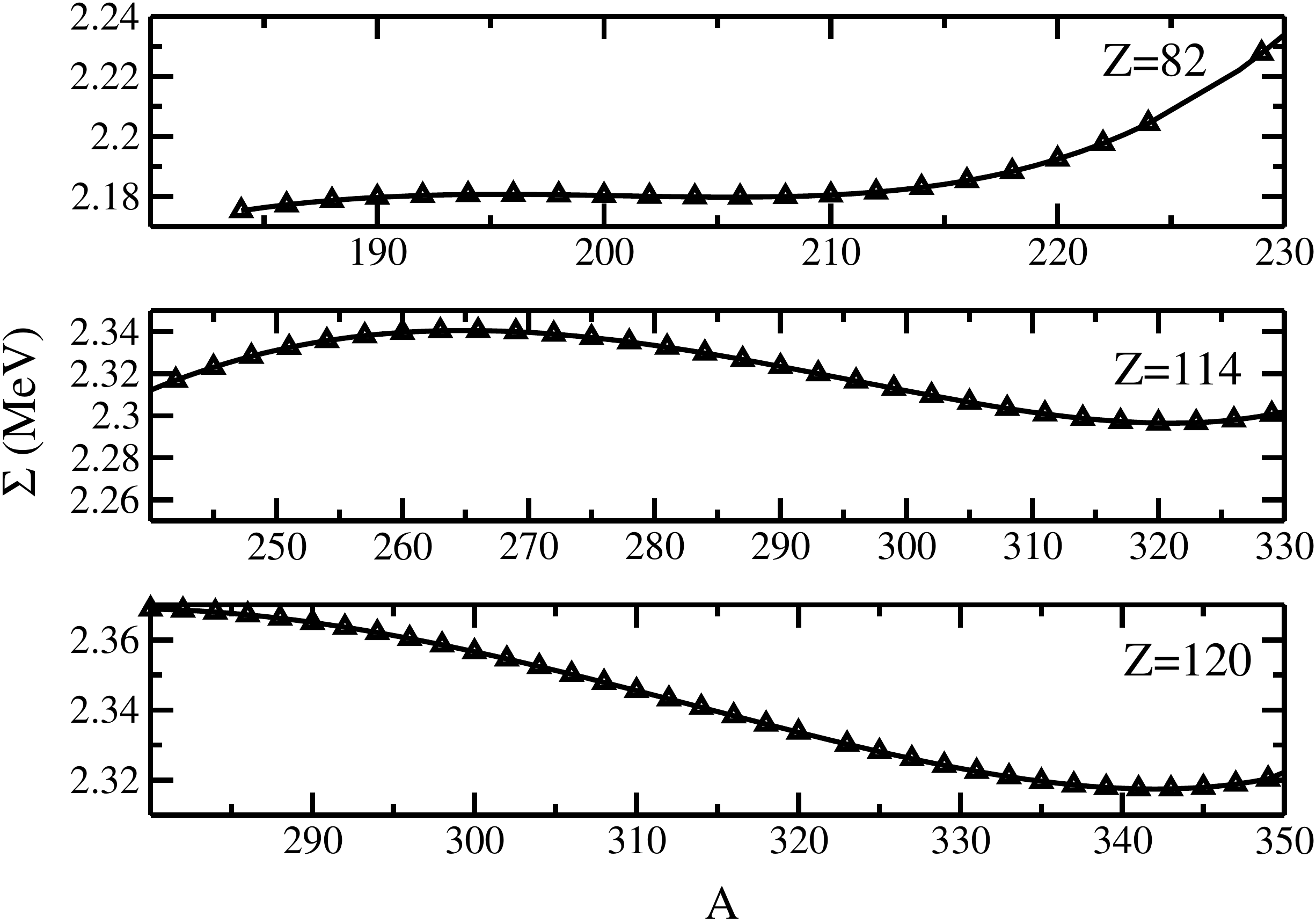}
\caption{\label{fig4:sigma} Same as Fig. \ref{fig3:sigma}, but for Pb, Z=114, and Z=120.
}
\end{figure}
In table~\ref{table4}, we have shown the results obtained in the RETF formalism with the
FSUGold parameter set and compared with other predictions like
the random phase approximation (RPA) and constrained relativistic mean filed (CRMF)\cite{wei13}. The experimental data are also given, where
ever available, for comparison. A comparative study of these 
results shows that for light nuclei such as $^{16}{O}$, the RPA 
and CRMF give better results than the semi-classical RETF
 approximation. For  example, the difference between the excitation 
energies of $^{16}O$ in the RPA and RETF formalism is $\sim$ 
2 MeV and it is only $~$0.1 MeV for $^{208}Pb$. This implies, 
for heavy nuclei, that the RETF gives comparable results with RPA and 
CRMF. On the other hand, the RETF results obtained by constrained 
calculation are within the experimental error bar. In RETF, the quantal 
correction is averaged out. Thus, the RETF result of monopole 
excitation energy differs from the 
RPA prediction for light nuclei, where quantal correction has a
significant role in structural properties. In heavy nuclei, the 
number of nucleons are larger, so the quantal and semi-classical 
approaches are almost similar. This could be a reason for the accuracy 
of application of semi-classical calculation to heavy mass nuclei. We have also 
compared our results with Pairing+MEM (mutually enhanced magicity)
\cite{khana09} and experimental data\cite{dpatel13} for $Pb$
isotopes. The Pairing+MEM theory says that a magic nucleus like
$^{208}Pb$ has a little higher excitation energy than its neighboring
isotopes. But recent experimental data are not in favor of the manifestation
of such an effect. Our theoretical results also come to the same conclusion.

There is no direct way to calculate $\Sigma$ in the scaling or constrained
method as in RPA. If we compare the excitation energy
obtained from scaling calculations with the non-relativistic RPA result,
then it is evident that the scaling gives the upper limit of the energy
response function. On the other hand, the constrained calculation predicts
the lower limit \cite{bohigas79}. As a result, the resonance width 
$\Sigma$ is obtained from the root mean square difference of $E_x^s$ and $E_x^c$.
We have plotted $\Sigma $ for the light nuclei in Fig. \ref{fig3:sigma} 
and for super heavy in Fig. \ref{fig4:sigma}. For lighter nuclei, $\Sigma$ is 
larger both in the proton and neutron drip-lines. As one proceeds from proton to 
neutron drip-line, the value of $\Sigma$ decreases up to a zero isospin 
combination (N=Z or double closed) and then increases. For example, 
$\Sigma$= 10.92, 5.0 and 8.9 MeV for $^{12}{O}$, $^{16}O$, and $^{26}O$, 
respectively. Similar trends are also followed by the medium nuclei Z=20, 28, and 50 isotopic 
chains. This conclusion can be drawn from the results of the
excitation energy also (see Figs. \ref{fig1:gmr} and \ref{fig2:gmr}), i.e., 
the difference between the scaling and constrained excitation energies are 
more in proton and neutron drip-lines as compared to the Z=N region.
The value of $\Sigma$ in an isotopic chain depends very much on the proton 
number. It is clear from the isotopic chains of $\Sigma$ for O, Ca, Ni, Sn, 
Pb, and Z=114, 120. All of the considered series have their own behavior and 
show various trends. Generally, for lighter elements, it decreases
initially to some extent and again increases monotonically.
On the other hand, for heavier nuclei like Pb, Z=114, or Z=120 this characteristic 
of $\Sigma$ differs with different mass number and can be seen in Fig. 
\ref{fig4:sigma}.
\subsection{Incompressibility of finite nuclei}
\begin{figure}
\includegraphics[scale=0.55]{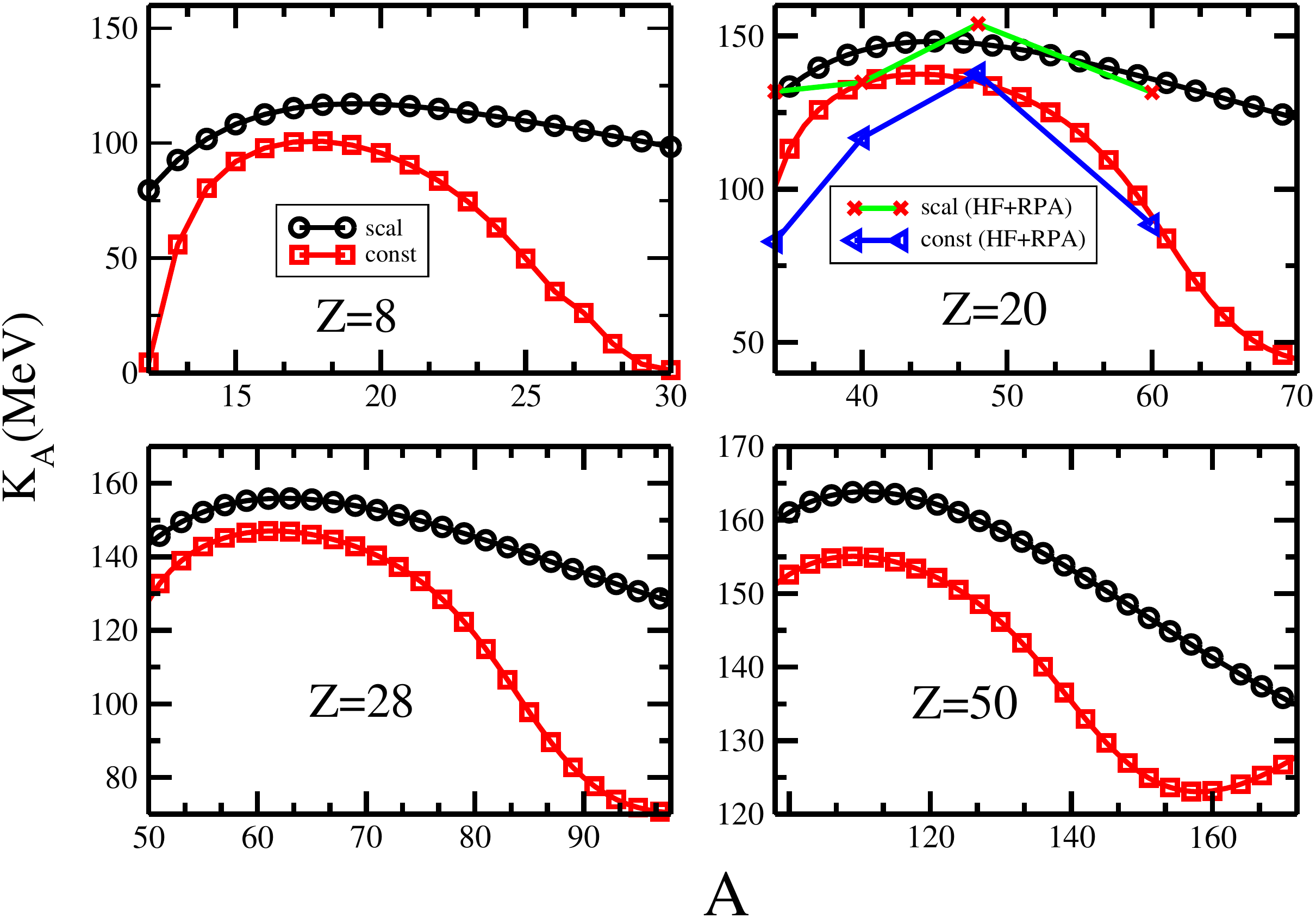}
\caption{\label{fig5:comp} The incompressibility modulus obtained by both
scaling and constrained approaches in the isotopic series of O, Ca, Ni, and Sn.
}
\end{figure}

\begin{figure}
\includegraphics[scale=0.55]{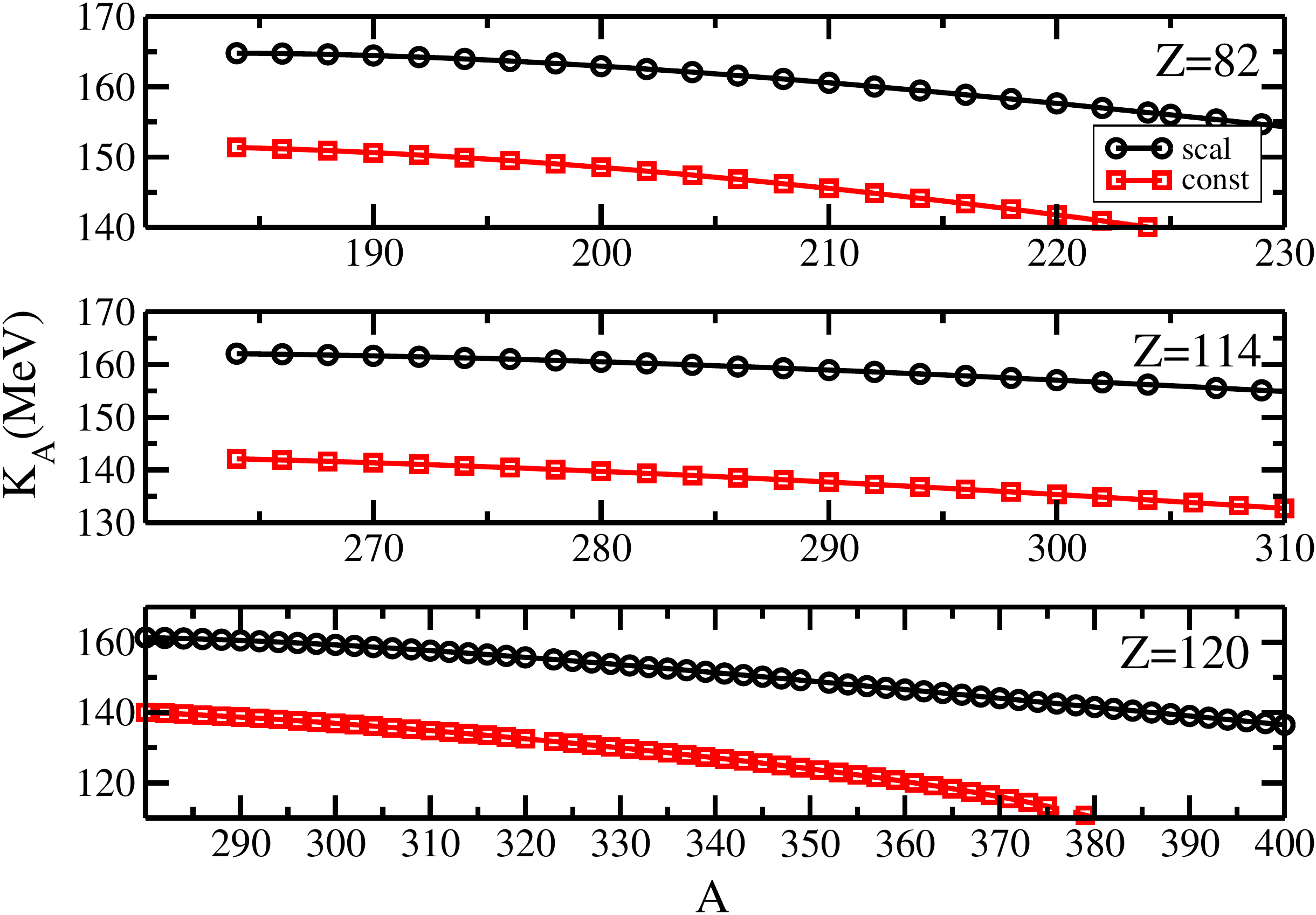}
\caption{\label{fig6:comp} Same as Fig. \ref{fig5:comp}, but for Pb, Z=114, and Z=120.}
\end{figure}
The nuclear matter incompressibility $K_{\infty}$ is a key quantity in the 
study of the equation of state. It is the second derivative of the energy 
functional with respect to density at the saturation and is defined as
$K_{\infty}=9\rho_0\frac{\partial^2\cal E}{\partial\rho^2}|_{\rho=\rho_0}$, 
which has a fixed value for a particular force parametrization. It is well 
understood that a larger $K_{\infty}$ of a parameter set gives a stiff EOS
 and produces a massive neutron star. It has also a direct 
relation to the asymmetry energy coefficient $J$ of the parameter set
\cite{estal01}. In the limit A approaches infinitely large, the finite 
nucleus can be approximated to an infinite nuclear matter system (N=Z for symmetry and $N\neq Z$ for asymmetry matter). Thus, it is instructive to study 
the nature of the incompressibility of a finite nucleus $K_A$ in the isotopic 
chain of super heavy nuclei. Here, we calculate the $K_A$ as a function of mass number for the
light nuclei considered in the present study (O, Ca, Ni, Sn) and then
extend the calculations to Pb, Z=114 and Z=120 in the super heavy region.
Our calculated results are shown in Figs. \ref{fig5:comp} and \ref{fig6:comp}.
The incompressibility of finite nuclei follows the same trend as the excitation 
energy. For light nuclei, the incompressibility has a small value for the proton 
and neutron drip-lines, whereas it is maximum in the neighborhood of double 
closed combinations.

It can be easily understood from Fig. \ref{fig5:comp} that, at some particular proton to neutron combination, the $K_A$ is high, i.e., at this combination of 
N and Z, the nucleus is more incompressible. In other words, the larger the 
incompressibility of a nucleus, the more compressible it will be. Here,
it is worthy to mention that the nuclear system becomes less compressible near both the neutron and proton drip-lines. This is because of the  
instability originating from the repulsive part of the nuclear force, 
revealing a large neutron-proton ratio, which progressively
increases with the neutron/proton number in the isotope without much 
affecting the density \cite{satpathy04}. Similar to the excitation 
energy, it is found that $ K_A$ obtained by the scaling method
is always higher than the constrained calculation. The decrease in
incompressibility near the drip-line regions is more prominent in constrained 
calculation than the scaling results. From leptodermous expansion 
\cite{blaizot}, we can get some basic
ideas about this decrease in the vicinity of drip-lines. The expression 
for finite nucleus incompressibility can be written as
\begin{eqnarray}
K_A=K_{\infty}+K_{Sur}A^{-{\frac{1}{3}}}+K_{\tau}{I^2}+K_{Coul}{Z}^2{A^{-\frac{4}{3}}},
\end{eqnarray}
where $ I={\frac{N-Z}{A}}$. The coefficient $K_{\tau}$ is negative, so 
incompressibility decreases with ${N-Z}$. For the Ca chain, the incompressibilities 
obtained by scaling and constrained calculations are compared with the 
Hartree-Fock plus RPA results \cite{blaizot} in Fig. \ref{fig5:comp}. From 
Fig. \ref{fig5:comp}, one can see that $K_A$ evaluated by the semi-classical
approximation deviates from RPA results for lighter isotopes, contrary to the
excellent matching for the heavier Ca nuclei. This is because 
of the exclusion of the quantal correction in the semi-classical
formalism. For higher mass nuclei, this correction becomes negligible and 
compares to the RPA predictions. This result is depicted in Fig. 
\ref{fig6:comp} for Pb and super heavy chain of nuclei. Here the results 
show completely different trends than for the lighter series.
The incompressibility has a higher value in the vicinity of the proton drip-line and
decreases monotonically towards the neutron drip-line. Because,
for high Z-series, the proton drip-line appears at a greater value of N in 
contrast to the lighter mass region. Again, the incompressibility decreases with
neutron number from the proton to neutron drip-lines.
\begin{figure}
\includegraphics[scale=0.55]{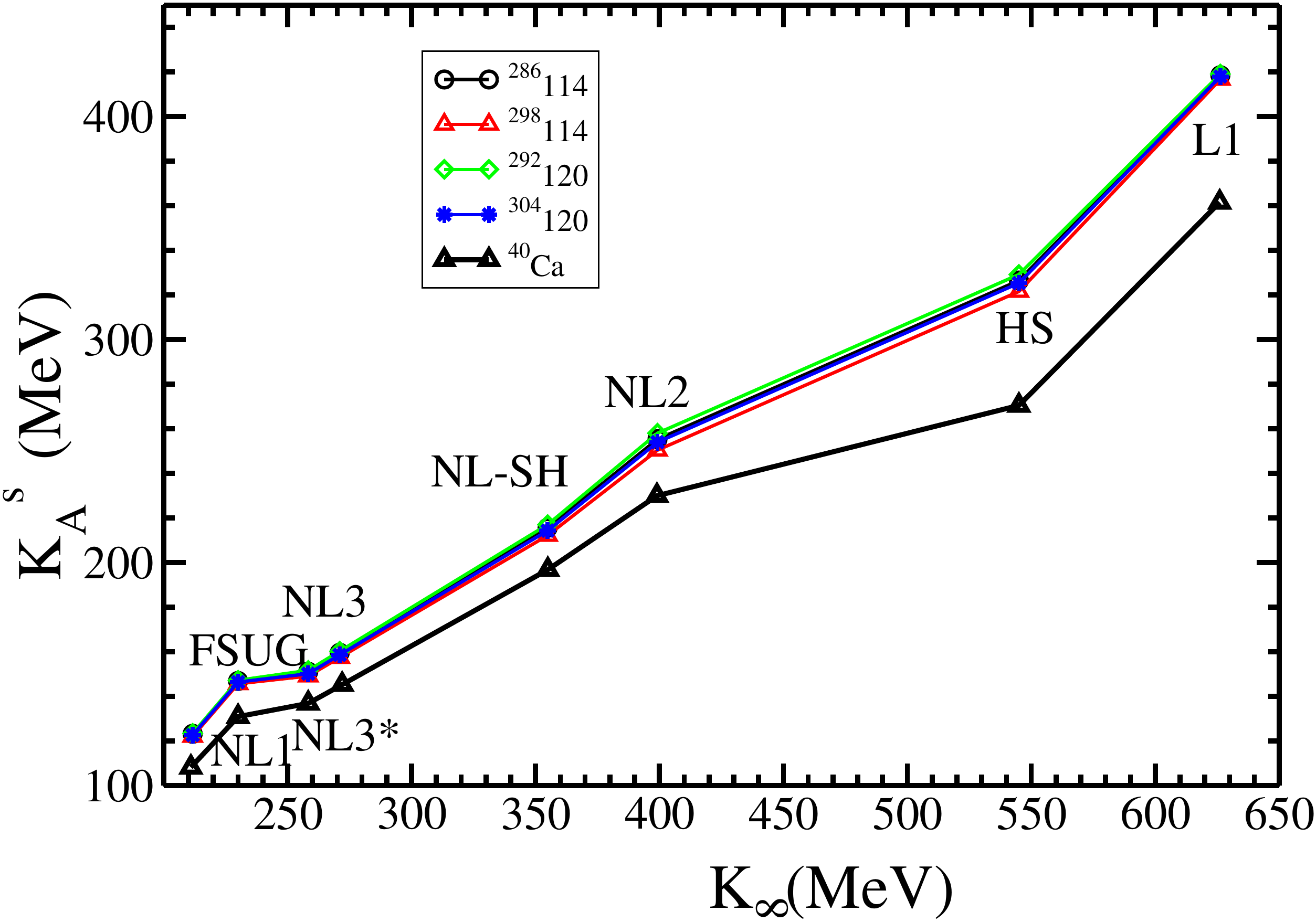}
\caption{\label{fig7:comp} Compressibility for finite nuclei obtained by scaling
calculation $K_A^{s}$ versus nuclear matter incompressibility $K_{\infty}$.}
\end{figure}
Finally, we would like to see the trend of $K_A$ with nuclear matter
incompressibility for various force parameters and also with the size of 
a nucleus which can reach the infinite nuclear matter limit. For this we 
choose  $^{286,298}114$, $^{292,304}120$, and $^{40}$Ca as the selected 
candidates as shown in Fig. \ref{fig7:comp}. Although, the super heavy
nuclei approach the nuclear matter limit, we can not reproduce the
$K_{\infty}$ from $K_A$. This may be due to the asymmetry needed to
form a bound nucleus, which is the reason for the deviation. That means the
asymmetry $I$ of $K_A$ and $K_{\infty}$  differs significantly
(where $I=\frac{N-Z}{N+Z}$), which is the main source
of deviation of $K_A$ from  $K_{\infty}$. Also, this deviation arises due to the
surface contribution of the finite nuclei.  For a quantitative estimation,
we have calculated the $K_A^{s}$ for different force parameters having various
$K_{\infty}$ at saturation. We find almost a linear variation of $K_A^{s}$
with $K_{\infty}$ for the considered nuclei as shown in Fig. \ref{fig7:comp}. 
For Ca isotopes also we find variation similar in nature, but with smaller $K_A$ than the
super heavy nuclei.

\section{Application of scaling and constrained formalism to Sn isotopes}
In this part of the chapter, we applied the scaling and constrained formalism to 
analyse the anamolous nature of Sn Isotopes. 
In the previous section, we have discussed the excitation energy and incompressibility of super heavy and drip line nuclei.  The recent experiment on the isotopic chain of Sn isotopes indicates the mismatch between the theoretical prediction and the experimental data of the excitation energy of isoscalar giant monopole resonance energy.  This  problem is addressed in the name of  " why the tin nucleus is so floppy ?".  So theoretical  physicists are  now more interested in the medium  heavy region (A$\sim$100).  In this section the excitation energy and the incompressibility  isotopes of Cd and  Sn  nuclei are discussed, within the framework of relativistic Thomas-Fermi and relativistic
extended Thomas-Fermi approximations. A large number of non-linear relativistic force parameters are used in the calculations that a parameter set is capable of reproducing the experimental monopole energy of Sn isotopes,
when its nuclear matter incompressibility lies within $210-230$ MeV, however, it fails to reproduce the GMR energy of other nuclei.
That means, simultaneously a parameter set can not
reproduce the GMR values of Sn and other nuclei.

                  Incompressibility of nuclear matter, also knows as compression modulus $K_{\infty}$ has a 
special interest in  nuclear and astro-nuclear physics, because of its 
fundamental role in guiding the equation of state (EOS) of nuclear matter.
The $K_{\infty}$ can not be measured by any experimental technique 
directly, rather it depends indirectly on the experimental measurement of 
isoscalar giant monopole resonance (ISGMR) for its confirmation \cite{blaizot}. 
This fact enriches the demand of correct measurement of excitation energy of 
ISGMR. The relativistic parameter with random phase approximation (RPA) 
constraints the 
range of the incompressibility $270\pm{10}$ MeV \cite{lala97,Vrete02} for 
nuclear matter. Similarly, the non-relativistic formalism with Hartree-Fock 
(HF) plus RPA allows the acceptable range of incompressibility  
$210-220$ MeV, which is less than the relativistic one. 
It is believed that the part of this discrepancy in the acceptable range of 
compressional modulus comes from the diverse behavior of the density 
dependence of symmetry energy in relativistic and non-relativistic formalism 
\cite{pieka02}. Recently, both the  formalism
come to a general agreement on the value of nuclear matter incompressibility 
i.e., $K_{\infty}=240\pm{10}$ MeV \cite{colo04,fsu,Agrawal05}. But the new experiment 
on Sn isotopic series i.e., $^{112}$Sn$-^{124}$Sn rises the question 
"why Tin is so fluffy ?"\cite{tli07,ugarg07,pieka07}. This 
question indicates toward the correct theoretical investigation 
of incompressibility. 
Thus, it is worthy to investigate the incompressibility  
in various theoretical formalisms. Most of the relativistic and non-relativistic
 models reproduce the strength distribution very well for medium 
and heavy nuclei, like $^{90}$Zr and $^{208}$Pb, respectively. But at the same 
time it overestimates the excitation energy of Sn around 1 MeV. This low value
 of excitation energy demands lower value of nuclear matter incompressibility. 
Till date, lots of effort have been devoted to solve this problem like 
inclusion of pairing effect 
\cite{citav91,gang12,jun08,khan09a,khan10,pieka14,vesel12}, mutually enhanced magicity (MEM) etc. \cite{khana09}. 
But the pairing effect reduces the theoretical excitation energy 
only by 150 KeV in Sn isotopic series, which may not be sufficient 
to overcome the puzzle. Similarly, new experimental data are not in favor 
of MEM effect \cite{dpatel13}. Measurement
on excitation energy of $^{204,206,208}$Pb shows that the MEM effect should 
rule out in manifestation of stiffness of the Sn isotopes. 
Here, we use the relativistic Thomas-Fermi (RTF) and  
relativistic extended Thomas-Fermi (RETF) 
\cite{acentel93,acentel92,speich93,centel98,centell93} 
with scaling and constraint approaches in the frame-work of non-linear 
$\sigma-\omega$ model \cite{boguta77}.

\section{Analysis of Sn results}

\label{mplsec3}
We calculate the GMR energy using both the scaling and constraint 
methods in the frame-work of relativistic
extended Thomas-Fermi approximation using various parameter sets
for Cd and Sn nuclei and compared with 
the excitation energy with moments ratio $({m_3/m_1})^{1/2}$ 
and $({m_1/m_{-1}})^{1/2}$ obtained  
from multipole-decomposition analysis (MDA). 
\begin{figure}
\includegraphics[scale=0.62]{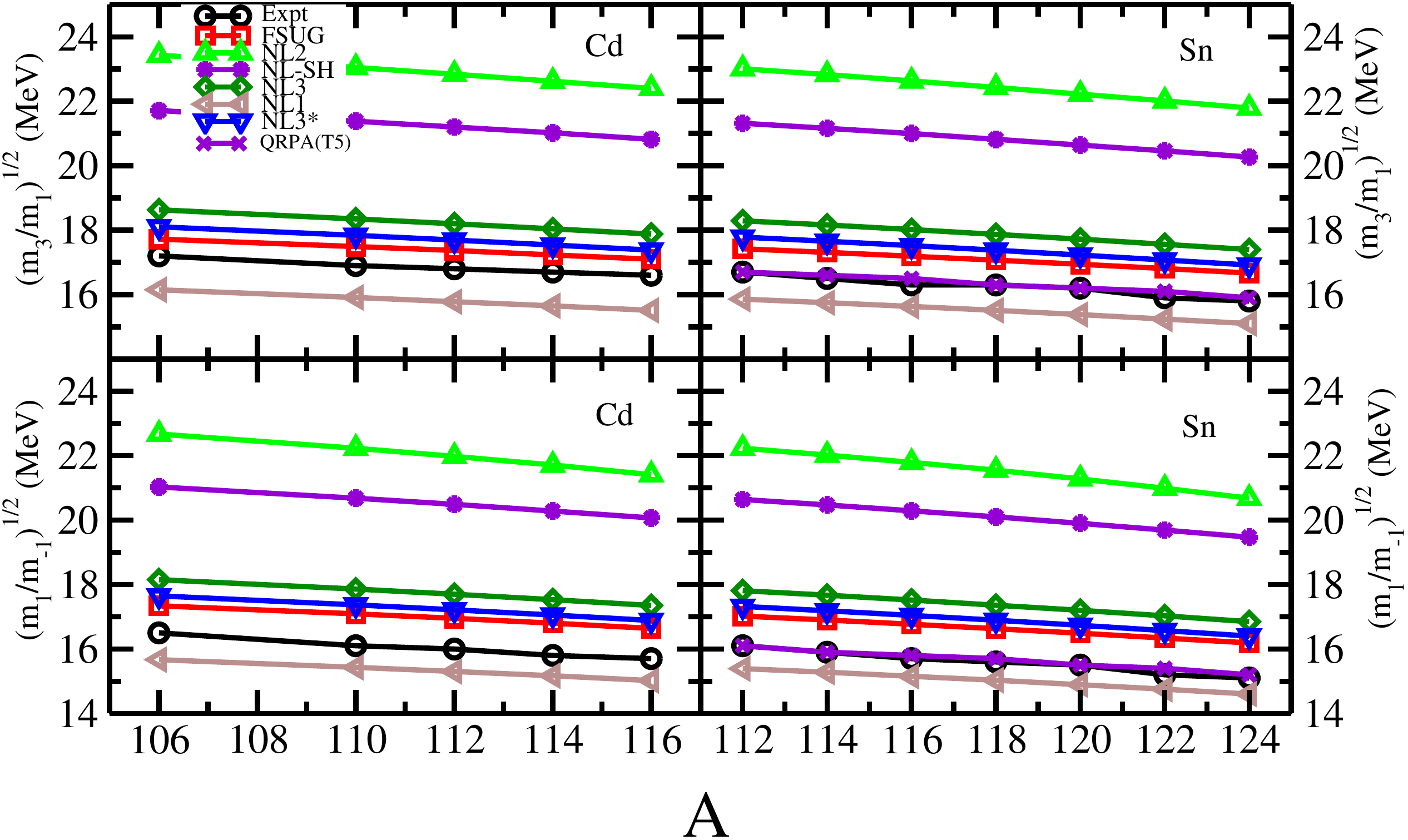}
\caption{\label{mplfi1}The isoscalar giant monopole excitation energy obtained
by scaling method with various parameter sets are compared with
experimental data of $\sqrt{{\frac{m_3}{m_1}}}$ and $\sqrt{{\frac{m_1}{m_{-1}}}}$
for Cd and Sn isotopes \cite{tli07,ugarg07}. The upper and lower panels are   
$\sqrt{{\frac{m_3}{m_1}}}$ and $\sqrt{{\frac{m_1}{m_{-1}}}}$, respectively.
}
\end{figure}

\begin{figure}
\includegraphics[scale=0.55]{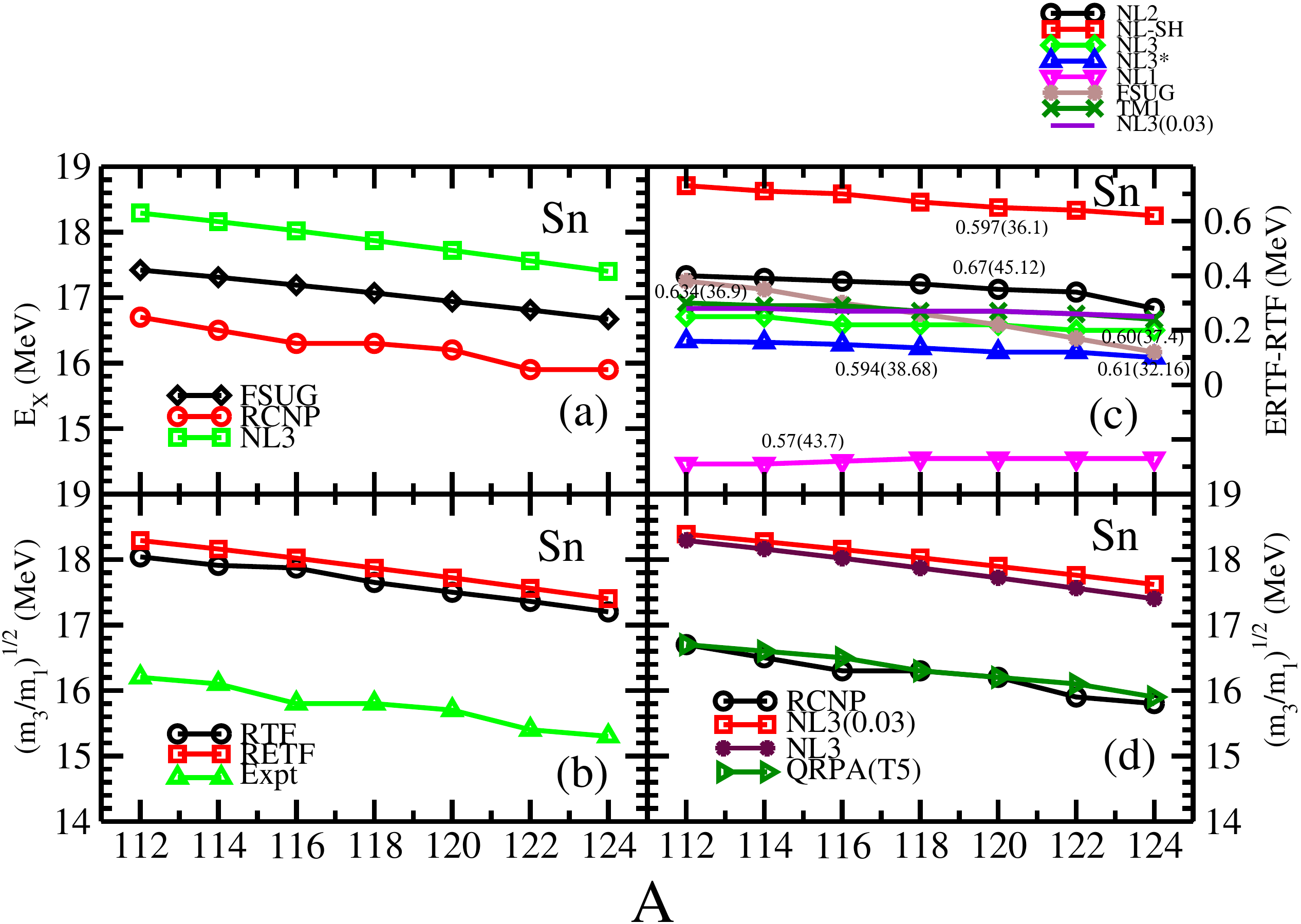}
\label{mplfi2}
\caption {(a) The variation of giant monopole excitation energy $E_x$ with 
mass number A for Sn isotopes, (b) the scaling monopole excitation energy within
RETF and RTF formalisms compared with the experimental moments
ratio ${\sqrt{{m_3/m_1}}}$ \cite{dpatel13}, (c) variation of the difference
of giant monopole excitation energy obtained from RETF and RTF
($\triangle{E}=RETF-RTF$) formalisms with various parameter sets for Sn 
isotopic chain, (d) moments ratio ${\sqrt{{m_3/m_1}}}$ for Sn isotopes obtained 
with NL3+$\Lambda_V$ is compared with NL3, QRPA(T5) and experimental 
data.}
\end{figure}

The basic reason 
to take a number of parameter sets is that the infinite nuclear matter 
incompressibility of 
these forces cover a wide range of values. For example, NL-SH has 
incompressibility 399 MeV, while that of NL1 is 210 MeV. From MDA analysis, we 
get different moments ratio, such as $m_3/m_1$, $m_0/m_1$ and $m_1/m_{-1}$.
These ratios are connected to scaling, centroid and constraint 
energies, respectively. 
In Fig.~\ref{mplfi1},  we have shown the $(m_3/m_1)^{1/2}$ and 
$(m_1/m_{-1})^{1/2}$ ratio for isotopic chains of Cd and Sn. The results are 
also compared with experimental data obtained from Research Center for 
Nuclear Physics (RCNP) \cite {tli07,ugarg07}. From 
the figures, it is cleared that the experimental values lie between the 
results obtained from FUSG (FSUGold) and NL1  force parameters. 
If one compares the experimental 
and theoretical results for $^{208}$Pb, the FSUG set gives better
results among all. For example, the experimental data and theoretical 
result for monopole energies are 14.17$\pm$0.1 and 14.04 MeV, respectively. 
From this, one could conclude that the infinite 
nuclear matter incompressibility lies nearer to that of FSUG (230.28 MeV) 
parametrization. But experimental result on ISGMR in RCNP shows that the predictive 
power of FUSG is not good enough for the excitation energy of Sn 
isotopes. This observation is not only confined to RCNP data, 
but also persists in the more sophisticated RPA. 
In Table~\ref{mplta1},  we have given the 
results for QRPA(T6), RETF(FSUG) and RETF(NL1). The experimental data are also
given to compare all these theoretical results.

\bigskip
\begin{table}
\caption{\label{mplta1}Moments ratio for Sn isotopes using RETF
approximation with FSUGold and NL1 sets are compared with QRPA(T6)
predictions \cite{tsel09}.}
\begin{tabular}{ccccccccc}
Nucleus &\multicolumn{4}{c}{$({m_3/m_1})^{1/2}$(MeV)}&\multicolumn{4}{c}{$({m_1/m_{-1}})^{1/2}$(MeV)}\\ \hline
 &QRPA&RETF&RETF& Expt. & QRPA&RETF&RETF& Expt. \\ 
& (T6)    & (FSUG) & (NL1) &  &   T6)    & (FSUG) & (NL1) & \\
$^{112}$Sn&17.3& 17.42&15.86 &16.7& 17.0&17.2&15.39&16.1\\
$^{114}$Sn&17.2& 17.32&15.75 &16.5& 16.9&16.9&15.28&15.9 \\
$^{116}$Sn&17.1& 17.19&15.63&16.3&16.8&16.77&15.15&15.7 \\
$^{118}$Sn&17.0&17.07&15.51&16.3&16.6&16.63&15.03&15.6 \\
$^{120}$Sn&16.9&16.94&15.38 &16.2& 16.5&16.44&14.89&15.5 \\
$^{122}$Sn& 16.8&16.81 &15.24 &15.9& 16.4& 16.34&14.75&15.2\\
$^{124}$Sn&16.7&16.67&15.1&15.8&16.2&16.19&14.6&15.1\\ 
\hline
\end{tabular}
\end{table}
The infinite nuclear matter incompressibility $K_{\infty}$ with T6 parameter set 
\cite{tsel09} is 236 MeV and that of FSUG is 230.28 MeV. 
The similarity in incompressibility (small difference of 6 MeV in $K_{\infty}$)
may be a reason for their prediction
in equal value of GMR. The table shows that, there is only 0.1 MeV difference 
in QRPA(T6) and RETF(FSUG) results in the GMR values for 
$^{112}$Sn$-$$^{116}$Sn  isotopes, but the results  exactly match 
for the $^{118}$Sn$-$$^{124}$Sn isotopes. This implies 
that for relatively higher mass nuclei, both the QRPA(T6) and RETF(FSUG) 
results are almost similar. If some one consider the experimental values 
of Sn isotopic series, then QRPA(T5) gives better result. For example,
experimental value of $(m_3/m_1)^{1/2}$ for $^{112}$Sn is 16.7$\pm$ 0.2 MeV 
and that for QRPA(T5) is 16.6 MeV. 
It is shown by  V. Tselyaev et al. \cite{tsel09} that the T5 parameter set 
with $K_{\infty}=202$ MeV, better explains the excitation energy of Sn 
isotopes, but fails to predict the excitation energy of $^{208}$Pb.
The experimental data of 
ISGMR energies for $^{90}$Zr and $^{114}$Sn lies in between the calculated 
values of T5 and T6 forces. 
In brief, we can say that the RPA analysis predicts the 
symmetric nuclear matter incompressibility within $K_{\infty}=202-236$ 
MeV and our 
semi-classical calculation gives it in the  range $210-230$ MeV.
These two predictions almost agree with each other in the acceptable
limit.
 
\begin{table}
\caption{\label{mplta2}Moments ratio ${\sqrt{{m_1/m_{-1}}}}$ for Pb 
isotopes within  RETF is compared  with pairing+ MEM results and 
experimental data.}
\begin{tabular}{cccccccccccc}
Nuclear Mass &\multicolumn{3}{c}{(${m_1/m_{-1}})^{1/2}$ (MeV)} 
&\multicolumn{2}{c}{$\Sigma$(MeV)}\\ \hline
 &pairing+MEM&our work&Expt.&our work & Expt.\\ \hline
$^{204}$Pb&13.4&13.6&13.7$\pm$0.1&2.02&3.3$\pm$0.2 \\
$^{206}$Pb&13.4&13.51&13.6$\pm$0.1&2.03& 2.8$\pm$0.2 \\
$^{208}$Pb&14.0&13.44&13.5$\pm$0.1&2.03&3.3$\pm$0.2  \\
\hline
\end{tabular}
\end{table}
In Table~\ref{mplta2}, we have displayed the data obtained from a recent experiment
\cite{dpatel13} and compared our results. Column two of the table is also devoted 
to the result obtained from pairing plus MEM effect \cite{khana09}. 
The data show clearly that our formalism (RETF) predicts the excitation energy 
more accurately than the result obtained by Pairing+ MEM prediction for Pb 
isotopes. For example, the difference between the 
pairing+MEM result and experimental observation is 0.3 MeV for $^{204}$Pb, 
which is away from the experimental error. It is 
only 0.1 MeV (within the error bar) in RETF calculation. 
This trend also follows for $^{206}$Pb and $^{208}$Pb. 
As it is mentioned earlier,
we have not included  the pairing correlation externally, yet the result is
good enough in comparison with MEM+pairing as pairing has a marginal role
for the calculation of collective excitation.  
Fig.~\ref{mplfi2}(a) shows the variation of excitation energy with mass number A for 
Sn isotopes. This results can also be treated as the monopole excitation energy $E_x$ with proton-neutron asymmetry ($I=\frac{N-Z}{N+Z}$). 
The graph shows that the variation of monopole excitation energy $E_x$ 
with both NL3 and FSUGold are 
following similar pattern as experimental one with a
different magnitude.
In Fig.~\ref{mplfi2}(b),  we have compared our results obtained from
RETF(NL3) and RTF(NL3) with the experimental data for Sn isotopes. 
The graph shows that there is only a small difference ($\sim 0.2$ MeV) in 
scaling monopole energies obtained with
RETF and RTF calculations. Interestingly, the RETF correction is additive 
to the RTF result instead of softening the excitation energy 
of Sn isotopes.  To see the behavior of different parametrization,
we have plotted Fig.~\ref{mplfi2}(c), where it has shown the
difference of ${\sqrt{m_3/m_1}}$ obtained with RETF and RTF results 
($\triangle{E}=RETF-RTF$) for various parameter 
sets. For all sets, except NL1, we find  $\triangle{E}$ as positive. Thus,
it is a challenging task to entangle the term which is the responsible
factor to determine the sign of RETF-RTF. Surprisingly, for most of 
the parameter sets, RTF is more towards experimental data. In spite 
of this, one cannot says anything about the qualitative behavior of 
RETF. Because, the variation of the density at the 
surface taken care by the RETF formalism, which is essential. 
One more interesting observation is that, when one investigate the 
variation of $\triangle{E}$ in the isotopic chain of Sn, it remains almost constant 
for all the parameter sets, except FSUGold. In this context, FSUGold behaves 
differently.

Variation of RETF-RTF with neutron-proton asymmetry for FSUGold set shows 
some possible correlation of RETF with the symmetry energy, which 
is absent in all other parameter sets. Now it is essential 
to know, in which respect the FSUGold parameter set is different from other. 
The one-to-one interaction terms for NL3, NL2, NL1 and NL-SH  
have similar couplings. However, the FSUGold is different from the 
above parameters in two aspects, i.e., two new coupling 
constants are added. One corresponds to the self-interaction of $\omega$ and
other one corresponds to the isoscalar-isovector meson coupling. 
It is known that self-interaction of $\omega$ is responsible for 
softening the EOS \cite{toki94,bodmer91} and the isoscalar-isovector
coupling takes care the softening of  
symmetric nuclear matter\cite{horow02}. The unique behavior shown by the FSUGold 
parametrization may be due to the following three reasons:
\begin{enumerate}
\item  introduction of isoscalar-isovector meson coupling $\Lambda_V$,
\item  introduction of self-coupling of $\omega-$meson, 
\item or simultaneous introduction of both these two terms with refitting 
       of parameter set with new constraint.
\end{enumerate}
In order to discuss the first possibility, we plotted NL3+$\Lambda_V$(0.03) 
in Fig.~\ref{mplfi2}(d). The graph shows that there is no difference between NL3 
and NL3+$\Lambda_V$, except the later set predicts a more positive 
RETF-RTF. It is well known that, the addition of $\Lambda_V$ coupling, i.e.,
NL3+$\Lambda_V$(0.03) gives a softer symmetry energy 
\cite{patra9}. This implies that, models with softer energy have greater 
difference in RETF and RTF. At a particular proton-neutron asymmetry, 
RETF-RTF has a larger value for a model with softer symmetry energy. This 
observation is not conclusive, because all the parameter sets do not 
follow this type of behavior. Quantitatively, the change of RETF-RTF 
in the Sn isotopic series is about $70 \%$ with FSUGold force, while this is 
only $20-30\%$ in NL3 and other parameter sets. 

In Table~\ref{mplta3}, we have 
listed the $\rho-$meson contribution to the total binding energy. 
From the analysis of our 
results, we find that only $\rho-$contribution  to the total binding energy 
change much more than other quantity, when one goes from RTF to RETF. But 
this change is more prominent in FSUGold parameter set than other sets 
like NL1, NL2, NL3 and NL-SH. Simple assumption says that the absent of 
$\Lambda_V$ coupling  may be the reason for this behavior in rest of the  
 parameter sets. But we have 
checked for the parameter NL3+$\Lambda_V$, which does not follow 
the assumption. This 
also shows similar behavior like other sets. In Table~\ref{mplta4}, we have given 
the results for FSUGold, NL3+$\Lambda_V$ and NL1. The data show  
a huge difference of monopole excitation energy 
in RETF and RTF with FSUGold parameter set. 
For example, the $\rho-$meson contribution to the GMR in RETF for 
$^{112}$Sn is 21.85 MeV, while in RTF it is only 0.00467 MeV.  
\begin{table}
\caption{\label{mplta3}Contribution of the $\rho-$meson to the total
binding energy in the RTF and RETF approximations with FSUGold, NL1 
and NL3+$\Lambda_V$ parameter set. RETF represent $\rho$ meson contribution 
to total energy from $R^2$ term and RETF$\Lambda_V$ represent contribution 
from the ${\Lambda_v}{R^2}{V^2}$. }
\begin{tabular}{ccccccccccc}
Nucleus &\multicolumn{4}{c}{FSUGold} &\multicolumn{2}{c}{NL3(0.03)}&
\multicolumn{2}{c}{NL1}\\ \hline
&RETF&RETF$\Lambda_V$&RTF$\Lambda_V$&RTF & RETF &RTF&RETF&RTF\\ \hline
$^{112}$Sn& 21.85&-6.66&-0.00130&0.00467&20.60 &20.12&17.91&16.73 \\
$^{114}$Sn& 28.72&-8.64&-0.00202&0.00664& 27.11&26.48&23.51&22.00  \\
$^{116}$Sn&36.42&-10.83&-0.00248&0.00829&34.40&33.62&29.73&27.90 \\
$^{118}$Sn&44.87&-13.23&-0.00298&0.01013&42.42&41.499&36.52&34.37 \\
$^{120}$Sn&54.04&-15.82&-0.00353&0.01213&51.13&50.05&43.84&41.37\\
$^{122}$Sn&63.88&-18.58&-0.00411&0.01429&60.48&59.24&51.63&48.85\\
$^{124}$Sn&74.33&-21.49&-0.00473&0.01660&70.43&69.03&59.84&56.76\\
\hline 
\end{tabular}
\end{table}
However, this difference is nominal in NL3+$\Lambda_V$  parameter set,
i.e., it is only 0.48 MeV. Similarly, this value is 1.18 MeV in 
NL1 set.  The contribution of $\rho-$meson to total 
energy comes from two terms: (i)  one from $\Lambda_V{R^2}{V^2}$ and 
other (ii) from $\rho^2$. We have explicitly shown that contribution comes from 
$\Lambda_V{R^2}{V^2}$ makes a huge difference between the GMR 
obtained from RETF and RTF formalisms. This type of contribution does not 
appear from NL3+$\Lambda_V$.  For example, in $^{112}$Sn the contribution 
of $\Lambda_V{R^2}{V^2}$ with RETF formalism is -6.0878 MeV, 
while with RTF formalism is -5.055 MeV. 

The above discussion gives us significant signature that the contribution of $\Lambda_V$ may 
be responsible for this anomalous behavior. But an immediate question arises, 
why NL3+$\Lambda_V$ parameter set does not show such type of effects, 
inspite of having $\Lambda_V{R^2}{V^2}$ term. This may be due to 
the procedure in which $\Lambda_V{R^2}{V^2}$ term is interpreted in these two 
parameter sets. In NL3+$\Lambda_V$(0.03), the $\Lambda_V{R^2}{V^2}$ term 
is not added independently. The $\Lambda_V$ and $g_\rho$ are interdependent to 
each other to fix the binding energy and difference in neutron and proton 
rms radii $R_n$-$R_p$. But in FSUGold, $\Lambda_V$ coupling 
constant is incorporate independently to reproduce the nuclear observable. 

\begin{table}
\caption{\label{mplta4}$^{S}K_A$ and $^{C}K_A$ are incompressibility for finite 
nuclei obtained from scaling and constraint methods, respectively are compared
with the values obtained from the equation of state (EOS).} 
\begin{tabular}{ccccccccc} 
Nucleus &\multicolumn{3}{c}{NL3} &\multicolumn{3}{c}{FSUGold}\\ \hline
 &$^{S}K $&$^{C}k$&$^{EOS}K$ &$^{S}K$ & $^{C}K$& $^{EOS}K$ \\ \hline
 $^{208}$Pb& 164.11& 149.96&145 &147.37& 134.57&138.42\\
 $^{116}$Sn&164.64& 155.39&131.57 &147.11& 139.71&127.64 \\
$^{40}$P&136.70&110.43&105&123.40&100.36&102.53 \\
$^{40}$Ca&145.32&134.47&105 &130.93& 123.15&102.53 \\
\hline 
\end{tabular}
\end{table}
In Table \ref{mplta4}, we have listed the incompressibility of 
some of the selected nuclei in scaling $^SK_A$ and 
constraint $^CK_A$ calculations. This results are compared  with the 
computed values obtained from the equation of state (EOS) model. 
To evaluate the incompressibility from the EOS, we 
have followed the procedure as discussed 
in Refs. \cite{centelles09,shailesh13}.  M. Centelles et al. \cite{centelles09},
parametrized the density for finite nucleus as $\rho_A=\rho_0-\rho_0/(1+c*{A^{1/3}})$ 
and obtained the asymmetry coefficient $a_{sym}$ of the nucleus with mass A 
from the EOS at this particular 
density. Here also, we have used the same parametric form of the density and obtained 
the incompressibility for finite nucleus from the EOS. For example, 
$\rho_A=0.099$ for $A=208$ in FSUGold parameter set, and the calculated 
incompressibility is $\sim 145$ MeV at this particular density. We have also 
calculated the incompressibility independently in Thomas-Fermi and extended 
Thomas-Fermi using scaling and constraint calculations, which are 161 MeV and 146.1 MeV, 
respectively.

\section{Summary and Conclusions}
\label{cejsec4}

In summary, we have calculated the isoscalar giant monopole resonance
for O, Ca, Ni, Sn, Pb, Z=114, and Z=120 isotopic series starting from 
the proton to the neutron drip-lines. We used four successful parameter sets,  NL1, NL3, NL3*, and FSUGold, with a wide range of incompressibility starting
from $211.7$ MeV to 271.76 MeV to see the dependency of the ISGMR on 
$K_{\infty}$. Also, we have analyzed the predictions of ISGMR with these
forces, which originate from various interactions and found that whatever
may be their origin, the differences in ISGMR predicted by them are found to be
marginal in the super heavy region. A recently developed scaling approach
in a relativistic mean field theory is used. A simple, but accurate 
constrained approximation is also performed to evaluate the isoscalar 
giant monopole excitation energy. From the scaling and constrained ISGMR 
excitation energies, we have evaluated the resonance width 
$\Sigma$ for the whole isotopic series. This is obtained by taking the 
root mean square difference of $E_x^s$ and $E_x^c$. The value of $E_x^s$ 
is always higher than the constrained result $E_x^c$. In a sum rule approach, 
the $E_x^s$ can be compared with the higher and $E_x^c$ as the 
lower limit of the resonance width.  In general, we found an increasing 
trend of $\Sigma$ for both the light and super heavy regions near the proton
 and neutron drip-lines. The magnitude of $\Sigma$ is predicted to be a minimum 
in the vicinity of N=Z or in the neighborhood of a double closed nucleus 
and it is a maximum for highly asymmetric systems. We 
have also estimated the incompressibility of finite nuclei. For some specific 
cases, the incompressibility modulus is compared with the nuclear matter 
incompressibility and found a linear variation among them. It is also concluded 
that the nucleus becomes less compressible with the increase of neutron or
proton number in an isotopic chain.  Thus neutron-rich matter, like 
neutron stars as well as drip-line nuclei, are less compressible than 
normal nuclei. In case of finite drip-line nuclei, the nucleus is incompressible, although it possess a normal density.
In brief, we analyzed the predictive power of various force parameters, 
like NL1, NL2, NL3, Nl-SH and FSUGold in the frame-work of relativistic 
Thomas-Fermi and relativistic extended Thomas-Fermi approaches for giant 
monopole excitation energy of Sn-isotopes. Then the calculation is extended
to some other relevant nuclei. 
The analysis shows that Thomas-Fermi approximation gives comparable 
results with
pairing+MEM data. It exactly reproduces the experimental data for Sn isotopes, when
the incompressibility of the force parameter is within $210-230$ MeV,
however, fails to reproduce the GMR data for other nuclei within
the same accuracy. 

We have qualitatively analyzed the difference in GMR energies RETF-RTF
using RETF and RTF formalisms in various force parameters. The FSUGold
parameter set shows different behavior from all other forces. Also, 
extended our calculations of monopole excitation energy for Sn isotopes
with a force parametrization having softer symmetry energy (NL3+ $\Lambda_V$). 
The excitation energy decreases with the increase of proton-neutron 
asymmetry agreeing with the experimental trend. In conclusion, after 
all these thorough analysis, it seems that the softening of Sn isotopes 
is an open problem for nuclear theory and more work in this direction are
needed.

\setcounter{equation}{0}
\setcounter{figure}{0}

\newpage
\newpage

\chapter{Perturbative constrained calculation of excitation energy of
ISGMR and IVGDR in semiclassical RMF theory.}
\label{chapter4}

In last two chapter,  we have extensively discussed the isoscalar giant monopole resonances in super heavy and medium-heavy nuclei. Semi-classical approximation like RTF and RETF are used, which are proved very successful to calculate the collective excitation. We have used constrained and scaling calculations for the excitation energy of ISGMR. In this chapter, we discuss a new constrained calculation, which we have developed to calculate the excitation energy of both 
ISGMR and IVGDR. From the theoretical point of view both the old and 
new constrained calculations are different from each other, but gives
similar results. In old constrained calculation, we  minimize 
the constrained energy functional with a constrained parameter $\lambda$. We get a different value of  constrained energy with a different value of  $\lambda$ and double derivative of  constrained energy with respect to $\lambda$ gives the constrained excitation energy. The new constrained calculation based on the perturbative approaches, in which we expand the energy functional in Taylor series around equilibrium state. We  calculated the excitation energy for some spherical (or double closed ) nuclei, where 
the experimental data are available.

\section{Introduction}
The nuclear matter incompressibility ($K_\infty$) is a measure of stiffness 
of the equation of state (EOS). It can be calculated from the curvature 
of the  EOS around the saturation density ($\rho_0$). This quantity is 
important in nuclear physics because it is related to many properties 
of finite nuclei (such as radii, masses and giant resonances), the 
dynamics of heavy-ion collisions,  neutron stars and supernovae 
collapses \cite{blaizot,harakeh01}. As $K_\infty$ is not an 
observable, its value must be deduced from a measurable quantity. In this respect,
an important source of information on $K_\infty$ is provided by the measurements of
compression modes, such as the isoscalar giant monopole resonance (ISGMR), in finite 
nuclei, in particular in $^{208}$Pb.
An accepted value $K_\infty=240\pm20$ MeV, predicted by the majority of mean field models, 
extracted from different analysis of experimental data and theoretical calculation\cite{stone14}.  


On the other hand, the existence of strong resonances in the photo-absorption 
cross sections was established almost seventy years ago \cite{baldwin47} and theoretically 
explained few years later \cite{goldhaber48,steinwedel50}. These resonances,
identified with the isovector giant dipole resonance (IVGDR), are correlated  
via the electrical polarizability with the neutron skin thickness of neutron rich nuclei
and with the properties of the nuclear symmetry energy around saturation
\cite{roca13,roca15}.

The ISGMR and IVGMR can be understood as small amplitude oscillations, which are the responses 
of the nucleus to an external field generated by electromagnetic or hadronic probes.
The standard theory to deal with these oscillations is the random-phase approximation
(RPA), which allows to study microscopically giant resonances. The key output provided
by the RPA calculations is the strength distribution $S_Q(E)$ associated to a given 
excitation operator ${\bf Q}$. If $S_Q(E)$ is concentrated in a narrow region of the energy 
spectrum, which is usually the case of the ISGMR and IVGMR at least for heavy stable 
nuclei, the knowledge of few low energy weighted moments of $S_Q(E)$ (sum rules)
allows to estimate the average excitation energy of the resonances. In some particular
cases it is possible to express some odd sum rules in a compact form involving only
ground-state properties, which avoids the calculation of the strength distribution 
\cite{bohigas79}. However, the full quantal calculation of these sum rules is still complicated
because the exact ground-state is, in general, unknown. Introducing additional approximations,
some sum rules can be calculated in a rather simple way, as for instance the cubic energy 
weighted sum rule, which can be computed by using the scaling method \cite{blaizot,bohigas79,jennings80} or the inverse energy weighed sum rule, which can be obtained through
constrained Hartree-Fock (HF) calculations \cite{blaizot,jennings80,treiner81,blaizot95}.

The theoretical study of giant resonances in the relativistic domain has been usually done
through relativistic RPA (RRPA) calculations. The RRPA approximation corresponds to the small
 amplitude limit of the time-dependent relativistic mean field (RMF) theory 
\cite{horowicz90,piekarewicz01,ma01}. 
On the other hand the time-dependent theory has also been directly 
used to study the isovector dipole \cite{vretenar95} as well as the isoscalar 
and isovector quadrupole \cite{vretenar95} and monopole
\cite{vretenar97} oscillations. As difference with the non-relativistic case, the sum rule 
theorems, which relate different moments of the RRPA strength with ground-state properties
computed at RMF level, have not been proved in the relativistic domain \cite{wei13}.
There are few relativistic constrained calculations available in the literature \cite{wei13,maru89,boer91,stoi94,ma01}. In Ref.\cite{ma01} the results of constrained RMF 
calculations of the $m_{-1}$ sum rule were compared with the corresponding RRPA values and it 
was suggested that the RRPA $m_{-1}$ moment could be estimated from a RMF calculation.
More recently, in Ref.\cite{wei13}, this suggestion was carefully analyzed for
several nuclei along the periodic table using two different RMF parameter sets.
It was found an excellent agreement between the average energies of ISGMR obtained 
through RMF constrained calculation and extracted from RRPA results.
In Ref.\cite{patra02} the scaling approach to the isoscalar giant monopole and quadrupole 
resonances in relativistic Thomas-Fermi (RTF) theory \cite{centelles90,centelles93}   
was discussed in detail. It was found that the semi-classical average excitation 
energies reproduced well the RRPA results. In the same reference the excitation energies 
of the ISGMR were also calculated performing RTF constrained calculations finding good 
agreement with the constrained RMF results of Ref.\cite{stoi94}. As a conclusion, these 
results together with the conjecture of Ref.\cite{ma01} points out that the scaling method and
 constrained calculations using the semi-classical RTF energy density functional may be a very 
efficient and simplified way to estimate the average excitation energy of some giant resonances. Although the constrained RTF approach has been used in 
\cite{patra02} to compute the average excitation energy of the ISGMR, in this 
chapter we want to use a new constrained
method applied together with the RTF approach, which is able to study not only the ISGMR but 
also the IVGDR. This new constrained calculation is based on a Taylor series expansion of 
the constrained energy density functional around the equilibrium. In Section~\ref{form1} we outline
the general formalism to calculate excitation energy of ISGMR and ISVGDR. In Section \ref{rmf1} we 
discussed the application of relativistic Thomas-Fermi approximation to calculate the energy and particle density. Section~\ref{pet} is dedicated to the explanation of perturbative approches to calculate the 
excitation energy of ISGMR and ISVGDR. Results of our new method with old one are given in 
section~\ref{res} and concluded in section~\ref{con1}.

\section{Formalism}
\label{form1}
The strength distribution associated to an excitation operator ${\bf Q}$ is defined as
\begin{eqnarray}
 {S_Q(E)} = {\sum_{n \ne 0}} {|\langle n|Q|0 \rangle|}^2 \delta(E - E_n),
\label{eq1}
\end{eqnarray}
where $|0\rangle$ and $|n\rangle$ are the ground and excited eigenstates, respectively,
of the exact Hamiltonian $H=T+V$ and $E_n$ is the excitation energy.

The energy weighted moments, also known as sum rules, are defined as 
\begin{eqnarray}
 {m_k}= \int_0^{\infty} {E^k S_Q(E)}dE = {\sum_{n\ne 0}} {E_n}^K {|\langle n|Q|0 \rangle|}^2,  
\label{eq2}
\end{eqnarray}
which allow to estimate the average excitation energy as   
\begin{equation}
\tilde{{E_K}}=\sqrt{\frac{m_k}{m_{k-2}}}.
\label{eq3}
\end{equation}
In this work we  concentrate in the study of the $\tilde{E}_1$ average
energy to estimate the excitation of the ISGMR and the IVGDR. A very 
useful approximation arises when the exact energy eigenstates of the  Hamiltonian
$H$  in (\ref{eq1}) is replaced by the one-hole-one particle ( 1p1h) RPA ones. 
In this case it is possible to show that some sum rules, in particular 
$m_1$ and $m_{-1}$, can be exactly obtained by
replacing the RPA ground-state by the uncorrelated Hartree-Fock (HF) or by
performing constrained HF calculations (see \cite{bohigas79} for more details).
In the monopole case, the energy weighted sum rule $m_1$ is almost a model 
independent quantity and according to the Thouless theorem it is given by    
\begin{equation}
{m_1}= {\sum_n}{E_n}|\langle n|\sum_{i=1}^A |0\rangle|^2 
= 2 A \frac{{\hbar}^2}{m} 
\langle {r^2} \rangle.
\label{eq4}
\end{equation}

For the isovector dipole, the previous simple approach provided in 
Eq.(\ref{eq4}) does not hold good because the excitation operator
${\bf Q}$ does not commute, in general, with the potential part $V$ of the
Hamiltonian. In the isovector case the full calculation of the $m_1$ 
sum rule is complicated and usually this moment is factorized as
\begin{equation}
{m_1}= {m_1}^0 ( 1 + \kappa),
\label{eq5}
\end{equation}
where ${m_1}^0$ is the sum rule computed assuming $[Q,V]=0$ and $\kappa$
is the so-called enhancement factor. In the dipole case, the ${m_1}^0$
is given by the kinetic Thomas-Reiche-Kuhn (TRK) term and therefore, the full 
energy weighted sum rule reads as 
\begin{equation}
{m_1}= \frac{\hbar^2}{2m}\frac{NZ}{A}( 1 + \kappa).
\label{eq6}
\end{equation}
For relativistic model the TRK enhancement factor in nuclear matter is given 
by \cite{bender03}  
\begin{equation}
\kappa(k_F)= \frac{m}{\sqrt{{m^*}^2 + {k_F}^2}}-1,
\label{eq7}
\end{equation}
where $k_F$  the Fermi momentum. Notice that in nuclear matter $\kappa$ is a function 
of $k_F$. In order to estimate the average enhancement factor in finite nuclei, we use 
(\ref{eq7}) together with a local density approximation:
\begin{equation}
\kappa = \frac{1}{A} \int{\kappa(k_F)}\rho(r){\bf dr}, 
\label{eq8}
\end{equation}
where the local Fermi momentum $k_F$ is related to the finite nucleus density by
$k_F(r)=(3 \pi^2 \rho(r))^{1/3}$.  

To obtain the inverse energy-weighted sum rule, we take into account the fact that this
sum rule computed at 1p1h RPA level, according to the dielectric theorem, 
can also be obtained from a constrained HF calculation
performed with the Hamiltonian $H + \lambda Q$, where $H$ is the Hamiltonian that describe the 
nucleus and $Q$ the one-body excitation operator as 
\begin{eqnarray}
{m_{-1}}= {\sum_n}\frac{|\langle n|Q|0\rangle|^2}{E_n} 
= \frac{1}{2} \frac{\partial \langle \lambda\vert Q \vert \lambda \rangle} 
{\partial \lambda} 
\bigg|_{\lambda=0 }
= \frac{1}{2} \frac{\partial^2 \langle \lambda\vert H\vert \lambda \rangle}
{\partial \lambda^2}\bigg|_{\lambda=0},
\label{eq9}
\end{eqnarray}
where $\vert \lambda \rangle$ is the HF ground state of the constrained Hamiltonian $H + \lambda Q$.

\section{Relativistic Thomas-Fermi approximation}
\label{rmf1}
In this work we describe the average excitation energy of ISGMR and IVGDR by using Thomas-Fermi approximation to the RMF theory discussed in early literature \cite{centelles90,centelles93,patra01,patra02}  
. A reason for using semi-classical technique is to study giant resonances 
is the fact that these oscillations are strong collective motion where shell effects have 
only a marginal impact \cite{patra02,mario10}. In the linear RMF model nucleons interact 
via the exchange of the effective $\sigma$, $\omega$ and $\rho$ mesons \cite{walecka74}.
However, due to the high incompressibility $K_{\infty}$ predicted by this model, 
self-interactions of the $\sigma$ meson, through non-linear terms, have been introduced
in the formalism to reduce $K_{\infty}$ to a more realistic value \cite{boguta77}.   
 The starting point of the semi-classical RMF theory is the nucleon-nucleon
interacting effective Lagrangian density, which can be written as
\begin{eqnarray}
{\cal H}= {\cal E} + g_{v} V \rho + g_{\rho} R \rho_3
+ e {\cal A} \rho_p + {\cal H}_f,
\label{eq10}
\end{eqnarray}
where $\rho=\rho_n + \rho_p$ is the baryon density, $\rho_3=(\rho_p - \rho_n)/2$ is the 
isovector density and ${\cal E}$ the nucleon energy density, which at RTF level can be written
as  \cite{centelles90,centelles93,patra01,patra02}
\begin{equation}
{\cal E} = \sum_q \frac{1}{8 \pi^2}[ k_{F_q}\epsilon_{F_q}^3 + k_{F_q}^3\epsilon_{F_q}
- {m^*}^4 \frac{k_{F_q} + \epsilon_{F_q}}{m^*}].
\label{eq11}
\end{equation}

For each kind of nucleon $(q=n,p)$, the local Fermi momentum $k_{F_q}$ and energy $\epsilon_{F_q}$
are defined as
\begin{equation}
k_{F_q}=(3 \pi^2 \rho_q)^{1/3} \quad ; \quad \epsilon_{F_q}=\sqrt{k_{F_q}^2 + {m^*}^2}. 
\label{eq12}
\end{equation}

In Eq. (\ref{eq10}), ${\cal H}_f$ stands for the free contribution of the meson fields
$\phi$, $V$ and $R$, associated to the $\sigma$, $\omega$ and  $\rho$ mesons, and for the 
Coulomb field ${\cal A}$. This contribution reads
\begin{eqnarray}
{\cal H}_f&=& \frac{1}{2} \left[ (\vec{\nabla}\phi)^2 + m_{s}^2 \phi^2 \right]
+\frac{1}{3} b \phi^3 + \frac{1}{4} c \phi^4
-\frac{1}{2} \left[ (\vec{\nabla} V)^2 + m_{v}^2 V^2 \right] 
\nonumber\\
&-&\frac{1}{2} \left[ (\vec{\nabla} R)^2 + m_\rho^2 R^2 \right]
- \frac{1}{2} \left(\vec{\nabla}  {\cal A}\right)^2, 
\label{eq13}
\end{eqnarray}
where $m_s$, $m_v$ and $m_{\rho}$ are the masses of the mesons and 
$m^*$, entering in Eqs.(\ref{eq11}) and (\ref{eq12}, is the nucleon effective mass
defined by $m^*=m-g_s\phi$. In these equations $g_s$, $g_v$, $g_{\rho}$ and $e$ are the 
coupling constants for the ${\sigma}$, $\omega$ and ${\rho}$ mesons and for the photon, 
respectively. 
In Eq.(\ref{eq13}) $b$ and $c$ stand for the 
coupling constants associated to the ${\sigma}$ meson non-linear terms. 
These self-interactions of the $\sigma-$ meson 
generates analogous effect of three body interaction
due to its {\it off-shell} meson couplings. 
These self-interactions  are also essential for the saturation properties of infinite nuclear 
matter \cite{fujita57,steven01,schiff50}. 

The semi-classical ground-state densities and meson fields are obtained by solving
the variational equations derived from the energy density (\ref{eq10}) constrained 
by the condition of fixed neutron ($N$) and proton ($Z$) numbers, which read
\begin{equation}
\epsilon_{F_n} + g_v V - \frac{1}{2}g_{\rho}R - \mu_n =0, 
\label{eq14}
\end{equation}
\begin{equation}
\epsilon_{F_p} + g_v V + \frac{1}{2}g_{\rho}R + e{\cal A} - \mu_p =0,
\label{eq15}
\end{equation}
\begin{equation}
\Delta V - {m_v}^2 V +{ g_v }\rho = 0,
\label{eq16}
\end{equation}
\begin{equation}
\Delta R - {m_\rho}^2 {R} +{ g_\rho }\rho_3 = 0,
\label{eq17}
\end{equation}
 \begin{equation}
\Delta \phi- m_s^2 {\phi} + g_s \rho_s - b \phi^2 - c \phi^3 = 0, 
\label{eq18}
\end{equation}
and
\begin{equation}
\Delta {\cal A} + e \rho_p = 0.
\label{eq19}
\end{equation}

In Eqs.(\ref{eq14}) and (\ref{eq15}) $\mu_n$ and $\mu_p$ are the neutron and proton
chemical potentials, respectively, introduced to ensure the right $N$ and $Z$ values.
In Eq.(\ref{eq18}) $\rho_s$  is the semi-classical scalar density given by
\begin{equation}
\rho_s = \frac{\partial {\cal E}}{\partial m^*} = 
\sum_q \frac{m^*}{2 \pi^2}[ k_{F_q}\epsilon_{F_q} - 
{m^*}^2 \frac{k_{F_q} + \epsilon_{F_q}}{m^*}].
\label{eq20}
\end{equation} 

\section{The constrained perturbative approach}
\label{pet}
We apply now the constrained method to estimate the excitation energy of the ISGMR and IVGDR.
In the semi-classical context one has to minimize the constrained energy density
functional \cite{patra02}
\begin{equation}
\int [{\cal H} - \mu_n \rho_n - \mu_p \rho_p - \lambda Q] d {\bf r} =
E(\lambda) - \lambda < Q > .
\label{eq21}
\end{equation} 

In previous studies of the ISGMR using the constrained method \cite{patra02,wei13,mario10}
one minimizes Eq. (\ref{eq21}) for several values of $\lambda$ and then computes numerically
the derivatives respect to $\lambda$, which define the $m_ {-1}$ sum rule (see Eq.(\ref{eq9}))
, using three- or five-point formulas. This technique may also be applied, in principle, 
to estimate the excitation energy  of the IVGDR using a deformed code. We present here an alternative 
route which avoids the explicit use of deformation. Keeping in mind that the derivatives 
respect to the parameter $\lambda$ in Eq.(\ref{eq9}) are computed  at $\lambda=0$, we expand 
the constrained energy density functional (\ref{eq21}) around equilibrium in powers of $\lambda$ 
up to quadratic terms, which according to  Eq.(\ref{eq9}), is needed to compute the 
$m_{-1}$ sum rule. To this end, we write the nuclear densities and meson fields as the sum of
the unperturbed solutions, obtained by solving self-consistently Eqs.(\ref{eq14})-(\ref{eq20}),
plus a perturbative contribution, which is linear in $\lambda$ and has the same angular 
dependence as the excitation operator $Q$. Notice that, as far as the right number of particles
is obtained by the unperturbed density, the additional constraint
$\int \delta \rho_n({\bf r}) d{\bf r}= \int \delta \rho_p({\bf r}) d{\bf r}=0$ must be imposed.
This can be achieved by introducing additional contributions to the neutron and proton 
chemical potentials, $\delta \mu_n$ and $\delta \mu_p$, respectively, which are also linear
in the parameter $\lambda$. 

The independent term of the constrained energy functional ($\ref{eq21}$) in powers of $\lambda$ 
is just the unperturbed energy $E_0$ of the ground-state of the 
nucleus. The linear term in $\lambda$  are proportional to the equation of motion of the  the unperturbed problem
Eqs.(\ref{eq14})-(\ref{eq20}), while the quadratic contribution provides the set of equations of
motion corresponding to the pertubative part to the nuclear densities and meson fields
 by applying the variations principle. According to Eq. ~\ref{eq9},
 $m_{-1}$ sum rule is given by one half of the second derivative of the constrained
energy with respect to the parameter $\lambda$. At this point two important comments are in 
order. First, the equations of motion of the perturbative part of the nuclear densities and
fields can also be obtained by performing a second order variation with  respect to the densities and 
fields in the equations of motion of the unperturbed problem, Eqs. (\ref{eq14})-(\ref{eq20}). Second, 
by replacing the solutions of the perturbative contribution to densities and fields in the 
second derivative of the energy density functional respect to the parameter $\lambda$, one can
recast this expression as one half of the first derivative of the excitation operator $Q$
respect to $\lambda$, in agreement with Eq.(\ref{eq9}). These two comments point out that
this perturbative approach to the calculation of the $m_ {-1}$ sum rule is properly formulated.
We will now apply this perturbative constrained method to estimate the excitation energies
of the ISGMR and IVGDR.

\subsection{Isoscalar giant monopole resonance}
The ISGMR is a collective excitation in which proton and neutron vibrate in a phase to each 
other. It corresponds to the radial oscillation of nucleus around its equilibrium radius. 
In order to calculate excitation energy we write the excitation operator as 
$Q = r^2 - <r^2>_0$, where  ${<r^2>_0}^{1/2}$ is the rms radius computed using the unperturbed
equilibrium density $\rho_0$. We apply the constrained perturbative method by writing the 
perturbative nuclear densities and fields as
$\lambda \delta \rho_n$, $\lambda \delta \rho_p$, $\lambda \delta V_0$ , 
 $\lambda \delta \phi$,  $\lambda \delta R$ and $\lambda \delta {\cal A}$.
Due to the spherical symmetry of the problem, we shall also introduce the corrections
$\lambda \delta \mu_n$ and $\lambda \delta \mu_p$ to the neutron and proton chemical
potentials. With all these ingredients we expand Eq. (\ref{eq21}) up to order $\lambda^2$
obtaining
\begin{eqnarray}
{\tilde E} &=& \int dr \bigg[ {\cal H}^0
+\lambda \bigg[\left({\epsilon^0_{F_n}} 
+{g_v} V^0-\frac{1}{2}g_\rho R^0\right)\delta \rho_n \nonumber \\ 
&+& \left({\epsilon^0_{F_p}} + {g_v} {V^0} + \frac{1}{2}g_\rho R^0
+ e{\cal A}^0\right)\delta \rho_p  
+ \left( \Delta {\cal A}^0 + e \rho^0_p \right) \delta {\cal A}\nonumber \\
&+& \bigg( \Delta {V^0} - {m_v}^2 {V^0} + g_v {\rho^0}\bigg) \delta{V}
+\bigg( \Delta R^0 - {m_\rho}^2 R^0 + g_\rho {\rho^0_3}\bigg) \delta R\nonumber \\
&-& \bigg( \Delta \phi - {m_s}^2 \phi^0
+ g_s {\rho^0_s}-b{\phi^0}^2 -c{\phi^0}^3 \bigg) \delta \phi \bigg] 
\nonumber \\ 
&+&\frac{\lambda^2}{2}
\bigg[ \bigg(\frac{\partial {\epsilon^0_{F_n}}}{\partial \rho^0_n}\delta \rho_n
-{g_s} \frac{\partial {\epsilon^0_{F_n}}}{\partial m_0^*}\delta \phi
+ g_v \delta V - \frac{1}{2}g_{\rho}\delta R\bigg)\delta \rho_n\nonumber \\
&+& \bigg(\frac{\partial {\epsilon^0_{F_p}}}{\partial \rho^0_p}\delta \rho_p
-{g_s} \frac{\partial {\epsilon^0_{F_p}}}{\partial m_0^*}\delta \phi + e\delta {\cal A}
+ g_v \delta V + \frac{1}{2}g_{\rho}\delta R\bigg)\delta \rho_p\nonumber \\
&+&\bigg(g_v \delta \rho +\Delta \delta V - m_v^2 \delta V\bigg)\delta V
+ \bigg(g_\rho \delta \rho_3 +\Delta \delta R - m_\rho^2 \delta R\bigg)\delta R\nonumber \\
&+&\bigg(e \delta \rho_p +\Delta \delta {\cal A} \bigg) \delta {\cal A}
+\bigg(\bigg({g_s}^2 \frac{\partial \rho^0_s}{\partial {m^*}^0}+ 2b \phi^0 + 3c {\phi^0}^2
\bigg)\delta \phi \nonumber \\
&-&{g_s}\frac{\partial {\epsilon^0_{F_n}}}{\partial {m^*}^0} \delta \rho_n
-{g_s} \frac{\partial {\epsilon^0_{F_p}}}{\partial {m^*}^0} \delta \rho_p
-\Delta \delta \phi + m_s^2 \delta \phi\bigg)\delta \phi\bigg]\nonumber \\
&-& \mu^0_n \rho^0_n -  \mu^0_p \rho^0_p 
-\bigg(\mu^0_n +\lambda \delta \mu_n \bigg)\lambda \delta \rho_n
-\bigg(\mu^0_p +\lambda \delta \mu_p \bigg)\lambda \delta \rho_p \nonumber \\
&-&\lambda (r^2 - <r^2>^0)( \delta \rho_n + \delta \rho_p) \bigg].
\label{eq22}
\end{eqnarray}

In the above equation the superscript 0 indicates quantities which are calculated using the 
unperturbed densities and fields. The linear contributions of the perturbative components
of densities and fields in Eq.(\ref{eq22}) vanish, because their prefactors are just the motion
equations of the unperturbed densities and fields Eqs.(\ref{eq14})-(\ref{eq20}), which
are zero for the self-consistent solutions. Therefore, the constrained energy Eq.(\ref{eq21})
becomes a quadratic function of the perturbative corrections to the nuclear densities and 
meson fields, from where the motion equations for $\delta\rho_n$, $\delta\rho_p$,$\delta V$,
$\delta R$, $\delta {\cal A}$ and $\delta \phi$ are easily obtained: 

\begin{equation}
\frac{\partial {\epsilon^0_{F_n}}}{\partial \rho^0_n}\delta \rho_n
-{g_s} \frac{\partial {\epsilon^0_{F_n}}}{\partial m_0^*}\delta \phi
+ g_v \delta V - \frac{1}{2}g_{\rho}\delta R - \delta {\tilde \mu}_n - r^2 =0,
\label{eq23}
\end{equation}

\begin{equation}
\frac{\partial {\epsilon^0_{F_p}}}{\partial \rho^0_p}\delta \rho_p
-{g_s} \frac{\partial {\epsilon^0_{F_p}}}{\partial m_0^*}\delta \phi + e\delta {\cal A}
+ g_v \delta V + \frac{1}{2}g_{\rho}\delta R - \delta {\tilde \mu}_p - r^2 =0,
\label{eq24}
\end{equation}

\begin{equation}
g_v \delta \rho +\Delta \delta V - m_v^2 \delta V = 0,
\label{eq25}
\end{equation}

\begin{equation}
g_\rho \delta \rho_3 +\Delta \delta R - m_\rho^2 \delta R = 0,
\label{eq26}
\end{equation}

\begin{equation}
(e \delta \rho_p +\Delta \delta {\cal A} = 0,
\label{eq27}
\end{equation}
and
\begin{eqnarray}
&&\bigg({g_s}^2 \frac{\partial \rho^0_s}{\partial {m^*}^0}+ 2b \phi^0 + 3c {\phi^0}^2\bigg)
\delta \phi -{g_s} \frac{\partial {\epsilon^0_{F_n}}}{\partial {m^*}^0} \delta \rho_n
-{g_s} \frac{\partial {\epsilon^0_{F_p}}}{\partial {m^*}^0}\delta \rho_p 
-\Delta \delta \phi + m_s^2 \delta \phi = 0. \nonumber\\
\label{eq28}
\end{eqnarray}

In Eqs.(\ref{eq23}) and (\ref{eq24}) $\delta {\tilde \mu}_q=\delta \mu_q - <r^2>^0$ 
($q=n,p$). The self-consistent solution of the set of equations  Eqs.(\ref{eq23})-(\ref{eq24})
gives the perturbed neutron and proton densities, $\delta \rho_n$ and  
$\delta \rho_p$, respectively, as well as the corrections to the fields
$\delta V$ ,$\delta \phi$, $\delta R$ and $\delta {\cal A}$, where $\delta_n$ and $\delta_p$ 
are the transition densities. The $m_{-1}$ sum rule is given by the 
second order derivative of the semi-classical energy $E(\lambda)$ (see Eq.(\ref{eq21})) 
with respect to the parameter $\lambda$. Combining Eq.(\ref{eq22}) with the motion equations
(\ref{eq23})-(\ref{eq24}) for the perturbative nuclear densities and meson and photon fields are 
easily obtained  
\begin{equation}
m_{-1}= \int \bigg(\delta {\tilde \mu}_n \delta \rho_n + 
\delta {\tilde \mu}_p \delta \rho_p + r^2 \delta \rho \bigg) d{\bf r},
\label{eq29}
\end{equation}
where $\delta \rho = \delta \rho_n + \delta \rho_p$. Taking into account
the constraint $\int \delta \rho_q d{\bf r} =0$, we get the 
final form of the $m_{-1}$, 
\begin{equation}
{m_{-1}}= \int r^2 \delta \rho d{\bf r}.
\label{eq30}
\end{equation}

In order to solve the motion equations (\ref{eq23})-(\ref{eq24}) for the perturbative 
corrections to the nuclear densities and meson fields we need to know, in the ISGMR case, the
corrections to the neutron and proton chemical potentials. To this end  we first isolate 
 these in equations $\delta \rho_n$ and $\delta \rho_p$ and next, using  explicitly the
constraints $\int \delta \rho_n d{\bf r}=\int \delta \rho_p d{\bf r}=0$, we obtain after 
little algebra the corrections $\delta {\tilde \mu}_n$ and $\delta {\tilde \mu}_p$ as
\begin{equation}
\delta {\tilde \mu}_n =\frac{\int {k^0_{F_n}} \bigg[{\epsilon^0}_{F_n}\bigg( 
g_v \delta V -\frac{1}{2} g_\rho \delta R - r^2 \bigg)- {m^0}^* g_s {\delta \phi}\bigg] 
d{\bf r}}{\int {k^0_{F_n}} {\epsilon^0}_{F_n} d{\bf r}},
\label{eq31}
\end{equation}
 and
\begin{equation}
\delta {\tilde \mu}_p =\frac{\int {k^0_{F_p}} \bigg[{\epsilon^0}_{F_p}\bigg( 
g_v \delta V + \frac{1}{2} g_\rho \delta R + e \delta A - r^2 \bigg)- 
{m^0}^* g_s {\delta \phi}\bigg] d{\bf r}}{\int {k^0_{F_p}}{\epsilon^0}_{F_p} d{\bf r}}.
\label{eq32}
\end{equation}
The self-consistent solution of the coupled equations (\ref{eq23})-(\ref{eq24}) 
allows to obtain the perturbative contribution to the nuclear densities and fields. From
Eq.(\ref{eq30}) one can obtain the inverse energy weighted sum rule ${m_{-1}}$, which combined 
with the $m_1$ sum rule given by Eq.(\ref{eq4}) provides an estimate of the excitation
energy of the ISGMR given by $E_1=\sqrt{m_1/m_{-1}}$.

\subsection{Isovector giant dipole resonance}
In case of IVGDR protons and neutrons vibrate in opposite phase 
to each other in such a way that the center of mass of the whole system 
remains unchange. This constraint is introduced through the excitation operator
$Q = z - <z> = rY_{10}(\Omega)$, where $<z>$ is the z-coordinate of the center of mass of the
nucleus. As in the case of ISGMR, we write the neutron and proton densities as the
sum of  unperturbed solution plus a perturbative term, which follows the geometry of the
excitation operator, i.e. $\rho_n({\bf r}) = \rho^0_n(r) + \lambda \delta \rho_n(r) 
Y_{10}(\Omega)$ and $\rho_p({\bf r}) = \rho^0_p(r) + \lambda \delta \rho_p(r)Y_{10}(\Omega)$.
With this choice of the nuclear densities
the right normalization of neutrons and protons is ensured and, consequently, the
corrections $\delta \mu_n$ and $\delta \mu_p$ to the corresponding chemical potentials
is not needed in the case of the IVGDR. Using this parametrization of the nuclear densities
we compute now the expectation values of the dipole operator $Q$ defined before. 
It is easy to show that
\begin{eqnarray}
&&<z> = \int\big({\rho_n({\bf r})+\rho_p({\bf r})\big)rY_{10}(\Omega)d{\bf r}}\nonumber \\
&=& \frac{1}{A}\int{r^3(\delta \rho_n(r)+\delta \rho_p(r))dr} =
\frac{N}{A}<z>_n + \frac{Z}{A}<z>_p,
\label{eq33}
\end{eqnarray}
and 
\begin{eqnarray}
&&<Q> = \int{\big(\rho_p({\bf r}) - \rho_n({\bf r})\big)\big(rY_{10}(\Omega)-<z>\big))d{\bf r}}
\nonumber \\
&=&2\bigg[\frac{N}{A}\int{r^3\delta \rho_p(r)dr}-\frac{Z}{A}\int{r^3\delta \rho_n(r)dr}\bigg]
\nonumber \\
&=& \frac{2NZ}{A}\big(<z>_p - <z>_n\big).
\label{eq34}
\end{eqnarray}

We assume, as in the ISGMR case, that the fields also split into an unperturbed part plus 
a corrective contribution that follows the excitation field, i.e.  
${V}={V}^0 +\lambda \delta {V}Y_{10}$, ${R}={R}^0 +\lambda \delta {R}Y_{10}$, 
${\cal A}={\cal A}^0 +\lambda \delta {\cal A}Y_{10}$ and 
${\phi}={\phi}^0 +\lambda \delta {\phi}Y_{10}$. Therefore, one can write the energy density 
functional (\ref{eq21}) in powers of $\lambda$ and derive from this expansion the motion 
equations of the nuclear densities and fields, as we have done in the case of  ISGMR
starting from Eq.(\ref{eq22}). However, as we have pointed out before, it is also possible
derive the variational equations for $\delta \rho_n$, $\delta \rho_p$, $\delta V$, $\delta R$,
$\delta \phi$ and $\delta {\cal A}$ expanding the motion equations associated to the 
constrained energy density functional (\ref{eq21}). In case of the IVGDR, these equations 
for the meson and photon fields are formally given by Eqs.(\ref{eq16})-(\ref{eq19}), keeping in
mind, however, that in this case densities and fields are, the ones associated to the 
constrained problem. The variations equations for neutron and proton are
\begin{equation}
\epsilon_{F_n} + g_v V - \frac{1}{2}g_{\rho}R - \mu_n  
+ \lambda rY_{10}\frac{N}{A}=0, 
\label{eq35}
\end{equation}
and
\begin{equation}
\epsilon_{F_p} + g_v V + \frac{1}{2}g_{\rho}R + e{\cal A} - \mu_p 
- \lambda rY_{10}\frac{N}{A}=0,
\label{eq36}
\end{equation}
which correspond to Eqs.(\ref{eq14}) and (\ref{eq15}) modified by the contribution of the 
excitation field. Expanding the densities and fields in these variations equations in
powers of $\lambda$, the independent term gives the motion equations of  unperturbed
densities and fields and the linear terms in $\lambda Y_{10}$ correspond to the variations
equations for the perturbations of densities and fields, which read
\begin{equation}
\frac{\partial {\epsilon^0_{F_n}}}{\partial \rho^0_n}\delta \rho_n
-{g_s} \frac{\partial {\epsilon^0_{F_n}}}{\partial m_0^*}\delta \phi
+ g_v \delta V - \frac{1}{2}g_{\rho}\delta R + \frac{Z}{A} r =0,
\label{eq37}
\end{equation}

\begin{equation}
\frac{\partial {\epsilon^0_{F_p}}}{\partial \rho^0_p}\delta \rho_p
-{g_s} \frac{\partial {\epsilon^0_{F_p}}}{\partial m_0^*}\delta \phi + e\delta {\cal A}
+ g_v \delta V + \frac{1}{2}g_{\rho}\delta R - \frac{N}{A}r = 0,
\label{eq38}
\end{equation}

\begin{equation}
g_v \delta \rho Y_{10} + \Delta(\delta V Y_{10}) - m_v^2 \delta V Y_{10} = 0,
\label{eq39}
\end{equation}

\begin{equation}
g_\rho \delta \rho_3 Y{10} + \Delta(\delta R Y_{10})- m_\rho^2 \delta R Y_{10} = 0,
\label{eq40}
\end{equation}

\begin{equation}
(e \delta \rho_p Y_{10} +\Delta(\delta {\cal A} Y_{10}) = 0,
\label{eq41}
\end{equation}
 and  

\begin{eqnarray}
&&\bigg({g_s}^2 \frac{\partial \rho^0_s}{\partial {m^*}^0}+ 2b \phi^0 + 3c {\phi^0}^2\bigg)
\delta \phi Y_{10} -{g_s} \frac{\partial {\epsilon^0_{F_n}}}{\partial {m^*}^0} \delta \rho_n
Y_{10} \nonumber \\&-&{g_s} \frac{\partial {\epsilon^0_{F_p}}}{\partial {m^*}^0}\delta \rho_p
Y_{10} - \Delta(\delta \phi Y_{10}) + m_s^2 \delta \phi Y_{10} = 0.
\label{eq42}
\end{eqnarray}

As in the case of ISGMR, we can also calculate the $m_{-1}$ sum rule for the IVGDR from the 
second order derivative of the constrained energy with respect to the $\lambda$,
which after little algebra and taking into account Eqs.(\ref{eq37})-(\ref{eq42})
become
\begin{equation}
\label{eq43}
m_{-1}= \frac{\partial^2 E }{\partial \lambda^2}.  
 \int r^3 dr \bigg[ \frac{N}{A}\delta \rho_p - \frac{Z}{A} \delta \rho_n \bigg].
\end{equation}

\section{Results and Discussions}
\label{res}

Before going to discussions of the predictions of our model and compare with the results obtained applying
another theoretical models as well as with the experimental data, we outline briefly our
numerical procedure to estimate semi-classically the ISGMR and IVGDR excitation energies in finite
nuclei. First, and assuming spherical symmetry, we compute for each nucleus the unperturbed
densities and fields by solving self-consistently the set of equations
(\ref{eq14})-(\ref{eq20}). 
using the so-called imaginary time-step method \cite{davi80,lev84,dali85}. 
 and these unperturbed values we can be  use to compute  the $m_{1}$ sum
rules given by Eqs.(\ref{eq4}) and (\ref{eq6}), for the ISGMR and IVGDR, respectively.
Next, we solve iteratively the sets of coupled linear equations (\ref{eq23})-(\ref{eq27}) for the
ISGMR and (\ref{eq37})-(\ref{eq42}) for the IVGDR, which allow to obtain the perturbative
corrections to the nuclear densities and meson and photon fields. Finally the $m_{-1}$
sum rules can be obtained through Eqs.(\ref{eq30}) and (\ref{eq43}) for the ISGMR and 
IVGDR, respectively. All these calculations are performed at RTF level paying special attention
 to the convergence of the $m_{-1}$ sum rule. The convergence of this quantity actually depends 
on the size of the box where the calculations are performed as well as on the number of iterations 
used in the unperturbed calculation.   

\bigskip
\begin{table*}
\begin{center}
\caption{\label{table1}{
Excitation energy of the ISGMR in MeV of some spherical nuclei calculated as
$\sqrt{m_1/m_{-1}}$ using the NL3 model. ${E_c}(RTF-P)$ is the estimte
of the present work, ${E_c}(RETF)$ is the RETF result computed as in
Ref.\cite{patra02},${E_c}(RMF)$ and ${E_c}(RPA)$ are the constrained
Hartree and RPA results, respectively, reported in \cite{wei13}.}} 
\begin{tabular}{cccccccccccc}
\hline
Nucleus&${E_c}(RTF-P)$&${E_c}(RETF)$&${E_c}(RMF)$&${E_c}(RPA)$& Expt. \\ \hline
$^{16}$O&26.26&25.98&23.34&23.35&21.13$\pm$0.49\\
$^{40}$Ca&22.89&23.20&21.55&21.57&19.18$\pm$0.37\\
$^{90}$Zr&18.74&19.08&18.58&18.55&17.89$\pm$0.20\\
$^{116}$Sn&17.29&17.57&16.98&17.06&16.07$\pm$0.12\\
$^{144}$Sm&16.05&16.33&16.08&16.16&15.39$\pm$0.28\\
$^{208}$Pb&13.84&13.91&14.07&14.10&14.17$\pm$0.28\\
\hline
\end{tabular}
\end{center}
\end{table*}

\begin{table*}
\begin{center}
\caption{\label{table2}{
Excitation energy of the IVGDR in MeV of some spherical nuclei calculated as
$\sqrt{m_1/m_{-1}}$ using the NL3 model. ${E_c}(RTF-P)$ is the estimte
of the present work and ${E}(RPA)$ are the RPA results reported in 
\cite{cao03}.}}
\begin{tabular}{cccccccccccc}
\hline
Nucleus&${E_c}(RTF-P)$&${E_c}(RPA)$&Expt.\cite{berman75} \\ \hline
$^{16}$O&22.05&21.1&22.3-24\\
$^{40}$Ca&19.90&19.57&19.8$\pm$0.5\\
$^{90}$Zr&17.18&17.19&16.5$\pm$0.2\\
$^{116}$Sn&16.25&15.77&15.7$\pm$0.2\\
$^{208}$Pb&13.54&13.16&13.3$\pm$0.1\\
\hline
\end{tabular}
\end{center}
\end{table*}

\begin{table*}
\begin{center}
\caption{\label{table3}{
Semi-classical $m_{-1}$ sum rule in fm$^2$ MeV$^{-1}$ of some spherical nuclei computed as explained in this 
work $(RTF-P)$ and using the DM approach (\ref{eq44}) ($(RTF-LDM)$ input). 
All these calculation are performed with the NL3 parameter set. }}
\begin{tabular}{cccccccccccc}
\hline
Nucleus&$m_{-1}(RTF-P)$&$m_{-1}(RTF-DM)$\\ \hline
$^{16}$O&0.219&0.192\\
$^{40}$Ca&0.699&0.657\\
$^{68}$Ni&1.499&1.403\\
$^{90}$Zr&2.130&2.076\\
$^{116}$Sn&3.061&3.012\\
$^{120}$Sn&3.250&3.173\\
$^{208}$Pb&7.685&7.256\\
\hline
\end{tabular}
\end{center}
\end{table*}
Experimental information about the excitation energy of the ISGMR in medium and 
heavy nuclei is obtained from inelastic scattering of $\alpha$ particles measured 
at forward angles \cite{young99,young01,youngblood01a,youngblood01b}. As a first
 application of our perturbative constrained approach developed in previous sections, we compute
the excitation energy of the ISGMR for several nuclei which value is
experimentally known  \cite{young99,young01,youngblood01a,youngblood01b} using the
RMF NL3 parametrization \cite{lala97}. Although the sum rule approach, which assumes a well 
defined peak, for light nuclei may be questionable, we display in table ~\ref{table1} our semi-classical
RTF estimate of the excitation energy of ISGMR of the nuclei $^{16}$O, $^{40}$Ca, 
$^{90}$Zr, $^{116}$Sn, $^{144}$Sm and $^{208}$Pb calculated with our perturbative 
constrained approach. In the same table we also show the constrained values 
obtained using the Relativistic Extended Thomas-Fermi Approach (RETF), which includes
$\hbar^2$ corrections, computed as described in Ref.\cite{patra02} together with the
corresponding experimental values. For a sake of completeness, we also display in the same table
the constrained Hartree and RPA values of these excitation energies reported in Ref.\cite{wei13}.  
From this table some comments are in order. First, as we mentioned previously, the constrained
Hartree predictions of the excitation energy of the ISGMR practically concides with the values
extracted from RRPA calculations, which numerically proves the conjecture of Ref.\cite{ma01}.
Second, as it was pointed out in \cite{patra02}, the importance of the $\hbar^2$
corrections in the excitation energy of the ISGMR is actually small as it can be seen 
from the comparison between the RTF and RETF results. Third, the excitation 
energies of the ISGMR decreases with increasing mass number, as expected, in both semi-classical 
and quantal (constrained Hartree and RRPA) predictions. However, the slope is larger for the 
semi-classical than for the quaintly  estimates. For light and medium nuclei the semi-classical 
values overestimate the quantal ones while the opposite is true for $^{208}$Pb, which is 
the heaviest nucleus considered in this work. 

The experimental values of the excitation energies of the IVGDR 
basically come from the analysis of the measured photo-absorption cross sections 
as mentioned before \cite{varlamov99}. From a theoretical point of view there are many studies of the 
IVGDR at RPA and RRPA levels using different mean field models in both non-relativistic
and relativistic frames \cite{vretenar03}. However, from long ago the IVGDR
has also been analyzed from a semi-classical point of view. Let us mention in this respect
the estimate of the $m_{-1}$ sum rule on the basis of the Droplet  Model (DM) 
\cite{meyer82} reported more than thirty years ago. More recently, the excitation energies 
of the IVGDR has been estimated using the scaling method \cite{guo90,gleissl90} and performing 
constrained calculations \cite{gleissl90}. In table ~\ref{table2} we display our constrained perturbative 
estimate of the excitation energy of some selected spherical even-even nuclei computed with the 
NL3 parameter set \cite{lala97} as well as the RRPA predictions obtained with the same
NL3 model \cite{todd03} and the experimental values extracted from Ref.\cite{berman75}. 
From this table it can be seen that our semi-classical RTF estimate reproduce remarkably 
well the RRPA values and the experimental data. This result point out, as discussed 
in earlier literature \cite{meyer82,guo90,gleissl90}, that semi-classical approaches, 
as the constrained perturbative RTF method discussed in this work, are well suited for 
describing also the average excitation energy of the IVGDR due to the fact that shell corrections 
in this collective oscillation are, actually, small \cite{meyer82}.

As mentioned before, the $m_{-1}$ sum rule can also be estimated in the framework of the 
LDM as \cite{meyer82}
\begin{equation}
m_{-1} = \frac{A <r^2>}{48 J} \big( 1 + \frac{15}{4 A^{1/3}}\frac{J}{Q}\big),
\label{eq44}
\end{equation}
where $\sqrt{<r^2>}$ stands for the mass rms radius, $J$ for the symmetry energy 
and $Q$ is the so-called surface stiffness coefficient. This coefficient $Q$ is
obtained from a semi-classical RTF calculation in semi-infinite nuclear matter as 
explained in Ref.\cite{mario98} and its corresponding  value is $J/Q$=1.46.
The DM estimates of the $m_{-1}$ sum rule at RTF  level for several nuclei
are reported in table ~\ref{table3}. To obtain these values, in addition to the $J/Q$ ratio and
the symmetry energy of the NL3 model ($J=$37.40 MeV), we also use in Eq.(\ref{eq44}) the 
$<r^2>$ value obtained from a self-consistent RTF calculation for each considered
nucleus. From table ~\ref{table3} we see that the semi-classical DM and the perturbative constrained RTF
estimates of the $m_{-1}$ sum rule are in a good agreement. 
Our RTF predictions of  $m_{-1}$ are also in good agreement with the values, computed 
semi-classically with the Skyrme Skm$^*$ force, displayed  in Table V of \cite{gleissl90}.
For example, $m_{-1}$ values of 0.215, 0.682, 2.054 and 7.047 fm$^2$ MeV$^{-1}$ are reported in
this reference for the nuclei $^{16}$O, $^{40}$Ca,$^{90}$Zr and $^{208}$Pb, respectively. 
In the same reference the experimental values of the  $m_{-1}$ sum rule, obtained
as the integral up to the pion threshold of the measured total nuclear photoabsortion
cross section \cite{meyer82}, are also reported in the same  Table V. These experimental values,
0.215$\pm$0.004, 0.682$\pm$0.016 and 7.35$\pm$0.51 fm$^2$MeV$^{-1}$ for $^{16}$O, $^{40}$Ca and
 $^{208}$Pb, respectively, are in harmony with our semi-classical predictions reported in table ~\ref{table3}.
However, I should be pointed out that recent measurements of the electric polarizability in
$^{208}$Pb, $^{120}$Sn and $^{68}$Ni give a smaller experimental value for the $m_{-1}$ sum rule 
\cite{roca15}. 
\section{Summary and outlook}
\label{con1}
We have estimated the average excitation energy of the isoscalar monopole and 
isovector dipole giant resonances for several nuclei within a semiclassical sum 
rule approach in the relativistic mean field framework with the NL3 parameter set. 
We are aware that a sum rule approach only gives information about some selected 
moments of the RRPA strength function and that in this approach a precise prediction of the excitation 
energy of the reasonance is only possible if the strength is concentrated in a single peak, 
which is the usual scenario for medium and heavy nuclei but not for light nuclei where 
the resonance broadens and fragments. The $m_1$ sum rules are evaluated at RPA level using  the 
semi-classical expectation values calculated with the relativistic Thomas-Fermi approximation . 
The $m_{-1}$ sum rule is obtained through a new constrained perturbative calculation also 
within the relativistic Thomas-Fermi theory. In this calculation the nuclear densities and 
meson and photon fields are splitted into a part corresponding to the unperturbed solution plus a 
perturbative correction, chosen in such a way that it follows the excitation field. The energy 
density corresponding to the constrained problem is expanded up to quadratic terms in these perturbations. 
The application of the variational principle to this energy density allows to obtain the equations 
of motion of the perturbative corrections to nuclear densities and fields.
With this simple semi-classical method the average excitation energies of the isoscalar monopole
and isovector dipole giant resonances is estimated reasonably well as compared with the more 
fundamental but also more cumbersome relativistic RPA calculations. The differences between both
calculations are basically due to shell effects and are less than 5\% in medium and hevay nuclei
where the sum rule can be considered more confidently. This result is in agreement with previous findings 
in earlier literature pointing out that shell effects have little impact on the average excitation energies 
of the giant resonances studied in this chapter. Our constrained perturbative approach has been obtained at 
relativistic Thomas-Fermi level, therefore, it seems reasonable, to improve it by including the $\hbar^2$
corrections which give a more realistic calculation at the nuclear surface. As far as our method is able
to describe collective oscillations with excitation operators without spherical symmetry, as the isovector 
dipole, it seems also appealing to apply our method to another isoscalar and isovector oscillations of
higher multipolarities.

\setcounter{equation}{0}
\setcounter{figure}{0}
\newpage
\newpage
\chapter{Effects of self interacting $\omega$-meson
on finite and infinite nuclear system.}
\label{chapter6}


The most mysterious and the debatable subject in nuclear physics is the nucleon-nucleon interaction.  A lots of theories have been proposed for the nucleon-nucleon interaction but still it is not clear. The nucleon-nucleon interaction affects the nuclear structure in a more prominent way.  So without the study of the nucleon -nucleon interaction nuclear structure will remain incomplete.  In this chapter, a detailed study is made for the nucleon-nucleon interaction based on relativistic mean field theory in which the potential is explicitly expressed in terms of masses and the coupling constants of the meson fields.  An unified treatment for self-coupling of isoscalar-scalar $\sigma-$, isoscalar-vector $\omega$-mesons and their coupling constants are given in an analytic form. The present investigation is focused on the effects of  self-interacting higher order $\sigma$ and $\omega$ fields
on nuclear properties. An attempt is made to explain the many-body effects by higher order  couplings $\sigma-$ and $\omega-$fields, which generally occurs in the high-density region. Both infinite nuclear matter and the
finite nuclear properties are included in the present study to observe
the behavior and sensitivity of this self-interacting terms.




\section{Introduction}
The Nucleon-Nucleon (NN) interaction  has been investigated for over half a 
century \cite{yakawa35}. Probably this is a
long standing question in history of nuclear physics. In fact,
describing the nuclear properties in terms of the interactions
between the nucleon pair is indeed the main goal for nuclear
physicists. The NN-interaction in terms
of mediated mesons was put forwarded by Yukawa \cite{yakawa35}
in 1935. Although the meson theory is not fundamental from the 
QCD point of view, it  has improved our understanding
of the nuclear forces as well as highlighted some good quantitative
results \cite{fuji86,fuji86a}. The modern theory of NN potential
in term of particle exchanges was made possible by the development
of quantum field theory \cite{fuji86a}. However, at low-energy,
one can assume that the interactions are instantaneous and
therefore the concept of interaction potential becomes useful.
The derivation of a potential through particle exchange is
important to understand the nuclear force as well as structural
properties.

Now-a-days, there are number of developments in the nuclear theory by
introducing quark and gluon in connection with the NN-potential
\cite{gross11,mech11}. These  models give the fundamental
understanding of NN interaction at present. Here, we are not
addressing all these rich and long standing problems about
NN-potential. Our aim is to highlight some basic features  of the
NN-interactions arising from relativistic mean field (RMF)
Lagrangian \cite{bhu12,bidhu14,love70,love72}. The behavior of
this potential gives an idea about the saturation properties of nuclear
force at high density limit.

This chapter is organized as follows. In Section ~\ref{bjp2}, we briefly discuss
the theoretical formalism of NN-interaction based on relativistic mean
field (RMF) theory. The general forms of the NN potentials are expressed
in the coordinate space ({\it r-space}) in term of masses  and coupling
constants of the force parameters. In Section ~\ref{bjp3}, we review 
the effects of modified term in the Lagrangian and their effects on 
the finite nuclei and in infinite nuclear matter. 
In Section ~\ref{bjp4}, we make few comments about the current 
form of the NN-interaction to saturation condition of nuclear systems.

\section{Theoretical frameworks}
\label{bjp2}
The nuclear potential in relativistic mean field (RMF) is possible via
various mesons interaction with nucleons. The linear relativistic
mean field (RMF) Lagrangian density for a nucleon-meson many-body system
\cite{miller72,walecka74,horo81,ser86} is given as:
\begin{eqnarray}
\label{bjpeq1}
{\cal L}&=&\overline{\psi_{i}}\{i\gamma^{\mu}
\partial_{\mu}-M\}\psi_{i}
+{\frac12}\partial^{\mu}\sigma\partial_{\mu}\sigma
-{\frac12}m_{\sigma}^{2}\sigma^{2} 
-g_{s}\overline{\psi_{i}}\psi_{i}\sigma \nonumber\\
&&-{\frac14}\Omega^{\mu\nu}
\Omega_{\mu\nu}+{\frac12}m_{w}^{2}V^{\mu}V_{\mu} 
-g_{w}\overline\psi_{i} \gamma^{\mu}\psi_{i} V_{\mu}
-{\frac14}\vec{B}^{\mu\nu}.\vec{B}_{\mu\nu} \nonumber\\
&&+{\frac12}m_{\rho}^{2}{\vec
R^{\mu}} .{\vec{R}_{\mu}} -g_{\rho}\overline\psi_{i}
\gamma^{\mu}\vec{\tau}\psi_{i}.\vec
{R^{\mu}}-{\frac14}F^{\mu\nu}F_{\mu\nu}\nonumber\\
&&-e\overline\psi_{i}
\gamma^{\mu}\frac{\left(1-\tau_{3i}\right)}{2}\psi_{i}A_{\mu}. 
\end{eqnarray}
If, we neglect the $\rho-$ meson, it corresponds 
to the Walecka model in its original
form \cite{walecka74,horo81}. From the above relativistic Lagrangian, we
obtain the field equations for the nucleons and mesons as,
\begin{eqnarray}
\Bigl(-i\alpha.\bigtriangledown+\beta(M+g_{\sigma}\sigma)
+g_{\omega}\omega+g_{\rho}{\tau}_3{\rho}_3 
){\psi}_i={\epsilon}_i{\psi}_i,\\
\nonumber \\
(-\bigtriangledown^{2}+m_{\sigma}^{2})\sigma(r)=-g_{\sigma}{\rho}_s(r),\\
\nonumber\\
(-\bigtriangledown^{2}+m_{\omega}^{2})V(r)=g_{\omega}{\rho}(r),\\
\nonumber \\
(-\bigtriangledown^{2}+m_{\rho}^{2})\rho(r)=g_{\rho}{\rho}_3(r).
\end{eqnarray}
In the limit of one-meson exchange and mean-field (the fields are replaced
by their expectation values or c-number), for a heavy and static baryonic medium, the solution of
single nucleon-nucleon potential for scalar ($\sigma$) and vector ($\omega$,
$\rho$) fields are given by \cite{ser86,brock78},
\begin{eqnarray}
\label{bjpeq6}
V_{\sigma}(r)&=&-\frac{g_{\sigma}^{2}}{4{\pi}}\frac{e^{-m_{\sigma}r}}{r}, 
\end{eqnarray}
and
\begin{eqnarray}
\label{bjpeq7}
V_{\omega}(r)&=&+\frac{g_{\omega}^{2}}{4{\pi}}\frac{e^{-m_{\omega}r}}{r},\quad
V_{\rho}(r)=+\frac{g_{\rho}^{2}}{4{\pi}}\frac{e^{-m_{\rho}r}}{r}.
\end{eqnarray}
The total effective nucleon-nucleon potential is obtained from the scalar
and vector parts of the meson fields. This can be expressed as \cite{bhu12},
\begin{eqnarray}
\label{bjpeq8}
v_{eff}(r)&=&V_{\omega} +V_{\rho} +V_{\sigma} 
=\frac{g_{\omega}^{2}}{4{\pi}}\frac{e^{-m_{\omega}r}}{r}
+\frac{g_{\rho}^{2}}{4{\pi}}\frac{e^{-m_{\rho}r}}{r}\nonumber\\
&&-\frac{g_{\sigma}^{2}}{4{\pi}}\frac{e^{-m_{\sigma}r}}{r}.
\label{eq:10}
\end{eqnarray}
\subsection{Non-linear case}
The Lagrangian density in the above Eq.~\ref{bjpeq1} contains only linear
coupling terms, which is able to give a qualitative description of the
nuclear system \cite{ser86,brock78}. The essential nuclear matter 
properties like incompressibility and the surface properties 
of  finite nuclei cannot
be reproduced quantitatively within this linear model. The suppression of
the two-body interactions within a nucleus in favor of the interaction
of each nucleon with the average nucleon density, means that the
non-linearity acts as a smoothing mechanism and hence leads in the
direction of the one-body potential and shell structure
\cite{boguta77,bodmer91,sgmuca92,suga94}. The replacement of mass
term $\frac{1}{2} m_{\sigma}^2\sigma^2$
of $\sigma$ field by $U (\sigma)$ and
$\frac{1}{2} m_{\omega}^2 V^{\mu} V_{\mu}$ of $\omega$ field by
$U (\omega)$. This can be expressed as
\begin{eqnarray}
\label{bjpeq2}
U (\sigma) = \frac{1}{2}m_{\sigma}^2 \sigma^2+ \frac{1}{3}g_{2}\sigma^{3} 
+\frac{1}{4} g_{3}\sigma^{4}, 
\end{eqnarray}
\begin{eqnarray}
\label{bjpeq3}
U (\omega) = \frac{1}{2}m_{\omega}^2 V_{\mu}V^{\mu} +\frac{1}{4}c_3 (V_{\mu}V^{\mu})^{2}.
\end{eqnarray}
The terms on the right hand side of Eqs.(~\ref{bjpeq2})- (\ref{bjpeq3}), except the first term,
from the non-linear self coupling of the $\sigma$ and
$\omega$ mesons, respectively \cite{boguta77,bodmer91}. Here, the
non-linear parameter $g_2$ and $g_3$ due to $\sigma-$ fields are adjusted
to the surface properties of finite nuclei \cite{schif51,schiff051}.
In general most of the successful fits like NL1 and NL3 sets yield, 
the $+ve$ and $-ve$ signs for $g_2$ and
$g_3$, respectively. The negative value of $g_3$ is a serious problem in
quantum field theory and  responsible for the divergence of a solution 
in the lighter mass region of periodic table i.e. for higher density region. As, we are dealing within the mean field level and
with normal nuclear matter density, the corresponding $\sigma$ field is
very small and the $-ve$ value of $g_3$ is still allowed
\cite{schif51,brock90}. Again, $c_3=\frac{1}{6}\zeta_0$ is the non-linear
coupling constant for self-interacting $\omega$-mesons. With the
addition of the non-linear terms in the Eqs. (~\ref{bjpeq2})- (\ref{bjpeq3}) 
to the Lagrangian, the field equations for $\sigma$ and $\omega$- 
fields (in Eq.\ref{bjpeq6}-\ref{bjpeq7} ) are modified
as:
\begin{eqnarray}
(-\bigtriangledown^{2}+m_{\sigma}^{2})\sigma(r)=-g_{\sigma}{\rho}_s(r)
-g_2 r \sigma^2 (r) - g_3 \sigma^3 (r), \nonumber \\
\nonumber \\
(-\bigtriangledown^{2}+m_{\omega}^{2})V(r)=g_{\omega}{\rho}(r)
-c_3 W^3 (r).
\end{eqnarray}
Here, $W(r)=g_{\omega}V_0 (r)$ is the non-linear self-interacting
$\omega$-fields. Because of the great difficulty in solving the above
nonlinear differential equations, it is essential to have a variational
principle for the estimation of the energies \cite{schif51,schiff051}. 
In the static case, the negative
sign of the third term in the Lagrangian is computed with the correct
source function and an arbitrary trial wave function. The limit on the
energy has a stationary value equal to the proper energy when the
trial wave-function is in the infinitesimal neighborhood of the correct
one. Now, the solution for the modified $\sigma$ and $\omega$
fields are given as \cite{schif51}
\begin{eqnarray}
V_{\sigma}=-\frac{g_{\sigma}^{2}}{4{\pi}}\frac{e^{-m_{\sigma}r}}{r}
+\frac{g_{2}^{2}}{4{\pi}}r \frac{e^{-2m_{\sigma}r}}{r}
+\frac{g_{3}^{2}}{4{\pi}}\frac{e^{-3m_{\sigma}r}}{r}, \nonumber \\
\nonumber \\
V_{\omega}=\frac{g_{\omega}^{2}}{4{\pi}}\frac{e^{-m_{\omega}r}}{r}
-\frac{c_{3}^{2}}{4{\pi}}\frac{e^{-3m_{\omega}r}}{r^2}.
\label{bjpeq10}
\end{eqnarray}
The new NN-interaction analogous to $M3Y$ form and is able to improve
the incompressibility and deformation of the finite nuclei results
\cite{brock90}. In addition to this, the non-linear self coupling
of the $\sigma$ and $\omega$-mesons help to generate the repulsive
and attractive part of the NN potential at $long$ and at $short$
distance, respectively to satisfy the saturation properties (Coester-band
problem) \cite{coester70}. We are dealing with two type of mesons, 
one is scalar ($\sigma$) and other is vector $(\omega)$. The range of
their interactions are also different due to their different masses. 
 Consider the case of $\sigma-$meson, where the range of interaction
is $\sim \frac{\hbar}{m_{\sigma}c}$ fm. In this range the attractive 
part of the potential comes from the exchange of the $\sigma-$meson.
The density dependent many-body effect demands a repulsive part in this
region. This is given by the self interacting terms like $\sigma^3$ and
$\sigma^4$ \cite{fujita57}. A suitable adjustment of the
parameter able to reproduces the proper potential satisfying the Coester
band problem. Generally, the exchange of $\omega-$meson gives the 
repulsive potential in the short-range part of the hard core region. 
Diametrically, opposite phenomenon is also occurs in case of $\omega-$meson 
coupling. Contrary to the non-linearity of $\sigma-$meson nature, the
self-coupling of $\omega-$meson gives an attractive component at very short 
distance ($\sim 0.2$ fm) of the nuclear potential.
The non-linear terms also generate the most discussed $3-body$
interaction \cite{fujita57}. The modified effective nucleon-nucleon 
interaction is defined as \cite{bhu12}:
\begin{eqnarray}
\label{bjpeq13}
v_{eff}(r)&=&V_{\omega}+V_{\rho}+V_{\sigma}\nonumber \\
&& =\frac{g_{\omega}^{2}}{4{\pi}}\frac{e^{-m_{\omega}r}}{r}
+\frac{g_{\rho}^{2}}{4{\pi}}\frac{e^{-m_{\rho}r}}{r}
-\frac{g_{\sigma}^{2}}{4{\pi}}\frac{e^{-m_{\sigma}r}}{r} \nonumber \\
&& +\frac{g_{2}^{2}}{4{\pi}}r\frac{e^{-2m_{\sigma}r}}{r}
+\frac{g_{3}^{2}}{4{\pi}}\frac{e^{-3m_{\sigma}r}}{r}
-\frac{c_{3}^{2}}{4{\pi}}\frac{e^{-3m_{\omega}r}}{r}.
\label{eq:10}
\end{eqnarray}


\section{Results and Discussions}
\label{bjp3}
The  Eq. (~\ref{bjpeq13}) shows that the 
effective NN-potential  in terms of the well known inbuilt RMF
theory parameters of $\sigma$, $\omega$ and $\rho$ meson fields. Here,
we have used RMF (NL3) force parameter along with varying $c_3$ for
$\omega$-self interactions to determine the nuclear properties. The
values of the parameters for NL3-force are listed in Table~\ref{bjpta1}.
Although, the $\omega^4$ term is already there in the FSU-Gold parameter
\cite{pika01,fsu}, here we are interested to see the effect
of non-linear self coupling of $\omega$ meson. Thus, we have added
the self-interaction of $\omega$ with coupling constant $c_3$ on top
of NL3 set and observed the possible effects.


\begin{figure}
\includegraphics[width=0.8\columnwidth,clip=true]{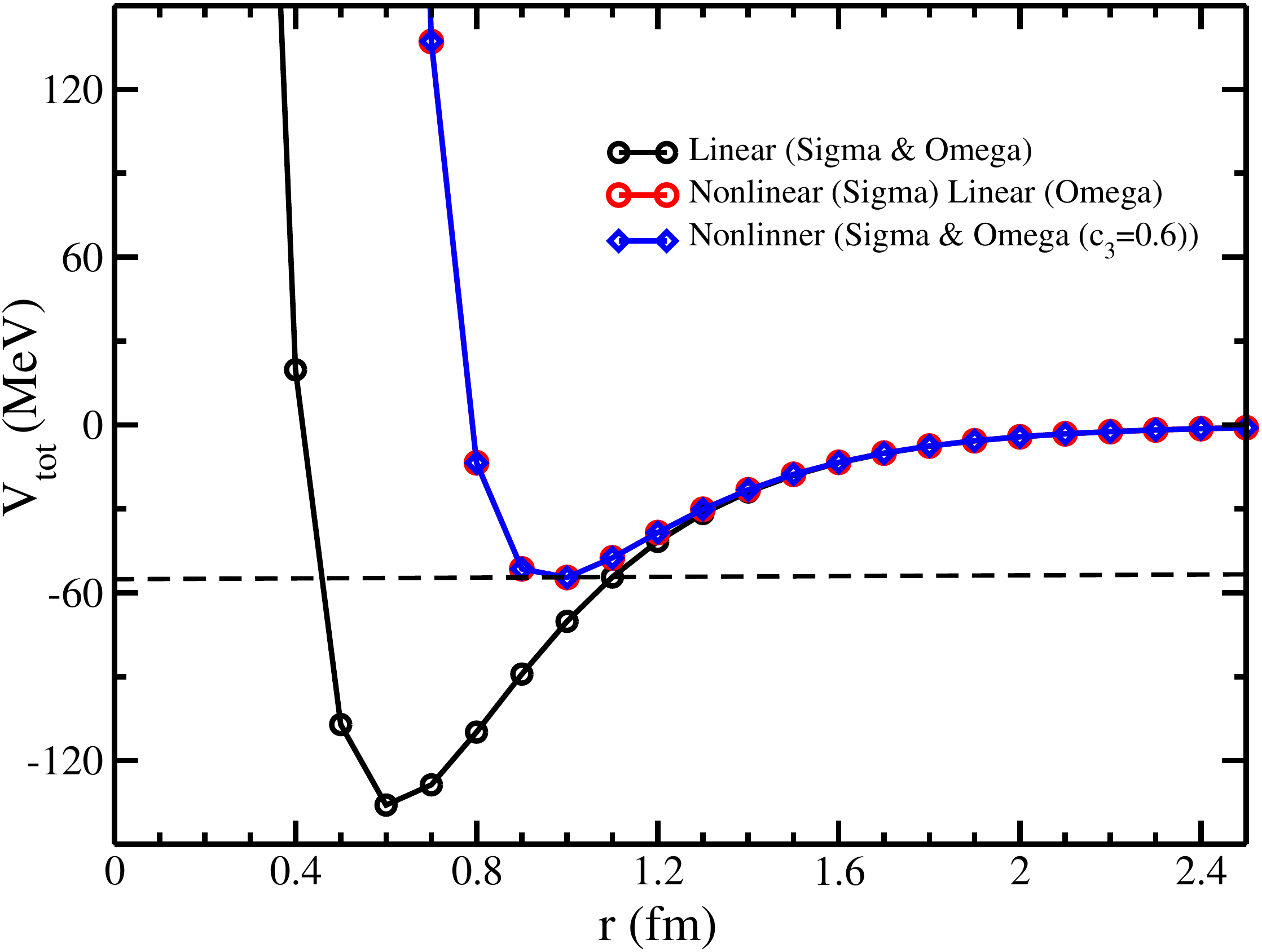}
\caption{The effective NN interaction potentials as a function of distance $r$ from Eq.~\ref{bjpeq6}-~\ref{bjpeq10} for NL3 parameter set.}
\label{bjpfi1}
\end{figure}
First of all, we have calculated the NN-potential for linear and
non-linear cases using Eq.~\ref{bjpeq8} and ~\ref{bjpeq13}, respectively. The obtained
results for each cases are shown in Fig.~\ref{bjpfi1}. From the figure, it is
clear that without taking the non-linear coupling for RMF (NL3), one
cannot reproduce a better NN-potential. In other word, the depth of
the potential for linear and non-linear are $\sim$ 150 MeV and 50 MeV,
respectively. Thus, the magnitude of the depth for linear case is not
reasonable to fit the NN-data. Again, considering the values of $c_3$,
there is no significant change in the total {\it nucleon-nucleon}
potential. For example, the NN-potential does not change at all for
$c_3$ $\simeq$ $\pm$ 0.6, which can be seen from Fig.~\ref{bjpfi1}.

\begin{figure}
\includegraphics[width=0.8\columnwidth,clip=true]{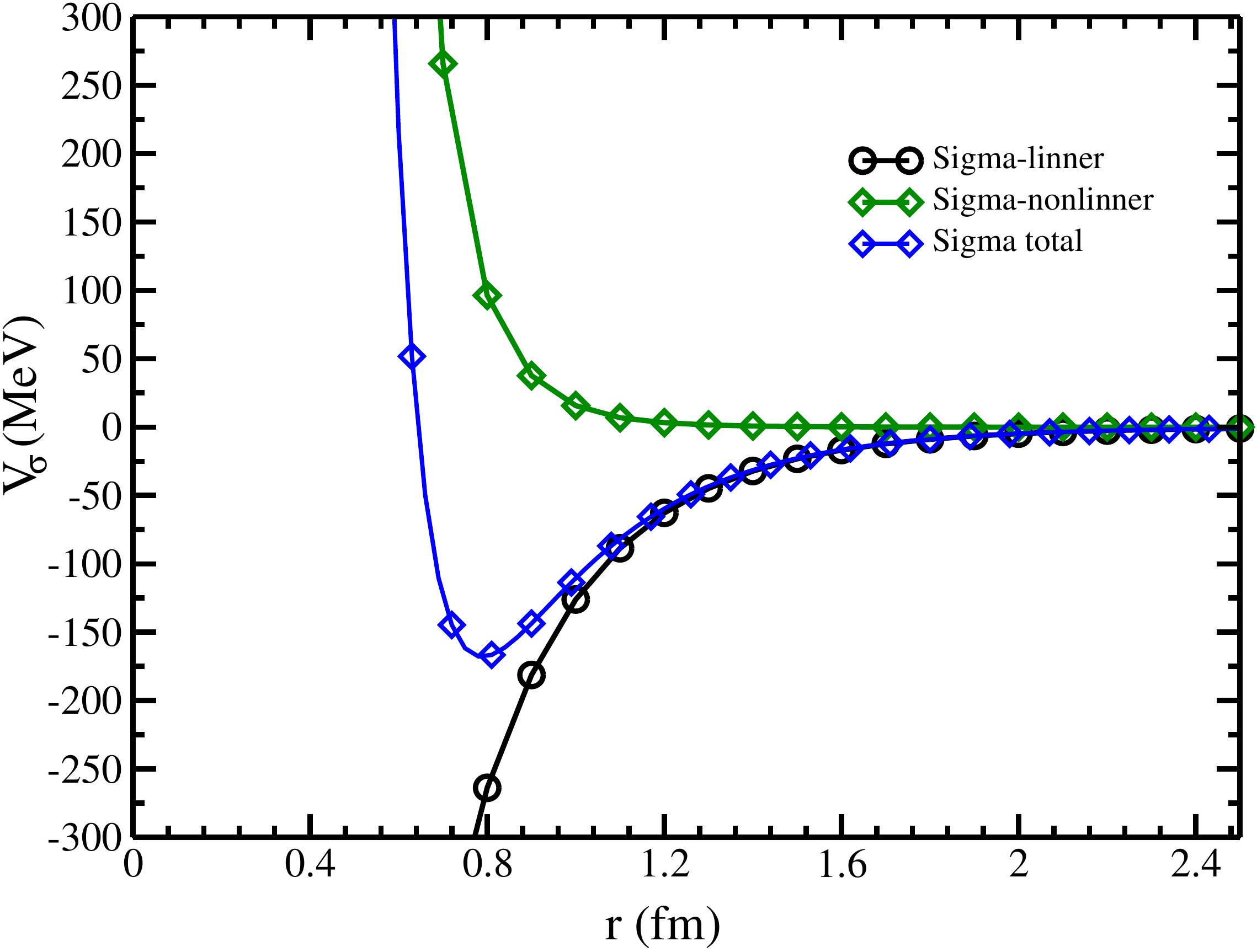}
\caption{ The contribution of $\sigma$-potential from linear, non-linear
and total as a function of distance $r$ for NL3 parameter set.}
\label{bjpfi2}
\end{figure}


\begin{figure}
\includegraphics[width=0.8\columnwidth,clip=true]{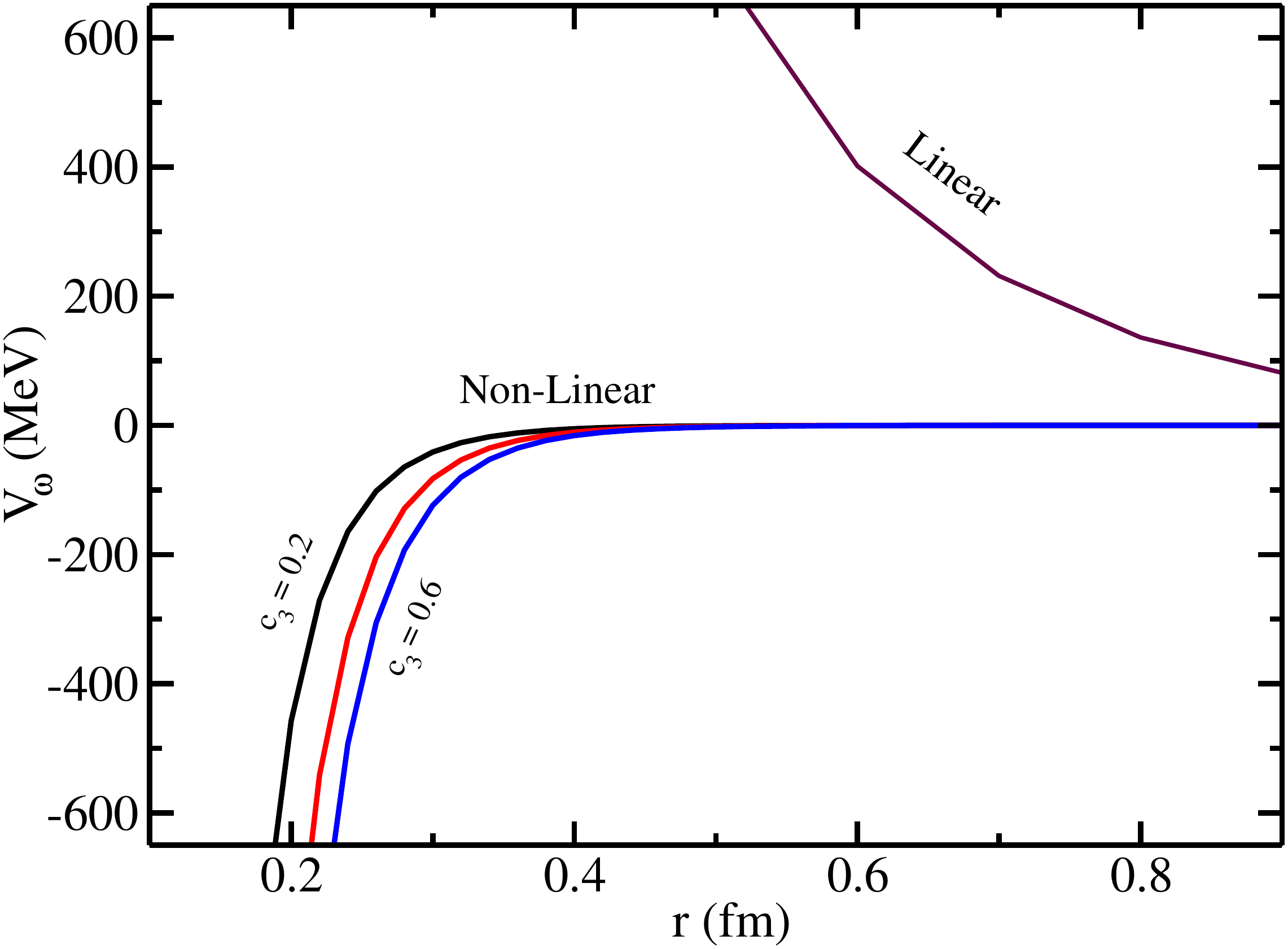}
\caption{ The contribution of $\omega$-potential from linear, non-linear
and total as a function of distance $r$ for for NL3 parameter set.}
\label{bjpfi3}
\end{figure}

Further, we have calculated the individual contribution of meson
fields to the NN-potential in particular case of $\sigma$ and
$\omega$-mesons. In case of $\sigma$-field, we have calculated
the linear and non-linear contributions separately, and combined
to get the total $\sigma$-potential as shown in Fig.~\ref{bjpfi2}.
From the figure, one can find the non-linear self-interacting
terms in the $\sigma$-field play an important role (contributing 
a repulsive interaction) in the attractive part of the $\sigma-$meson 
domain, giving rise to a repulsive potential complementing the 3-body
effect of the nuclear force in the total NN-potential \cite{schif51}. The
linear and non-linear contribution of the $\omega$-field at
various $c_3$ are shown in Fig.~\ref{bjpfi3}. The important feature in this
figure is that the linear term give an infinitely large repulsive
barrier at $\sim 0.5$ fm, in which range, the influence of the non-linear
term of the $\omega-$meson is zero. However, this non-linear
terms is extremely active at very short distance ($\sim 0.2$ fm), which
can be seen from the figure. 


\begin{figure}
\includegraphics[width=0.8\columnwidth,clip=true]{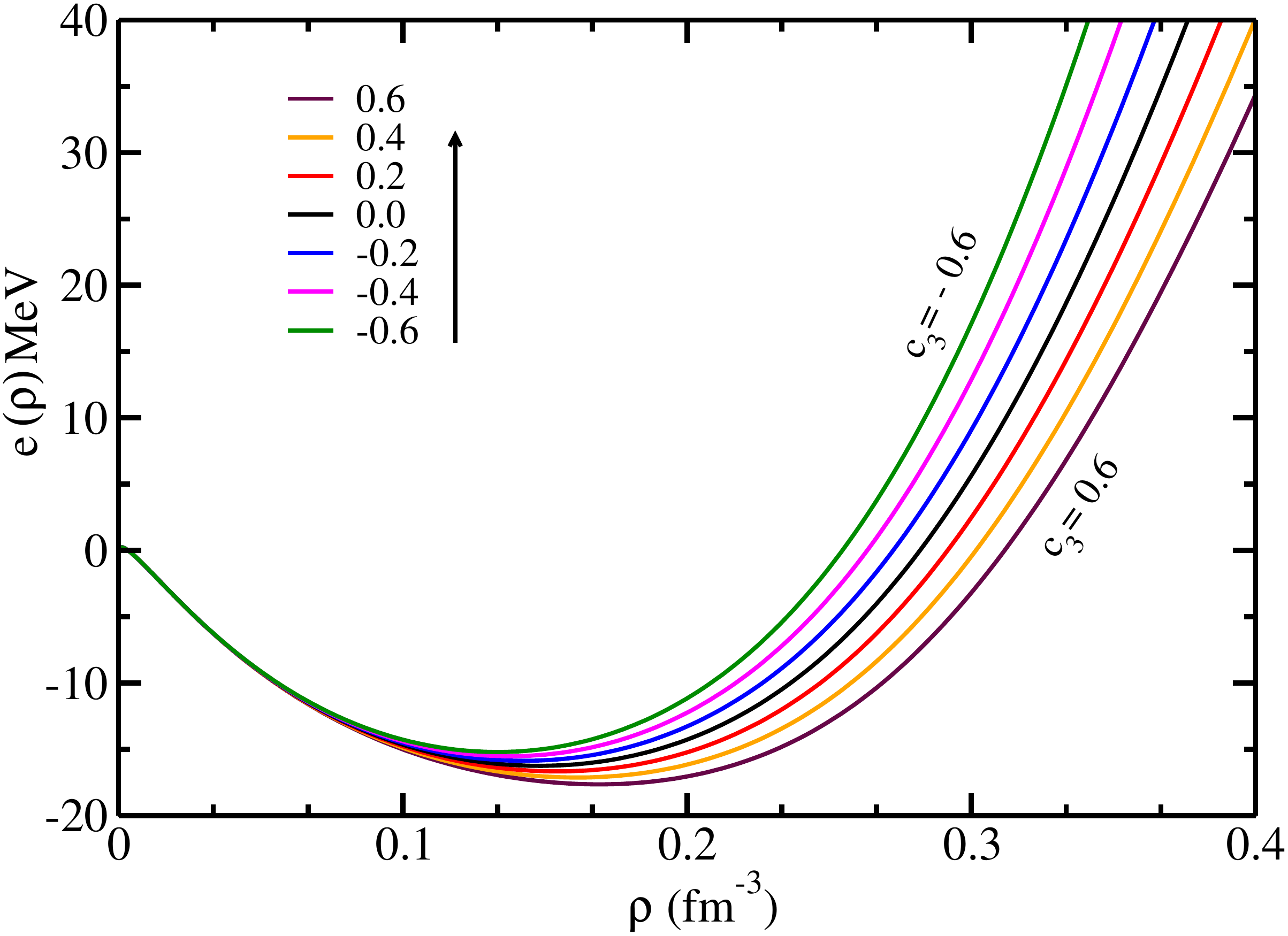}
\caption{The energy per particle of symmetric nuclear matter as a
function of baryon density for various values of $c_3$.}
\label{bjpfi4}
\end{figure}
That means, mostly the (i) linear term of the $\sigma-$ meson is responsible
for the attractive part of the nuclear force (nuclear binding energy) (ii)
the non-linear terms are responsible for the repulsive part of the nuclear
force at long distance, which simulate the 3-body interaction of the nuclear
force \cite{schif51}, which also help to explain the Coester band problem.
(iii) similarly, the linear term of the $\omega-$ meson is restraint for
the repulsive part of the nuclear force ({\it known as hard core}) and
(iv) the non-linear self-coupling of the $\omega-$ meson
($\frac{1}{4}c_3 V_{\mu}V^{\mu}$) is responsible for the attractive part
in the very shortest ($\sim 0.2 fm$) region of the NN-potential. 
It is worthy to
mention that the values of these constants are different for
different forces of RMF theory. Hence, the NN-potential somewhat
change a little bit in magnitude by taking different forces,
but the nature of the potential remains unchanged.
\subsection{Energy density and Pressure density}
In the present work, we study the effect of the additional term
on top of the $NL3$ force parameter to the Lagrangian \cite{lala97}, 
which comes
from the self-interaction of the vector fields with $c_3$ as it is
done in the Refs. \cite{pika01,fsu,furn87,furn89}. The inclusion 
of this term is
not new, it is already taken into account for different forces
of RMF and effective field theory motivated relativistic mean
field theory (E-RMF). Here, our aims is to see the effect of $c_3$ to
the nuclear system and the contribution to the attractive part of
the hard core of NN-potential. We have solved the mean field equations
self-consistently and estimated the energy and pressure density as
a function of baryon density. The NL3 parameter set along with the 
additional $c_3$ is used in the calculations\cite{furn87,furn89}. 
The obtained results for different values of $c_3$ are shown 
in Figs.~\ref{bjpfi4} and ~\ref{bjpfi5},
respectively. From the figure, it is clearly identify that the $-ve$
value of $c_3$ gives the {\it stiff} equation of state (EOS), meanwhile
the $+ve$ value shows the {\it soft} EOS. It is to be noted that mass
and radius of the neutron star depends on the softness and stiffness
of EOS. Here, in our investigation, we observed that the softening
of the EOS depends on the non-linear coupling of the $\omega-$ meson
\cite{suga94,fsu}. The recent measurement of Demorest {\it et. al.}
\cite{demo10} put a new direction that the NL3 force needs slightly
softer EOS. However, when we deals with G2 (E-RMF) model, the results
of Ref. \cite{singh13} demands a slightly stiffer EOS. This implies
that, the value of $c_3$ should be fixed according to solve the above
discussed problem.

\begin{figure}
\includegraphics[width=0.8\columnwidth,clip=true]{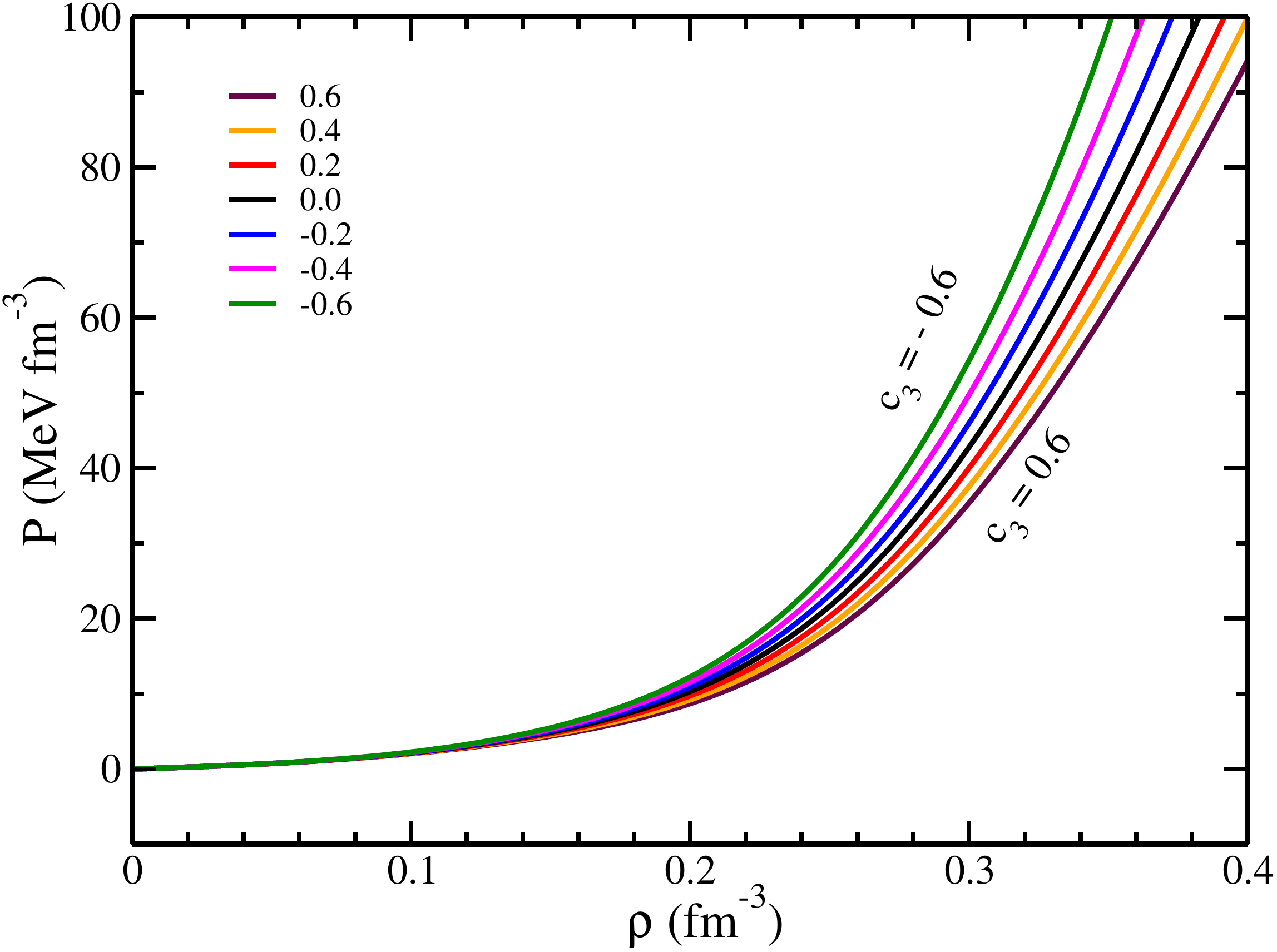}
\caption{The pressure density of symmetric nuclear matter
as a function of baryon density for various values of $c_3$.}
\label{bjpfi5}
\end{figure}


\begin{figure}
\includegraphics[width=0.8\columnwidth,clip=true]{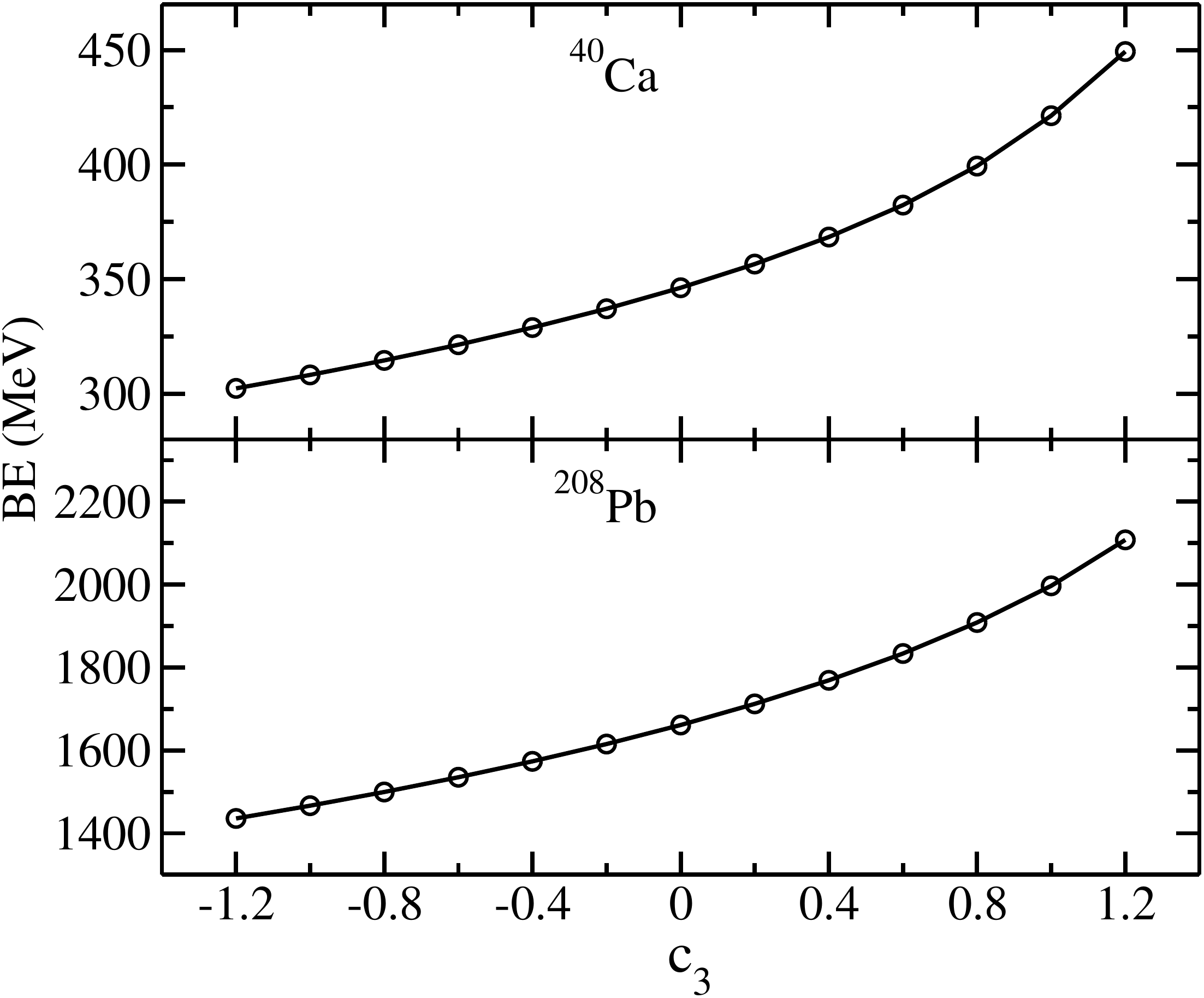}
\caption{The binding energy of $^{40}$Ca and $^{208}$Pb
in their ground state for different values of $c_3$.}
\label{bjpfi6}
\end{figure}

\begin{figure}
\includegraphics[width=0.8\columnwidth,clip=true]{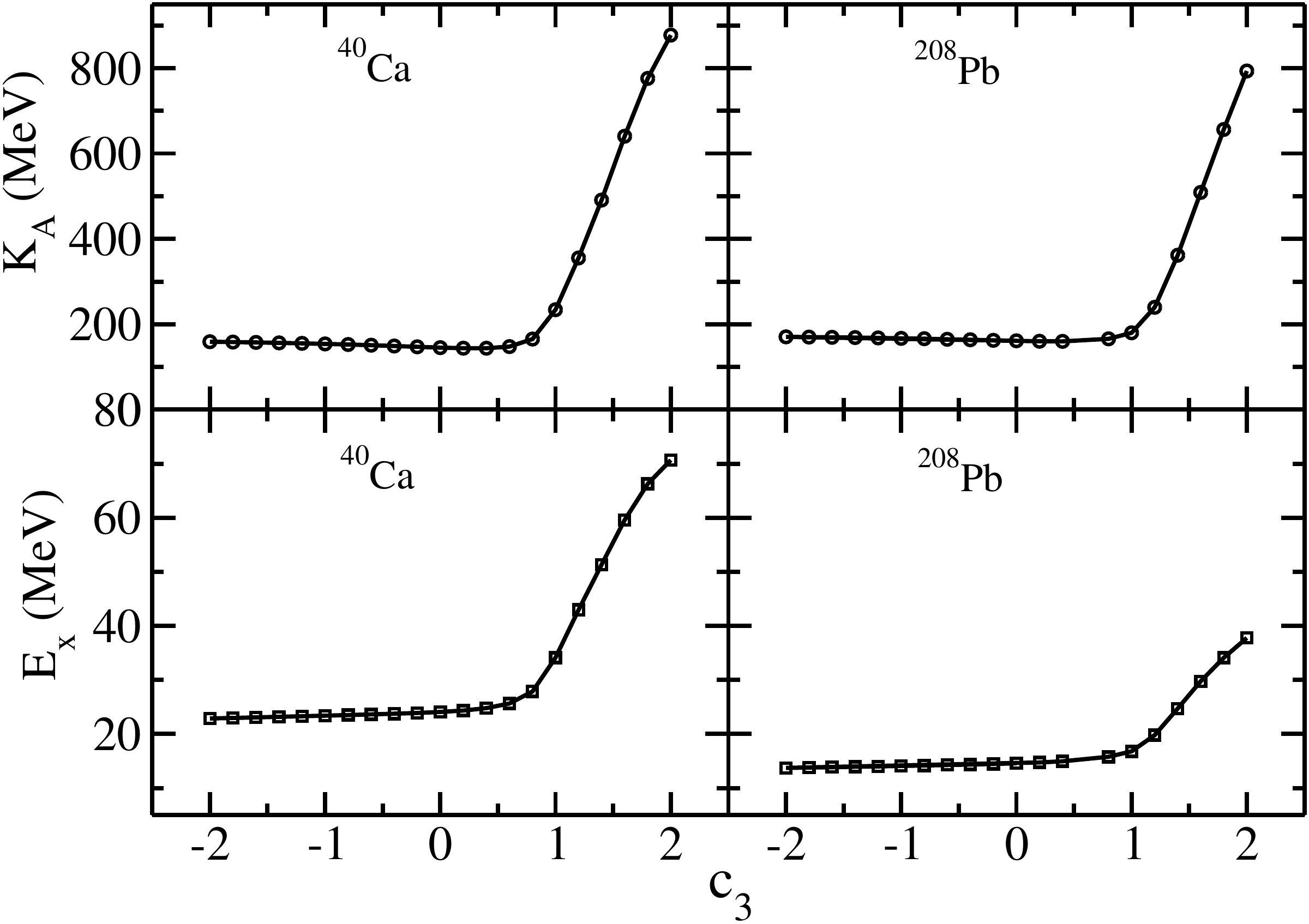}
\caption{The excitation energy as a function of
$c_3$ for $^{40}$Ca and $^{208}$Pb. (b) The incompressibility of $^{40}$Ca
and $^{208}$Pb as a function of $c_3$.}
\label{bjpfi7}
\end{figure}


\begin{table*}
\begin{center}
\caption{\label{bjpta1}The values
of $m_{\sigma}$, $m_{\omega}$, $m_{\rho}$
(in MeV) and $g_{\sigma}$, $g_{\omega}$, $g_{\rho}$ for RMF (NL3)
force, along with the self-interacting $\omega-$ field with coupling
constant $c_3$ .}
\begin{tabular}{cccccccccc}
\hline
 set & ${m}_\sigma$ & ${m}_\omega$ &$m_{\rho}$&$g_{\sigma}$&$g_{\omega}$&$
g_{\rho}$&$g_{2}(fm^{-1})$&$g_{3}$&$c_{3}$ \\
\hline
NL3&508.194&782.5&763.0&08.31&13.18&6.37&-10.4307&-28.8851&0.0$\pm$0.6\\
 \hline
\end{tabular}
\end{center}
\end{table*}

\subsection {Binding energy, Excitation energy and Compressibility}
To see the sensitivity of $c_3$ on the finite nuclei, we calculated
the binding energy (BE), giant monopole excitation energy ($E_x$)
for $^{40}$Ca and $^{208}$Pb nuclei as representative cases as
a function of $c_3$. The excitation energy and incompressibility are 
calculated by scaling method within the framework of extended Thomas-Fermi 
approximation. The obtained results are shown in upper and
lower panel of the Fig.~\ref{bjpfi6} and ~\ref{bjpfi7}. 
The experimental and empirical data are also displaced for comparison.

       From the figure, one observes a systematic variation
of binding energy by employing the isoscalar-vector selfcoupling
parameter $c_3$. For example, the binding energy
monotonically changes for all values of $c_3$. When $c_3 = 0$,
the NL3 set reproduces the original binding energies for
both $^{40}Ca$ and $^{208}Pb$, which are fitted with the data while
constructing the force parameters. As soon as the self-coupling
constant is non-zero, the calculated BE deviates
from the data, because of the influence of $c_3$. Again, analyzing
the excitation energy, we find reasonable match of $E_x$
with the observation (lower panel of Fig. ~\ref{bjpfi7}). These values
of $E_x$ remain almost constant for a wide range of $c_3$ ($\sim$ −2
to $\sim$1), beyond which $E_x$ increases drastically for positive
value only. Further, we analyze the variation of compressibility
modulus with c3 for $^{40}$Ca and $^{208}$Pb (upper panel of
Fig. ~\ref{bjpfi7}). We include both positive and negative values of $c_3$
(−2 to +2) to know the effects on the sign of $c_3$. Similar
to the monopole excitation energy, we find that the compressibility
modulus does not change with the increase of
c3 up to some optimum value. The empirical (nuclear matter
compressibility modulus $K_\infty$) data of ${K_\infty}^{emp}= 210\pm30$ MeV 
and the nuclear matter incompressibility for NL3 set (271.76 MeV) are given 
in the figure to have an idea about the bridge between the nuclear matter 
limit for finite nuclei. In general, these two values
along with the finite nuclei compressibility modulus
gives an overall estimation about a possible link among
them in finite nuclei and nuclear matter limit. It is interesting
to notice that although we get a stiff equation of state
with negative value of $c_3$ for infinite nuclear matter system,
this behavior does not appear in finite nuclei, i.e. the
$K_A$ and $E_x$ do not change with sign of $c_3$. May be the
density with which we deal in the finite nucleus is responsible
for this discrepancy. However, $K_A$ and $E_x$ increase
substantially after certain value of $c_3$, i.e. the finite nucleus
becomes too much softer at about $c_3$ $\sim$1.0 resulting a larger
incompressibility.

\section{Summary and Conclusions}
\label{bjp4}
In summary, we analyzed the effects of the non-linear self-coupling
of the $\sigma-$scalar and $\omega-$vector mesons. At long range, i.e.,
more than 0.5 fm, the self coupling of the 
$\sigma-$meson gives a repulsive component contrary to the
attractive part of the linear term. This repulsive nature
of the nuclear potential originated from the nonlinear terms of the
$\sigma-$meson couplings simulate the 3-body
force. This 3-body force is mostly responsible to solve the Coester band
problem in RMF formalism. On the other hand, at extremely short distance, 
the nonlinear term of the $\omega-$meson coupling gives
a strongly attractive potential for both $positive$ and $negative$
value of $c_3$. This short range distance is about
$0.2 fm$, beyond (more than 0.2 fm) which the $vector-$meson interaction itself
shows a strong repulsion due to its linear interaction of the $\omega-$meson 
 and responsible for the saturation of nuclear force. Thus, one conclude 
that the effects of the vector self-coupling is crucial for the attractive 
nature of the nuclear force at the extremly short range region and should 
be taken in equal footing while constructing the force parameter in  
relativistic field theory.










\setcounter{equation}{0}
\setcounter{figure}{0}

\newpage
\newpage
\chapter{Effects of $NN$ potentials on $p$ nuclides in 
the A$\sim$100-120 region}
\label{chapter7}
In the previous chapter, we  discussed the effects of self-interacting $\omega$
-meson ($\omega^4$) coupling on  various properties starting from the 
binding energy of finite nucleus  to equation of state of the  infinite 
nuclear matter. A special attention was given on the effect of 
$\omega^4$ coupling on  newly proposed R3Y nucleon-nucleon potential. 
In this chapter, we will discuss some more applications of our 
proposed nucleon-nucleon interaction, R3Y  in the calculation of 
astrophysical S-factor. Microscopic optical potentials for low energy 
proton reactions have been obtained by folding density dependent M3Y and R3Y 
interaction derived from nuclear matter calculation with densities from mean 
field approach to study astrophysically important proton-rich nuclei in  
mass 100-120 region. We compare S factors for low-energy $(p,\gamma)$ 
reactions with available experimental data and further calculate 
astrophysical reaction rates for $(p,\gamma)$ and $(p,n)$ reactions 
using both R3Y and M3Y interaction. Along with the linear R3Y interaction, 
we choose some nonlinear R3Y (NR3Y) interactions from RMF calculation 
and folded them with corresponding RMF densities to reproduce 
experimental S factor values in this mass region. Impact of the  
non-linearity of NR3Y interaction on S-factor of proton rich nuclei 
is discussed in detail.
\begin{figure}
\center
\resizebox{9cm}{8cm}{\includegraphics{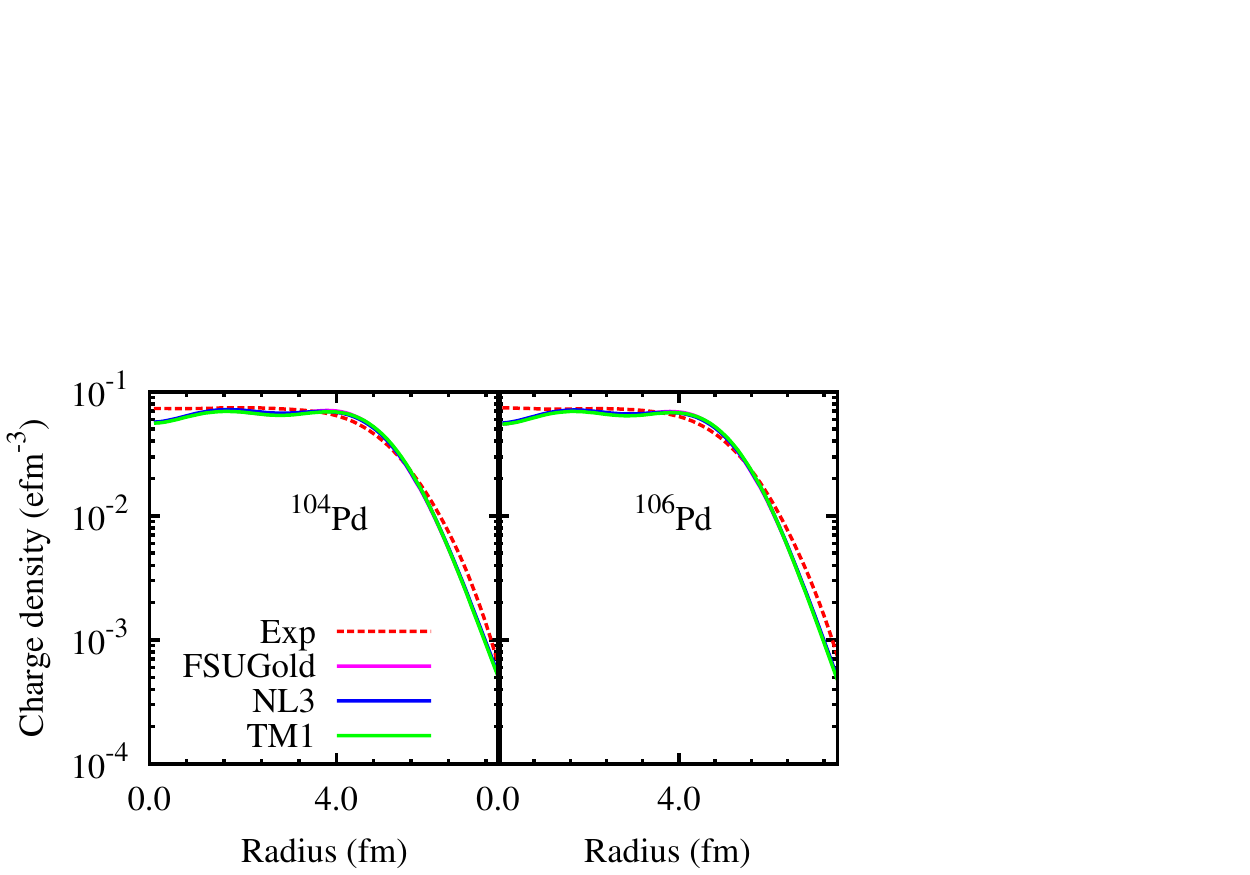}}
\caption{\label{cden} Comparison of charge density from our calculation with  Fourier-
Bessel analysis of experimental electron scattering data\cite{de}
}
\end{figure}
\section{Introduction}
In nature 35 nuclei, commonly termed as $p$ nuclei,  can be found on the proton-rich side of the nuclear landscape ranging 
between $^{74}$Se to $^{196}$Hg. As they are neutron deficit, the astrophysical reactions involved in the synthesis of these 
elements do not correspond to the slow($s$) or  fast($r$) neutron capture processes. It mainly includes
reactions such as proton capture, charge exchange and photo-disintegration.  One can find a detailed study related to 
the $p$  process
in standard text books 
[for example, Illiadis\cite{ili}] and reviews\cite{prev}.

The natural abundances for $p$ nuclei are very low in
the order of 0.01\% to 1\%. In general, the calculation of isotopic abundances 
require a network calculation typically involving 2000 nuclei
and approximately 20000 reaction and decay channels and 
one  major problem with this $p$ network is that most of the nuclei involved in the reaction network are very shortly lived.
As a consequence, it is  
very difficult to track the $p$ process nucleosynthesis network experimentally.  
However, recent radioactive ion beam facilities are giving new prospects, still we are far away from measuring astrophysical
reaction rates for the main reactions involved in the $p$ process. Thus, one often has to depend on theoretical models
to study these reactions. 
These type of calculations acutely exploit the Hauser-Feshbach formalism where the optical model potential, in a local or a
global form, is a key ingredient. Rauscher et.al. substantially calculated astrophysical reaction rates and cross 
sections in a global approach\cite{rate}. They further made a comment that the statistical model 
calculations may be improved by using locally tuned parameterization. 

Here, we perform a fully microscopic calculation. The framework is
based on microscopic optical model utilizing the theoretical  density profile of a nucleus. In presence of a stable target,
electron scattering experiment can be performed to avail nuclear charge density distribution data. However, in absence of a stable target, 
theory remains a sole guide to describe the density. Therefore, in this work we employ
relativistic mean field (RMF) approach to extract the density information of a nucleus. This has the advantage 
of extending it to  unknown mass regions. In some earlier works\cite{gg,clah1,clah2,clah3,saumi}, this method has been used
to study low energy proton reactions in the A $\sim$ 55-100 region. Therefore we use this method in A $\sim$ 100-120 region as an
extension of previous works.

The non-linearity in the scalar field\cite{rein,bidhu14} in a RMF theory 
has been proved very successful in reproducing various observable 
like nuclear ground state including nuclear matter properties and the 
surface phenomena like proton radioactivity etc. In this chapter, 
we intend to study the effects of  microscopic optical potentials obtained from nonlinear $NN$ interactions also in 
addition to the conventional linear $NN$ interactions in the A$\sim$ 100-120 region. We concentrate mainly on 
the region relevant to the $p$ network and therefore mainly proton rich and stability region of the nuclear landscape is our 
main concern. 

\section{Procedure}
The RMF approach has successfully explained various features of stable 
and exotic nuclei like ground state binding energy, radius, deformation, 
spin-orbit splitting, neutron halo etc.
\cite{walecka74,pring96,bender03,soret86,miller72}.
The RMF theory is nothing but the relativistic generalization of the 
non-relativistic effective theory like Skyrme and Gogny. This theory 
does the same job, what the non-relativistic theory can do, with an additional guarantee that
 it works in a better way in high density region\cite{compare}. We have used the RMF formalism in both direct and indirect 
way. Directly we have calculated the nuclear density, which is an essential 
quantity to calculate the optical potential. Indirectly we used  RMF Lagrangian to derive $NN$ 
interactions also along with the phenomenologically availed $NN$ interaction model. 
Here we have used different types of $NN$ interactions, namely 
the density dependent M3Y interaction (DDM3Y) and 
nonlinear R3Y interactions(NR3Y). The concept of the NR3Y was originally developed from 
basic idea of the RMF formalism\cite{bhuyan12ab} and will be discussed later in this section. 

In order to calculate the nuclear density, different forms of Lagrangian densities can be used from RMF approach. The chosen form of the interaction Lagrangian density is given by
\begin{eqnarray}
{\cal L}_{int}&=&\bar{{\psi}}\bigg[{g_\sigma}{\phi}-\bigg({g_\omega}{V_\mu}+\frac{g_\rho}{2}
{\tau}.{\mathbf b_\mu}+\frac{e}{2}(1+\tau_3){A_\mu}\bigg)\gamma_{\mu}\bigg]{\psi}
-\frac{g_2}{3}{\phi^3}-\frac{g_3}{4}\phi^4+\frac{\xi}{4}
({V_\mu}{V^\mu})^2\nonumber\\
&+&\Lambda(\mathbf R_\mu. \mathbf R^\mu)(V_\mu V^\mu).
\end{eqnarray}
The values of $g_\sigma$, $g_\rho$ and $g_\omega$ are the coupling constants for sigma, rho, and omega mesons respectively, given in Table \ref{nl3}.
The coupling constants for nonlinear terms of sigma are $g_2$ and $g_3$, that for omega meson is given by $\xi$ and $\Lambda$ denotes the 
cross coupling strength between rho and omega meson.




For example, in case of FSUGold parameter set\cite{fsu}, one can see that, apart from the usual
nucleon-meson interaction terms, it contains two additional nonlinear meson-meson 
self interaction  terms including isoscalar ($\omega$) meson self interactions, and mixed isoscalar-isovector ($\omega^2 R^2$) coupling, whose main aim is to softening the equation of state (EOS) of symmetric nuclear matter. 
As a result, the new parameterization becomes more effective in reproducing 
quite a few nuclear collective modes, namely the breathing modes in $^{90}$Zr and $^{208}$Pb, 
and the isovector giant dipole resonance in $^{208}$Pb\cite{fsu}. 

Again there are many other
parameter sets 
in RMF which are different 
from each other in various ways like inclusion of new interaction or different 
value of masses and coupling constants of the mesons etc. For the comparison and 
better analysis we have included different parameter sets (NL3, TM1) as it is a matter of great concern 
 to check their credibility in astrophysical prediction. 
 Therefore, for the  astrophysical calculations we have used nuclear densities from different sets of parameters like NL3 and TM1
and folded them with corresponding $NN$ interactions respectively. In case of DDM3Y interaction, which is not obtained 
 from the RMF theory, we folded it with RMF density obtained from FSUGold. This FSUGold folded DDM3Y interaction have been used
 in earlier works\cite{gg,clah1,clah2,clah3,saumi} and successfully reproduced some astrophysically important cross sections 
 and reaction rates
 in A $\sim$ 55-100 region. 
 Therefore, it will be interesting to see the behavior of such potential in A $\sim$ 100-120 region.


Typically, a microscopic optical model potential is obtained by folding an  
effective interaction, derived either from the nuclear matter calculation, in the 
local density approximation, {\em i.e.} by substituting the nuclear matter 
density with the density distribution of the finite nucleus (for example DDM3Y), or directly by folding different R3Y interactions 
using different sets of parameters 
from RMF with corresponding density distributions. 
 The folded potential therefore takes the form

\begin{equation}
 V(E,\vec R)=\int\rho(\vec {r'})v_{eff}(r,\rho,E) \vec{dr'},
\end{equation}
with $\vec r=\vec {r'}-\vec R$ in fm. These effective interactions ($v_{eff}(r,\rho,E)$) are described below in more details.


The  density dependent M3Y (DDM3Y)  
interaction\cite{ddm3y} is obtained from a finite range energy independent G-matrix 
elements
of the Reid potential by adding a zero range energy dependent pseudo-potential 
and introducing a density dependent factor.
The interaction is given by

\begin{equation}
 v_{eff}(r)=t^{M3Y}(r,E)g(\rho).                 
\end{equation} 
Here $v_{eff}(r)$ is a function of r, $\rho$ and $E$, where $E$ is the  incident energy and $\rho$, the nuclear density. 
The $t^{M3Y}$ interaction is defines as
\begin{equation}
t^{M3Y}=7999\frac{e^{-4r}}{4r}-2134\frac{e^{-2.5r}}{2.5r}+J_{00}(E)\delta(r)
\end{equation} 
with the zero range pseudo potential $J_{00}(E)$ given by, 
\begin{equation}
\label{pnueq5}
J_{00}(E)=-276\left( 1-0.005\frac{E}{A}\right) {\rm MeV} fm^{3}\end{equation} 
and $g(\rho)$ is the density dependent  factor expressed as,
\begin{equation}
g(\rho)=C(1-b\rho^{2/3}),
\end{equation} 
with $C=2.07$ and $b=1.624$ $fm^2$ \cite{ddm3y}.



 In 2014, Sahu et.al.\cite{bidhu14} introduced a simple form of nonlinear self-coupling of the scalar meson field and suggested 
a new $NN$ potential from relativistic mean field theory (RMFT) analogous to the M3Y interaction. Rather than using usual
phenomenological prescriptions, the authors derived the microscopic $NN$ interaction from the RMF Lagrangian. 
Starting with the nonlinear relativistic mean field Lagrangian density  for a nucleon-meson many-body system they solved the 
nuclear system under the mean-field approximation using the  Lagrangian and obtained the field equations for the nucleons and mesons.
It is necessary here to mention that the authors \cite{bidhu14} had taken the nonlinear part of the scalar meson $\sigma$ proportional
to $\sigma^3$ and $\sigma^4$ in account. 
 Finally for a normal nuclear medium
the resultant effective nucleon-nucleon interaction, obtained from the summation of the scalar and vector meson fields takes the form. \footnote{There is a typographical error in the expression of $v_{eff}$ in Sahu et.al \cite{bidhu14} and the corrected 
 form is given in this thesis.}

\begin{eqnarray}
\label{pnueq7}
 v_{eff}(r) = \frac{g_{\omega}^2}{4\pi}\frac{e^{-m_\omega r}}{r}+\frac{g_{\rho}^2}{4\pi}\frac{e^{-m_\rho r}}{r}
 -\frac{g_{\sigma}^2}{4\pi}\frac{e^{-m_\sigma r}}{r}\\\nonumber
 +\frac{g_2^2}{4\pi}{r}{e^{-2m_\sigma r}}+\frac{g_3^2}{4\pi}\frac{e^{-3m_\sigma r}}{r}-\frac{\xi^2}{4\pi}\frac{e^{-3m_\omega r}}{r}\\\nonumber
 +J_{00}(E)\delta(r).
\end{eqnarray}
Here $m_\sigma$, $m_\rho$, $m_\omega$ are the masses of sigma, rho, and omega mesons respectively, whereas the zero
range pseudo potential $J_{00}(E)$ is given in Eqs.(\ref{pnueq5}).
\begin{table}
\begin{center}
\caption{\label{nl3} Model parameters for the Lagrangian FSUGold\cite{fsu}, NL3\cite{lala97} and TM1\cite{toki94}.}
\vspace{.5cm}
\begin{tabular}{|c|c|c|c|}
\hline
{}&FSUGold&NL3&TM1\\
\hline
$M$ (MeV)&939&939&938\\

$m_\sigma$ (MeV) &491.500&508.194&511.198\\

$m_\omega$ (MeV)&782.500&782.501&783.000\\

$m_\rho$ (MeV)&763.000&763.000&770.000\\

$g_\sigma$&10.592&10.2170&10.0290\\

$g_\omega$&14.298 &12.8680&12.6140\\

$g_\rho$&11.767&4.4740&4.6320\\

$g_2$ (fm$^{-1}$)&-4.2380&-10.4310&-7.2330\\

$g_3$&-49.8050&-28.8850&0.6180\\

$\xi$&2.0460&{-}&71.3070\\

$\Lambda$&0.0300&{-}&{-}\\


\hline

\end{tabular}
\end{center}

\end{table}

 Using NL3 parameters from Table \ref{nl3}, Eqs. (\ref{pnueq7}) becomes\cite{bidhu14}
 
 \begin{eqnarray}
 v_{eff}(r) = 10395\frac{e^{-3.97r}}{4r}+1257\frac{e^{-3.87r}}{4r}
 -6554\frac{e^{-2.58r}}{4r}\\\nonumber
 +6830{r}\frac{e^{-5.15r}}{4}+52384\frac{e^{-7.73r}}{4r}+J_{00}(E)\delta(r).
\end{eqnarray}

The authors of Ref. \cite{bidhu14} denoted this $NN$ interaction potential as NR3Y(NL3). 
Further, putting parameter sets from TM1 (Table \ref{nl3}),
one can obtain $v_{eff}$ for  NR3Y(TM1).






Since the DDM3Y folded potential described above do not include any spin-orbit term, the
spin-orbit potential from the Scheerbaum prescription\cite{SO} has been coupled with the
phenomenological complex potential depths $\lambda_{vso}$ and $\lambda_{wso}$ 
. The spin-orbit potential is given by
\begin{equation} U^{so}_{n(p)}(r)=(\lambda_{vso} +i\lambda_{wso})
\frac{1}{r}\frac{d}{dr}(\frac{2}{3}\rho_{p(n)}+\frac{1}{3}\rho_{n(p)}).
\end{equation}
The depths are functions of energy, given by 
\begin{displaymath}\lambda_{vso}=130\exp(-0.013E)+40,\end{displaymath}
and 
\begin{displaymath}\lambda_{wso}=-0.2(E-20),\end{displaymath}
where $E$ is in MeV. These standard values have been used in the 
present work. However, in case of nonlinear $NN$ folded potentials from RMF (NR3Y(NL3), NR3Y(TM1)), 
one need not require to add spin-orbit term from outside, as it is contained within the RMF\cite{bidhu14}.  

Finally reaction cross-sections and astrophysical reaction rates are calculated in the Hauser-Feshbach 
formalism using the computer package TALYS1.2\cite{talys}. 

\section{Results and discussions}
For simplicity, this section is divided in three subsections. In the 
Sec.~\ref{rmf}, results from RMF calculations are given. We will concentrate on the reaction cross-sections and astrophysical S factors in the 
Sec. ~\ref{sfac}. Furthermore, results for reaction rates for astrophysically 
important nuclei are provided. Sec.~\ref{opt} is devoted to the 
effects of different $NN$ potentials in this mass region.

\subsection{RMF calculations}
\label{rmf}

\begin{table*}
\caption{\label{be} Calculated  binding energy per nucleon (B.E/A) \cite{adi} and charge radii $r_{ch}$ \cite{angel} for some selected $p$ nuclei
compared with experimental values.}
\begin{center}
\begin{tabular}{lcccccccc}\hline
&\multicolumn{4}{c}{B.E./A(MeV)}
&\multicolumn{4}{c}{$r_{ch}$(fm)} \\
{}&FSUGold&TM1&NL3&Exp&FSUGold&TM1&NL3&Exp\\\hline
$^{102}$Pd&{8.480}&{8.537}&8.572&8.580& 4.460&{4.476}&{4.483}&4.483 \\
$^{106}$Cd& 8.494&{8.518}&8.532&8.539 &4.525&{4.535}&{4.535}&4.538  \\
$^{108}$Cd& {8.498}&{8.529}&8.537&8.550 &4.537&{4.549}&{4.552}&4.558 \\
$^{113}$In&{8.507}&{8.461}&{8.523}& 8.523&4.480&{4.575}&{4.588}&4.601 \\
$^{112}$Sn&{8.514}&{8.520}&8.502&8.514&4.595&{4.598}&{4.594}&4.594 \\
$^{114}$Sn&{8.534}&{8.526}&8.490&8.523&4.636&{4.611}&{4.662}&4.610 \\
$^{115}$Sn&{8.530}&{8.527}&8.494&8.514&4.607&{4.611}&{4.617}&4.615\\
$^{120}$Te&{8.461}&{8.461}&8.460&8.477&4.682&{4.688}&{4.735}&4.704 \\
\hline
\end{tabular}
\end{center}

\end{table*}

In some earlier works \cite{gg,clah1,clah2,clah3,saumi}, FSUGold has been proved to be successful in reproducing experimentally obtained 
binding energy, charge radius and charge density data in the A$\sim$55-100 region. Again in 1997, NL3 parameter set had been 
introduced by Lalazissis et.al\cite{lala97} with a aim to provide a better description not only for the properties of stable nuclei 
but also for those far from the $\beta$ stability line and during last two decades, this parameter set successfully reproduces 
binding energy, charge radius etc. for various elements throughout the periodic table\cite{lala97,p2}.   
In order to confirm the applicability of RMF calculations in A$\sim$100-120 region, in Table \ref{be},
we compare nuclear binding energy per nucleon and charge radii of $p$ nuclei 
in the concerned mass region  with different sets of parameters of RMF formalism with existing experimental data\cite{adi,angel}. 
We find that, in most cases, our calculations with
different sets of parameters 
match quite well with the experimental data.   
In Fig. \ref{cden} charge density from our calculations are compared with 
existing electron scattering data\cite{de} for Pd isotopes and here also, the agreement is well enough to confirm the credibility
of RMF models in this mass region.



\subsection{Astrophysical S factor and reaction rates}
\label{sfac}

In the present case, our calculations, being more
microscopic, are more restricting. In general, phenomenological models are usually fine-tuned for nuclei
near the stability valley, but not very successful in describing elements near the proton and neutron rich regions.
Microscopic models, in contrary, can be extended to the drip line regions and 
therefore, this method can be used to study the reaction rates of nuclei involved in $p$ process 
nucleosynthesis network ($\sim$ 2000 nuclei are present in the total $p$ network). 
However, only a few number of stable $p$ nuclides are available in nature that can be accessed by the experiment and therefore
we are restricted to those nuclei for the purpose of comparison. 

Let us first take the case of DDM3Y folded potential. As a first test of the optical model potential, 
we have calculated elastic proton scattering at low energies where experimental
data are available. As the elastic scattering process
involves the same incoming and outgoing channel for the
optical model, it is expected to provide the easiest way
to constrain various parameters involved in the calculation.
Here we are mainly interested in the energy region between 2-8 MeV as the
astrophysical important Gamow window lies within this energy range in the concerned mass region.
However, scattering experiments are very difficult at
such low energies, because the cross sections are extremely
small, and hence no experimental data are available. Therefore we have
compared the cross sections from our calculations with the lowest energy experimental data available in
the literature.

\begin{figure}
\center
\resizebox{10cm}{9cm}{\includegraphics{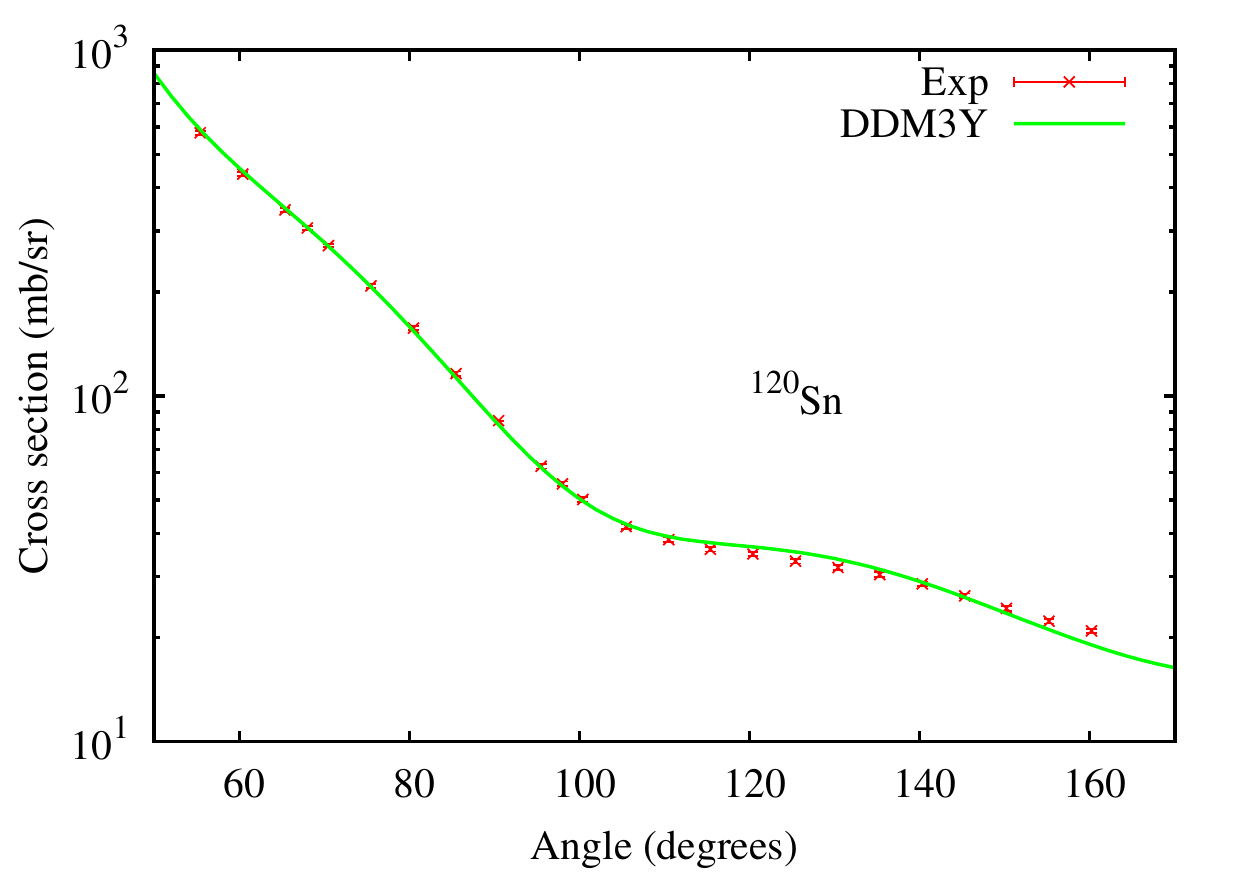}}
\caption{ \label{120sn} Experimental and calculated cross sections for elastic
proton scattering at 9.7 MeV proton energy.
}
\end{figure}

In Fig. \ref{120sn}, we present the result of our calculation with DDM3Y folded potential for $^{120}$Sn with 
available experimental data\cite{diff}.
To fit the experimental data, at first, the folded DDM3Y
potential is multiplied by factors 0.3 and 0.7 to obtain the
real and imaginary parts of the optical potential, respectively. 
However better fits for individual
reactions can be possible by varying different parameters. But if
the present calculation has to be extended to an unknown
mass region, this approach is clearly inadequate. Therefore,
we have refrained from fitting individual reactions. A detailed description of these normalizing constants are available in 
references\cite{gg,clah1,clah2,clah3,saumi}.

Yet, the astrophysical reaction rates depend on the proper choice of the level density and the E1 gamma strength. Therefore,
we have calculated all of our
results with microscopic level densities in Hartree-Fock (HF)
and Hartree-Fock-Bogoliubov (HFB) methods, calculated for
TALYS database by Goriley and Hilaire\cite{talys,gori} on the basis of Hartree-Fock calculations\cite{gori1}
. We have also
compared our results using phenomenological level densities
from a constant-temperature Fermi gas model, a back-shifted
Fermi gas model, and a generalized super-fluid model from
TALYS. All these model parameters can be availed from TALYS database. 
We find that the cross sections are very sensitive to the
level density parameters.
We therefore analyzed, in most of the cases, the HF level densities fit the
experimental data better in this mass region.
Again, for E1 gamma strength functions, results derived from
HF + BCS and HFB calculations, available in the TALYS
database, are employed.  In this case also, the results for HF+BCS calculations describe the experimental data reasonably well 
and we present our results for that
approach only.

We now calculate some $(p,\gamma)$ cross sections relevant to $p$ nuclei in A$\sim$100-120 region, where experimental
data are available.  At such low energies, 
reaction cross-section varies  rapidly making comparison between theory and
experiment rather difficult. Therefore the usual practice in low-energy
nuclear reaction is to compare another key observable, viz. the
S factor. It can be expressed as\cite{clah1}

\begin{equation}
S(E)=E\sigma(E)e^{2\pi\eta},
\end{equation}
where E is the energy in center of mass frame in keV, which factorises out the pre-exponential low energy dependence of reaction
cross-section $\sigma(E)$, and $\eta$ indicates
 the Sommerfeld parameter with  
\begin{equation}
 2\pi\eta=31.29 Z_{p}Z_{t}\sqrt{\frac{\mu}{E}}.
\end{equation}

The factor exp$(2\pi\eta)$ is inversely proportional to the transmission 
probability through the Coulomb barrier with zero angular momentum(s-wave) 
and therefore removes exponential low energy dependence of $\sigma(E)$. 
Here $\sigma(E)$ is in  barn, $Z_{p}$ and $Z_{t}$ are the charge numbers 
of the projectile and the target, respectively and $\mu$ is the reduced 
mass (in amu) of the composite system. This S factor varies much slowly 
than reaction cross-sections and for this reason, we calculate this quantity
and compare it with experimentally obtained values.

\begin{figure}
\center
\includegraphics{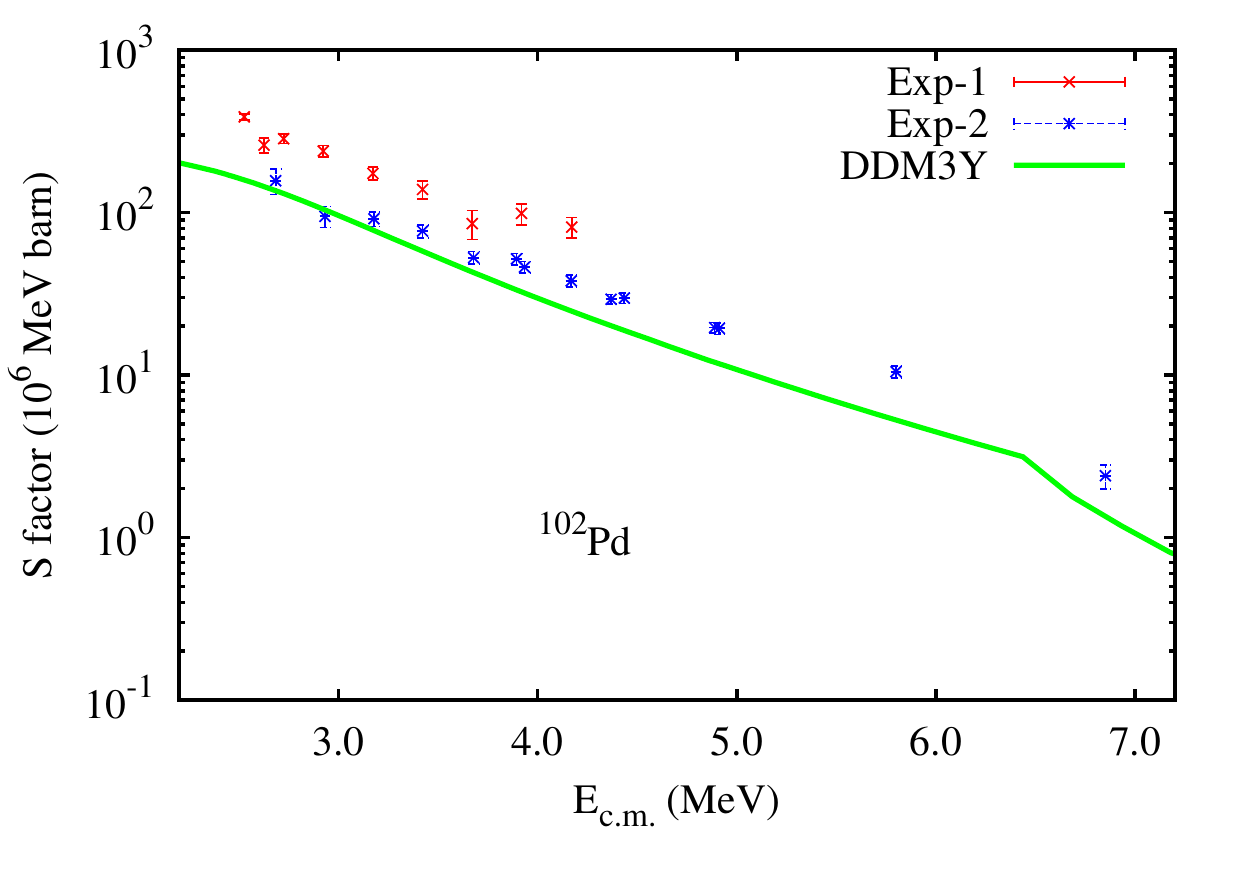}
\caption {\label{102pd} S factors from two different microscopic potentials are compared with  experimental measurements
  for $^{102}$Pd. Here \textquotedblleft Exp-1\textquotedblright   is the experimental data from 
  reference\cite{pd102r1}, \textquotedblleft Exp-2\textquotedblright from reference\cite{pd102r2} and 
  \textquotedblleft DDM3Y\textquotedblright is for the DDM3Y-folded potential.}
\end{figure}

\begin{figure}
\center
\resizebox{10cm}{9cm}{\includegraphics{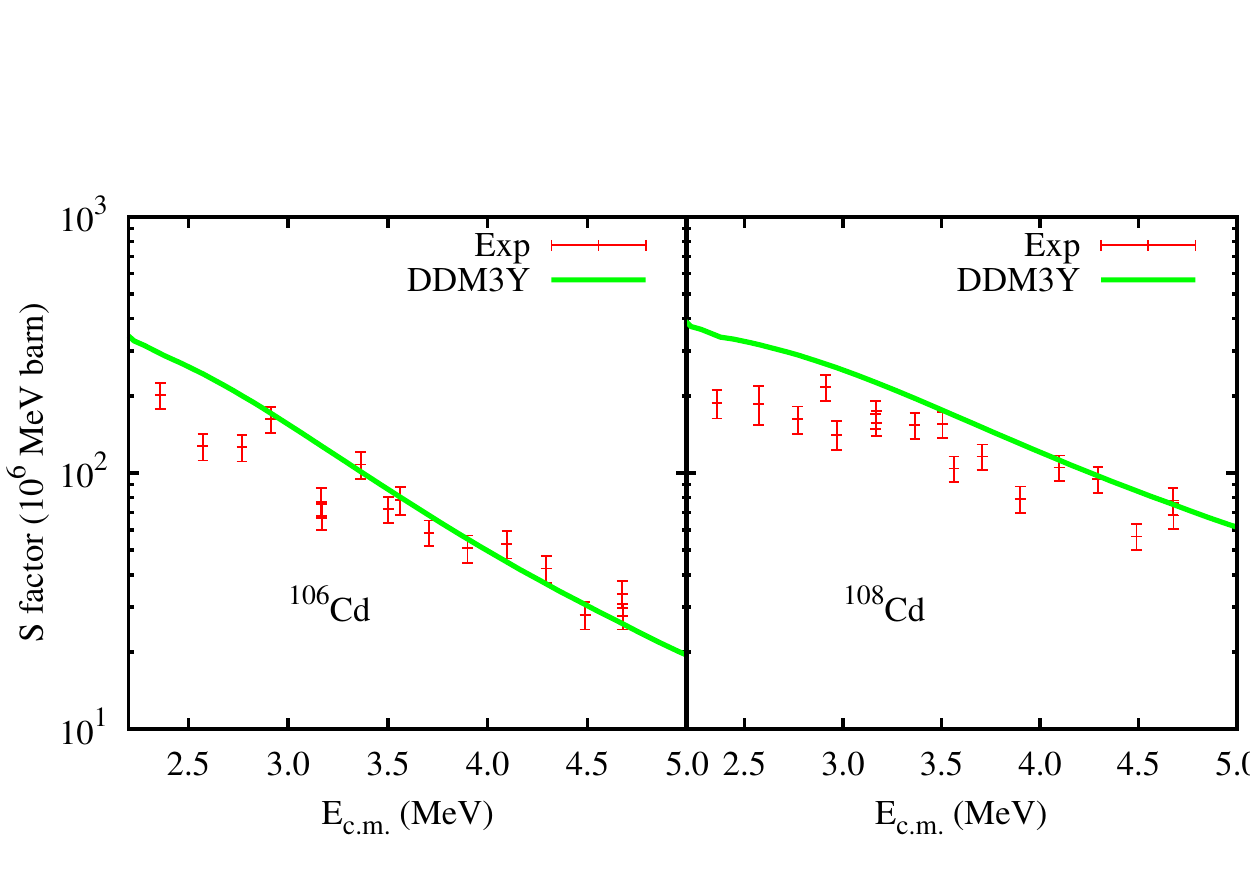}}
\caption {\label{106cd} S factors extracted from theory compared with experimental measurements
 for $^{106,108}$Cd. Here \textquotedblleft Exp\textquotedblright is the experimental data from reference\cite{cdref}. 
 }
\end{figure}

\begin{figure}
\center
\resizebox{10cm}{9cm}{\includegraphics{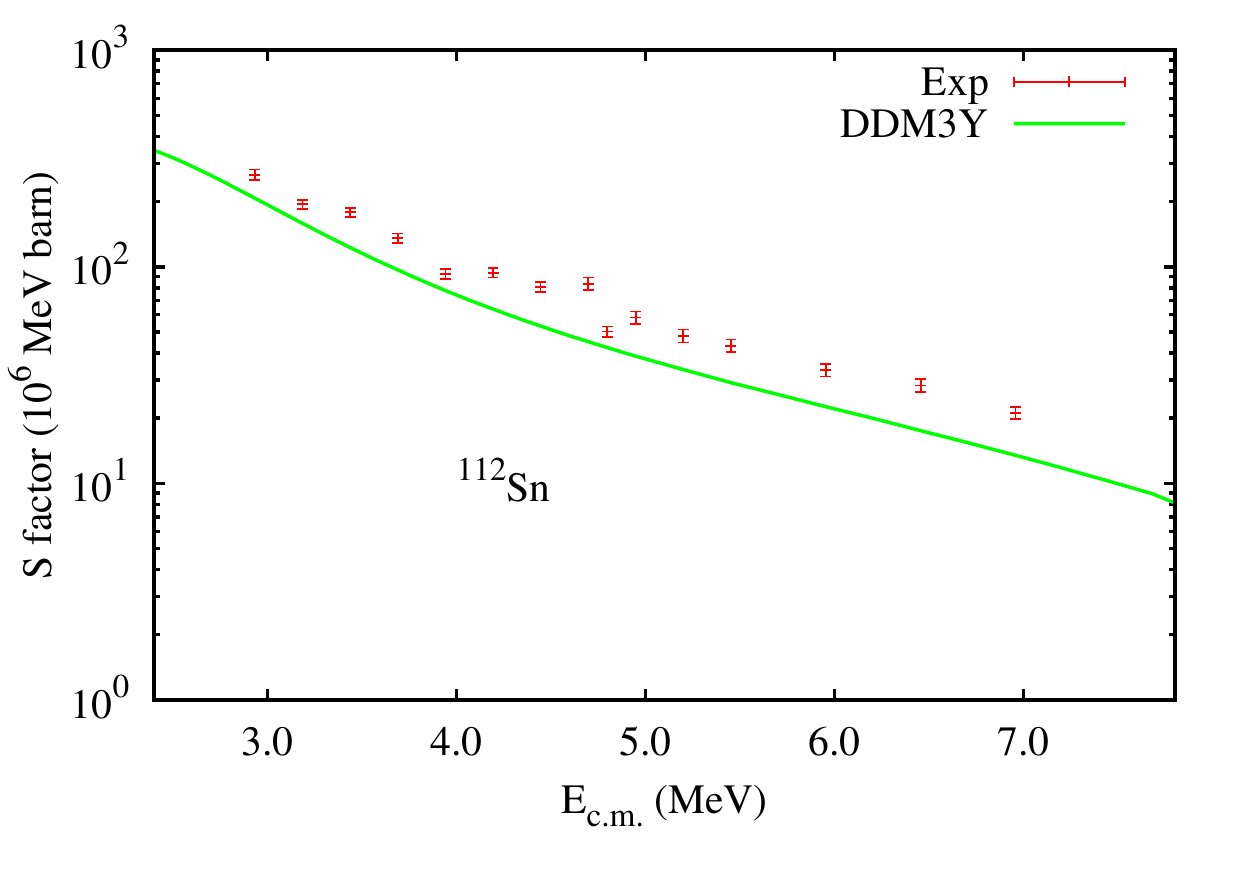}}
\caption{ \label{112sn} S factors extracted from theory compared with experimental measurements
 for for $^{112}$Sn.  }
\end{figure}

In Figs. \ref{102pd}-\ref{112sn} we present the results of some 
of our calculations with folded DDM3Y potential for Pd, Cd and Sn isotopes, respectively,
along with the corresponding experimental results. 
The experimental values for $^{102}$Pd are from Ref. \cite{pd102r1}(red point) and \cite{pd102r2}(blue cross), 
$^{106,108}$Cd from Gy. Gy\"urky et.al\cite{cdref} and $^{112}$Sn from Ref. \cite{sn112ref}.
 In case of $^{102}$Pd in Fig. \ref{102pd}, theoretical prediction 
is in a good agreement, mainly in the low energy regime, 
with the experimental data from Ref. \cite{pd102r2} but under 
estimates the data obtained from the Ref. \cite{pd102r1}. In Ref. \cite{pd102r1}, an activation technique was used 
in which gamma
rays from decays of the reaction products were detected off-line by two hyper-pure germanium
detectors in a low background environment, whereas in Ref. \cite{pd102r2}, cross-section measurements have been carried out at
the cyclotron and Van de Graaff accelerator by irradiation of thin sample
layers and subsequent counting of the induced activity. However, we can not comment on the individual merits of these experiments.

In case of $^{106,108}$Cd (in Fig. \ref{106cd}), one can find that the agreement of theory with experimental
values are good enough, however there is a slight over estimation for $^{108}$Cd in the low energy regime. In case of $^{112}$Sn 
in Fig. \ref{112sn}, our calculation follows the experiment in a fairly good fashion.    

The success of this microscopic optical potential (DDM3Y interaction folded with FSUGold density) in reproducing S-factor data for
the above $p$ nuclei leads us to calculate reaction rates
of some astrophysically important reactions. In Fig. \ref{pgrate}, we compare  
$(p,\gamma)$ reaction rates for some important $p$ nuclei 
 with NONSMOKER rates\cite{rate} obtained from statistical model calculation with a global approach.
 Again in Fig. \ref{pnrate}, reaction rates for charge exchange reactions $(p,n)$ for some nuclei,
 however not astrophysically significant enough, in this mass region are compared with existing NONSMOKER calculations. One
can see that the present calculation is very similar to the NONSMOKER values in almost all cases. Therefore, it is expected that all the results can
also be reproduced with commonly used NONSMOKER rates.

\begin{figure}
\vspace{-3.0 cm}
\center
\resizebox{10cm}{10cm}{\includegraphics{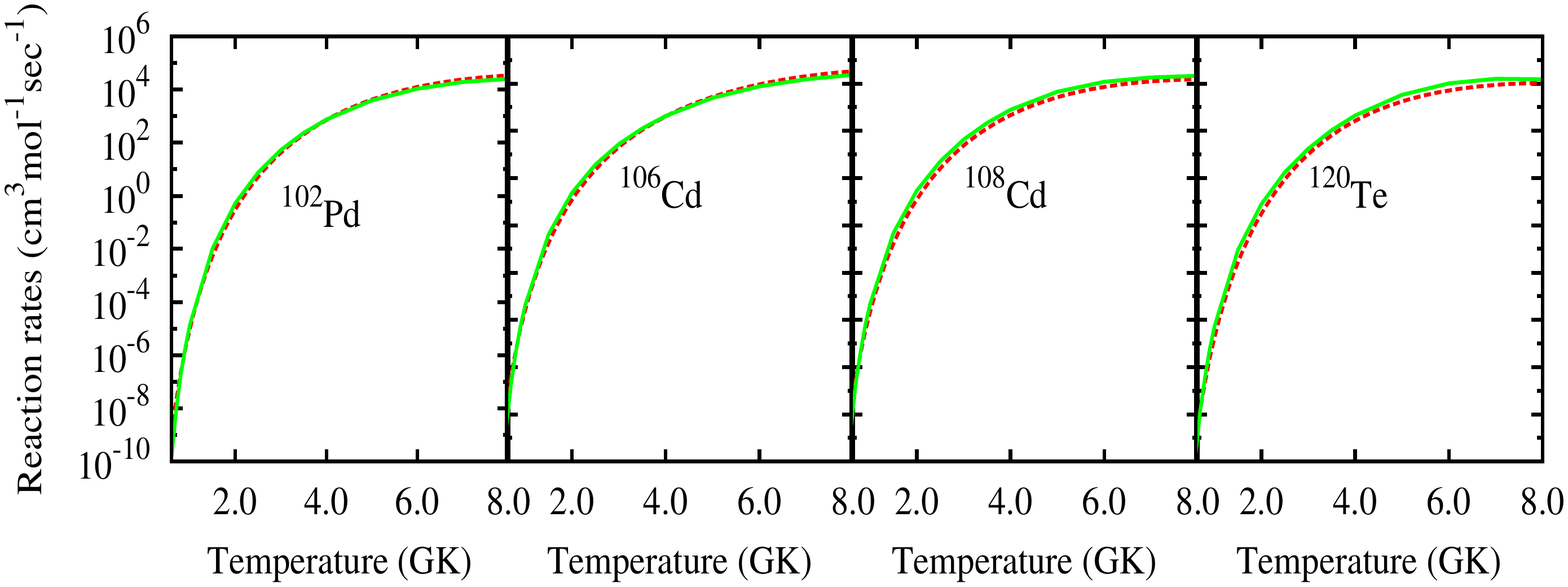}}
\caption { \label{pgrate} Astrophysical reaction rates for $(p,\gamma)$ reactions of some important $p$ nuclei 
compared with NONSMOKER rates\cite{rate}. Here Green Continuous line: Present calculation, Red Dotted line: NONSMOKER calculation.
 }
\end{figure}

\begin{figure}
\vspace{-3.0 cm}
\center
\resizebox{10cm}{9cm}{\includegraphics{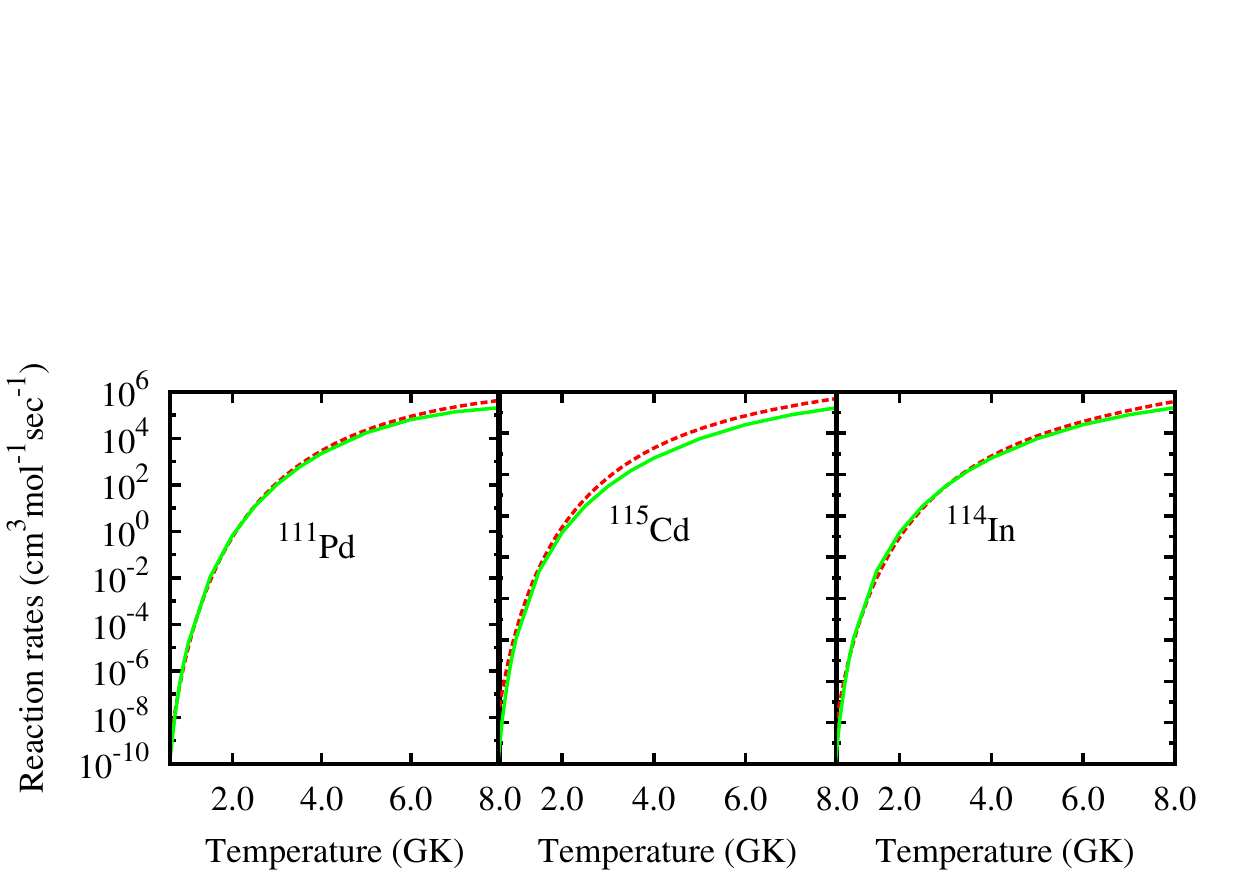}}
\caption {\label{pnrate} Astrophysical reaction rates for $(p,n)$ reactions
compared with NONSMOKER rates\cite{rate}. Here Green Continuous line: Present calculation, Red Dotted line: NONSMOKER calculation.
 }
\end{figure}

\begin{figure}
\center
\resizebox{10cm}{9cm}{\includegraphics{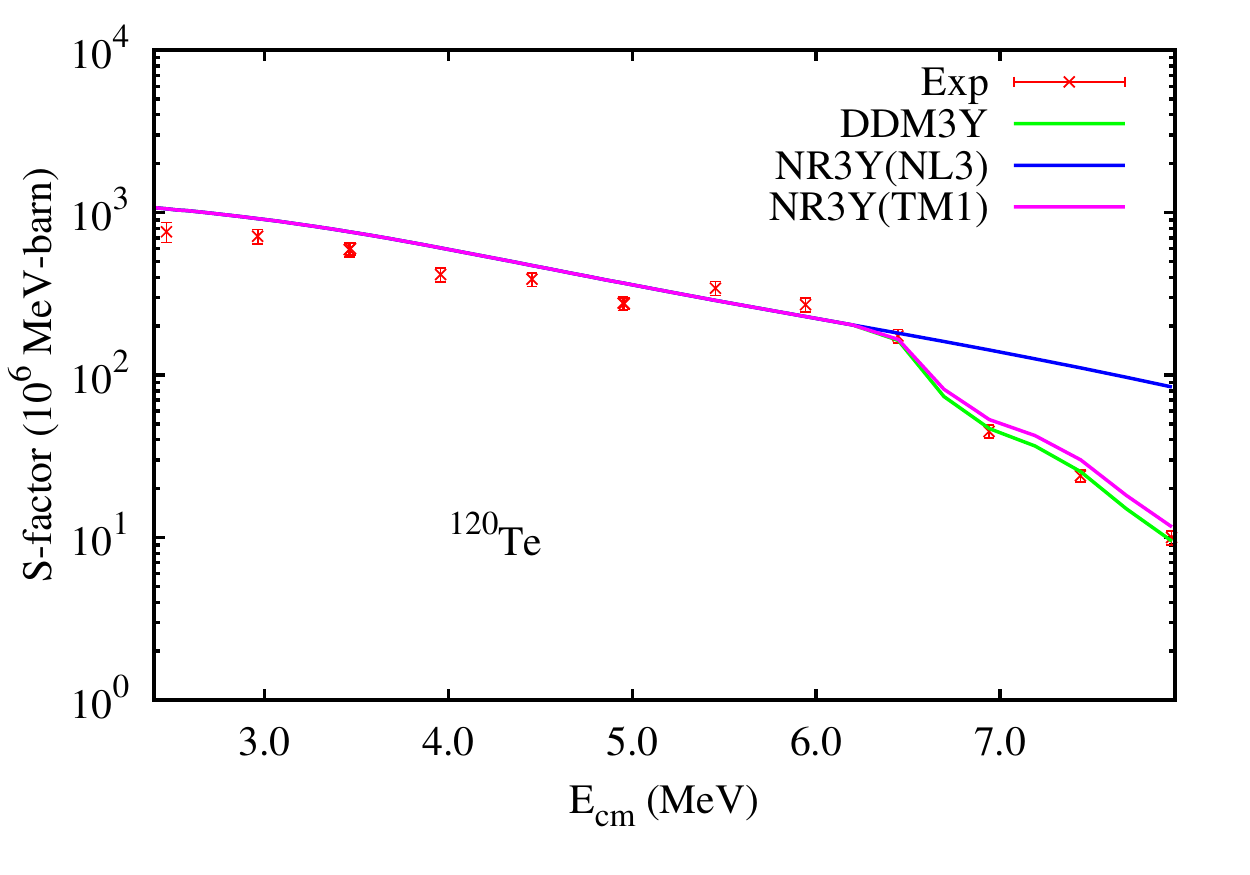}}
\caption {\label{120te} S factors extracted from our calculations compared with experimental measurements
 for $^{120}$Te. Here \textquotedblleft Exp\textquotedblright is the experimental data from reference\cite{te120ref}. 
 For other details, see the text. }
\end{figure}

In the remaining part, we mainly concentrate on the effects of optical potentials obtained by folding nonlinear interactions from RMF (NR3Y). 
In Fig. \ref{120te}, S factors for $^{120}$Te  obtained from NR3Y(NL3) and
NR3Y (TM1) potentials are compared with the experimental data taken from Ref. \cite{te120ref}.
The S- factor with  DDM3Y interaction folded with  FSUGold density is also given 
for comparison. 
 
 One can see that our calculation with 
 folded DDM3Y potential shows a very nice agreement with experimental values throughout the energy range. 
 In contrary, in case of 
 NR3Y(NL3) folded potential, there is a wide deviation of the theory with experimental data after 6 MeV whereas 
 the TM1 folded potential NR3Y(TM1) shows a decrease in S factor value around 6 MeV energy unlike the NR3Y(NL3) case.  
 
 The rapid drop of S factor
 values with increasing energy actually takes place due to the increasing contribution of higher angular momentum channels (l$>$0).
 Therefore, if the center of mass energy E$_{c.m.}$ becomes larger than the Coulomb barrier for a specific set of nucleon-nucleus
 reaction (E$_{c.m.}>$E$_c$), as a result the S factor will decrease rapidly with the growth of energy (E$_{c.m.}$)\cite{sf}.
 In the next subsection,
 we illustrate this physics in detail and show how this phenomenon is associated with different form of potentials.

\subsection{Optical potentials and effects of non-linearity}
\label{opt}
\begin{figure}
\center
\resizebox{10cm}{9cm}{\includegraphics{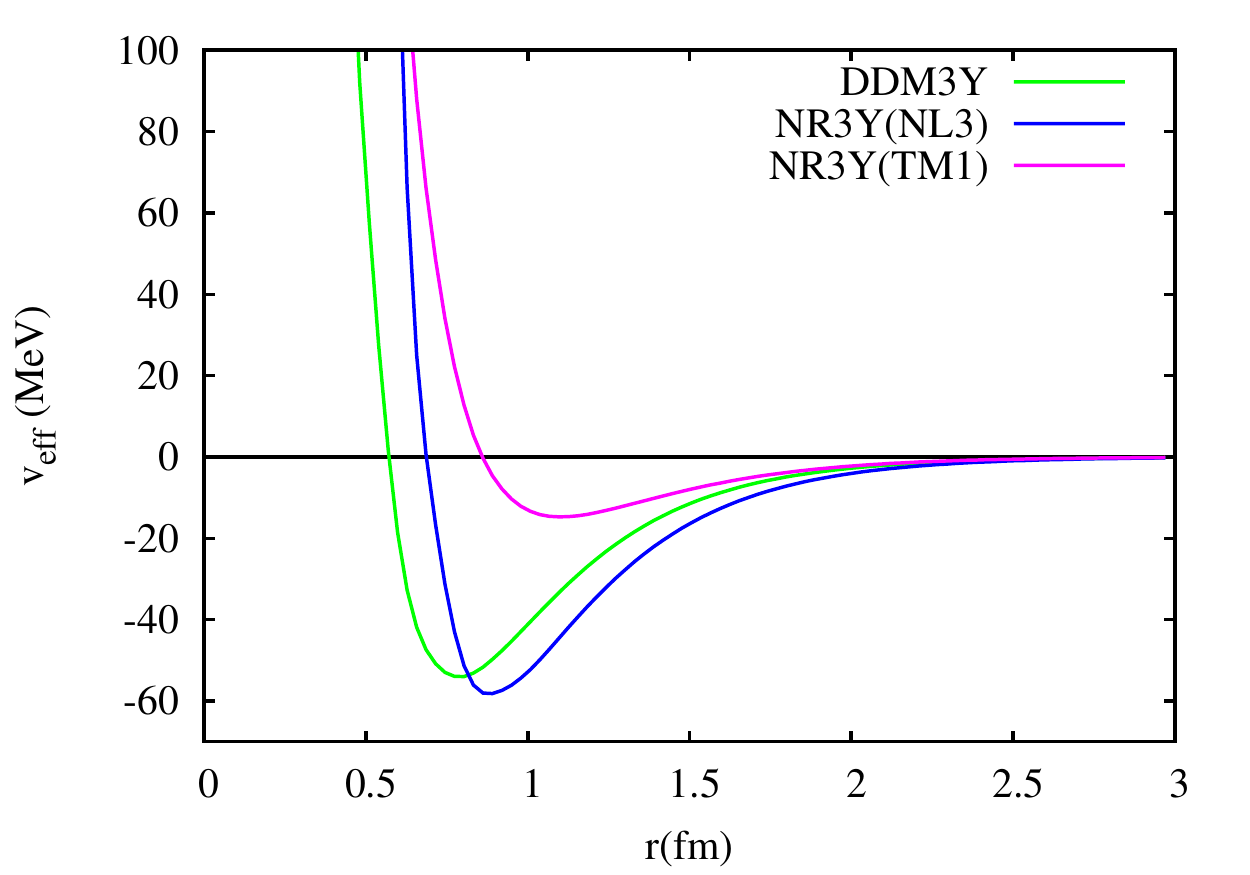}}
\caption{(Color online) Effective interaction potential for $^{120}Te$.
 \label{int}}
\end{figure}

We now interpret the above results (for example, see Figs. \ref{120te}) 
with help of the microscopic potentials obtained from different $N-N$ interactions. In Fig. \ref{int}, the effective $NN$ interaction potentials (in MeV) are plotted with the radius r (fm) for $^{120}$Te. The DDM3Y interaction,
being dependent on the density, 
is different for different elements of the periodic table, whereas in contrary, other interactions remain unaltered for different
elements. In Fig. \ref{int} different forms of $NN$ interactions are given.
We find that the curves from DDM3Y and NR3Y(NL3) interactions generated from two different formalisms show almost similar
trend.

\begin{figure}
\center
\resizebox{10cm}{9cm}{\includegraphics{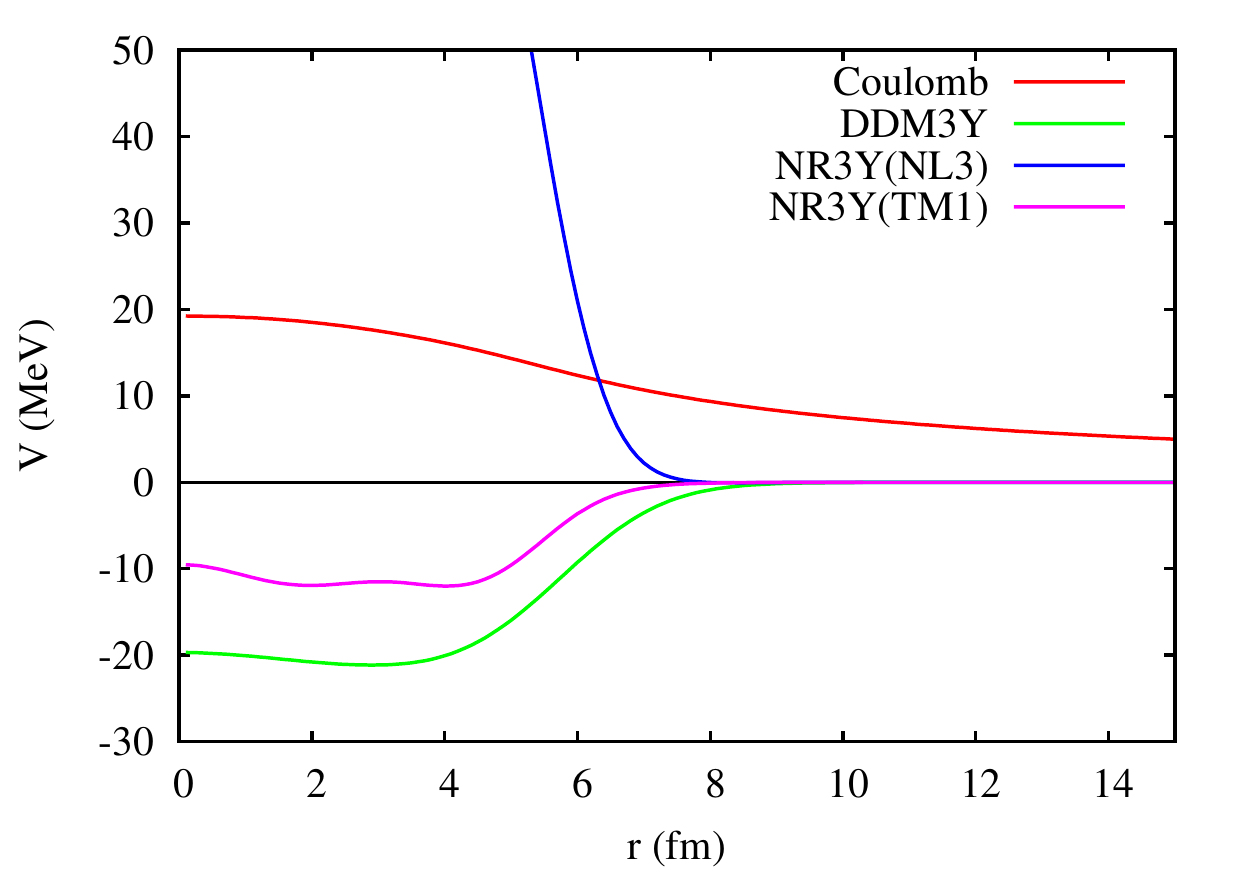}}
\caption{(Color online) Real central part of folded potentials and Coulomb potential for 7 MeV proton(Lab) incident on $^{120}$Te \label{pot}}
\end{figure}

A graphical representation of microscopic potentials for  $^{120}$Te after folding the interactions is represented
in Fig. \ref{pot}. Here the real central part of the optical potential is plotted with the radius. 
In the figure, one can see that the DDM3Y folded potential provides an attractive potential whereas in
case of NR3Y(NL3) folded potential, the repulsive part overpowers the attractive part, as well as the Coulomb part of the potential. 
As a result, the resultant repulsive barrier becomes greater than the Coulomb barrier almost upto 
a range for a nuclear reaction to occur. Therefore the
penetrability of the higher angular momentum channels get reduced and as a obvious consequence, the desirable sharp drop in S factor 
(Fig. \ref{120te}) has not been achieved. In case of TM1 folded potential, we can see that the effective contribution
of the optical potential is attractive in nature similar to the DDM3Y  potential and therefore, the Coulomb energy serves as 
the only repulsive barrier. As a result the 
penetration probability for higher angular momentum channels becomes higher than that of the NR3Y(NL3) case. This reason is replicated
as a drop of S factor
values at higher energies in Fig. \ref{120te}. In case of the imaginary part of the potential, the curves follow exactly the similar trend 
as that of the real part, i.e., apart from NR3Y(NL3) potential, rest of them gives attractive contribution. 
 One can explain the above scenario
from the numerical value of the nonlinear coupling
constant $g_3$ of TM1 parameter set,
as given in Table \ref{nl3}, which is much less than that of the NL3 parameter set. Therefore it can be understood that with 
 decreasing values of the nonlinear coupling constants $g_2$ and $g_3$,
the repulsive
component of the optical potential also gets reduced and one point is attained when only the effect of 
Coulomb barrier remains as a dominating repulsive contributor 
and we will get patterns like TM1,
DDM3Y as shown in Fig. \ref{pot} and we can find the expected drop of S factor at higher
energies due to the opening of 
 higher angular momentum 
channels. So from the above observations we can comment that there should be an 
upper cut-off for the coupling constants values of the nonlinear components.   

\section{Summary }
To summarize, cross section for low energy $(p,\gamma)$ reactions for 
a number of $p$ nuclei in A$\sim$100-120 region have been calculated 
using microscopic optical model potential with the Hauser Feshbach reaction code TALYS. Mainly, microscopic potential 
is obtained by folding DDM3Y interaction with densities from RMF approach. Astrophysical reaction rates for $(p,\gamma)$ and $(p,n)$
reactions are compared with standard NONSMOKER results. Finally, the effect of microscopic optical potential obtained by folding 
nonlinear NR3Y(NL3) and NR3Y(TM1) interactions with corresponding RMF densities are employed to fit the experimental S factor data
for $^{120}$Te. The reason of the deviation of theoretical 
prediction with nonlinear NR3Y(NL3) potential from experiment at higher 
energies has been discussed and finally we made a comment on magnitude of 
the coupling terms of the nonlinear components that an upper cut-off 
 for $g_2$ and $g_3$ should be fixed to get proper repulsive component of 
the $NN$ interaction.

 \setcounter{equation}{0}
 \setcounter{figure}{0}


\newpage
\chapter{Effects of hyperon on both static and rotating neutron star}
\label{chapter8}

In last few chapters, we discussed the structure of finite nuclei by using relativistic as well as non-relativistic formalisms.  Along with the finite nuclear system, an infinite nuclear system also plays very crucial role in understanding the nucleon-nucleon interaction and hence the nuclear structure. In this chapter, we have discussed the structure of the neutron star, which is a perfect example of the infinite nuclear matter system.   Particularly , we study the effects of  isovector-scalar  ($\delta$)-meson on neutron and hyperon stars. Influence of $\delta$-meson on both static and rotating stars  are discussed. The $\delta$-meson in a neutron star  consisting of protons, neutrons, and electrons makes the equations of state stiffer at a higher density and consequently increases the maximum mass of the star. But induction of $\delta$-meson in the hyperon star decreases the maximum mass. This is due to the early evolution of hyperons in the presence of $\delta$-meson.

\section{Introduction}\label{sec1}
Neutron star is a venerable candidate to discuss the physics at 
high density. We can not create such a high density in a terrestrial
 laboratory, so a
neutron star is and the only object, which can provide much information on 
high-density nature of the matter\cite{prakash04,prakash07}. But it is not an 
easy task to deal with 
the neutron star for it's complex nature, as all the 
four fundamental forces (strong, weak, gravitational and electromagnetic) 
are active.  High gravitational field makes  
mandatory to use general theory of relativity for the study of neutron star 
structure. Equations of states (EOS) are the sole ingredient that must be 
supplied to the equation of stellar structure,
Tolman-Oppenheimer-Volkoff (TOV) equation, whose output 
is the mass-radius profile of the dense neutron star. 
In this case, the nuclear EOS plays an intimate 
role in deciding the mass-radius of a neutron star. Its indispensable 
importance attracts the attention of physicists to have an anatomy of
the interactions Lagrangian.
As the name suggests, a neutron star is not completely made up 
neutrons, a small fraction of protons and electrons  are also present, 
which is the consequence of the $\beta-$equilibrium and charge neutrality 
condition\cite{glen85}. Also, the presence of exotic degrees of freedom like 
hyperons and kaons can not be ignored in such  high dense matter. 
It is one among the most asymmetric and dense nuclear object  
in  nature. 

From last three decades \cite{rein89,ring96}, the relativistic mean field (RMF)
approximation, generalized by Walecka \cite{walecka74} and later on developed by Boguta and 
Bodmer \cite{boguta77} is one amongst the most reliable theory 
to deal with the infinite nuclear matter and finite nuclei. The original RMF 
formalism starts with an effective Lagrangian, whose degrees of freedom are
nucleons, $\sigma-,$ $\omega-$, $\rho-$ and $\pi-$mesons. To reproduce proper
experimental observable, it is extended to the self-interaction of 
$\sigma-$meson.  Recently, all other self- and crossed interactions including
the baryon octet are also introduced keeping in view the extra-ordinary 
condition of the system, such as highly asymmetric system or extremely high- 
density medium \cite{bharat07}. 
Since the RMF formalism is an effective nucleons-mesons model, the
coupling constants for both nucleon-meson and hyperon-meson are fitted
to reproduce the properties of selected nuclei and infinite nuclear matter
properties \cite{walecka74,boguta77,rein86,ring90}. In this case, it is
improper to use the parameters obtained from the free nucleon-nucleon 
scattering data. 
The parameters, with proper relativistic kinematics and with the
mesons and their properties are already known or fixed from the properties of a 
small number of finite nuclei, the method gives excellent results 
not only for spherical nuclei but also  well-known deformed cases.
The same force parametrization can be used both for $\beta-$stable 
and $\beta-$unstable nuclei through-out the periodic table
\cite{rein88,sharma93,suga94,lala97}. 

The importance of the self- and crossed- interactions  are significant for
some specific properties of nuclei/nuclear-matter in certain conditions. 
For example, self-interaction of  $\sigma$-meson takes care  the reduction
of nuclear matter incompressibility $K_{\infty}$ from an unacceptable high
value of $K_{\infty}\sim 600$ MeV to a reasonable number of $\sim 270$ MeV
 \cite{boguta77,bodmer91}, while the self-interaction 
of vector meson $\omega$ soften the equations of states \cite{suga94,fsu}. 
Thus, it is imperative to include all the mesons and their possible interactions
with nucleons and hyperons, self- and crossed terms in the effective Lagrangian density. 
However, it is not necessary to do so, because of the symmetry reason and 
their heavy masses \cite{mechl87}. For example, to keep the spin-isospin 
and parity symmetry in the ground state, the contribution of  $\pi-$meson 
is ignored \cite{green72} and also the effect of heavier mesons are neglected for their
negligible contribution. Taking into this argument, in many versions of the
RMF formalism, the inclusion of isovector-scalar ($\delta$) meson is neglected 
due to its small contribution. 
But  recently it is seen \cite{sing14,shailesh14b,kubis97,bunta04,ortel15,roca11}
that the endowment of the $\delta$-meson goes on increasing with  density 
and  asymmetry of the nuclear system. 
Thus, it will be impossible for us to justify the abandon of $\delta-$meson
both conceptually and practically, while considering the high asymmetry 
and dense nuclear systems, like the neutron star and relativistic heavy ion 
collision. Recent observation of neutron star like  PSR J1614-2230 with 
mass of (1.97$\pm$0.04)$M_\odot$ \cite{demo10} and the PSR J0348+0432 
with mass of (2.01$\pm$0.04)$M_\odot$ \cite{antoni13} re-open the challenge 
in the dense matter physics. 
The heavy mass of PSR J0348+0432 (M=2.01$\pm$0.04$M_\odot$) forces the nuclear 
theorists to re-think the composition and interaction inside the 
neutron star. Therefore, it is important to establish the effects of the 
$\delta$-meson and all possible interactions of other mesons for such
compact and asymmetry system.

The paper is organized as follows: In Sec.~\ref{form}, we have outlined a brief
theoretical formalism. The necessary steps of the RMF model and the
inclusion of $\delta-$meson is explained. The results and discussions are
devoted in Sec.~\ref{resu}. Here, we have attempted to explain 
the effects of  $\delta$-meson on the nuclear matter system like 
hyperon and  proton-neutron stars. This analysis is done for both
static and rotating neutron and neutron-hyperon stars. In this
calculations, the E-RMF Lagrangian (G2 parameter set) is used 
to take care of all possible self- and crossed interactions \cite{furn97}. 
On top of the G2 Lagrangian, the $\delta-$meson interaction is added to 
take care of the isovector channel. 
The concluding remarks are given in section ~\ref{conc}.

\section{Theoretical formalism}\label{form}

From last one decade a lot of work have been done to emphasize the role 
of  $\delta-$meson on both finite and infinite nuclear matter
\cite{hofmann01,liu02,mene04,sula05}. It is seen that the contribution of 
$\delta$-meson to the symmetry energy is negative \cite{shailesh12}. To fix 
the symmetry energy around the empirical value ($\sim$30 MeV ) we need a 
large coupling constant of the $\rho-$meson ($g_\rho$) value 
in the absence of the $g_\delta$. The proton and neutron effective masses 
split due to inclusion of  $\delta$-meson and consequently it affects 
the transport properties of neutron star\cite{kubis97}. The addition of
$\delta$-meson not only modify the property of 
infinite nuclear matter, but also enhances the spin-orbit 
splitting in the finite nuclei\cite{hofmann01}.  A lot of mysteries 
are present in the effects of $\delta$-meson till date. The motivation
of the present chapter is to study such information. It is to be noted that
both the $\rho-$ and $\delta-$mesons correspond to the isospin asymmetry, and
a careful precaution is essential while fixing the $\delta$-meson coupling
in the interaction.

The effective field theory and naturalness of the parameter are 
described in \cite{furn97,furn96,muller96,furn00,mach11}. The Lagrangian 
is consistent with underlying symmetries of the QCD. The G2 parameter is 
motivated by E-RMF theory. The terms of the Lagrangian are taken into 
account up to $4^{th}$ order in meson-baryon coupling. For the study of 
isovector channel, we have introduced the isovector-scalar $\delta$-meson. 
The baryon-meson interaction is given by \cite{bharat07}:   

\begin{eqnarray}
{\cal L}&=&\sum_B\overline{\psi}_B\left(
i\gamma^{\mu}D_{\mu}-m_B+g_{\sigma B}\sigma+g_{\delta B}\delta.\tau \right)
\psi_B 
+ \frac{1}{2}\partial_{\mu}\sigma\partial_{\mu}\sigma
\nonumber \\
&&-m_{\sigma}^2\sigma^2\left(\frac{1}{2}+\frac{\kappa_3}{3!}\frac{g_{\sigma}\sigma}{m_B}
+\frac{\kappa_4}{4!}\frac{g_{\sigma}^2\sigma^2}{m_B^2}\right) - \frac{1}{4}\Omega_{\mu\nu}\Omega^{\mu\nu}\nonumber \\
&&+\frac{1}{2}m_{\omega}^2
\omega_{\mu}\omega^{\mu}\left(1+\eta_1\frac{g_{\sigma}\sigma}{m_B}
+\frac{\eta_2}{2}\frac{g_{\sigma}^2\sigma^2}{m_B^2}\right)-\frac{1}{4}R_{\mu\nu}^aR^{\mu\nu a}\nonumber \\
&&+\frac{1}{2}m_{\rho}^2
\rho_{\mu}^a\rho^{a\mu}\left(1+\eta_{\rho}
\frac{g_{\sigma}\sigma}{m_B} \right)
+\frac{1}{2}\partial_{\mu}\delta.\partial_{\mu}\delta-m_{\delta}^2\delta^2
+\frac{1}{4!}\zeta_0 \left(g_{\omega}\omega_{\mu}\omega^{\mu}\right)^2
\nonumber \\&&+\sum_l\overline{\psi}_l\left(
i\gamma^{\mu}\partial_{\mu}-m_l\right)\psi_l. 
\end{eqnarray}
The co-variant derivative $D_{\mu}$ is defined as:
\begin{eqnarray}
D_{\mu}=\partial_{\mu}+ig_{\omega}\omega_{\mu}+ig_{\rho}I_3\tau^a\rho_{\mu}^a,
\end{eqnarray}
where $R_{\mu\nu}^a$ and $\Omega_{\mu\nu}$ are field tensors and
defined as follow
\begin{eqnarray}
R_{\mu\nu}^a=\partial_{\mu}\rho_{\nu}^a-\partial_{\nu}\rho_{\mu}^a
+g_{\rho}\epsilon_{abc}\rho_{\mu}^b\rho_{\nu}^c,
\end{eqnarray}
\begin{eqnarray}
\Omega_{\mu\nu}=\partial_{\mu}\omega_{\nu}-\partial_{\nu}\omega_{\mu}.
\end{eqnarray}
Here, $\sigma$, $\omega$ , $\rho$ and $\delta$ are the sigma, omega, rho and 
delta meson fields, respectively and in real calculation, we ignore the non-abelian term from the $\rho-$field. All symbols are carrying their own usual 
meaning~\cite{bharat07,sing14}.

The Lagrangian  equation for different mesons are given by \cite{bharat07}:
\begin{flushleft}
\begin{eqnarray}
m_{\sigma}^2 \left(\sigma_0+\frac{g_{\sigma}\kappa_3\sigma_0}{2m_B}
+\frac{\kappa_4 g_{\sigma}^2\sigma_0^2}{6m_B^2} \right) \sigma_0 
-\frac{1}{2}m_{\rho}^2\eta_{\rho}\frac{g_{\sigma}\rho_{03}^2}{m_B}\nonumber \\
-\frac{1}{2}m_{\omega}^2\left(\eta_1\frac{g_{\sigma}}{m_B} 
+\eta_2\frac{g_{\sigma}^2\sigma_0}{m_B^2}\right)\omega_0^2 
=\sum g_{\sigma}\rho^s_B,
\end{eqnarray}
\end{flushleft}
\begin{flushleft}
\begin{eqnarray}
m_{\omega}^2\left(1+\eta_1\frac{g_{\sigma} \sigma_0}{m_B}
+\frac{\eta_2}{2}\frac{g_{\sigma}^2\sigma_0^2}{m_B^2}\right)\omega_0 
+\frac{1}{6}\zeta_0g_{\omega}^2\omega_0^3 
=\sum g_{\omega}\rho_B,
\end{eqnarray}
\end{flushleft}
\begin{eqnarray}
 m_{\rho}^2\left(1+\eta_{\rho}\frac{g_{\sigma}\sigma_0}{m_B}\right)
=\frac{1}{2}\sum g_{\rho}\rho_{B3},
\end{eqnarray}
\begin{eqnarray}
m_{\delta}^2{\delta^3}&=&g_{\delta}^2 \rho^s_{3B},
\end{eqnarray}
with $\rho^s_{3B}=\rho^s_{p}-\rho^s_{n}$, $\rho^s_{p}$ and $\rho^s_{n}$ are
scalar densities for the proton and neutron, respectively. The total scalar 
density is expressed as the sum of the proton and neutron densities 
$\rho^s_{B} = \rho^s_{p}+\rho^s_{n}$, which is given by 
\begin{eqnarray}
\rho^s_{i}=\frac{2}{(2\pi)^3}\int_0^{k_i}\frac{M_i^*d^3k}{E_i^*} ,
i=p,n
\end{eqnarray}
and the vector (baryon) density 
\begin{eqnarray}
\rho_B=\frac{2}{(2\pi)^3}\int_0^{k_i}d^3k,
\end{eqnarray}
where, $E_i^*=(k_i^2+M_i^{*2})^{1/2}$ is the effective energy, $k_i$ is
the Fermi momentum of the baryons. $M_p^*$ and $M_n^*$ are the proton and 
neutron effective masses written as\\
\begin{eqnarray}
M_p^*=M_p-g_{\sigma}\sigma_0-g_{\delta}{\delta^3}\\
M_n^*=M_n-g_{\sigma}\sigma_0+g_{\delta}{\delta^3},
\end{eqnarray}
which are solved self-consistently. $I_3$ is the third component of
isospin projection and $B$ stands for baryon octet. The energy and pressure
density depends on the effective mass $M_B^*$ of the system, which first
needed to solve these self-consistent equations and obtained the fields
for mesons. Using the Einstein's energy-momentum tensor, the total energy 
and pressure density are given as \cite{bharat07}:

\begin{eqnarray}\label{energy}
\cal{E}&=&\sum_B\frac{2}{(2\pi)^3}\int_0^{k_B}d^3kE_B^*(k)
+\frac{1}{8}\zeta_0g_{\omega}^2\omega_0^4 
+ m_{\sigma}^2\sigma_0^2\left(\frac{1}{2}+\frac{\kappa_3}{3!}
\frac{g_{\sigma}\sigma_0}{m_B}+\frac{\kappa_4}{4!}
\frac{g_{\sigma}^2\sigma_0^2}{m_B^2}\right) \nonumber \\
&& + \frac{1}{2}m_{\omega}^2 \omega_0^2\left(1+\eta_1
\frac{g_{\sigma}\sigma_0}{m_B}+\frac{\eta_2}{2}
\frac{g_{\sigma}^2\sigma_0^2}{m_B^2}\right)
+ \frac{1}{2}m_{\rho}^2 \rho_{03}^2\left(1+\eta_{\rho}
\frac{g_{\sigma}\sigma_0}{m_B} \right) \nonumber \\ 
&&+\frac{1}{2}\frac{m_{\delta}^2}{g_{\delta}^{2}}(\delta^3)^2
+\sum_l\varepsilon_l,
\end{eqnarray}

and

\begin{eqnarray}\label{pressure}
\cal{P}&=&\sum_B\frac{2}{3(2\pi)^3}\int_0^{k_B}d^3kE_B^*(k)
+\frac{1}{8}\zeta_0g_{\omega}^2\omega_0^4 
- m_{\sigma}^2\sigma_0^2\left(\frac{1}{2}+\frac{\kappa_3}{3!}
\frac{g_{\sigma}\sigma_0}{m_B}+\frac{\kappa_4}{4!}
\frac{g_{\sigma}^2\sigma_0^2}{m_B^2}\right) \nonumber \\
&& + \frac{1}{2}m_{\omega}^2 \omega_0^2\left(1+\eta_1
\frac{g_{\sigma}\sigma_0}{m_B}+\frac{\eta_2}{2}
\frac{g_{\sigma}^2\sigma_0^2}{m_B^2}\right)
+ \frac{1}{2}m_{\rho}^2 \rho_{03}^2\left(1+\eta_{\rho}
\frac{g_{\sigma}\sigma_0}{m_B} \right) \nonumber \\
&&-\frac{1}{2}\frac{m_{\delta}^2}{g_{\delta}^{2}}(\delta^3)^2+\sum_l P_l,
\end{eqnarray}
where $P_l$ and $\varepsilon_l$ are lepton's pressure and energy 
density, respectively.  

\section{Results and discussions}\label{resu}
Before going to the discussions of our results, we give 
a brief description of the parameter fitting procedure for 
g$_\rho$ and g$_\delta$.  As it is commonly known, the symmetric energy
$E_{sym}$, is an important quantity to select the equation of states. This
value of $E_{sym}$ determines the structure of both static and rotating 
neutron stars. On the other hand, an arbitrary combination of g$_\rho$ and g$_\delta$ with a fixed value of E$_s$ can affect the ground state properties
of finite nuclei. Thus, to have a clear picture on the effect of g$_\delta$
on hyperon star structure, we have chosen two different prescriptions for the
selection of g$_\delta$ in our present calculations. (1) In the first method,
we have constructed various sets of g$_\rho$ and g$_\delta$ keeping E$_s$ fixed. 
Here,  all the other parameters of G2 set are remained unchanged. The G2 set and 
the combination of  g$_\rho$ and g$_\delta$ are displayed in Table ~\ref{tab1}.
(2) In the second procedure, we have chosen the g$_\rho$, g$_\delta$ pairs
keeping the binding energy constant (experimental binding energy) for finite 
nuclei. The values of these g$_\rho$ and g$_\delta$ are given in 
Table ~\ref{tab3}  with other properties of infinite nuclear matter. It is 
worthy to re-emphasized here that we are not looking for a full-fledged 
parameter set including the $\delta-$meson coupling,
but our aim in this paper is to study the effects of $\delta-$meson coupling on hyperon star
and the production of baryon octet. Therefore, after splitting the g$_{\rho}$ coupling
constant into two parts (g$_{\rho}$, g$_{\delta}$) using the first prescription, 
the results on hyperon star along with the neutron star structures 
both for static and rotating cases under $\beta-$equilibrium condition 
are discussed in the subsequent subsections ~\ref{resub}, ~\ref{resuc},~\ref{resud}, ~\ref{resue}  and ~\ref{resuf}. 
In Sec. IV, we follow the second procedure to get the (g$_{\rho}$, g$_{\delta}$) pairs
and applied these to some selective cases.

\subsection{Parametrization of $g_\rho$ and $g_\delta$ with constant symmetry energy}\label{resua}

It is important to fix $g_\delta$  value to see the effects of the 
$\delta$-meson. The isovector channels in  RMF theory come to exist 
through both the $\rho-$ and $\delta-$mesons couplings. 
While considering the effects of the $\delta$-meson, we have to 
take the $\rho$-meson into account. Since both the isovector channels are
related to isospin, one can not optimize the $g_\delta$ coupling
independently. Here, we have followed a more reliable procedure by fixing
the symmetry energy $E_{sym}$ with adjusting simultaneously different values of 
$g_\rho$ and $g_\delta$ \cite{kubis97}. In general, for most of the non-relativistic formalism, the
symmetry energy $E_{sym}$ is around 30-33 MeV. However, in some specific 
parametrization like GS4, $E_{sym}=$12.83 MeV and for PRC45 set it is 
51.01 MeV \cite{chen07,dutra12}.  On the other hand, in non-linear, 
density-dependent and point-coupling
relativistic mean field forces, the $E_{sym}$ varies from 26.1 to 44.0 MeV. Here we have used the well
known G2 parametrization, which has a moderate symmetry energy $E_{sym}=36.4$ MeV. It is
to be noted that the symmetry energy plays a crucial role both in finite nuclei and in the equation
of state, which
include the neutron distribution radius in the nucleus and the mass and radius of a neutron star,
respectively. For a smaller value of $E_{sym}$, both the relativistic
and non-relativistic forces predict a smaller neutron star mass contrary to the recent observation of
about 2M$_{\odot}$. A detail variation of symmetry energies for Skyrme effective interaction and
non-linear relativistic mean field formalism is available in Refs.
\cite{chen07,dutra12}. Recently,
a large number of papers have been devoted to $E_{sym}$ for a definite value, but till it is
under active discussions.

As it is mentioned earlier, 
we have added $g_\delta$ on top of the G2 parameter set. Thus, the symmetry 
energy of G2 parameter $E_{sym} = 36.4 $ MeV is kept constant at the time of
re-shuffling $g_\rho$ and $g_\delta$.  It is to be noted that, we do not want
to change the value of $E_{sym}$ of the original G2 parameter set with the 
addition of $\delta$-meson. The G2 parameters and the $g_{\delta}$
and $g_{\delta}$ combinations are displayed in Table ~\ref{tab1}. The
nuclear matter properties are also listed in the table (middle panel).
\begin{figure}
\includegraphics[width=0.50\textheight,clip=true,angle=0]{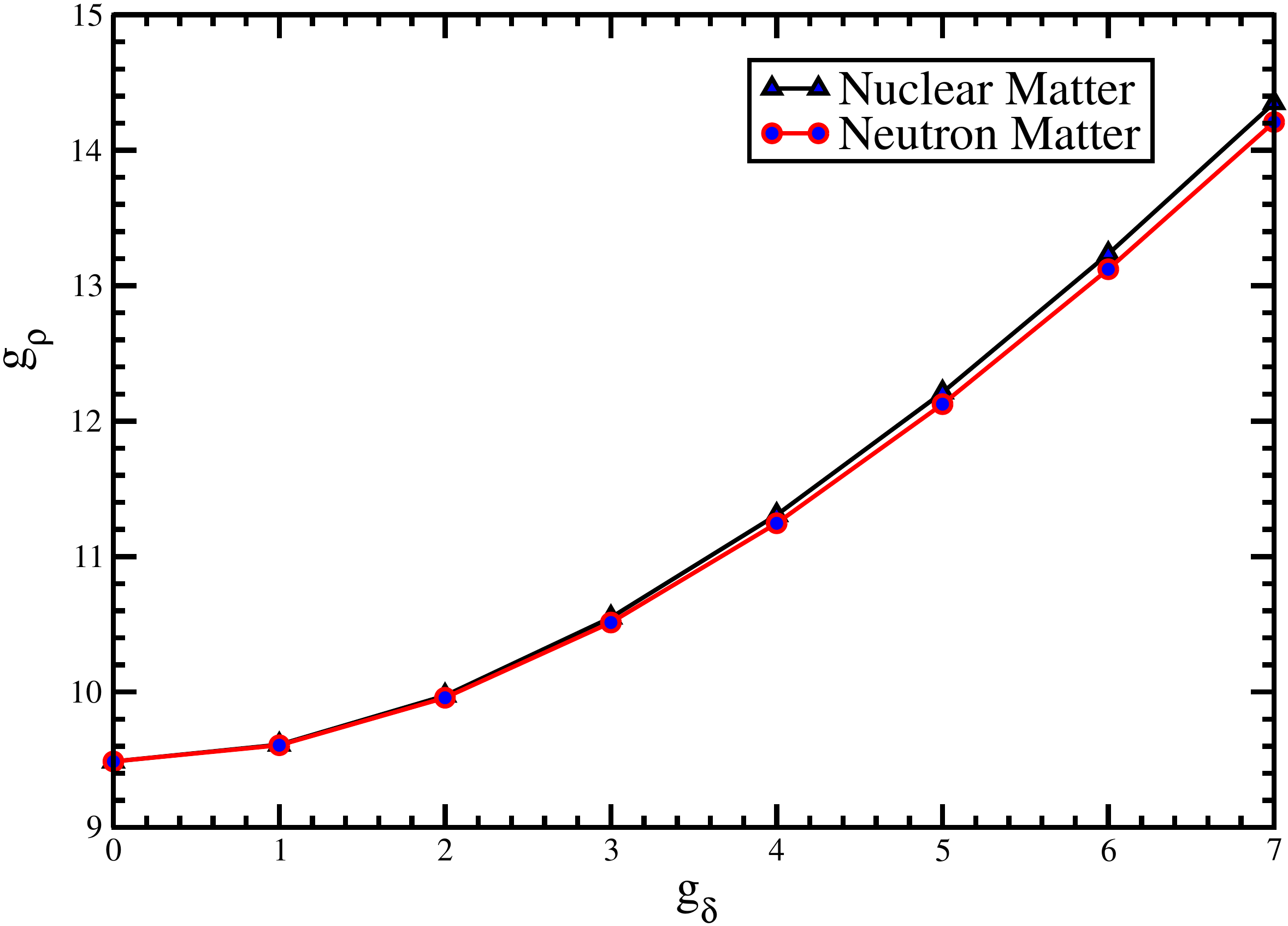}
\caption{\label{fig1}Variation of $g_\rho$ and $g_\delta$ at
a constant value of symmetry energy $E_{sym} = 36.4 $ MeV for both nuclear and neutron matter.}
\end{figure}
\begin{table*}
\caption{\label{tab1}{The parameters for G2 set are in the upper panel of 
the Table. The nuclear matter saturation properties are in the middle
panel and various $g_{\rho}$ and $g_{\delta}$ combinations are in the
lower panel, keeping symmetry energy E$_{sym}$ = 36.4 MeV fixed.
}}
\bgroup
\def\arraystretch{1.7}
\scalebox{0.56}{
{\begin{tabular}{|c|c|c|c|c|c|c|c|c|c|c|c|c|c|c|}
\hline
$m_{n}$ = 939.0 MeV& $m_{\sigma}$ = 520.206 MeV&$m_{\omega}$ = 782.0 MeV &
$m_{\rho}$ = 770.0 MeV& $m_{\delta}$ = 980.0 MeV &$\Lambda$= 0.0&
$\zeta_{0}$ = 2.642&$\eta_{\rho}$ = 0.39&\\ 
\hline
$g_{\sigma}$ = 10.5088&$g_{\omega}$ = 12.7864& $g_{\rho}$ = 9.5108&$g_{\delta}$ = 0.0&$k_{3}$
 = 3.2376&$k_{4}$ = 0.6939&$\eta_{1}$ = 0.65& $\eta_{2}$ = 0.11& \\
\hline
$\rho_{0}$ = 0.153$fm^{-3}$&E/A = -16.07 MeV& $K_{\infty}$ = 215 MeV  
&$E_{sym}$ = 36.4 MeV& $m^*_{n}/m_{n}$ = 0.664 &&&&  \\
\hline
\hline
$(g_{\rho}, g_{\delta})$&(9.510, 0.0)&(9.612, 1.0)&(9.973, 2.0)&(10.550, 3.0)&(11.307, 4.0)&
(12.212, 5.0)&   
(13.234, 6.0) &  
(14.349, 7.0)\\
\hline
\end{tabular}
}}
\egroup
\end{table*}
\begin{table*}
\hspace{3.0 cm}
\caption{Binding energy (MeV) and charge radius (fm) are  calculated with various combination of 
$g_{\rho}$ and $g_{\delta}$ in G2+$\delta$. The results are compared with experimental data
{\cite{nndc}}.}
\renewcommand{\tabcolsep}{0.13 cm}
\renewcommand{\arraystretch}{1.}
\scalebox{0.7}{
{\begin{tabular}{|c| c| c| c| c| c| c| c| c| c| c| c| c| c| c|}
\hline
($g_{\rho}$,$g_{\delta}$) &(9.510, 0.0)&(9.612, 1.0)&(9.973, 2.0)&(10.550, 3.0)&(11.307, 4.0)& (12.212, 5.0)& (13.234, 6.0) & (14.349, 7.0)&\\
\hline
\multicolumn{1}{|c|}{Nucleus}&\multicolumn{8}{c}{Theory}& \multicolumn{1}{|c|}{Expt. }\\
\hline 
$^{16}$O (BE)&  127.2  & 127.2 & 127.2  & 127.2  & 127.2 & 127.2   &127.2 & 127.2 &127.6 \\
r$_{ch}$& 2.718        &2.718  &  2.718 &2.718  & 2.718 &2.717 &  2.717& 2.716   &2.699 \\
\hline
$^{40}$ Ca (BE)& 341.1 & 341.1 &  341.1 &  341.1 & 341.1 &  341.1 & 341.1 & 341.1 &342.0 \\
r$_{ch}$& 3.453        & 3.453  & 3.453   & 3.453   &  3.452& 3.451 &  3.450 & 3.449&3.4776 \\
\hline
$^{48}$Ca (BE)& 416.0  & 415.8 &  415.2 & 414.1  & 412.6 & 410.7&  408.4 & 405.7  &416.0 \\
r$_{ch}$& 3.440        & 3.439 &  3.438 &  3.437   & 3.435& 3.432  & 3.430 & 3.427  & 3.477\\
\hline
$^{56}$Ni (BE)& 480.4  & 480.3 &  480.3 & 480.3  & 480.3 &  480.3 & 480.3 & 480.2   &484.0 \\
r$_{ch}$&  3.730       & 3.730   & 3.730  & 3.730  & 3.730& 3.730 & 3.730 & 3.724    &    \\
\hline
$^{58}$Ni (BE)& 497.2     & 497.2 &  497.1& 497.0  & 496.9 & 496.7  & 496.5 & 496.3  &506.5 \\
r$_{ch}$& 3.765        & 3.765  &  3.763   & 3.762  & 3.761& 3.758 &  3.756 &  3.753 & 3.775  \\
\hline
$^{90}$Zr (BE)& 781.6  & 781.2 &780.6   &  779.4 & 777.9 &  775.9  & 773.6 & 770.8&783.9    \\
r$_{ch}$& 4.238        & 4.238 &  4.237   & 4.235  & 4.233& 4.230 & 4.228 & 4.225 & 4.269   \\
\hline
$^{116}$Sn (BE)& 981.2 & 980.7 &979.4   & 977.2  & 974.1 & 970.2 & 965.6 & 960.3  &988.7   \\
r$_{ch}$&  4.604       & 4.603 &  4.601  & 4.598  & 4.594&  4.589 & 4.584 & 4.579  & 4.625   \\
\hline
$^{118}$Sn (BE)& 997.6 & 997.1 & 995.4  & 992.7  & 989.0 & 984.3  & 978.7& 972.2 &1004.9   \\
r$_{ch}$& 4.620        & 4.619  &  4.617  &  4.613  & 4.610& 4.604 &  4.599 & 4.594 & 4.639   \\
\hline
$^{120}$Sn (BE)&1013.9  & 1013.2 & 1011.2  & 1008.0 & 1003.5  & 998.0  & 991.3 & 983.8  &1020.5   \\
r$_{ch}$& 4.627         &4.626  & 4.624  & 4.620 & 4.616& 4.610 & 4.605 & 4.600   &4.652         \\
\hline
$^{208}$Pb (BE)& 1633.3 & 1631.4 & 1625.7   &  1616.2 &  1603.0 & 1586.4 &  1566.4 &  1543.5&1636.4  \\
r$_{ch}$& 5.499        & 5.498 & 5.497   &  5.494 &  5.492&  5.489 & 5.487&  5.485   &5.501        \\
\hline
\end{tabular}\label{tab2}}
}
\end{table*}
\begin{table}
\hspace{5.0 cm}
\caption{ Mass and radius of the neutron star are calculated at different 
 values of  $g_{\rho}$ and $g_{\delta}$  keeping  binding energy of $^{208}$Pb (1633.296 MeV) constant.
The calculated results of $E_{sym}$, $L_{sym}$ and $K_{sym}$ are for nuclear matter at different
combinations of ($g_{\rho}$, $g_{\delta}$) pairs.
}
\renewcommand{\tabcolsep}{0.13 cm}
\renewcommand{\arraystretch}{1.}
\scalebox{0.85}{
{\begin{tabular}{|c| c| c| c|c|c|c|}
\hline
$ (g_\rho, g_\delta)$ & $\frac {M}{M_\odot}$ & Radius (Km)&E$_{sym}$ (MeV)& L$_{sym}$ (MeV)& K$_{sym}$ (MeV) \\
\hline
(9.510, 0.0)  & 1.980 & 11.230& 36.4& 101.0& -7.58\\

(9.588, 1.746) & 1.993 & 11.246& 35.3& 98.3& -0.60\\

(9.896, 3.543) & 1.997 & 11.262 & 31.7& 90.2& 20.90\\

(10.518, 5.742) & 2.004 & 11.294 &23.8& 72.5& 67.07\\

(11.774, 8.834) & 2.018 & 11.510  & 6.35 & 30.6& 169.03\\
\hline
\end{tabular}\label{tab3}}
}
\end{table}
For a particular value of $E_{sym} = 36.4 $ MeV, the variation of $g_\rho$ and $g_\delta$ are plotted in Fig.~\ref{fig1}. From Fig.~\ref{fig1},  it is clear that 
as the $g_\delta$ increases 
the $g_\rho$ value also increases, almost linearly, to fix the symmetry energy
unchanged. This implies that $\rho-$ and $\delta$-mesons have opposite effect on 
$E_{sym}$ contribution, i.e.,  the $\delta$-meson has negative contribution 
of the symmetry energy contrary to the positive contribution of 
$\rho$-meson.

We feel that it is instructive to check the finite nuclear properties 
with these combinations of $g_\rho$ and $g_\delta$. We have tabulated the binding 
energy and charge radius of some spherical nucleus in Table~\ref{tab2}.  From 
the table, it is clear that binding energy for asymmetric nucleus goes on 
decreasing with increasing $\delta-$meson and decreasing $\rho-$meson couplings. 
However, it is well understood that the scalar $\delta-$meson gives a positive
contribution to the binding energy. Thus, the binding energy of asymmetric
nuclei should go on increasing with g$_{\delta}$ contrary to the observation
seen in Table ~\ref{tab2}. This happens, because of the simultaneous change of
($g_\rho$, $g_\delta$) pair to keep the constant symmetry energy, i.e., g$_{\rho}$ is
decreasing and $g_\delta$ is increasing. As a result, the contribution of $\rho-$ meson, 
which is negative to the binding energy dominates over the $\delta-$meson 
effect on binding energy. But in case of symmetric nucleus, like $^{16}$O etc.
the effects of both $\rho-$ and $\delta-$mesons are absent due to iso-spin
symmetry. A further inspection of Table ~\ref{tab2} reveals a slight change in 
binding energy and charge radius even for symmetric nuclei because of the
slight different in density distribution of protons and neutrons, although it is small.
 
\subsection{Fields of $\sigma, \omega, \rho$ and $\delta$ mesons}\label{resub}
The fields of the meson play a crucial role to construct the nuclear 
potential, which is the deciding factor for all type of calculations in
the relativistic mean field model. In Fig.~\ref{fig2}, we have plotted 
various meson fields included in the present calculations, such as 
$\sigma$, $\omega$, $\rho$ and $\delta$ with $g_{\delta}$ on top of 
G2 parameter set $(G2+g_{\delta})$. 
\begin{figure}
\includegraphics[width=0.60\textheight,clip=true,angle=0]{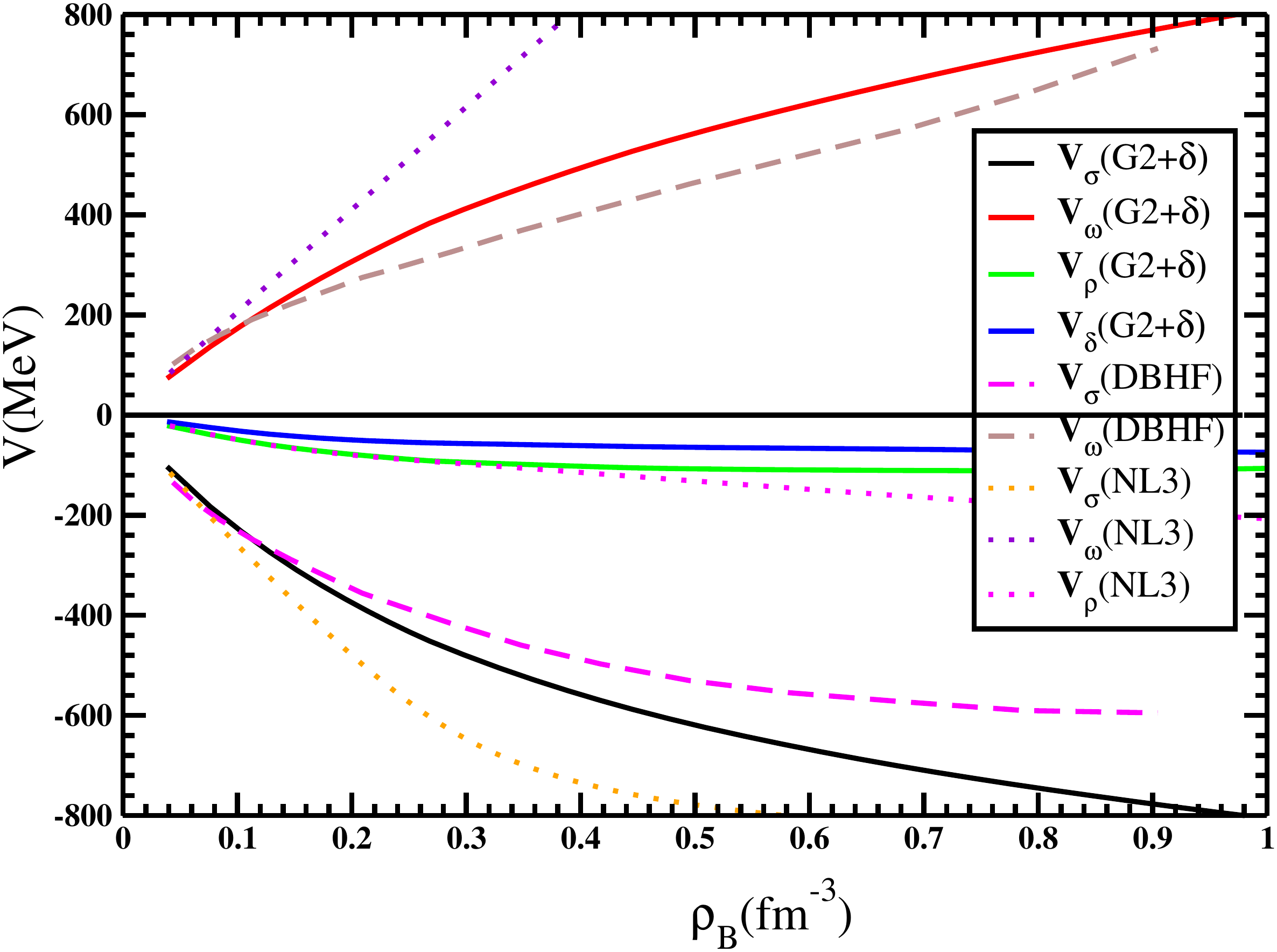}
\caption{Various meson fields are obtained from the 
RMF theory with $G2+g_{\delta}$ and NL3 parameter sets. The $\sigma-$ 
meson field $V_\sigma$ and $\omega-$meson field  $V_\omega$ from 
$G2+g_{\delta}$ calculations are compared with the results of DBHF 
theory \cite{brok90} and NL3 set.}
\label{fig2}
\end{figure}
It is obvious that $V_\sigma$ and $V_\omega$ are opposite to each
other, which is also reflected in the figure. This means, the positive
value of $V_{\omega}$ gives a strong repulsion, which is compensated by the
strongly attractive potential of the $\sigma-$meson field $V_\sigma$.
The nature of the curves for $V_\sigma$ and $V_\omega$ are almost similar 
except the sign. The magnitude of $V_\sigma$ and $V_\omega$ looks
almost equal. However, in real (it is not clearly visible in the curve,
because of the scale), the value of $V_\sigma$ is slightly larger
than $V_\omega$, which keeps the overall nuclear potential strongly attractive.
The attractive $V_\sigma$ and repulsive $V_\omega$ potentials  combinely 
give the saturation properties of the nuclear force. It is worthy to mention
that the contributions of self-interaction terms are taken care
both in $V_\sigma$ and $V_\omega$, which are the key quantities to solve the
Coester band problem \cite{bidhu14} and the explanation of quark-gluon-plasma 
(QGP) formation within the relativistic mean field formalism \cite{subrat15}.
The self-interaction of the $\sigma-$meson gives a repulsive force at long
range part of the nuclear potential, which is equivalent to the 3-body 
interaction and responsible for the saturation properties of nuclear force.
The calculated results of $V_\sigma$ and $V_\omega$ are compared with the 
results obtained from DBHF theory with Bonn-A potential\cite{brok90} and
NL3 \cite{lala97} force. 

Fig.~\ref{fig2} clearly shows that in the low density region 
(density $\rho_B$ $\sim 2\rho_0$) both RMF and DBHF theories well matched. 
But as it increases beyond density $\rho_B$ $\sim 2\rho_0$ ($\rho_0$ is the
nuclear saturation density) both the calculations 
deviate from each other. 
The possible reason may be the fitting procedure of parameters in Bonn-A potential 
is up to $2-3$ times of saturation density $\rho_0$, beyond that the DBHF
data are simple extrapolation of the DBHF theory. Again, the $V_{\omega}$ and
$V_{\sigma}$ fields of NL3 are very different from
$G2+\delta$ results. The $V_{\omega }$ for NL3 follows a linear
path contrary to the results of $G2+\delta$ and Bonn-A. This could be due to the
absence of self- and crossed couplings in NL3 set. 
The contribution of both $\rho-$ and $\delta-$ mesons correspond to the 
isovector channel. The $\delta-$meson gives different effective masses 
for proton and neutron, because of their opposite iso-spin of the third
component. The nuclear potential generated by the $\rho-$ and $\delta-$mesons
are also shown in  Fig~\ref{fig2}. We noticed that although their contributions
are small, but non-negligible. These non-zero values of $V_{\rho}$ and 
$V_{\delta}$ to the nuclear potential has a significant  consequence, mostly
in compact dense object like neutron or hyperon stars, which will be
discussed later in this paper.

\subsection{Energy per particle and pressure density}\label{resuc}
 
The energy and pressure densities as a function of baryonic density
$\rho_B$ are known as equations of states (EOS). 
These quantities are the key ingredients to describe the structure of 
neutron/hyperon stars. To see the sensitivity of the EOS, we have plotted  
energy per particle ($E/\rho_B -M$)  as a function of density for pure
neutron matter in  Fig~\ref{fig3}. Each curve corresponds to a particular 
combination of $g_\delta$ and $g_\rho$ (taken from Table ~\ref{tab1}), 
which reproduce the symmetry energy $E_{sym} = 36.4 $ MeV without destabilizing 
other parameters of G2 set. The green line represents for $g_\delta =0$, i.e., 
with pure G2 parameter set. Both the binding energy per particle as well as 
the pressure density increase with the value of $g_\delta$.
This process continues till the value of $g_\delta$ reaches, at which 
$E/\rho_B -M$ equals the nuclear matter binding energy per particle. An
unphysical situation arises beyond this value of $g_\delta$ because the
binding energy of the neutron matter will be greater than $E/\rho_B -M$ for
the symmetric nuclear matter. In the case of G2+$\delta$ parametrization, this limiting
value of $g_{\delta}$ reaches at $g_\delta$= 0.7, after which
we do not get a convergence solution in our calculations.
\begin{figure}
\includegraphics[width=0.60\textheight,clip=true,angle=0]{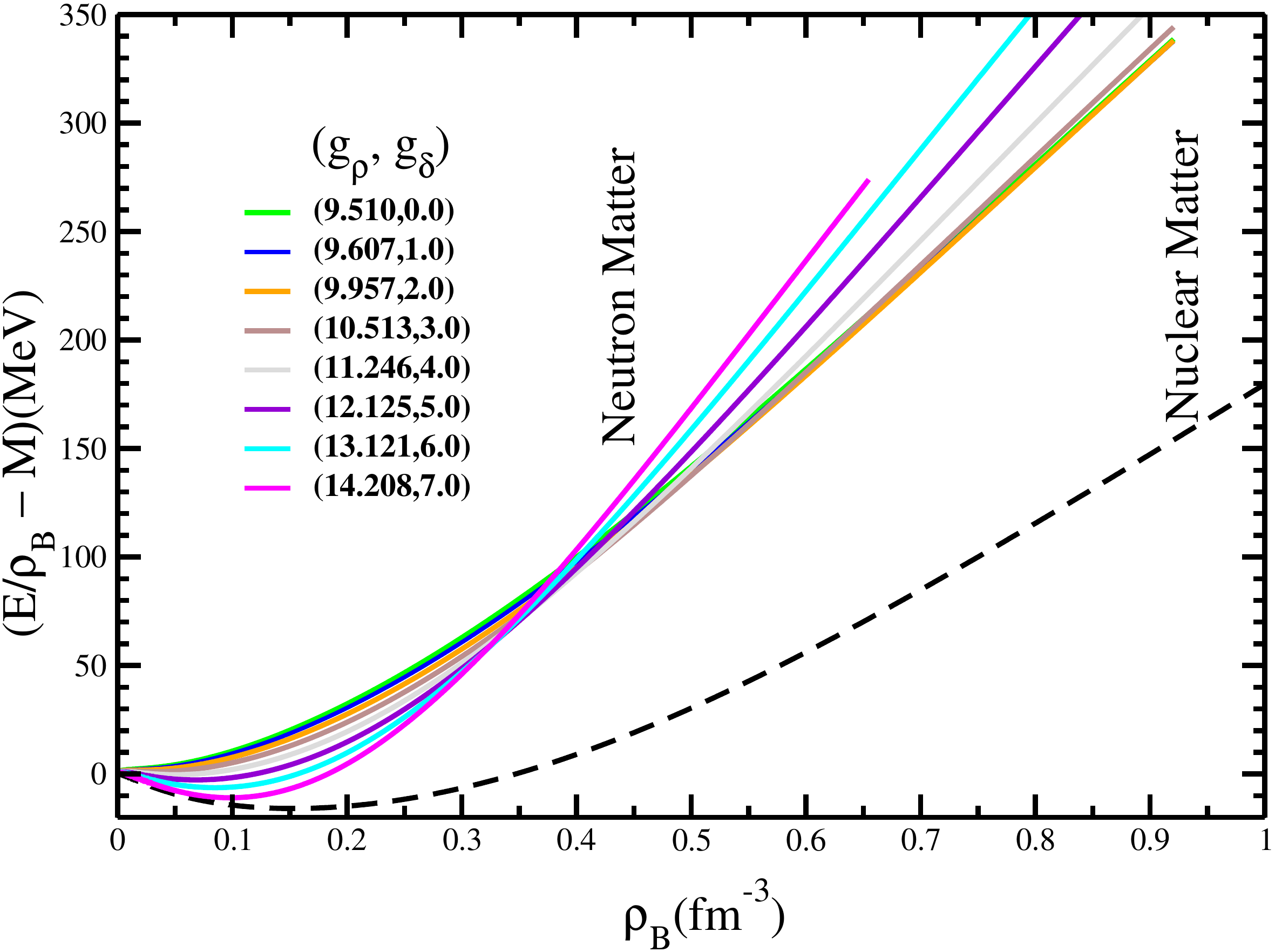}
\caption{(Color online)  Variation of binding energy per particle
with density at various $g_\rho$ and $g_\delta$. }
\label{fig3}
\end{figure}

\begin{figure}
\includegraphics[width=0.60\textheight,clip=true,angle=0]{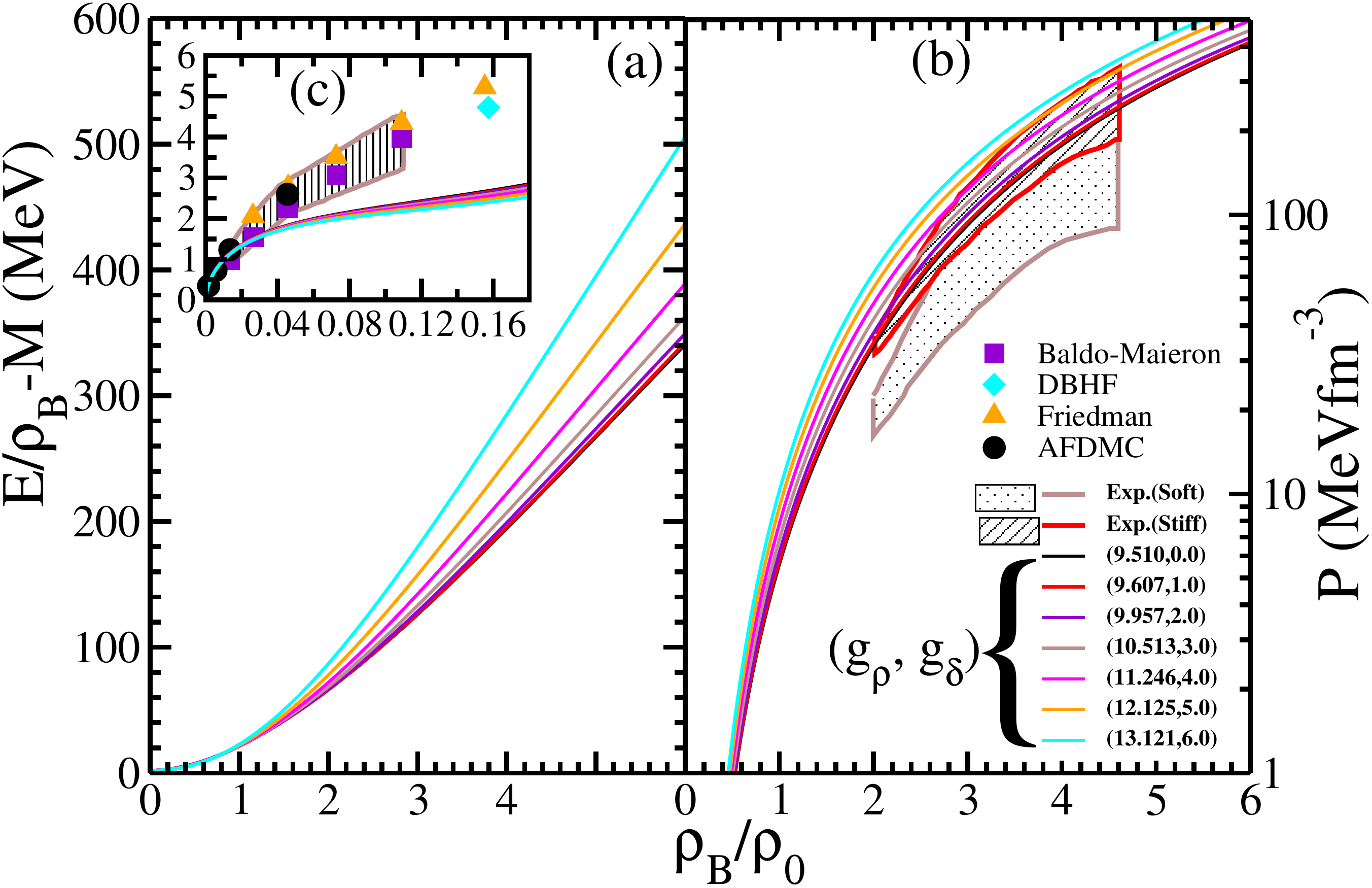}
\caption{ Variation of energy per particle (panel (a)) and pressure density (panel (b))
with $\rho_B/\rho_0$ at different values of $g_\rho$ and $g_\delta$. The enlarged version of energy 
per particle for sub-saturation region is in panel (c). The results of other theoretical models 
like Baldo-Maieron \cite{baldo}, DDHF \cite{sagawa08}, Firedman \cite{fried81} and AFDMC \cite{gando08}
are also given for comparison.
 }
\label{fig4}
\end{figure}

In Fig~\ref{fig4} we have plotted the variation of energy and pressure densities as
a function of ${\rho_B/\rho_0}$ for different combinations of $g_\rho$ and $g_\delta$.
The enlarge version of energy density in the sub-saturation region is shown in panel (c) 
of the figure. Similar to other parameter sets of RMF formalism, the G2+$\delta$ set also
deviates from the experimental data. It is to be recalled here that special attentions
are needed to construct nucleon-nucleon interaction to fit the data at sub-saturation
density. For example, the potentials of Friedman and Pandharipande \cite{fried81},  Baldo-Maieron \cite{baldo},
DDHF \cite{sagawa08} and AFDMC \cite{gando08} are designed to fit the data in this region. The three-body effect
also can not be ignored in this sub-saturation region of the density \cite{pieper01}.  Although, the
non-linear interactions fulfilled this demand to some extent \cite{bidhu14,subrat15,fujita57},
like Coester band problem \cite{coester70}, till some further modification of the couplings are needed. In this regard, the relativistic mean field 
calculations with density dependent meson-nucleon coupling \cite{typel99} 
and constraining the RMF models of the nuclear matter equation of 
state at low densities \cite{voskresenskaya12} are some of the attempts.
The mean field approximation is also a major limitation in the region of 
sub-saturation density. This is because, the assumption of classical meson field
is not a proper approximation in this region to reproduce precisely the data. 
In higher density region, most of the RMF forces reproduces the experimental
data quite well and the predictive power of these forces for finite nuclei
is in excellent agreement both for $\beta-$stable and $\beta-$unstable
nuclei.  The energy and pressure densities with G2 set reproduce 
the experimental data satisfactorily \cite{aru04}. The variation of 
pressure density as a function of $\rho_B$ is shown in panel (b) of
Fig. 4, which passes inside the stiff flow data at higher density \cite{dane02}. 
Also, the $\delta-$meson coupling has  significant effect in supersaturation density 
than the sub-saturation region.
All the EOS with different $g_\rho$ and $g_\delta$ remain inside the stiff 
flow data (Fig. 4, panel (b)). In the present investigation, we are more concerned for highly
dense neutron and hyperon stars, which are considered to be 
super-saturated nuclear objects.

\subsection{Stellar properties of static and rotating neutron stars}\label{resud}
The $\beta$-equilibrium and  charge neutrality  
are two important conditions to justify the structural composition 
of the neutron/hyperon stars. Both these conditions force the stars to 
have $\sim$90$\%$ of neutron and $\sim$10$\%$ proton. With the inclusion of  
baryons, the $\beta-$equilibrium conditions between chemical potentials
 for different particles: 
\begin{eqnarray}\label{beta}
\mu_p = \mu_{\Sigma^+}=\mu_n-\mu_{e}, \nonumber\\
\mu_n=\mu_{\Sigma^0}=\mu_{\Xi^0}=\mu_{n},\nonumber \\
\mu_{\Sigma^-}=\mu_{\Xi^-}=\mu_n+\mu_{e},\nonumber \\
\mu_{\mu}=\mu_{e}, \nonumber\\
\end{eqnarray}
and the charge neutrality condition is satisfied by  
\begin{eqnarray}\label{charge}
n_p+n_{\Sigma^+}=n_e+n_{\mu^-}+n_{\Sigma^-}+n_{\Xi^-}.
\end{eqnarray}

To calculate the mass and radius profile of the static (non-rotating) and 
spherical neutron star, we solve the general relativity
Tolmann-Oppenheimer-Volkov (TOV)\cite{tolm39} equations which are written as: 
\begin{eqnarray}
\frac{d P(r)}{d r}=-\frac{G}{c^2}\frac{[{\cal E}(r)+P(r)][M(r)+\frac{4\pi r^3 P(r)}{c^2}]}{r^2(1-\frac{2GM(r)}{c^2 r})}
\end{eqnarray} 
and
\begin{eqnarray}
\frac{d M(r)}{d r}=\frac{4\pi r^2 {\cal E}(r)}{c^2},
\end{eqnarray}
with G as the gravitational constant, $\cal E$$(r)$ as the energy density, 
$P(r)$ as the pressure density and $M(r)$ as the gravitational mass inside 
radius $r$. We have used c=1. For a given EOS, these equations can be 
integrated from the origin as an initial value problem for a given choice 
of the central density $\cal E$$_c$($r$). The value of r (= R) at which the 
pressure vanish defines the surface of the star. 
In order to understand the effect of $\delta-$meson coupling on neutron
star structure, we must also look, what happens to massive objects as they 
rotate and how this affects the space-time around them. 
\begin{figure}
\includegraphics[width=0.60\textheight,clip=true,angle=0]{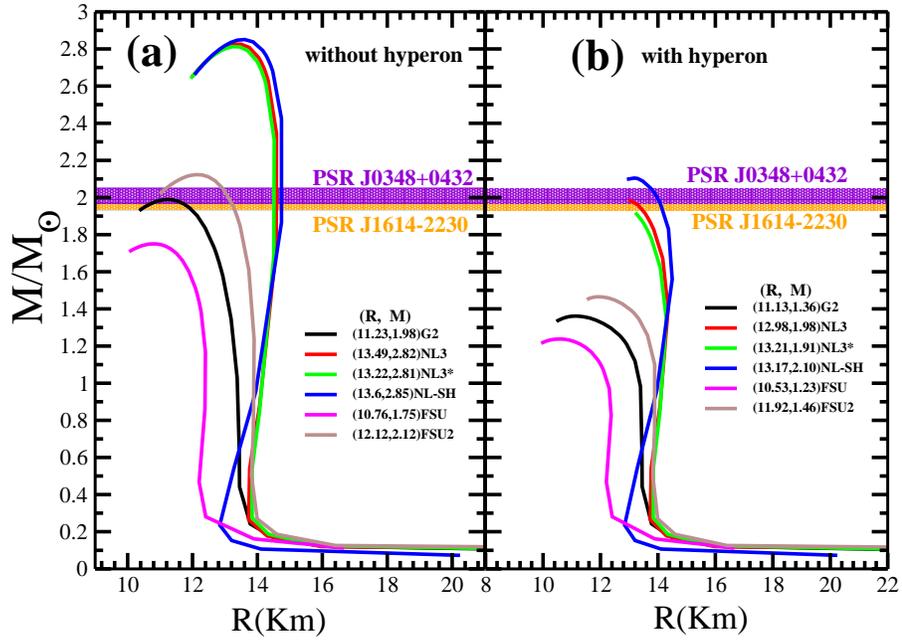}
\caption{(Color online) The mass-radius profile for static star with different 
parametrization like G2\cite{furn97}, NL3\cite{lala97}, NL3*\cite{lala09}, 
NL-SH\cite{sharma93}, FSU\cite{fsu}and FSU2\cite{fsu2}. (a) is for 
proton-neutron star and (b) is for the hyperon star. The maximum mass M and the
corresponding radius obtained by various parameter sets are given in the parenthesis.
}
\label{fig5}
\end{figure}
For this, we use the code written by Stergioulas\cite{ster95} based on Komastu, 
Eriguchi, and Hachisu (KEH) method (fast rotation)\cite{kom89,koma89} to 
construct mass-radius of the uniform rotating star. One should note that the
maximum mass of a static star is less than the rotating stars. Because, 
when the massive objects rotate they flatten at their poles. The forces of 
rotation, namely the effective centrifugal force,  pulls the mass farthest 
from the center further out, creating the equatorial bulge. This pull away 
from the center will, in part, counteract  gravity, allowing the star to be 
able to support more mass than its  non-rotating star.

We know that the core of neutron stars contain hyperons at very high density
($\sim$7-8 $\rho{_0}$) matter. As it is mentioned before, with the presence of 
baryons, the EOS becomes softer and stellar properties will change. The 
maximum mass of hyperon star decreases about 10-20$\%$ depending on the choice 
of the meson-hyperon coupling constants. The 
hyperon couplings are expressed as the ratio between the meson-hyperon and 
meson-nucleon couplings as:
\begin{eqnarray}
\chi_{\sigma} = \frac {g_{Y\sigma}}{g_{N\sigma}}, \chi_{\omega} = \frac {g_{Y\omega}}{g_{N\omega}},\chi_{\rho} = \frac {g_{Y\rho}}{g_{N\rho}}, \chi_{\delta} = \frac {g_{Y\delta}}{g_{N\delta}}.
\end{eqnarray}
In the present calculations, we have taken $\chi_{\sigma} = \chi_{\rho} = 
\chi_{\delta}$ = 0.6104 and $ \chi_{\omega}$ = 0.6666\cite{glen91}. One can
find similar calculations for stellar mass in Refs. \cite{glen20,weis12,lope14}. 
Now we present the star properties like mass and radius in Figs. 
~\ref{fig5} and ~\ref{fig6}. In Fig.~\ref{fig5} we  have plotted 
the mass-radius profile for the proton-neutron star as well as 
for the hyperon star using a wide variation of parameter sets starting
from the old parameter like NL-SH\cite{sharma93} to the new set of
FSU2 \cite{fsu2}. The mass-radius profile varies 
to a great extend over the choice of the parameter. For example, in FSU
parameter set \cite{fsu}, the maximum possible mass of the proton-neutron 
star is $\sim$ 1.75 $M_\odot$, while the maximum possible mass for the 
NL3 set \cite{lala97} is $\sim$ 2.8 $M_\odot$. 
These results are shown in the left panel of the Fig.~\ref{fig5}, while 
right panel shows the same things for the hyperon star (the maximum mass and the 
corresponding radius for different forces are given in the parenthesis). 

\subsection{Effects of $\delta-$meson on static and rotating stars}\label{resue}

The main aim of  this paper is to understand the effects of $\delta$-meson 
on  neutron stars both with and without hyperons. 
Fig.~\ref{fig6} represent the mass-radius profiles for 
non-rotating and rotating stars taking into account the presence of with 
and without hyperons.  These profiles are shown for various combinations 
of $g_{\rho}$ and $g_{\delta}$ (see Table ~\ref{tab1}), which we have 
obtained by fitting the symmetry energy $E_{sym}$ of pure nuclear matter.  
\begin{figure*}
\includegraphics[width=0.60\textheight,clip=true,angle=0]{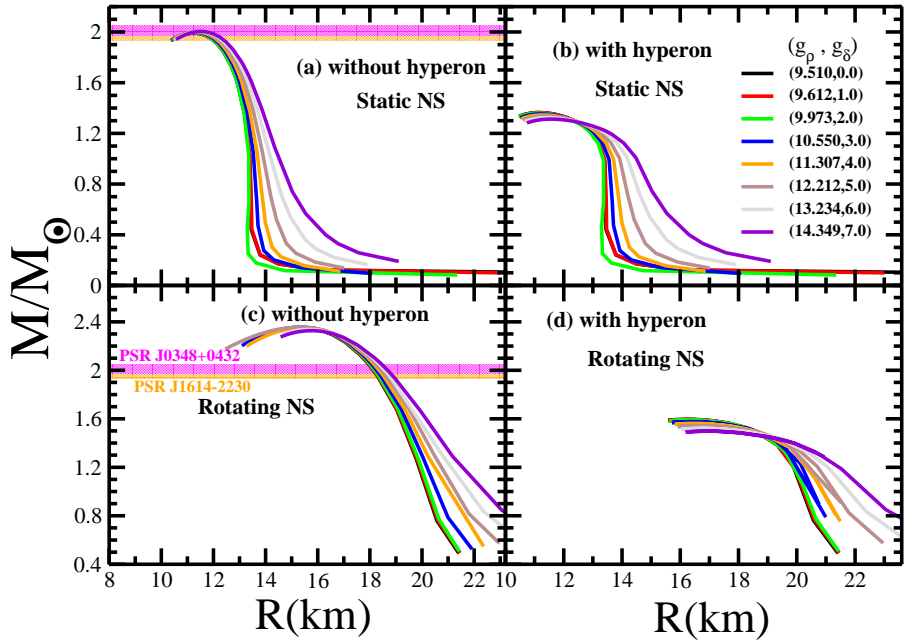}
\caption{(Color online) The mass-radius profile of the static and rotating
 proton-neutron and hyperon stars with various combination of $g_{\delta}$ 
and $g_{\rho}$ in G2+$\delta$. (a,c) is for proton-neutron star and (b,d) 
is for the hyperon star. (a,b) for static and (c,d) for rotating cases.}
\label{fig6}
\end{figure*}
\begin{figure}
\includegraphics[width=0.60\textheight,clip=true,angle=0]{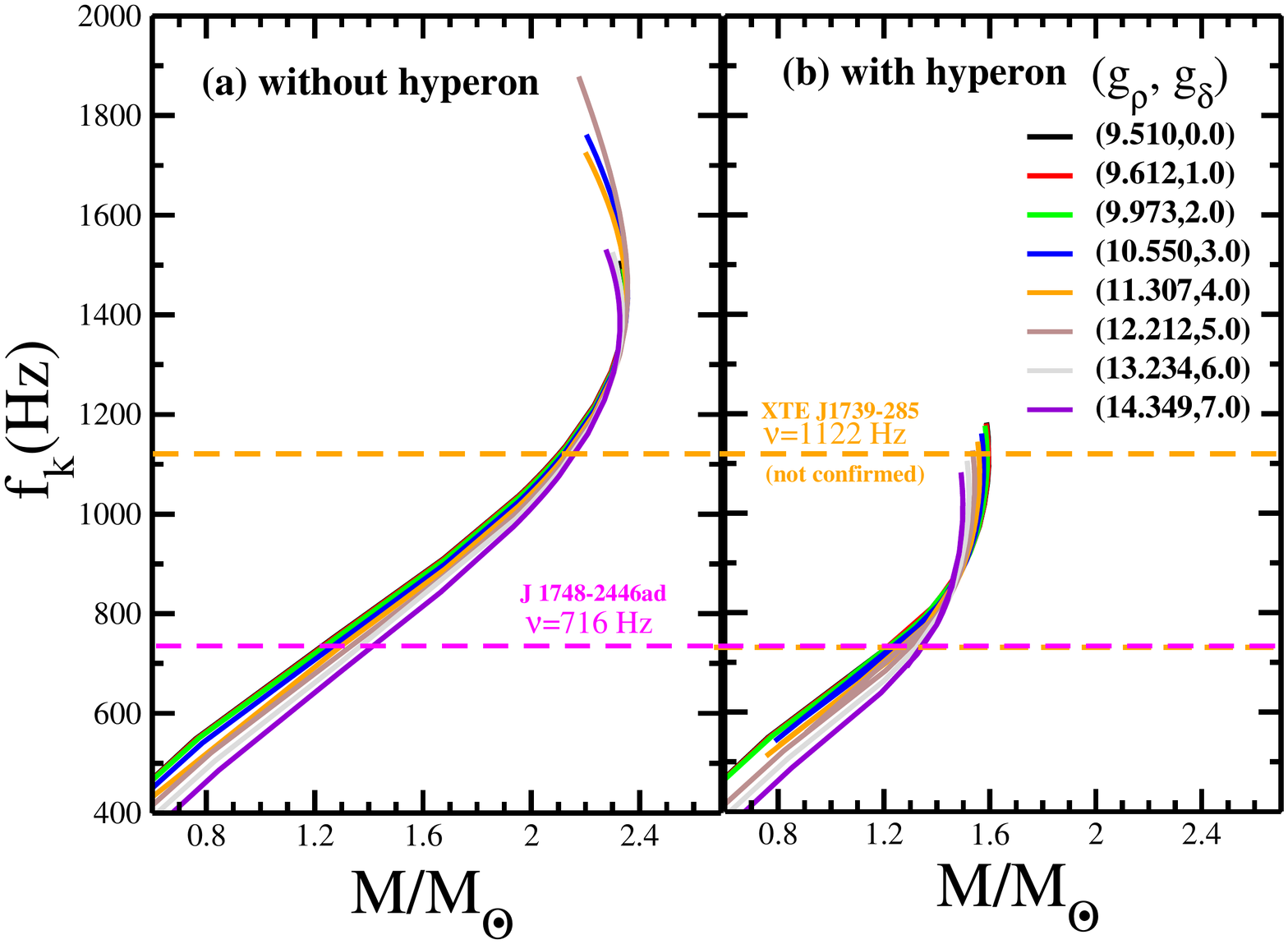}
\caption{(Color online) Keplerian frequency of the rotating proton-neutron
and hyperon stars with various combination of $g_{\delta}$ and $g_{\rho}$ in
G2+$\delta$. (a) is for proton-neutron star and (b) is for the hyperon star.
The results are obtained from RNS code \cite{ster95}.}
\label{fig7}
\end{figure}
Analyzing the graphs, we notice a slight change in the maximum mass with 
g$_\delta$ value. That means, the mass of the star goes on decreasing with
an increase value of the $\delta$-meson coupling in hyperon star. A further 
inspection of the
results reveals that, although the  $\delta$-meson coupling has a 
nominal effects on the maximum mass of the proto-neutron stars, we get an 
asymptotic  increase in the mass. This asymptotic nature of the curves is more
prominent in presence of hyperons inside the stars. Similar phenomena are
also observed in case of rotating stars. 

The empirical formula for the relation between maximum frequency $f_{max}$
with mass of the neutron star for a given EOS is given as
\cite{haensel16,haensel07}
\begin{equation}
f_{max}\approx 1.22 KHz \sqrt{M_{max}^{static}/M_{\odot}} 
({R_{max}^{static}/10Km})^{-3/2},
\end{equation}
 where $M_{max}^{static}$=maximum static mass and
$R_{max}^{static}$=maximum allowed radius for a neutron star. In actual, the neutron stars
have a wide range of frequencies due to the fluid of the stars oscillate in various modes
\cite{benhar04}. Among them, the most important modes are the first pressure mode
($p_I$-mode) and the fundamental mode of the fluid oscillation
(f-mode). The empirical formula for the frequencies of these two modes are
$f_f=(0.79\pm{0.09})+(33\pm2)\sqrt{M/R^3}$ and $f_p=1/M(-1.5\pm{0.8})+(79\pm{4}){M/R}$,
respectively, where M and $f$ in Km and KHz. The above two relations are obtained by
using a wide sample of EOSs \cite{benhar04}. 

In the present calculations, we assume the frequency of the rotating
neutron star is within the Keplerian frequency limit. At this limit
the spin frequency of the neutron star is equal to the orbital frequency
$f_{or}$ (along a circular path on the equator of the NS) \cite{haensel16}.
If the orbital frequency $f_{or} > f_{K}$ (Kepler frequency), then the
hydrostatic equilibrium of the NS does not hold good. To make it clear,
the Kepler frequency as a function of NS mass is shown in Fig. 7 with and
without considering hyperon into account. The results of Fig. 7 is obtained
from the RNS code and the expression for the Keplarian frequency, i.e. the 
maximum frequency obtained from the general theory of relativity
can be found in Refs. \cite{glen94,ster95}. In this figure, the variation
of $f_K$ is shown as a function of $M/M_{\odot}$ with various combinations
of $g_{\rho}$ and $g_{\delta}$ which we have already fixed (see Table 1).  
We noticed a finite effect of $g_{\rho}$ and $g_{\delta}$ variation on the mass
and Keplerian frequency of the pure neutron and hyperon stars.

For a quite some time  pulsar B1937+21 with frequency 642 Hz was considered 
as the fasted spinning NS. However, Hessels et. al.\cite{hess} found even 
more faster spinning NS pulsar J1748-2446ad at frequency 716.356 Hz. This NS 
has a mass of 0.14$M_\odot$ companion. It is difficult to obtain 0.14$M_\odot$ 
from the equation of state  at supra-nuclear densities. Our calculations 
suggest that if pulsar has a mass less than 1.4$M_\odot$ than the larger 
density slope of the symmetry energy at saturation would be excluded. 
If we consider the neutron star mass to be greater than 2.0$M_\odot$ and 
 hyperons are present in it, then the star mass  will be 1.6$M_\odot$,
 within the pulsar  XTE J1739–285 NS\cite{kaar}.

Here, we analyzed the effects of $\delta-$meson
coupling on neutron and hyperon stars. We observed that, the mass of
the star decreases when hyperons are included in the calculations, as
a result, the maximum mass of the star with  G2+$\delta$
set become much less than 2M$_{\odot}$, the latest observation of a
massive neutron star. In summary, the following possibilities are in order:

(i) Since the mass obtained without hyperon for static case is 
$\sim 2M_{\odot}$, in this situation one do not need to reduce the mass any 
more by adding hyperons into it. This can be justify by assuming that in 
massive neutron star,
there is no hyperon. The absence of hyperons in massive neutron star
may not be a convincing explanation, because of the highly dense matter 
in the core of the NS, which favor the production of hyperon. 
(ii) The rotation of a NS increases the maximum allowed mass. On the other hand,
the inclusion of hyperon deceases the mass. In the present case, even if consider the rotation of the star, it is not sufficient to get the maximum mass$\sim$
2$M_\odot$ (see Fig. 6(d)). (iii) The 3$^{rd}$ possibility  is the effect 
of $\delta-$meson coupling, which may increase the mass of the hyperon star 
after its insertion into the model. Although, its effect is finite, it is not sufficient to increase the hyperon
mass to two solar unit. Thus, the addition of $\delta-$meson may not 
be sufficient to explain the heavier mass of the NS. (iv) Probably, the
fourth possibility may be the most acceptable explanation in which
we suggest for the modification of the EOS, such that after the
addition of hyperon, the mass of the static neutron-hyperon star
will be $\sim 2M_{\odot}$. In particular, the hyperon-meson coupling should 
be re-investigated to get a proper coupling constants, which allowed the 
maximum mass$\sim 2M_{\odot}$ with hyperon. Work in this direction is in progress\cite{skp16}.

\subsection{Effects of $\delta-$meson on baryon production}\label{resuf}
Finally, we want to see the effects of $\delta-$meson coupling on the 
particle production for the whole baryonic family at various densities
in nuclear matter system.  
\begin{figure}
\includegraphics[width=0.60\textheight,clip=true,angle=0]{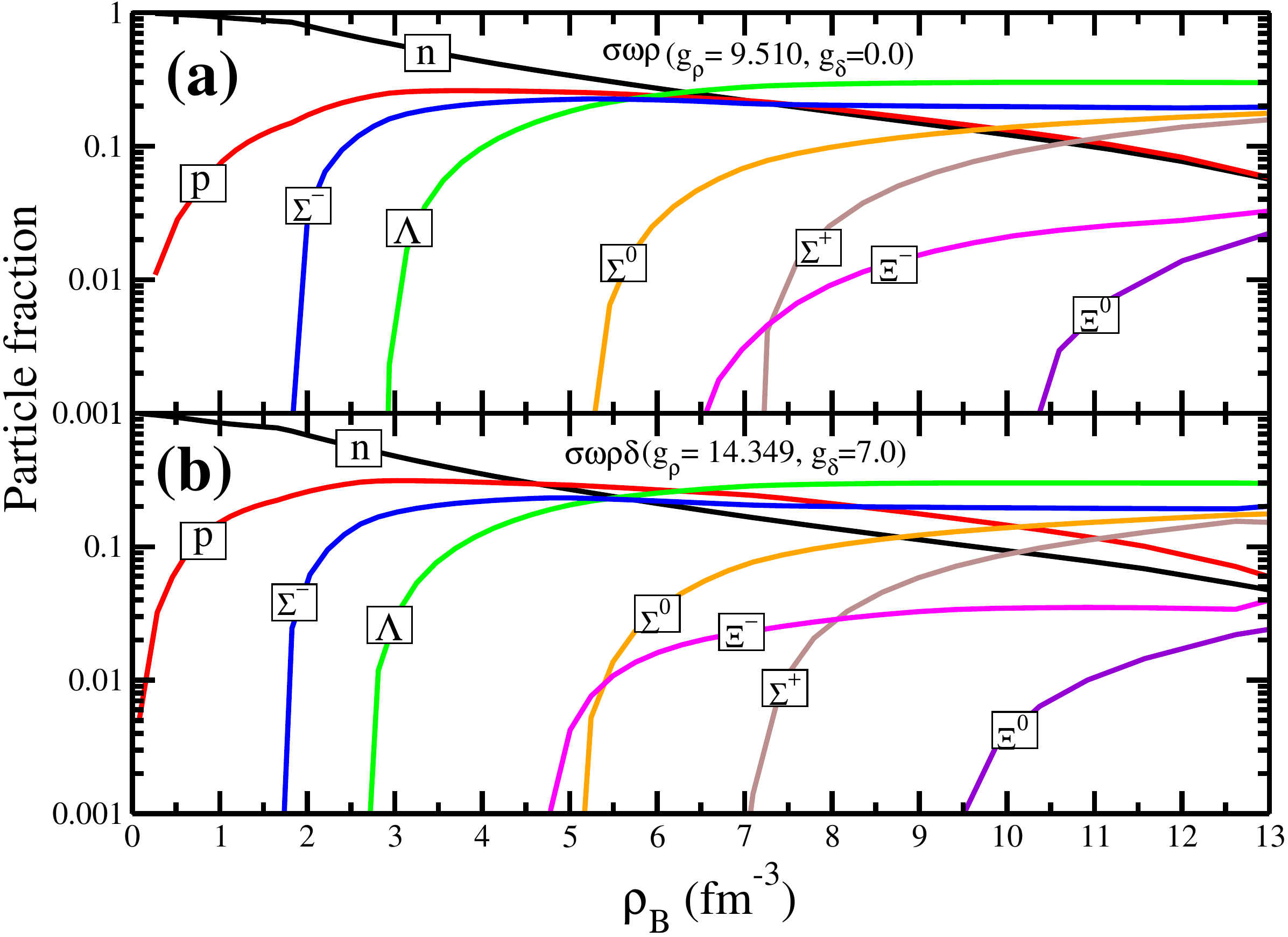}
\caption{(Color online) Yield of strange particles as a function of density. 
The upper panel (a) is with G2 parameter set (without taking $\delta-$meson 
coupling) and the lower panel (b) is with $\delta-$meson coupling. }
\label{fig8}
\end{figure}
The Fermi energy of both proton and neutron increases with density 
for their Fermionic nature. After a certain density, the Fermi energy of the
nucleon exceeded the rest mass energy of the nucleon ($\sim$1000 MeV), 
and strange particles ($\Sigma, \Lambda,\Xi$) are produced. As a result, the
equations of state of the star becomes soft and gives a smaller star mass 
compare to the neutron star containing only protons, neutrons and electrons. 
The decrease in star mass in the presence of whole baryon octet can be
understood from the analysis of Fig.~\ref{fig8}. From the figure, it is clear 
that $\delta$-meson has a great impact on the production of hyperons. The
inclusion of $\delta-$meson accelerate the strange particle production. 
For example, the evolution of $\Sigma^-$ takes place at density 
$\rho_B=1.75\rho_0$ in absence of $\delta-$meson. However, it produces at
$\rho_B=1.67\rho_0$ when $\delta-$meson is there in the system. Similarly,
analyzing the evolution of other baryons, we notice that although the
early production of baryons in the presence of $\delta-$meson is not 
in a definite proportion to each other, in each case the yield is faster. A significant
shifting towards lower density is maximum for heaviest hyperon ($\Xi^0$) and
minimum for nucleon (see Fig. ~\ref{fig8}). For example,  $\Xi^{-}$ evolves
at $\rho_{B}$ = 6.5 $\rho_{0}$ for a non-$\delta$ system and $\rho_{B}$$\sim$5.0 $\rho_{0}$ for medium when $\delta-$meson is included. 
Thus, the $\delta-$coupling has a sizable impact on the production of 
hyperons like  $\Xi^{-}, \Xi^{0}$ and $\Sigma^{+}$.

\begin{figure}
\includegraphics[width=0.60\textheight,clip=true,angle=0]{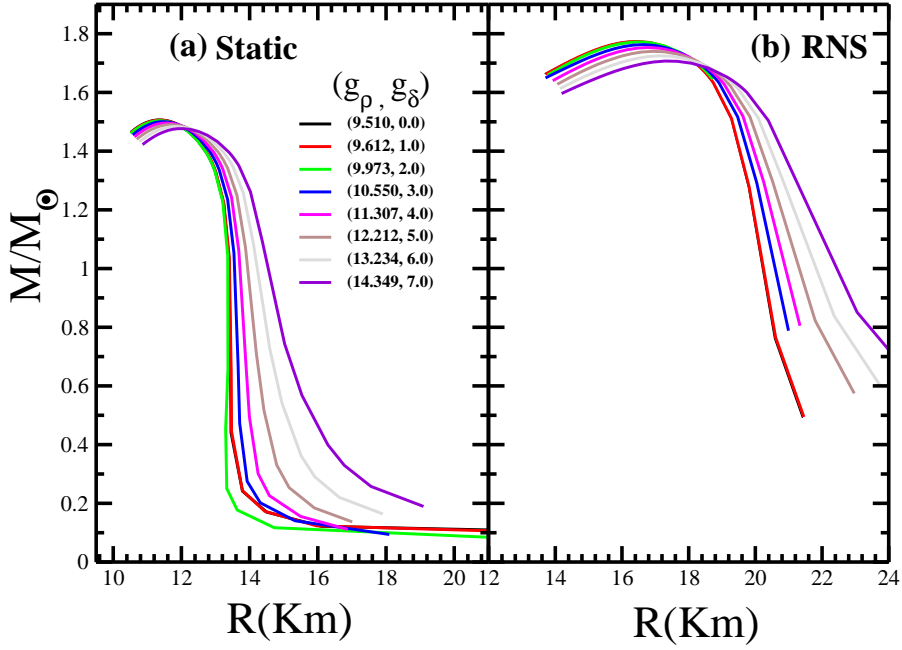}
\caption{(Color online) Mass and radius profile of hyperon star with 
G2+$\delta$ parameter set, but with different meson-hyperon coupling of 
Ref.\cite{glen20}. }
\label{fig9}
\end{figure}

\subsection{Fitting of $g_\rho$ and $g_\delta$ with fixed binding energy and charge radius}

In previous sub-sections we have seen the effects of ($g_\rho$, $g_\delta$) 
pair with a constant symmetry energy on the maximum mass and radius of the  neutron 
and hyperon stars. The effects of the ($g_\rho$, $g_\delta$) pairs are not prominent
on the  star structure in this method. On the other hand,  it affects the bulk properties, like binding energy and  root mean square radius considerably 
for asymmetric finite nucleus.  In Table~\ref{tab2}, we have given the mass and charge radius 
for some of the selected nuclei. Although, all the combination of $g_\rho$ and $g_\delta$ 
are fixed at a constant symmetry energy, the binding of  $^{208}$Pb differ by 90 MeV 
in the first and last combination of $g_\rho$ and $g_\delta$.  In this sub-section, we 
would like to change the strategy to select the ($g_\rho$, $g_\delta$) pairs.
Here, we have followed the second procedure as we have discussed in the previous sub-section,
i.e., we find the values of $g_\rho$ and $g_\delta$ by adjusting 
the binding energy and charge radius of $^{208}$Pb. Once we get the ($g_\rho$, $g_\delta$),
we used the pair for the calculations of
other nuclei of Table~\ref{tab2}. Surprisingly, the outcome of binding energy and
charge radius matches pretty well with the original calculations. The $g_\rho$ and 
$g_\delta$ combination along with the corresponding mass and radius of a neutron 
star is given in the Table ~\ref{tab3}. 
From the table it is clear that these combinations are also not affecting much 
to the maximum mass and radius of the neutron star. However, the $E_{sym}$,
$L_{sym}$ and $K_{sym}$ calculated from the corresponding ($g_\rho$, $g_\delta$) combinations
for nuclear matter changes a lot (See Table ~\ref{tab3}).
We used the hyperon-meson coupling constants of Ref. \cite{glen20} to evaluate 
the hyperon star structure. The calculated results for static and rotating hyperon star
are plotted in Fig. ~\ref{fig9}. The maximum mass increases and the radius
decreases slightly with the addition of $\delta-$meson to the star 
system.

\section{Summary and Conclusions}\label{conc}

In summary, using the effective field theory approach, we discussed the effects
 of isovector scalar meson on hyperon star. The inclusion of $\delta$-meson with
G2 parameter set, we have investigated the static and rotating stellar 
properties of neutron star with hyperons. We fitted the parameters and see the
variation of g$_\rho$ and g$_\delta$ at a constant symmetry energy for 
both the nuclear and neutron matter. We also used these (g$_{\rho}$, g$_{\delta}$)
pairs to finite nuclei and find a large change in binding energy for asymmetric
nuclei. Then we re-fitted the (g$_{\rho}$, g$_{\delta}$) pairs keeping
binding energy and charge radius fixed for $^{208}$Pb and tested the effects
for some selected nuclei and able to reproduced the data similar to the original
G2 set.  
With the help of G2+$\delta$ model, for static and rotating stars without 
hyperon core, we get the maximum mass of $\sim$2$M_\odot$ and 
$\sim$2.4$M_\odot$, respectively. This prediction of masses is in agreement 
with the recent observation of $M\sim$$2M_\odot$ of the stars. However, with 
hyperon core the maximum mass obtained are $\sim$1.4$M_\odot$ and 
$\sim$1.6$M_\odot$ for static and rotating hyperon stars, respectively. 
In addition, we have also calculated the production of whole
baryon octet with variation in density. We find that the particle fraction 
changes a lot in the presence of $\delta-$meson coupling. When there is
$\delta-$meson in the system the evolution of baryons are faster compare
to a non-$\delta$ system. This effect is significant for heavier masses and
minimum for lighter baryon. Hence, one can conclude that the yield of 
baryon/hyperons depends very much on the mesons couplings. One important information
is drawn from the present calculations is that the effect of g$_{\delta}$ is
just opposite to the effect of g$_{\rho}$. As a consequence, many long standing 
anomaly, such as the comparable radii of $^{40}$Ca and $^{48}$Ca  be
resolved by adjusting the (g$_{\rho}$, g$_{\delta}$) pairs properly.  Keeping in 
view the importance of $\delta-$meson coupling and the reverse nature of 
g$_{\rho}$ and g$_{\delta}$, it is necessary to get a new parameter
set including proper values of $g_{\delta}$ and $g_{\rho}$, and the work is under progress.

\setcounter{equation}{0}
\setcounter{figure}{0}
\newpage

\newpage
\chapter {Summary and Conclusions}
\label{chapter9}

The relativistic mean field formalism provides a common platform to 
study both the finite and infinite nuclear matter systems.  Its relativistic nature defined a pathway from the less   dense finite nuclear system to highly dense neutron star. We studied various aspects of  finite  nuclear structure like  magicity of proton and neutron number and collective excitation like giant resonance and the nucleon-nucleon interaction .   For the infinite nuclear system, we applied RMF to study the mass and radius of the neutron star. In brief, we have applied RMF formalism to study both finite and infinite nuclear systems.

Recently the advent of  radioactive ion beam (RIB) facility  inspires the nuclear physicists to look forward to the structure of  drip-lines and super-heavy nuclei more seriously.  Nuclei away from the $\beta$-stability lines are far different from the nuclei on the $\beta$-stability line. In Ch.~\ref{chapter2}, we have calculated BE, $S_{2n}$ energy, single particle levels, pairing gaps and chemical potential, in the isotopic chain of Z = 82, 114, 120 and 126. All our calculations are done in the framework of  non-relativistic SEI  and relativistic RMF interactions. We have compared our results with  FRDM and other theoretical predictions.  Overall, the discussions and analysis of all possible  evidence of shell closure property with SEI interaction and RMF show that one can take Z = 120 and N = 182 or 184 as the next magic combination beyond Z = 82 and N = 126, which is consistent with other theoretical models.  

Not only the single particle properties, which are discussed in first part of the thesis but also the collective excitation plays an important role in nuclear structure physics. In Ch.\ref{chapter3},  we have calculated the excitation energy  of isoscalar giant monopole resonance  and incompressibility for O, Ca, Ni, Sn, Pb, Z=114, and Z=120 isotopic series starting from  proton to  neutron drip lines. We used four successful parameter sets, NL1, NL3, NL3*, and FSUGold, with a wide range of nuclear matter incompressibility starting from 211.7 MeV to 271.76 MeV to see the dependency of the ISGMR on $K_\infty$. Also, we have analyzed the predictions of ISGMR with these forces, which originate from various interactions and found that whatever may be the parameter set, the differences in excitation energy ISGMR predicted by them are found to be marginal in the super heavy region. A recently developed scaling approach in a relativistic mean field theory is used. A simple, but accurately constrained approximation is also performed to evaluate the isoscalar giant monopole excitation energy. From the scaling and constrained ISGMR excitation energies, we have evaluated the resonance width $\Sigma$ for the whole isotopic series. This is obtained by taking the root mean square difference of ${E^s}_x$ and ${E^c}_x$ . The value of  ${E^s}_x $is always higher than the constrained result ${E^c}_x$ . In a sum rule approach, the ${E^s}_x $ can be compared with the higher and $ {E^c}_x $ as the lower limit of the resonance width. In general, we found an increasing trend of resonance width $\Sigma$ for both the light and super heavy regions near the proton and neutron drip lines. The magnitude of $\Sigma$ is predicted to be minimum in the vicinity of N=Z or in the neighborhood of a double closed nucleus and it is maximum for highly asymmetric system. In the present thesis, we have also estimated the incompressibility of finite nuclei. For some specific cases, the  incompressibility is compared with the nuclear matter incompressibility and found a linear variation among them. It is also concluded that the nucleus becomes less compressible with the increase of neutron or proton number in an isotopic chain. Thus neutron-rich matter, like neutron star as well as drip-line nuclei, are less compressible than normal nuclei. In the case of exotic (drip-line nuclei) system, the nucleus is incompressible, although it possesses a normal density.

The recent experiment on isoscalar monopole excitation in the isotopic chain of the Sn isotopes indicates a new problem in the medium heavy mass range (A$\sim$ 100). In the second part of Ch.~\ref{chapter3},  we analyzed the predictive power of various force parameters, like NL1, NL2, NL3, Nl-SH and FSUG in the framework of RTF and RETF approaches for giant monopole excitation energy of Sn-isotopes. Then the calculation is extended to some other relevant nuclei in the mass range A$\sim$100.  The analysis shows that relativistic Thomas-Fermi  and extended relativistic Thomas-Fermi approximation  give comparable results with pairing+MEM prediction. It can be exactly reproduced the experimental data for Sn isotopes, when the incompressibility of the force parameter is within 210-230 MeV, however, fails to reproduce the GMR data for other nuclei within the same accuracy. We have qualitatively analyzed the difference in GMR energies ($\triangle E_{GMR}$=RETF-RTF) using RETF and RTF formalisms in various force parameters. The FSUGold parameter set shows different behavior from all other forces. Also, we extended our calculations for monopole excitation energy of Sn isotopes with a force parametrization having softer symmetry energy (NL3 + $\Lambda_V$ ). The excitation energy decreases with the increase of proton-neutron  asymmetry agreeing with the experimental trend. In conclusion, after all, from these thorough analyses, it seems that the softening of Sn isotopes is an open problem in nuclear theory and more work in this direction is needed. In Ch.~\ref{chapter4} we discussed 
about the new constrained caiculation developed by us. This new method is 
based on the Taylor series expansion. Although this method is very simple, 
it gives reasonable results which are comparable with other microscopic 
calculation like RRPA. For simplicity, we applied the constrained method 
to Thomas-Fermi approximation. But this method can be extended to extended 
Thomas-Fermi approach without loosing any generality.

     Nucleon-Nucleon interaction has a very crucial role in the nuclear structure as well as the other branch of nuclear physics.  So it is important to study more about the nucleon-nucleon interaction and its effects on various properties of the nuclear system.  In Ch.~\ref{chapter6}, 
we extensively discussed a new approach R3Y,  for the N-N interaction, which is first suggested by Patra et at. We have added a new self-interacting $\omega $ meson contribution to this new formalism and checked its contribution to finite and infinite nuclear matter system.  In Ch.~\ref{chapter7}, we used the density dependent M3Y and R3Y to study the p-$\gamma$ reaction for the proton-rich nuclei.  We have folded the M3Y and R3Y interaction with RMF densities to obtain the cross-section of p-$\gamma$ reaction.  From the cross-section of these reaction, we can study the astrophysical- S factor, which is 
important in the study of the r-process of nuclear synthesis.

In Ch.~\ref{chapter8}, using the effective field theory approach, we discussed the effects of isovector scalar meson on hyperon star with the inclusion of $\delta$-meson with G2 parameter set and then investigated the static and rotating stellar  properties of neutron star with hyperons. We fitted the parameters and see the variation of g$_\rho$ and g$_\delta$ at a constant symmetry energy for  both the nuclear and neutron matter. We also used these (g$_{\rho}$,   $g_{\delta}$) pairs to finite nuclei and find a large change in binding energy for asymmetric nuclei. Then we re-fitted the (g$_{\rho}$, g$_{\delta}$) pairs keeping binding energy and charge radius fixed for $^{208}$Pb and tested the effects for some selected nuclei and able to reproduce the data similar to the original G2 set.  With the help of the G2+$\delta$ model, for static and rotating stars without  hyperon core, we get the maximum mass of $\sim$2$M_\odot$ and  $\sim$2.4$M_\odot$, respectively. This prediction of masses is in agreement 
with the recent observation of $M\sim$$2M_\odot$ of the stars. However, with  hyperon core the maximum masses obtained are $\sim$1.4$M_\odot$ and  $\sim$1.6$M_\odot$ for static and rotating hyperon stars, respectively.  In addition, we have also calculated the production of whole baryon octet with variation in density. We find that the particle fraction  changes a lot in the presence of $\delta-$meson coupling. When there is $\delta-$meson in the system the evolution of baryons is faster compared to a non-$\delta$ system. This effect is significant for heavier masses and minimum for lighter baryon. Hence, one can conclude that the yield of  baryon/hyperons depends very much on the mesons couplings. One important information is drawn from the present calculations is that the effect of g$_{\delta}$ is just opposite to the effect of g$_{\rho}$. As a consequence, many long-standing anomalies, such as the comparable radii of $^{40}$Ca and $^{48}$Ca  can be resolved by adjusting the (g$_{\rho}$, g$_{\delta}$) pairs properly.  Keeping in  view the importance of $\delta-$meson coupling and the reverse nature of 
g$_{\rho}$ and g$_{\delta}$ with each other, it is necessary to get a new parameter
set including proper values of $g_{\delta}$ and $g_{\rho}$, and the work is in progress.

\setcounter{equation}{0}
\setcounter{figure}{0}
\newpage
\addcontentsline{toc}{chapter}{REFERENCES}
\renewcommand{\bibname}{REFERENCES}
\bibliographystyle{utcaps}
\bibliography{mythesis}

\end{document}